\documentclass[12pt]{article}
\usepackage{amsmath,graphicx,amssymb,amsfonts,amsthm}
\usepackage{verbatim,enumitem,lipsum,float}
\usepackage{color,epstopdf,epsfig,lscape,longtable,mathabx,natbib}
\usepackage{xr}
\usepackage{mathrsfs}
\usepackage{placeins}
\usepackage{multirow}

\usepackage{mathtools}

\newtheorem{Lem}{\underline{\bf Lemma}}
\newtheorem{Th}{\underline{\bf Theorem}}
\newtheorem{Mth}{\underline{\bf Main Theorem}}

\newtheorem{Cor}{\underline{\bf Corollary}}

\def\wh{\widehat}
\def\wt{\widetilde}
\def\wc{\widecheck}
\def\n{\nonumber}

\def\pr{\mbox{pr}}

\def\expit{{\mbox{expit}}}
\def\trans{^{\rm T}}

\def\sumI{\sum_{i=1}^N}

\def\0{{\bf 0}}

\def\A{{\bf A}}
\def\a{{\bf a}}
\def\B{{\bf B}}
\def\b{{\bf b}}

\def\g{{\bf g}}

\def\h{{\bf h}}

\def\k{{\bf k}}

\def\U{{\bf U}}
\def\S{{\bf S}}

\def\u{{\bf u}}
\def\V{{\bf V}}

\def\X{{\bf X}}
\def\x{{\bf x}}
\def\Y{{\bf Y}}
\def\Z{{\bf Z}}
\def\z{{\bf z}}

\def\fyx{f_{Y\mid \X}}

\def\fx{f_{\X}}
\def\fyxrone{f_{Y\mid \X,R=1}}

\def\ba{{\boldsymbol\alpha}}
\def\bb{{\boldsymbol\beta}}

\def\bt{{\boldsymbol\theta}}

\def\bphi{\boldsymbol\phi}

\def\bmu{\boldsymbol\mu}

\def\bz{{\boldsymbol \zeta}}

\def\btau{\boldsymbol\tau}

\def\mR{\mathbb{R}}
\def\eff{_{\rm eff}}

\def\bse{\begin{eqnarray*}}
\def\ese{\end{eqnarray*}}
\def\be{\begin{eqnarray}}
\def\ee{\end{eqnarray}}
\def\bsq{\begin{equation*}}
\def\esq{\end{equation*}}
\def\bq{\begin{equation}}
\def\eq{\end{equation}}

\def\squarebox#1{\hbox to #1{\hfill\vbox to #1{\vfill}}}
\def\boxit#1{\vbox{\hrule\hbox{\vrule\kern6pt\vbox{\kern6pt#1\kern6pt}\kern6pt\vrule}\hrule}}

\allowdisplaybreaks

\begin{document}
\begin{center}
{\LARGE{\bf
Avoid Estimating the Unknown Function in a Semiparametric Nonignorable Propensity Model
}}
\end{center}

\baselineskip=15pt

\vskip 2mm
\begin{center}
	Samidha Shetty, Yanyuan Ma and Jiwei Zhao\\
	Department of Statistics, Pennsylvania State University,
        University Park, PA 16802 \\
     Department of Biostatistics \& Medical Informatics, University of Wisconsin-Madison,
    Madison, WI 53726\\
    sss269@psu.edu\\
	 yzm63@psu.edu\\
    jiwei.zhao@wisc.edu
\end{center}

\begin{abstract}
We study the problem of estimating a functional or a parameter
in the context where
outcome is subject to nonignorable missingness. We completely avoid
modeling the regression
relation, while allowing the propensity to be modeled by a
semiparametric logistic relation where the dependence on
covariates is unspecified. We discover a surprising phenomenon in
that the estimation of the parameter in the propensity model as well
as the functional estimation can be carried out without assessing
the missingness dependence on covariates. This allows us to propose a
general class of estimators for both model parameter estimation and
functional
estimation, including estimating the outcome mean. The robustness of
the estimators are nonstandard and are established
rigorously through theoretical derivations, and are supported by
simulations and a data application.
\end{abstract}


\section{Introduction}\label{sec:intro}

Handling missing data is an inevitable issue in many empirical
studies, especially when the data are directly collected from human beings or strongly linked to the subjects' behaviors.
Missing data are termed missing at random or ignorable
\citep{little2019statistical} if its propensity only depends on the
observed data.
In applications, however, missing data are usually nonignorable in the
sense that its propensity not only depends on the observed data but
also on the unobserved ones.
Nonignorable missing data are prevalent in surveys in social sciences
as well as in patient reported outcomes in biomedical studies \citep{fielding2008simple, li2014standards, gomes2016addressing,
  leurent2018sensitivity}.

In applications, the outcome variable  $Y$ can be, for
example,  the pain
score of a
patient suffering from arthritis, or the depression score of a
breast cancer patient.
In these scenarios, the scientific interest is to estimate some
unknown quantities about the marginal distribution of $Y$ such as
$E(Y)$.
In these problems, it is not a good idea to impose assumptions on
$\fyx(y,\x)$, the conditional probability distribution function of $Y$
given $\X$, a vector of covariates. This is because an incorrect assumption of $\fyx(y,\x)$ will
easily jeopardize the estimation of $E(Y)$ if we compute it through using
$E(Y)=E\{E(Y\mid \X)\}$ and  incorporating $\fyx(y,\x)$.
On the other hand, \cite{robins1997toward} pointed out that the whole
model would be non-identifiable if there are no parametric
assumptions on either $\fyx(y,\x)$ or the propensity score model.
To enable model identification, one has to impose some assumption,
which would include some parametric component, on the propensity model.
Then a key question is: what type of assumption is appropriate on the
propensity model for nonignorable missing data?

A simplistic approach is to directly impose a fully parametric assumption,
such as logistic regression, see, e.g.,
\cite{ibrahim1996parameter,
  rotnitzky1997analysis, qin2002estimation, chang2008using,
  wang2014instrumental, morikawa2016semiparametric}.
Since any model on the propensity contains variable $Y$, which is
subject to missing, it is difficult to empirically verify the
adequacy of the model.
Therefore, it is often preferred to have a flexible model as long as it
guarantees identifiability.

To this end,
probably the most successful endeavor thus far is
\cite{shao2016semiparametric},
who assumed that the logit of the propensity model is a sum of a
parametric component of $Y$ indexed by known function $h(\cdot)$ and
unknown parameter $\bb$, $h(Y,\bb)$, and an unknown nonparametric
component of $\U$, $g(\U)$. Here $\X=(\U\trans,\Z\trans)\trans$ and
the variable $\Z$ is termed nonresponse instrument
\citep{wang2014instrumental} or shadow variable.
The use of the nonresponse instrument is to enable model
identifiability in their context.
This model is certainly more flexible than a purely parametric model
mentioned in the previous paragraph. The parameter $\bb$ in the model,
characterizing the dependence of the propensity on outcome $Y$,
indicates how severe the nonignorability is. If the parametric
component does not depend on $Y$ as a special case, the model
degenerates to a standard ignorable one.
However, in the estimation method presented in
\cite{shao2016semiparametric},
an estimate to $g(\U)$ is required.
Although $g(\U)$ itself does not have missing data, its estimation
cannot be stand-alone and would have to involve some missing data
techniques or other unknown components of the model.
In \cite{shao2016semiparametric}, they adopted the profiling approach,
and estimated $\wh g(\U,\bb)$ via the standard kernel estimation for
every fixed value of $\bb$.
Additionally, since $\wh g(\U,\bb)$ has to be repeatedly estimated in
the algorithm, the whole procedure is computationally very expensive.
Clearly this approach brings tremendous complexity for end-users, and
it limits the applicability of this semiparametric nonignorable
propensity model to broader practice.

In this paper, we propose a
completely novel estimation framework which consistently estimates the
unknown quantities of interest, without either estimating or modeling the
nonparametric component $g(\cdot)$. In other words, our method has
certain robustness property against $g(\cdot)$. Importantly, we find
that such robustness does not inherit from standard robustness in
semiparametric literature. Instead, it benefits from the specific
modeling structure and our treatment, as will be detailed later.
We first consider the estimation of $\bb$. Through
extensive and careful derivations, we find that our framework can
always consistently estimate $\bb$ without a correct estimator of
$g(\cdot)$.
We then extend our framework to the estimation of
$\bt$ that satisfies $E\{\bz(\X, Y,\bt)\}=\0$ for a
given function $\bz(\cdot)$. We show that, once $\bb$ can be
consistently estimated without estimating $g(\cdot)$,
$\bt$ can also be consistently estimated in a
similar fashion.
Note that although in most literature, the
quantity of interest is $\bt$, such as $E(Y)$, we single out
$\bb$ as a quantity of interest as well because its estimation is
crucial for the latter. In other words, $\bb$
is of interest due to its supporting role.

The key idea of our framework is to view $g(\cdot)$ as a nuisance
component in a semiparametric model, and to project the effect of $g(\cdot)$ to an orthogonal direction
via semiparametric treatment
\citep{bickel1993efficient, tsiatis2006semiparametric}.
Our procedure only needs a working model of $g(\cdot)$ for the
implementation, and this working model does not have to contain the
true $g(\cdot)$.
Although the estimation procedure is elegant,
the theoretical development is complex and involved. The proof
involves various nonparametric and semiparametric analyses,
in combination with heavy usage of the U-statistic techniques. The technical details are given in the Supplementary Material.

\section{Semiparametric Nonignorable Propensity Model}\label{sec:SW}

Throughout the paper, we
consider the case that $\X$ is fully observed but $Y$ is subject to
nonignorable missingness. We encode $R$ as the indicator of observing
$Y$ in that $R=1$ if $Y$ is observed and $R=0$ otherwise.
In reality we observe $N$ independent and identically distributed samples of $(R,RY,\X)$.
The propensity
of the missing data is the conditional probability distribution function
$\pi(y,\x)=\pr(R=1\mid y,\x)$. Because of the nonignorable missingness,
$\pr(R=1\mid y,\x)\neq \pr(R=1\mid \x)$, but $\pi(y,\x)$ does not
necessarily depend on all variables in $\X$.

The semiparametric nonignorable propensity model proposed in
\cite{shao2016semiparametric} is
\be\label{eq:mechanism}
\pi(y,\x)=\pi(y,\u,\bb,g)=\expit\{h(y,\bb)+g(\u)\},
\ee
where $\expit(\cdot)=\exp(\cdot)/\{1+\exp(\cdot)\}$, $\bb$ is an unknown
$d$-dimensional parameter, $h(\cdot)$ is a
known function, and $g(\cdot)$ is an arbitrary unspecified function.
We denote the dimensions of $\U$ and $\Z$ be $q$ and $p-q$, respectively where $p$ is the dimension of $\X$.
Note that \cite{shao2016semiparametric} simply wrote $h(y,\bb)=\beta
y$, but all their methods also apply to this more general possibly nonlinear
function $h(y,\bb)$.

As introduced in Section~\ref{sec:intro},  (\ref{eq:mechanism}) is
 one of the most widely adopted models if one would like to
leap forward from the pure parametric models and pursue a more
flexible semiparametric model for the nonignorable propensity.
Unlike the parametric models in the literature, model
(\ref{eq:mechanism}) is semiparametric which contains the
nonparametric component $g(\u)$ that provides flexibility to protect
against model misspecification.

The main idea of estimation in \cite{shao2016semiparametric} is to
construct sufficient estimating equations for $\bb$. Their
estimating equations are based on
\bse
E[w_l(\Z)\{R/\pi(Y,\u,\bb,g)-1\}\mid \u] = 0, l=1,\ldots,L,
\ese
where $w_l(\z)$ is an arbitrary function of $\z$.
If $\Z$ is discrete with the number of categories $L$ satisfying
$L\geq d+1$, the authors simply advocated using $w_l(\z)=I(\z=l)$;
while when $\Z$ is continuous, the authors suggested replacing them by
some moment functions.
Inserting the form of $\pi(Y,\u,\bb,g)$, we obtain that
\bse
\exp\{-g(\u)\}E[w_l(\Z)R \exp\{-h(Y,\bb)\}\mid \u] = E\{w_l(\Z)(1-R)\mid \u\}.
\ese
Further, to estimate $\exp\{-g(\u)\}$, the authors simply chose $w_l(\z)$ as a constant and then adopted the profiling technique, i.e., for
 each fixed $\bb$, they estimated $\exp\{-\wh g(\u,\bb)\}$ by
\bse
\exp\{-\wh g(\u,\bb)\} = \frac{\sumI (1-r_i)K_t(\u-\u_i)}{\sumI r_i \exp\{-h(y_i,\bb)\}K_t(\u-\u_i)},
\ese
where $K$ is a kernel function and $t$ is a bandwidth.
In other words, the authors estimated the nonparametric component
$g(\u)$ via the kernel estimation. In their method, this step of
estimating $g(\u)$ is important and
inevitable. Given that $g(\u)$ is not at the center of our interest,
we would like to explore the possibility of avoiding estimating it.

The overarching goal of this paper is to show, by rigorously
characterizing the geometric structure of model (\ref{eq:mechanism}), one can avoid
modeling or estimating $g(\u)$ completely.

\section{Geometric Structure of the Model}\label{sec:structure}
\subsection{Likelihood of the Model}
Let $\fx(\x)$ denote the probability distribution function (pdf) of $\X$
and define $\fyxrone(y,\x)$ to be the conditional pdf of $Y$ given
$\X$ and $R=1$.
Incorporating the relation
\bse\label{eq:relation}
\fyx(y,\x)=\frac{\fyxrone(y,\x)/\pi(y,\u,\bb,g)}{\int \fyxrone(t,\x)/\pi(t,\u,\bb,g)dt},
\ese
the joint pdf of $(\X, R, RY)$, or the likelihood function from one observation, is
\be
&&f_{\X,R,RY}(\x,r,ry) \label{eq:model}\\
&=&\fx(\x)\left\{
\frac{\fyxrone(y,\x)}{\int \fyxrone(t,\x)/\pi(t,\u,\bb,g)dt}
\right\}^r \left\{1-\frac{1}{\int \fyxrone(t,\x)/\pi(t,\u,\bb,g)dt} \right\}^{1-r}.\n
\ee
Here,
we view $\bb$ as the parameter of interest and $\fx(\cdot),
\fyxrone(\cdot), g(\cdot)$ as
nuisance components.
Note that the models $\fx(\cdot)$ and
$\fyxrone(\cdot)$ do not involve missing data and their estimates could be stand-alone.

\subsection{Likelihood Identifiability}\label{sec:iden}

As pointed out in \cite{shao2016semiparametric} and the references
therein, the likelihood (\ref{eq:model}) may not be fully identifiable
without some other assumptions or conditions.
Our result below imposes a condition on $\fyxrone(y,\x)$, which does
not involve missing data hence can be empirically tested, and under
which the likelihood (\ref{eq:model}) is fully 
identifiable.

\begin{Lem} \label{lem:one}
  Assume that for any function $\kappa(Y,\U)$ with finite mean,
  $E\{\kappa(Y,\U)\mid \X,R=1\}=\0$ implies $\kappa(Y,\U)=\0$ almost
  surely.
Assume also that $h(y,\bb)$ is suitably parameterized so that
$h(y,\bb_1)\ne h(y,\bb_2)+c$ for any nonzero constant $c$ and any
$\bb_1\ne\bb_2$.
Then, all unknown components $\bb$, $g(\u)$, $\fyxrone(y,\x)$ and $\fx(\x)$ in (\ref{eq:model}) are identifiable.
\end{Lem}

Requiring $h(y,\bb_1)\ne h(y,\bb_2)+c$ for any nonzero constant $c$ and any
$\bb_1\ne\bb_2$ is very mild. It simply assumes the parametrization of
$h(y,\bb)$ is reasonable. For example, it excludes the pathological
case $h(y,\bb)=\beta_1y+\beta_2+\beta_3$.
The proof of Lemma~\ref{lem:one} is in Supplement S.1.
The condition presented in Lemma~\ref{lem:one} is usually referred to
as the completeness condition in the literature.
The completeness condition is satisfied for many commonly-used models.
For example, in exponential families where
$\fyxrone(y,\x)=s(y,\u)t(\x)\exp\{\bmu(\x)\trans \btau(y,\u)\}$
with $s(y,\u)>0$, $t(\x)>0$, $\btau(y,\u)$ is one-to-one in $y$ 
and the support of $\bmu(\x)$ is an open set, the completeness
condition holds. This is documented in classic textbooks such as
\cite{lehmann2006testing} (Theorem 4.3.1).
Therefore, commonly-seen regression models, such as the linear
regression for continuous $Y$ and the logistic regression for binary
$Y$, satisfy the completeness condition.
The completeness condition has been well used in different disciplines
for investigating model identifiability, such as
\cite{newey2003instrumental, d2010new, hu2018nonparametric,
  miao2015identification, zhao2021versatile}.
It is worthwhile to mention that  in our context, the completeness
condition is imposed on the model $\fyxrone(y,\x)$, which is free of
missing data, therefore, it can be  adequately assessed empirically.

Note that the identifiability results presented here are different
from, but also relevant to those discussed in
\cite{shao2016semiparametric}.
In \cite{shao2016semiparametric}, they did not work on the likelihood
function (\ref{eq:model}) directly. Instead, they aimed to find enough
estimating equations to estimate $\bb$. Once their estimating
equations are identifiable, the parameter $\bb$ is estimable and hence
identifiable.
In a special situation when both $Y$ and $\Z$ are discrete with finite
support $\{y_1, \ldots, y_S\}$ and $\{\z_1, \ldots, \z_L\}$, the
completeness condition implicitly requires
\citep{newey2003instrumental} $L\geq S$, the support of the
nonresponse instrument has to be no smaller than that of the outcome
variable.
Similarly with a discrete instrument $\Z$ with $L$ categories, to
create enough estimating equations, \cite{shao2016semiparametric}
required that $L\geq d+1$, i.e., the support of the nonresponse
instrument has to be greater than the dimension of the unknown
parameter $\bb$. From this aspect, the identifiability requirements of
\cite{shao2016semiparametric}
 are similar to ours---both impose conditions on the support
of the instrument $\Z$.

\subsection{Nuisance tangent space $\Lambda$ and its complement
  $\Lambda^\perp$}\label{sec:space}

We take a geometric point of view for the semiparametric model with
(\ref{eq:model}) as the likelihood function from one single
observation.
The influence functions for regular asymptotically linear
estimators of $\bb$ lie in the Hilbert space $\mathcal{H}$ of all
 $d$-dimensional zero-mean measurable functions of the
observed data with finite variance.
The inner product of $\mathcal{H}$ is defined as
$\langle h_1, h_2\rangle = E\{h_1\trans(O)h_2(O)\}$.
For semiparametric models, the nuisance tangent space $\Lambda$ is
defined as the mean squared closure of
nuisance tangent spaces associated with  all parametric submodels.
According to the theory of semiparametrics \citep{bickel1993efficient,
  tsiatis2006semiparametric}, influence functions  belong to the
linear space orthogonal to the nuisance tangent space, denoted as
$\Lambda^\perp$ throughout.

For notational simplicity, in the following, we write
$E(\cdot\mid\cdot,R=1)$ as
$E(\cdot\mid\cdot,1)$
and let $w(\x)\equiv[E\{\pi^{-1}(Y,\u,\bb,g)\mid\x,1\}]^{-1}$.
The nuisance tangent space for $\fx(\x)$ in (\ref{eq:model}) is a
standard construction and is described as
\bse
\Lambda_1=\left[ \a(\x)\in{\mR^d}: E\{\a(\X)\}=\0 \right].
\ese
Taking into account that $\fyxrone(\cdot)$ is a conditional pdf,
it is also straightforward to derive that
the nuisance tangent space for $\fyxrone(y,\x)$ is
\bse
\Lambda_2=
\left[\frac{
\{w(\x) -r\}E\{\b(Y,\x)\mid\x\}}
{1-w(\x)}
+r \b(y,\x): \b(Y,\X)\in{\mR^d}, E\{\b(Y,\X)\mid\x,1\}=\0
\right].
\ese
Finally, the nuisance tangent space for $g(\u)$ can be verified to be
\bse
\Lambda_3
=\left[\a(\u)
  \{r-w(\x)\}: \forall
  \a(\u)\in\mR^{d}\right].
\ese
We can easily verify that $\Lambda_1\perp\Lambda_2$ and
$\Lambda_1\perp\Lambda_3$. We now project $\Lambda_2$ onto
$\Lambda_3$ to obtain its residual space to form
$\Lambda_2^*$. To this end,
after some algebraic derivation, we arrive at
\bse
\Lambda_2^*
&=&
\left\{r \b(y,\x)-\{r-w(\x)\}
\left(\frac{ E\{\b(Y,\x)\mid\x\}}
{1-w(\x)}
-\frac{
E\left[ E\{ \b(Y,\x)\mid\x\}
\mid\u,1\right]}{E\{1-w(\x)
\mid\u,1\}}\right) :\right.\\
&&\left.\b(Y,\x)\in{\mR^d}, E\{\b(Y,\x)\mid\x,1\}=\0
\right).
\ese
We can verify that $\Lambda_2^*\perp\Lambda_3$ and
$\Lambda_2+\Lambda_3=\Lambda_2^*+\Lambda_3$. Thus, the nuisance
tangent space is
$\Lambda=\Lambda_1\oplus\Lambda_2^*\oplus\Lambda_3$.

For convenience, we write a function in $\Lambda^\perp$ as
$\g(r,ry,\x)=r\g_1(y,\x)+(1-r)\g_0(\x)$.
It is easy to show that
\bse
\Lambda_1^\perp
=
\left[\g(r,ry,\x):
  \g_0(\x)=\frac{-E\{\g_1(\Y,\x)\mid\x,1\}}{w^{-1}(\x)-1}\right],
\ese
and
\bse
\Lambda_3^\perp
=
\left\{\g(r,ry,\x): E\left(
\{w(\x)-
w^2(\x)\}
[E\{\g_1(Y,\x)\mid\x,1\}-\g_0(\x)]\mid\u\right)=\0
\right\}.
\ese
Hence the intersection of $\Lambda_1^\perp$ and $\Lambda_3^\perp$ is
\bse
\Lambda_1^\perp\cap\Lambda_3^\perp
=
\left[\g(r,ry,\x):
  \g_0(\x)=\frac{-E\{\g_1(\Y,\x)\mid\x,1\}}{w^{-1}(\x)-1}, E\{\g_1(Y,\x)\mid\u,1\}=\0
\right].
\ese
To obtain $\Lambda^\perp$, we consider
any $\g(r,ry,\x)\in \Lambda_1^\perp\cap\Lambda_3^\perp$, and further
require $\g(r,ry,\x)\perp\Lambda_2^*$. This implies that for any $\b(Y,\x)$ such
that
$\b(Y,\X)\in{\mR^d}, E\{\b(Y,\X)\mid \x,1\}=\0$,
we must have
\bse
0&=&
E\left[
\{R\g_1(Y,\x)+(1-R)\g_0(\x)\}\trans\right.\\
&&\left.\times\left\{
r \b(y,\x)-\{r-w(\x)\}
\left(\frac{ E\{\b(Y,\x)\mid\x\}}
{1-w(\x)}
-\frac{
E\left[ E\{ \b(Y,\x)\mid\x\}
\mid\u,1\right]}{E\{1-w(\x)
\mid\u,1\}}\right)\right\}\right]\\
&=&
E\left(
E\left[\frac{\g_1(Y,\x)\trans \b(Y,\x)}{E\{\pi^{-1}(Y,\u,\bb,g)\mid\x,1\}}\mid\x,1\right]
+
E\left[\frac{\g_0(\x)\trans
\b(Y,\x)\pi^{-1}(Y,\u,\bb,g) }
{E\{\pi^{-1}(Y,\u,\bb,g)\mid\x,1\}}\mid\x,1\right]
 \right).
\ese
This is equivalent to that
\bse
&&\frac{\g_1(Y,\x)}{E\{\pi^{-1}(Y,\u,\bb,g)\mid\x,1\}}
+
\frac{\g_0(\x)\pi^{-1}(Y,\u,\bb,g) }
{E\{\pi^{-1}(Y,\u,\bb,g)\mid\x,1\}}\\
&=&\frac{\g_1(Y,\x)}{E\{\pi^{-1}(Y,\u,\bb,g)\mid\x,1\}}
-
\frac{
\pi^{-1}(Y,\u,\bb,g) }
{E\{\pi^{-1}(Y,\u,\bb,g)\mid\x,1\}}\frac{E\{\g_1(\Y,\x)\mid\x,1\}}{E\{\pi^{-1}(Y,\u,\bb,g)\mid\x,1\}-1}
\ese
nd is a function of $\X$ only. Combining with the requirement of
$\Lambda_1^\perp\cap\Lambda_3^\perp$, we obtain that the nuisance
tangent space is
\bse
\Lambda^\perp
=
\left(
\g_0(\x)\{1- r\pi^{-1}(Y,\u,\bb,g)\}:
E[\{1-w^{-1}(\x)\}\g_0(\x)\mid\u,1]=\0
\right).
\ese

\subsection{Efficient score for $\bb$}\label{sec:score}

We next derive the
efficient score $\S\eff$ for $\bb$, the residual of the score vector $\S_\bb$ after projecting it on to the nuisance tangent space $\Lambda$.
By the projection theorem for Hilbert spaces, the projection of $h\in\mathcal{H}$ on to a closed linear subspace $\Lambda$ of $\mathcal{H}$ is the unique element in $\Lambda$, denoted by $\Pi(h\mid \Lambda)$, such that $\|h-\Pi(h\mid\Lambda)\|$ is minimized and the residual $h-\Pi(h\mid\Lambda)$ is orthogonal to all $\lambda\in\Lambda$, that is, $E[\{h-\Pi(h\mid\Lambda)\}\trans\lambda]=0$ for all $\lambda\in\Lambda$.
Taking derivative with respect to $\bb$ of the logarithm of
(\ref{eq:model}),  we get the score function
\bse
\S_\bb(\x,r,ry)
=
E[\{\pi^{-1}(Y,\u,\bb,g)-1\}\h'_\bb(Y,\bb)\mid\x,1]\left\{
w(\x)-\frac{1-r}{w^{-1}(\x)-1}\right\}.
\ese
After some derivations, we obtain that the efficient score is
\be\label{eq:effscore}
\S\eff(\x,r,ry)=\g(\x)
[1-r \{1+e^{-g(\u)-h(y,\bb)}\}], \mbox{ where }
\ee
\bse
\g(\x)&=&\frac{\a(\u)
E\{e^{-h(Y,\bb)}\mid\x,1\}
-E\{e^{-h(Y,\bb)}\h'_\bb(Y,\bb) \mid\x,1\}
}{
E\{e^{-h(Y,\bb)}\mid\x,1\}
+e^{-g(\u)}E\{e^{-2h(Y,\bb)}\mid\x,1\}
},\\
\a(\u)
&=&
\frac{E\left[
E\{e^{-h(Y,\bb)}\h'_\bb(Y,\bb) \mid\x,1\}
E\{e^{-h(Y,\bb)}\mid\x,1\}/d(\x)
\mid\u,1\right]}
{E\left(
[E\{e^{-h(Y,\bb)}\mid\x,1\}]^2/
d(\x)
\mid\u,1\right)}, \mbox{ and }\\
d(\x)
&=&E\{e^{-h(Y,\bb)}\mid\x,1\}
+e^{-g(\u)}E\{e^{-2h(Y,\bb)}\mid\x,1\},
\ese

By defining $\b(y,\x)=\g(\x)[\pi^{-1}(y,\u,\bb,g)-w^{-1}(\x)]$, one can verify this result by checking that $\S_\bb=\S_2+\S_3+\S\eff$ as well as the facts that $\S_2(\x,r,ry)\in\Lambda_2^*$, $\S_3(\x,r,ry)\in\Lambda_3$, $\S\eff(\x,r,ry)\in\Lambda^\perp$, and $E\{\b(Y,\x) \pi^{-1}(Y,\u,\bb,g)\mid\x,1\}=
\g(\x)
[E\{\pi^{-2}(Y,\u,\bb,g)\mid\x,1\}-w^{-2}(\x)]$,
where
\bse
\S_2(\x,r,ry)
&=&r\b(y,\x)-\{r-w(\x)\} \left[\frac{E\{\b(Y,\x)\mid\x\}}{
1-w(\x)}-\frac{E[
E\{\b(Y,\x) \mid\x\}\mid\u,1]}
{E\{1-w(\x)\mid\u,1\}}\right],\\
\S_3(\x,r,ry)
&=&\frac{
E\left(w(\x) E[\{\pi^{-1}(Y,\u,\bb,g)-1\}\h'_\bb(Y,\bb)\mid\x,1]
\mid\u,1\right)}
{E\{1-w(\x)\mid\u,1\}}
\{r-w(\x)\}.
\ese

\section{Proposed Method for $\bb$}\label{sec:betamethod}
To implement the efficient estimator of $\bb$, one needs to evaluate
the conditional expectations of various functions of $\X$ given
$\U,R=1$, $E(\cdot\mid \u,1)$,  the conditional expectations of
various functions of $Y$ given $\X,R=1$, $E(\cdot\mid \x,1)$, as well
as to estimate $g(\u)$.

Since $\x$ is fully observed, the conditional expectations of functions
of $\x$ given $\u$ is a standard full data estimation problem and we
consider this as something feasible. For example, we could use results
based on exploration on the relation between $\X$ and $\U$, either
known or parametrically modeled, or even employing standard nonparametric
regression techniques if necessary.  The conditional expectations of
functions of $Y$ given $\x,1$ is also a full data estimation problem
hence we also consider this as something standard and feasible. Indeed, both
estimations can be done via the existing parametric, nonparametric or
semiparametric modeling and estimation procedures. So we do not
discuss these issues further.

The estimation of $g(\u)$ is however not a full data problem hence is
far from conventional. In investigating the effect of $g(\u)$,
we make an important discovery, in that
the difficult task of estimating $g(\u)$ can be completely avoided.
Indeed, we have the freedom of
adopting a working model, say, $g^*(\u)$, in using the
efficient score, and this still leads to a consistent
estimator of $\bb$. This discovery is a surprise to us since the
original model (\ref{eq:model}) does not promise such a feature. In
our analysis, we find that this property is a benefit from the
logistic propensity form and the additive form inside the logistic
link function, bearing the resemblance to the case-control study,
where the retrospective sampling feature can be ignored without
hampering the covariate effect estimation.

To see the details of the property, using a working model $g^*(\u)$, we would get
\be\label{eq:effscore*}
\S\eff^*(\x,r,ry)=\frac{\a^*(\u)
E\{e^{-h(Y,\bb)}\mid\x,1\}
-E\{e^{-h(Y,\bb)}\h'_\bb(Y,\bb) \mid\x,1\}
}{
E\{e^{-h(Y,\bb)}\mid\x,1\}
+e^{-g^*(\u)}E\{e^{-2h(Y,\bb)}\mid\x,1\}
}[1-r \{1+e^{-g^*(\u)-h(y,\bb)}\}],
\ee
where
\bse
\a^*(\u)
&=&\frac{
E\left[
E\{e^{-h(Y,\bb)}\h'_\bb(Y,\bb) \mid\x,1\}
E\{e^{-h(Y,\bb)}\mid\x,1\}/d^*(\x)
\mid\u,1\right]
}{
E\left(
[E\{e^{-h(Y,\bb)}\mid\x,1\}]^2/d^*(\x)
\mid\u,1\right)},\\
d^*(\x)
&=&E\{e^{-h(Y,\bb)}\mid\x,1\}
+e^{-g^*(\u)}E\{e^{-2h(Y,\bb)}\mid\x,1\}.
\ese
Although at the first glance, $\S\eff^*(\x,r,ry)$ does not appear to have mean
zero, we find that the construction of $\a(\u)$ actually ensures a
certain adaptivity to $d^*(\x)$ hence to $g^*(\u)$ so that
$\S\eff^*(\x,r,ry)$ has conditional mean zero given $\u$. We write
this important discovery as a Main Theorem, while provide its proof in
Supplement S.2.
\begin{Mth} \label{th:seff*}
$E\{\S^*\eff(\X,R,RY)\mid\u\}=\0.$
\end{Mth}

With the Main Theorem \ref{th:seff*} at hand,
to obtain a consistent estimator of $\bb$, we can actually
bypass estimating $g(\u)$ completely. Instead, we can adopt an
arbitrary working function $g^*(\u)$, and construct
an estimating equation $\sumI\wh\S\eff^*(\x_i,r_i,r_iy_i)=\0$, i.e.,
\be\label{eq:effscore*est}
\sumI \frac{\wh\a^*(\u_i)
\wh E\{e^{-h(Y,\bb)}\mid\x_i,1\}
-\wh E\{e^{-h(Y,\bb)}\h'_\bb(Y,\bb) \mid\x_i,1\}
}{
\wh E\{e^{-h(Y,\bb)}\mid\x_i,1\}
+e^{-g^*(\u_i)}\wh E\{e^{-2h(Y,\bb)}\mid\x_i,1\}
}[1-r_i \{1+e^{-g^*(\u_i)-h(y_i,\bb)}\}]
=\0,
\ee
to solve for $\bb$,
where at any $\u, \x$,
\bse
\wh \a^*(\u)&=&
\frac{
\wh E\left[
\wh E\{e^{-h(Y,\bb)}\h'_\bb(Y,\bb) \mid\x,1\}
\wh E\{e^{-h(Y,\bb)}\mid\x,1\}/\wh d^*(\x)
\mid\u,1\right]
}{
\wh E\left(
[\wh E\{e^{-h(Y,\bb)}\mid\x,1\}]^2/\wh d^*(\x)
\mid\u,1\right)},\\
\wh d^*(\x)
&=&\wh E\{e^{-h(Y,\bb)}\mid\x,1\}
+e^{-g^*(\u)}\wh E\{e^{-2h(Y,\bb)}\mid\x,1\}.
\ese

In the above construction, we use $\wh E(\cdot\mid\x,1)$ and
$\wh E(\cdot\mid\u,1)$ to denote a general assessment of the
corresponding conditional expectations. Note that these are all
standard operations based on fully observed data, hence in practice one can use any traditional parametric or nonparametric methods or various machine learning methods.
In this paper, we study three options which lead to three different estimators:
\begin{enumerate}
  \item Estimator $\wh\bb^*$, which satisfies (\ref{eq:effscore*est}) where
$E(\cdot\mid\x,1)$ and
$E(\cdot\mid\u,1)$ are known. Although $\wh\bb^*$ is infeasible in practice, it is included here for comparison purposes.
    \item Estimator $\wc\bb^*$, which satisfies (\ref{eq:effscore*est}) where
$E(\cdot\mid\x,1)$ and
$E(\cdot\mid\u,1)$ are estimated parametrically, where the parameters
are collected as $\ba$.
    \item Estimator $\wt\bb^*$, which satisfies (\ref{eq:effscore*est}) where
$E(\cdot\mid\x,1)$ and
$E(\cdot\mid\u,1)$ are estimated via nonparametric kernel regression. More details of estimator $\wt\bb^*$ is contained in Supplement S.3.
\end{enumerate}

For estimators $\wh\bb^*$ and $\wc\bb^*$, we have the following results with their proofs in Supplement S.4 and S.5 respectively.

\begin{Th} \label{th:idealbetarob}
Estimator $\wh\bb^*$ satisfies $N^{1/2}(\wh\bb^*-\bb)=N^{-1/2}\sumI
\bphi_\bb(\x,r,ry,\bb,g^*)+o_p(1)$, where
\bse
\bphi_\bb(\x,r,ry,\bb,g^*)=
-\left[E\left\{\frac{\partial\S^*\eff(\X,R,RY,\bb) }{\partial \bb\trans}\right\}\right]^{-1} \S^*\eff(\x,r,ry,\bb),
\ese
and $\wh\bb^*$ satisfies
$
N^{1/2}(\wh\bb^*-\bb)\to N(\0, \A_\bb^{-1}\B_\bb{\A_\bb^{-1}}\trans)
$
in distribution  when $N\to\infty$, where
\bse
\A_\bb&=&E\left\{\frac{\partial\S\eff^*(\X_i,R_i,R_iY_i,\bb)}{\partial\bb\trans}\right\},\\
\B_\bb&=&E\left\{\S\eff^*(\X_i,R_i,R_iY_i,\bb)^{\otimes2}\right\}.
\ese
\end{Th}

\begin{Th}\label{th:parabetarob}
Estimator $\wc\bb^*$ satisfies $N^{1/2}(\wc\bb^*-\bb)=N^{-1/2}\sumI
\bphi_\bb(\x,r,ry,\bb,g^*)+o_p(1)$, where
\bse
\bphi_\bb(\x,r,ry,\bb,g^*)&=&
-\left[E\left\{\frac{\partial\S^*\eff(\X,R,RY,\bb)}{\partial \bb\trans}\right\}\right]^{-1} \\
&&\times
\left[
\S^*\eff(\x,r,ry,\bb)+
E\left\{\frac{\partial
  \S^*\eff(\X_i,R_i,R_i Y_i,\bb,\ba) }{\partial \ba\trans} \right\}
r_i\bphi_\ba(\x_i,y_i,\ba)\right],
\ese
and $\wc\bb^*$ satisfies
$
N^{1/2}(\wc\bb^*-\bb)\to N(\0, \A_\bb^{-1}\B_\bb{\A_\bb^{-1}}\trans)
$
in distribution  when $N\to\infty$, where
\bse
\A_\bb&=&E\left\{\frac{\partial\S\eff^*(\X_i,R_i,R_iY_i,\bb)}{\partial\bb\trans}\right\},\\
\B_\bb&=&E\left(
\left[
\S^*\eff(\x_i,r_i,r_i y_i,\bb)+
E\left\{\frac{\partial
  \S^*\eff(\X_i,R_i,R_i Y_i,\bb,\ba) }{\partial \ba\trans} \right\}
r_i\bphi_\ba(\x_i,y_i,\ba)\right]^{\otimes2}\right).
\ese
Here $\bphi_\ba(\x,y,\ba)$ is the influence function associated with
$\wh\ba$ in the full data parametric estimation of the conditional
expectations.
\end{Th}

To present the theoretical property of $\wt\bb^*$, we need the following regularity conditions.
\begin{enumerate}[label=C\arabic*,series=conditions]
 \item\label{con:kernel}
The univariate kernel function $K(u)$ is symmetric,  has order $m$ and
has smooth $m$th
derivative and is supported on $[-1, 1]$. The multivariate kernel
function $K(\u)$ is the product of the univariate kernel function of
each component, i.e. for a $q$-dimensional vector $\u$, the multivariate kernel function can be written as $K(\u)=\prod_{j=1}^q K(u_j)$.
Here, with a slight abuse of
notation, we use the notation $K$
for both univariate and multivariate kernel functions. 
Similar assumptions hold for a $p$-dimensional vector $\X$ as well.

\item\label{con:smooth1}
The functions  $\fyxrone(y,\x)$, $\fx(\x)$ and
$\pi(y,\u,\bb,g)$
have $m$th smooth derivatives with respect to $\x$.
\item\label{con:bandwidth1}
The bandwidth $h$ satisfies $N^{1/2}h^m\to0$ and $N^{1/2}h^p\to\infty$ as $N\to\infty$.
\end{enumerate}

\begin{Th}\label{th:betarob}
Under the regularity conditions \ref{con:kernel},
\ref{con:smooth1}
and \ref{con:bandwidth1},
the estimator $\wt\bb^*$ satisfies
$N^{1/2}(\wt\bb^*-\bb)=N^{-1/2}\sumI
\bphi_\bb(\x,r,ry,\bb,g^*)+o_p(1)$, where
\bse
\bphi_\bb(\x,r,ry,\bb,g^*)&=&
-\left[E\left\{\frac{\partial\S^*\eff(\X,R,RY,\bb)}{\partial \bb\trans}\right\}\right]^{-1}
\left\{
\S^*\eff(\x,r,ry,\bb)+r_i \k(\x, y,\bb,g^*)\right\},
\ese
and $\wc\bb^*$ satisfies
$
N^{1/2}(\wt\bb^*-\bb)\to N(\0, \A_\bb^{-1}\B_\bb{\A_\bb^{-1}}\trans)
$
in distribution  when $N\to\infty$, where
\bse
\A_\bb&=&E\left\{\frac{\partial\S\eff^*(\X_i,R_i,R_iY_i,\bb)}{\partial\bb\trans}\right\},\\
\B_\bb&=&E\left[\{\S\eff^*(\X_i,R_i,R_iY_i,\bb)+ R_i \k(\X_i, Y_i,\bb,g^*)\}^{\otimes2}\right].
\ese
Here
\bse
k(\x_i, y_i,\bb,g^*)&=& \frac{\{e^{-g(\u_i)}-e^{-g^*(\u_i)}\}[\a^*(\u_i)
E\{e^{-h(Y_i)}\mid\x_i,1\}
-E\{e^{-h(Y_i)}\h'_\bb(Y_i,\bb) \mid\x_i,1\}] }{ d^*(\x_i)}\n\\
&&\times\left[2E\{e^{-h(Y_i)}\mid\x_i,1\}-e^{-h(y_i)}-\frac{E\{e^{-h(Y_i)}\mid\x_i,1\} \{e^{-h(y_i)}+e^{-g^*(\u_i)}e^{-2h(y_i)}\}}{d^{*}(\x_i)} \right].
\ese
\end{Th}

The proof of Theorem
\ref{th:betarob} is given in Supplement S.6.
Under the special situation that $g^*(\u)=g(\u)$,
$\S\eff^*(\X_i,R_i,R_iY_i,\bb)$ becomes the true efficient score, hence \\
$E\{\partial
  \S^*\eff(\X_i,R_i,R_i Y_i,\bb,\ba)/\partial \ba\trans\}=\0$
and the function $\k(\x,y,\bb,g^*)$ also vanishes,
therefore the corresponding estimators $\wh\bb$, $\wc\bb$ and $\wt\bb$ are all efficient estimators.
\begin{Cor}\label{cor:knowng}
When $g^*(\u)=g(\u)$, 
the corresponding estimators $\wh\bb$,
$\wc\bb$  and $\wt\bb$ are identical to the first order.
In addition, they are all efficient and reach
the semiparametric efficiency bound.
\end{Cor}

Our main message in
Theorems \ref{th:idealbetarob}, \ref{th:parabetarob},
\ref{th:betarob} is that the difficult task of estimating $g(\u)$
can
be avoided due to a surprising robustness property discovered in the Main
Theorem \ref{th:seff*}.
Corollary \ref{cor:knowng} further reveals an
interesting fact that,  when we use a working model $g^*$ that is not the
same as $g$, the various parametric or nonparametric estimation
procedures of the conditional means bring an
efficiency loss. However, this efficiency loss disappears if the
working model $g^*(\u)$ happens to be identical to $g$.
This phenomenon is a direct consequence of the fact that $\S\eff^*$ is no
longer an element in $\Lambda^\perp$ unless $g^*=g$, which is a
major difference from many other more familiar robust semiparametric estimators.

\section{Proposed Method for $\bt$}\label{sec:bb+}

Once $\bb$ is estimable, it is then hopeful to proceed to estimate
$\bt$ without the need to estimate $g(\u)$. To this end, we
make use of the relation
\bse
E\{\bz(\X,Y,\bt)\}=E\{\bz(\X,Y,\bt)\mid R=1\}\pr(R=1) + E\{\bz(\X,Y,\bt)\mid R=0\}\pr(R=0).
\ese
Obviously, $E\{\bz(\X,Y,\bt)\mid R=1\}\pr(R=1)$
can be estimated by $N^{-1} \sumI r_i \bz(\x_i,y_i,\bt)$.
Further,
$E\{\bz(\X,Y,\bt)\mid R=0\} = \int E\{\bz(\X,Y,\bt)\mid R=0,\x\}f_{\X\mid R}(\x,0)d\x$ and
\bse
E\{\bz(\x,Y,\bt)\mid R=0,\x\} = \frac{E \{\bz(\x,Y,\bt)(\pi^{-1}-1)\mid \x,1\}}{E\{ (\pi^{-1}-1)\mid \x,1\}} = \frac{E[\bz(\x,Y,\bt)\exp\{-h(Y,\bb)\}\mid \x,1]}{E[\exp\{-h(Y,\bb)\}\mid \x,1]}.
\ese
Hence
$E\{\bz(\X,Y,\bt)\mid R=0)\pr(R=0)$ can be estimated by
\bse
N^{-1}\sumI (1-r_i)\frac{\wh E[\bz(\x_i,Y,\bt)\exp\{-h(Y,\bb^{\rm est})\}\mid \x_i,1]}{\wh E[\exp\{-h(Y,\bb^{\rm est})\}\mid \x_i,1]},
\ese
where $\bb^{\rm est}$ is a generic notation for an estimator of $\bb$.
Summarizing the above analysis, we propose to estimate $\bt$ by solving
\be\label{eq:bz}
\0 = \wh E\{\bz(\X,Y,\bt)\}=N^{-1} \sumI \left( r_i \bz(\x_i,y_i,\bt)+ (1-r_i)\frac{\wh
    E[\bz(\x_i,Y,\bt)\exp\{-h(Y,\bb^{\rm est})\}\mid \x_i,1]}{\wh E[\exp\{-h(Y,\bb^{\rm est})\}\mid \x_i,1]} \right).
\ee
Importantly, (\ref{eq:bz}) does not involve the unknown function
$g(\u)$, hence it shares the same robust property as the estimation
for $\bb$. The two expectations in (\ref{eq:bz}) are both full data
conditional expectations, hence can be assessed similarly as in
Section~\ref{sec:betamethod}. Also, any  estimator of $\bb$
obtained in Section~\ref{sec:betamethod} can be used as $\bb^{\rm est}$.

We simply denote the estimator from solving (\ref{eq:bz}) as
$\wh\bt$. In a special case, if ones interest is to estimate
$E\{\bz(\X,Y)\}$, then
\be\label{eq:bz-}
\wh E\{\bz(\X,Y)\}=N^{-1} \sumI \left( r_i \bz(\x_i,y_i)+ (1-r_i)\frac{\wh
    E[\bz(\x_i,Y)\exp\{-h(Y,\bb^{\rm est})\}\mid \x_i,1]}{\wh E[\exp\{-h(Y,\bb^{\rm est})\}\mid \x_i,1]} \right).
\ee

The
theoretical property of $\wh E\{\bz(\X,Y)\}$ is relatively simple to
obtain given the results in Theorems \ref{th:idealbetarob},
\ref{th:parabetarob}, and \ref{th:betarob}, as we summarize below.

\begin{Lem}\label{lem:zetarob}
When $N\to\infty$,
the estimator given in (\ref{eq:bz-}) satisfies
$N^{1/2}[\wh E\{\bz(\X,Y)\}-
E\{\bz(\X,Y)\}]=N^{-1/2}\sumI
\bphi_\bz(\x_i,r_i,r_iy_i,\bb,g^*)+o_p(1)$, where
\bse
\bphi_\bz(\x,r,ry,\bb,g^*)
&=&
r \bz(\x,y)+ (1-r)\frac{
     E[\bz(\x,Y)\exp\{-h(Y,\bb)\}\mid \x,1]}{
    E[\exp\{-h(Y,\bb)\}\mid \x,1]}
-
E\{\bz(\X,Y)\}\\
&&+E\left(
\frac{\partial}{\partial\bb\trans}\frac{
    E[\bz(\X,Y)\exp\{-h(Y,\bb)\}\mid \X,1]}{
    E[\exp\{-h(Y,\bb)\}\mid \X,1]} \right)
\bphi_\bb(\x,r,ry,\bb,g^*)\\
&&+\k_\ba(\x,r,ry,\bb).
\ese
Here, $\k_\ba(\x,r,ry,\bb)=\0$
 when the conditional expectations are known as in Theorem
\ref{th:idealbetarob},
\bse
\k_\ba(\x,r,ry,\bb)=E\left(\frac{\partial}{\partial\ba\trans}\frac{
    E[\bz(\X,Y)\exp\{-h(Y,\bb)\}\mid \X,1,\ba]}{
    E[\exp\{-h(Y,\bb)\}\mid \X,1,\ba]}\right)N^{-1/2}\sumI\bphi_\ba(\x,r,ry)
\ese
 when the conditional expectations are parametrically estimated  as in Theorem
\ref{th:parabetarob}, and
\bse
\k_\ba(\x,r,ry,\bb)=
\frac{r\{1-E(R\mid\x)\} \exp\{-h(y,\bb)\}}{
E[\exp\{-h(Y,\bb)\}\mid \x,1]E(R\mid\x)}
\left(
    \bz(\x,y)
-
\frac{
    E[\bz(\X,Y)\exp\{-h(Y,\bb)\}\mid \x,1]
}{
    E[\exp\{-h(Y,\bb)\}\mid \x,1]  }
\right)
\ese
when the conditional expectations are nonparametrically estimated  as in Theorem
\ref{th:betarob}. Further, $N^{1/2}[\wh E\{\bz(\X,Y)\}-
E\{\bz(\X,Y)\}]\to N(\0,\V_\bz)$,
where $\V_\bz=E\{\bphi_\bz(\X,R,RY,\bb,g^*)^{\otimes2}\}$.
\end{Lem}

Consequently, for $\wh\bt$, we have
\begin{Th}\label{th:thetarob}
When $N\to\infty$,
$\wh\bt$ satisfies
\bse
N^{1/2}(\wh\bt-\bt)=
-\A_\bt
^{-1}
N^{-1/2}\sumI
\bphi_\bz(\x_i,r_i,r_iy_i,\bb,g^*)+o_p(1),
\ese
therefore,
\bse
N^{1/2}(\wh\bt-\bt)
\to N(\0,\A_\bt^{-1}\V_\bz{\A_\bt^{-1}}\trans),
\ese
where $\bphi_\bz(\x_i,r_i,r_iy_i,\bb,g^*), \V_\bz$ are given in Lemma
\ref{lem:zetarob}, and
$
\A_\bt=\partial E\{\bz(\X_i,
      Y_i,\bt)\}/\partial\bt\trans.
$
\end{Th}

\section{Simulation Studies}\label{sec:simu}

\subsection{Estimators of $\bb$}\label{sec:simu1}
We conduct some simulation studies to examine the finite-sample
performance of our proposed estimators as well as their comparison
with some existing methods in the literature. We first present the
simulation results for the estimators of $\bb$, then for the
estimators of $\bt$.

We first compare the performance of the three proposed estimators of $\bb$.
We consider variable $Z$ with two
categories such that $\pr(Z=-1)=\pr(Z=1)=0.5$. Given
$Z=z$, $U \sim N(z,1)$. Given $Z=z$, $U=u$
and $R=1$,
$Y \sim N\{(u-z)^2,1\}$.
The propensity model is defined
as $\pi (y,u,\beta,g)=\expit\{h(y,\beta)+g(u)\}$ with $h(y)=-\beta y$
and $g(u)=-a_1-a_2u$ where $(\beta,a_1,a_2)=(-0.2,0.4,-0.3)$.

We implement the estimator $\wh\bb^*$ where all the conditional
expectations are replaced with their truth, the estimator $\wc\bb^*$
where the conditional expectations are evaluated parametrically with
the conditional densities estimated using the maximum likelihood
method, and the estimator $\wt\bb^*$ where the conditional
expectations are estimated via nonparametric
kernel regression with Nadaraya-Watson kernel estimator with bandwith
$1.5 N^{1/3}$.
We consider two different situations for $g^*(u)$ that it is correctly
specified with $g(u)=-0.4+0.3u$ and misspecified with $g^*(u)=-0.4 u$,
and three different situations for the sample size $N=200, 500,
1000$.

Based on 1000 simulation replicates, Table~\ref{table:simu1}
summarizes the estimation bias, the Monte Carlo standard deviation,
the estimated standard error, and the 95\% coverage probability. From
Table~\ref{table:simu1}, it is clear that all the estimators have very
small estimation bias, no matter the working model $g^*(u)$ is correct
or not. In all scenarios, the estimated standard errors calculated
through the asymptotic theory are fairly close to the Monte Carlo
standard deviations. Compared to the situation with the correct model
of $g(u)$, although not drastic at all, the misspecified situation has
a little bit efficiency loss, which matches with the theoretical
properties of these estimators presented in
Section~\ref{sec:betamethod}. The coverage probability, in almost all
scenarios, is quite close to the nominal level 95\%. The only
exception is the estimator $\wt\bb^*$, where it is slightly
under-covered but gets better when the sample size increases. The
potential reason is the relatively more restrictive requirement on the
sample size for nonparametric kernel estimation.
Overall the results presented in Table~\ref{table:simu1} are in
accordance to that found in our theoretical derivations. They are
convincing in that the proposed estimators of $\bb$ with an
arbitrarily misspecified working model $g^*(u)$ are still consistent
and asymptotically normally distributed.

\FloatBarrier
\begin{table}[!h]
\begin{center}
\caption{Estimation bias (Bias), standard deviation (SD), estimated
  standard error (SE) and 95\% coverage probability (CVP) of the
  proposed estimators of $\bb$. All values have been multiplied by 100}
\begin{tabular}{c c c r r r r }
\hline
$g^*(u)$ & $N$ & Estimator & Bias & SD & SE & CVP\\\hline\hline

Correct & 200 & $\wh\beta$ & -0.90 & 18.23   & 16.49 & 94.9\\
$g(u)=-0.4+0.3u$& & $\wc\beta$ & -0.91 & 18.16   & 16.31 & 94.9\\
& & $\wt\beta$ & -3.96 & 16.61 & 15.25  & 93.4\\
& 500 & $\wh\beta$ & -0.25 & 10.41   & 9.72 & 93.5\\
& & $\wc\beta$ & -0.25 & 10.39  & 9.73  & 93.4\\
& & $\wt\beta$ & -1.20 & 9.89  & 9.29  & 94.3\\
& 1000 & $\wh\beta$ & -0.23 & 6.94   & 6.80  & 95.7\\
& & $\wc\beta$ & -0.23 & 6.93  & 6.80 & 95.9\\
& & $\wt\beta$ & -0.69 & 6.71   & 6.54 & 95.8\\
\hline
Incorrect & 200 & $\wh\beta^*$ & -0.88 & 19.11 & 16.72   & 93.3\\
$g^*(u)=-0.4u$& & $\wc\beta^*$ & -0.79 & 18.96   & 16.53  & 93.4\\
& & $\wt\beta^*$ & -4.40 & 17.20 & 14.34 & 90.4\\
& 500 & $\wh\beta^*$ & -0.16 & 11.03   & 10.27 & 94.3\\
& & $\wc\beta^*$ & -0.10 & 10.94  & 10.14 & 94.0\\
& & $\wt\beta^*$ & -1.40 & 10.36  & 8.50  & 90.7\\
& 1000 & $\wh\beta^*$ & -0.24 & 7.45 & 7.22 & 94.9\\
& & $\wc\beta^*$ & -0.20 & 7.43 & 7.12 & 94.3\\
& & $\wt\beta^*$ & -0.80 & 6.90  & 5.95  & 91.4\\\hline
\end{tabular}
\label{table:simu1}
\end{center}
\end{table}
\FloatBarrier

\subsection{Estimators of $\theta=E(Y)$}\label{sec:simu2}

To compare the proposed estimators with Shao and Wang's method, we focus on the
estimation of the outcome mean, i.e., $\theta=E(Y)$. From
Section~\ref{sec:bb+}, the proposed estimators $\wh\theta^*$,
$\wc\theta^*$ or $\wt\theta^*$ estimate $E(Y)$ as
\bse
N^{-1} \sumI \left( r_i y_i+ (1-r_i)\frac{\wh
    E[Y\exp\{-h(Y,\bb^{\rm est})\}\mid \x_i,1]}{\wh
    E[\exp\{-h(Y,\bb^{\rm est})\}\mid \x_i,1]} \right),
\ese
where $\bb^{\rm est}=\wh\bb^*$, $\wc\bb^*$ or $\wt\bb^*$.
Besides Shao and Wang's method, we also compare with the oracle estimator
$\sumI y_i/N$ and the na\"ive estimator $\sumI r_i y_i/\sumI r_i$.

We consider two simulation settings. In the first setting, the data
generation is largely identical to that in Section~\ref{sec:simu1}
except that we use $g(u)=-a_1-a_2u^2$ where
$(\beta,a_1,a_2)=(-0.1,0.2,-0.3)$.
When we implement the estimator $\wt\beta^*$, we use the Gaussian
kernel with bandwidth $1.5 N^{-1/3}$. We consider the misspecified
model $g^*(u)=-0.4u$ in our implementations.
Based on 1000 simulation replicates, we report the estimation bias,
the Monte Carlo standard deviation, the mean squared error, the
estimated standard error based on 200 bootstrap samples, and the 95\%
coverage probability in Table~\ref{table:simu2}.

The second setting concerns a two-dimensional $\U$. Specifically, we
consider variable $Z$ with two
categories such that $\pr(Z=1)=\pr(Z=2)=0.5$. We generate
$U_1 |Z=z \sim N(z,1)$ and $U_2 |Z=z \sim N(z,1)$. Also, given $Z=z$, $\U=(u_1,u_2)$
and $R=1$,
$Y \sim N\{(u_1-z)^2+(u_2-z)^2,1\}$. The propensity model is defined
as $\pi (y,\u,\beta,g)=\expit\{h(y,\beta)+g(\u)\}$ with $h(y)=-\beta y$
and $g(\u)=-a_1-a_2u_1^2-a_3u_2^2$ where
$(\beta,a_1,a_2,a_3)=(-0.1,0.8,-0.2,-0.2)$.
When we implement the estimator $\wt\theta^*$, we use a fourth order
Tri-weight kernel with bandwidth $1.0N^{-1/6}$ when estimating
$\wt\bb^*$ and we use a second order Gaussian kernel with bandwidth
$1.0N^{-1/3}$ for the $E(Y)$ estimation.

We consider the misspecified model $g^*(\u)=-0.4u_1-0.6u_2$ in our experiments.
The corresponding results are summarized in Table~\ref{table:simu3}.

The results in Tables~\ref{table:simu2} and \ref{table:simu3} deliver
the following information. Firstly, the na\"ive method has very large
bias and unreliable coverage probability, hence should not be used in
practice.
Similar to Section~\ref{sec:simu1}, the three proposed methods all
perform well in terms of small bias, close Monte Carlo standard
deviation and estimated standard error, as well as close to nominal
coverage probability. Relatively, the nonparametric estimator
$\wt\theta^*$ has slightly larger bias, but the bias quickly vanishes
when the sample size increases. In terms of estimation efficiency, the
proposed estimators possess very close, albeit slightly larger, Monte
Carlo standard deviation and estimated standard error compared to the
oracle estimator, especially when the model $g(\u)$ is correct and the
sample size is large.

The comparison to Shao and Wang's method is also obvious. With a
one-dimensional $U$, although all the standard deviations are close,
the proposed estimators have a smaller bias hence a smaller MSE. With
a two-dimensional $\U$, the bias, standard deviation and the MSE are
all much smaller than Shao and Wang's method.
One may notice that we do not report the estimated standard error and
the coverage probability results for Shao and Wang's method. This concerns the computation
time. For the two-dimensional $\U$ case, our method with sample size
200 only takes less than a minute to finish 1000 simulation replicates
whereas Shao and Wang's method takes about 25 minutes. When the sample size
increases to 1000, our method takes about 10 minutes but Shao and Wang's method would need
about 16 hours, not even to mention that the standard error estimation
needs 200 bootstrap samples. The computational complexity of Shao and Wang's method comes
from the fact that their kernel estimation has to be conducted for
every value of $\bb$ in every iteration. Due to the facts that our
goal of the paper is not to replicate the results in
\cite{shao2016semiparametric} and that the superior performance of our
proposed methods is already very apparent without comparing the
estimated standard error, we decide not to report the standard error
estimates of Shao and Wang's method.

\FloatBarrier
\begin{table}[!h]
\begin{center}
\caption{Estimation bias (Bias), standard deviation (SD), mean squared
  error (MSE), estimated standard error (SE) and 95\% coverage
  probability (CVP) of the proposed estimators of $\theta=E(Y)$, with
  one-dimensional $U$. All values have been multiplied by 100}
\begin{tabular}{c c c r r r r r }
\hline
$g^*(u)$ & $N$ & Estimator & Bias & SD & MSE & SE & CVP\\\hline\hline

Correct & 200 & Oracle & 0.57  & 12.32 & 12.32 & 12.22           & 94.7           \\
$g(u)=-0.2+0.3u^2$& & Na\"ive & 23.46  & 17.41 & 29.21 & 17.14           & 74.6            \\
& & SW & -5.70 & 14.89 & 15.94 &                  &                  \\
& & $\wh\theta$ & -0.14 & 13.63 & 13.62 & 15.69          & 97.3            \\
& & $\wc\theta$ & -0.17 & 14.03 & 14.02 & 15.87           & 97.3            \\
& & $\wt\theta$ & -2.74 & 15.74 & 15.97 & 16.61           & 95.8            \\
& 500 & Oracle & 0.63  & 7.56 & 7.58 & 7.77           & 95.5\\
 &  & Na\"ive & 23.84  & 11.09 & 26.29 & 10.93           & 41.6\\
 &  & SW & -6.14 & 9.44 & 11.26 &                  &\\
 &  & $\wh\theta$ & 0.46  & 8.20 & 8.21 & 8.45           & 95.9\\
 &  & $\wc\theta$ & 0.48  & 8.49 & 8.50 & 8.70           & 95.7\\
 &  & $\wt\theta$ & -0.73 & 9.64 & 9.66 & 9.80           & 95.5\\
 & 1000 & Oracle & 0.29  & 5.59 & 5.60 & 5.49           & 94.3\\
 &  & Na\"ive & 23.60  & 7.82 & 24.86 & 7.72           & 12.0\\
 &  & SW & -6.28 & 6.59 & 9.10 &                  &  \\
 &  & $\wh\theta$ & 0.15  & 5.87 & 5.87 & 5.88          & 94.9\\
 &  & $\wc\theta$ & 0.16  & 6.05 & 6.05 & 6.05           & 95.1\\
 &  & $\wt\theta$ & -0.44 & 6.83 & 6.84 & 6.82           & 94.9\\\hline
 
Incorrect & 200 & Oracle & 0.57  & 12.32 & 12.32 & 12.22           & 94.4            \\
$g^*(u)=-0.4u$& & Na\"ive & 23.46  & 17.41 & 29.21 & 17.11           & 74.0            \\
& & SW & -5.70 & 14.89 & 15.94 &                  &                  \\
& & $\wh\theta^*$ & 0.31  & 14.66 & 14.67 & 18.22           & 97.6            \\
& & $\wc\theta^*$ & 0.09  & 14.47 & 14.46 & 17.48           & 97.8            \\
& & $\wt\theta^*$ & -4.35 & 15.97 & 16.54 & 16.62           & 95.0            \\
 & 500 & Oracle & 0.63  & 7.56 & 7.58 & 7.80           & 95.6            \\
 &  & Na\"ive & 23.84  & 11.09 & 26.29 & 10.95           & 41.8            \\
 &  & SW & -6.14 & 9.44 & 11.26 &                  &\\
 &  & $\wh\theta^*$ & 0.64  & 7.97 & 7.99 & 8.52           & 95.9            \\
 &  & $\wc\theta^*$ & 0.64  & 8.24 & 8.26 & 8.59           & 96.1            \\
 &  & $\wt\theta^*$ & -1.70 & 9.78 & 9.92 & 9.90           & 95.6            \\
 & 1000 & Oracle & 0.29  & 5.59 & 5.60 & 5.49           & 94.2            \\
 &  & Na\"ive & 23.60  & 7.82 & 24.86 & 7.72           & 12.2            \\
 &  & SW & -6.28 & 6.59 & 9.10 &                  &  \\
 &  & $\wh\theta^*$ & 0.24  & 5.71 & 5.71 & 5.73           & 94.6            \\
 &  & $\wc\theta^*$ & 0.23  & 5.88 & 5.88 & 5.90           & 94.8            \\
 &  & $\wt\theta^*$ & -1.10 & 6.86 & 6.95 & 6.88           & 94.4           \\\hline

\end{tabular}
\label{table:simu2}
\end{center}
\end{table}
\FloatBarrier

\FloatBarrier
\begin{table}[!h]
\begin{center}
\caption{Estimation bias (Bias), standard deviation (SD), mean squared error (MSE), estimated standard error (SE) and 95\% coverage probability (CVP) of the proposed estimators of $\theta=E(Y)$, with two-dimensional $\U$. All values have been multiplied by 100}
\begin{tabular}{c c c r r r r r }
\hline
$g^*(\u)$ & $N$ & Estimator & Bias & SD & MSE & SE & CVP\\\hline\hline

Correct & 200 & Oracle & -1.89 & 15.22 & 15.32 & 15.64           & 95.2                                                                          \\
$g(\u)=-0.8+0.2u_1^2+0.2u_2^2$& & Na\"ive & 25.37  & 19.87 & 32.22 & 20.87           & 77.9                                                                          \\
& & SW & -11.00 & 20.71 & 23.44 &                  &                                                                                \\
& & $\wh\theta$ & -1.63 & 16.21 & 16.29 & 16.82           & 94.7                                                                          \\
& & $\wc\theta$ & -2.11 & 16.70 & 16.82 & 17.28           & 94.4                                                                          \\
& & $\wt\theta$ & -5.44 & 17.00 & 17.84 & 18.09           & 95.6                                                                          \\
& 500 & Oracle & -2.09 & 9.85 & 10.07 & 9.96          & 94.6                                                                          \\
 &  & Na\"ive & 24.82  & 12.63 & 27.85 & 13.29           & 55.8                                                                         \\
 &  & SW & -8.60 & 13.88 & 16.32 &                  &                                                                                \\
 &  & $\wh\theta$ & -1.72 & 10.41 & 10.54 & 10.61           & 94.8                                                                          \\
 &  & $\wc\theta$ & -2.20 & 10.71 & 10.93 & 10.92          & 95.1                                                                          \\
 &  & $\wt\theta$ & -4.14 & 10.74 & 11.51 & 11.18           & 94.7                                                                          \\
 & 1000 & Oracle & -1.59 & 6.91 & 7.09 & 7.06           & 94.6                                                                          \\
 &  & Na\"ive & 25.65  & 9.16 & 27.23 & 9.42           & 22.5                                                                          \\
 &  & SW & -6.38 & 8.72 & 10.80 &                  &                                                                                \\
 &  & $\wh\theta$ & -1.22 & 7.33 & 7.43 & 7.48           & 95.2                                                                          \\
 &  & $\wc\theta$ & -1.65 & 7.55 & 7.73 & 7.70           & 94.7                                                                          \\
 &  & $\wt\theta$ & -2.50 & 7.72 & 8.11 & 7.81           & 94.3                                                                          \\\hline
 
Incorrect & 200 & Oracle & -1.82 & 15.27 & 15.37 & 15.66                & 95.5                                                                          \\
$g^*(\u)=-0.4u_1-0.6u_2$& & Na\"ive & 25.44  & 20.02 & 32.36 & 20.88                 & 78.4                                                                          \\
& & SW & -11.00 & 20.71 & 23.44 &                        &                                                                                \\
& & $\wh\theta^*$ & -2.33 & 15.69 & 15.85 & 16.43                 & 94.4                                                                         \\
& & $\wc\theta^*$ & -2.56 & 16.27 & 16.46 & 16.65                 & 94.0                                                                          \\
& & $\wt\theta^*$ & -5.48 & 17.36 & 18.19 & 18.24                 & 94.6                                                                          \\
 & 500 & Oracle & -2.10 & 9.84 & 10.05 & 9.95                 & 94.4                                                                          \\
 &  & Na\"ive & 24.84  & 12.59 & 27.85 & 13.28                & 55.8                                                                          \\
 &  & SW & -8.60 & 13.88 & 16.32 &                        &                                                                                \\
 &  & $\wh\theta^*$ & -2.13 & 10.43 & 10.64 & 10.60                 & 93.8                                                                          \\
 &  & $\wc\theta^*$ & -2.46 & 10.69 & 10.96 & 10.79                & 93.4                                                                          \\
 &  & $\wt\theta^*$ & -4.22 & 10.63 & 11.43 & 11.25                & 94.2                                                                          \\
 & 1000 & Oracle & -1.61 & 6.91 & 7.09 & 7.03                 & 94.3                                                                          \\
 &  & Na\"ive & 25.63  & 9.16 & 27.21 & 9.40                 & 21.9                                                                          \\
 &  & SW & -6.38 & 8.72 & 10.80 &                        &                                                                                \\
 &  & $\wh\theta^*$ & -1.35 & 7.48 & 7.59 & 7.43                 & 94.5                                                                          \\
 &  & $\wc\theta^*$ & -1.69 & 7.65 & 7.83 & 7.59                & 93.9                                                                          \\
 &  & $\wt\theta^*$ & -2.44 & 7.69 & 8.07 & 7.83                 & 94.7                                                                          \\\hline

\end{tabular}
\label{table:simu3}
\end{center}
\end{table}
\FloatBarrier

\section{Data Application}\label{sec:data}

\FloatBarrier
\begin{table}[!h]
\begin{center}
\caption{Estimates of monthly income in 2006 (standard error estimates in parentheses) from the Korean Labor \& Income Panel Study data set (SW; Shao and Wang)} \label{table:realdata}
\begin{tabular}{l c c c c }
\hline
Instrument $\z$            & na\"ive & parametric & SW & $\wt\theta^*$ \\\hline\hline
age,education,gender & \multirow{7}{*}{205.7(2.8)} & 183.9(2.6)    & 186.0 (2.4) & 187.4 (2.4)       \\
age,education        & & 185.4(3.0)    & 186.0 (2.4) & 187.4 (2.4)       \\
age,gender            & &183.1(3.8)   & 186.3 (2.4) & 187.8 (2.5)       \\
education,gender      & &183.2(4.5)   & 185.4 (2.1) & 186.8 (2.6)       \\
age                    &&196.2(6.3)  & 186.4 (2.3) & 188.2 (2.5)       \\
education              &&188.2(5.9)  & 185.1 (2.2) & 187.4 (2.8)       \\
gender                 &&186.1(5.5)  & 185.5 (2.3) & 185.7 (2.5)       \\\hline

\end{tabular}
\end{center}
\end{table}
\FloatBarrier

In this section, we analyze a data set from the Korean Labor \& Income
Panel Study with its details  at
\verb"https://www.kli.re.kr/klips_eng/index.do". This data set
contains $N=2506$ regular wage earners, and the scientific interest is
to estimate the mean of the monthly income in the year of 2006
(outcome $Y$), which has around 35\% missing values.
The fully observed covariates $\X$ in this data set include: monthly
income in 2005 (continuous), gender (discrete with two categories:
male, female), age (discrete with three categories: less than 35,
between 35 and 51, greater than 51), education (discrete with two
categories: up to high school and beyond high school).
This data set was analyzed in \cite{wang2014instrumental} and
\cite{shao2016semiparametric} where the former simply assumed a
parametric logistic regression model for the propensity.

According to \cite{shao2016semiparametric}, the missingness of the
2006 income will definitely depend on the 2006 income itself as well
as the 2005 income. Further, it is reasonable to assume that, given
2005 income and 2006 income, the missingness does not depend on some
or all of gender, age and education. We follow the strategy in
\cite{shao2016semiparametric} to consider different instruments $\Z$
in our analysis.

We implement the proposed estimator $\wt\theta^*$ presented in
Section~\ref{sec:bb+}. The working model of $g^*(u)$ was chosen as
$0.3+0.1u$ where $u$ is the 2005 income.
The
estimation results as well as the corresponding estimated standard
errors based on 200 bootstrap samples are summarized in
Table~\ref{table:realdata}. Table~\ref{table:realdata} also
outlines the results from the na\"ive method ($\sumI r_i y_i/\sumI
r_i$), the parametric method proposed in \cite{wang2014instrumental}
and Shao and Wang's method.

From Table~\ref{table:realdata}, the na\"ive method, which does not take into account the nonignorable missingness, clearly over-estimates the monthly income in 2006, therefore the na\"ive method is severely biased.
The parametric method proposed in \cite{wang2014instrumental} can largely alleviate the estimation bias, but it seems sensitive to the choice of the instrument, i.e., the model misspecification. Additionally, the estimated standard errors from this method is much larger than the methods that assume the semiparametric propensity. One potential reason is that the parametric propensity is very fragile to the misspecified $g(\u)$.
In general, Shao and Wang's method and the proposed estimator $\wt\theta^*$ give very similar results to both parameter estimation and standard error estimation. Interestingly, $\wt\theta^*$ always estimates the monthly income in 2006 just slightly larger than Shao and Wang's method.

\section{Discussion}\label{sec:disc}

In this paper, we propose a completely novel estimation procedure for
the parameter $\bb$ in the semiparametric nonignorable propensity
model as well as for a general functional $\bt$. In this model, there
are three nonparametric nuisance components: $\fyxrone(y\mid \x,1)$,
$\fx(\x)$ and $g(\u)$, where the first two do not involve any missing
data so their estimates are stand-alone and can be flexible, whereas
the estimation of $g(\u)$ is generally not straightforward.

We treat it as a semiparametric model and characterize its geometric
structure. This geometric representation is useful because it will
enable us to identify influence functions of regular asymptotically
linear estimators, which, in turn, will motivate various estimating
equations for the parameter of interest.
We derive the efficient score $\S\eff$ for estimating $\bb$.
Instead of directly employing this efficient score, we exploit
different forms of $\S\eff$ with a working model $g^*(\u)$ bearing in
mind that the estimation of $g(\u)$ is cumbersome.
Fortunately, our paper shows that, this ``working'' efficient score
function $\S\eff^*$ remains to be mean zero, for reasons pertaining to
the logistic propensity form and its additive form inside the link
function, which is nonstandard.
This allows us to use a simpler, but possibly incorrect, model for
$g(\u)$ to form estimating equations whose solution will still yield
consistent estimators of $\bb$.
We also find out that the similar spirit applies to the estimation of $\bt$ as well.
In other words, our estimation procedure does not need the estimation of $g(\u)$ at all.
It is exciting that, compared to other methods in the literature, our
proposal would potentially bring tremendous convenience for end-users,
and greatly broaden the applicability of this semiparametric
nonignorable propensity model for real applications.
This is the key innovation of our work.

\section*{Acknowledgment}

This research was partially supported by the National Science
Foundation under award number 2122074. 
The authors would like to thank Dr. Jun Shao and Dr. Lei Wang for
sharing their programming code for replications. 

\section*{Supplementary Material}
Supplementary material available online contains proof of Theorems and Lemmas.

\bibliographystyle{agsm}
\bibliography{reference}

\end{document}


\begin{center}
{\LARGE{\bf Supplementary Material to
``Avoid Estimating the Unknown Function in a Semiparametric Nonignorable Propensity Model''}}\\ 
	Samidha Shetty, Yanyuan Ma and Jiwei Zhao\\
	Department of Statistics, Pennsylvania State University,
        University Park, PA 16802 \\
     Department of Biostatistics \& Medical Informatics, University of Wisconsin-Madison,
    Madison, WI 53726\\
    sss269@psu.edu\\
	 yzm63@psu.edu\\
    jiwei.zhao@wisc.edu

\end{center}
\setcounter{equation}{0}\renewcommand{\theequation}{S.\arabic{equation}}
\setcounter{subsection}{0}\renewcommand{\thesubsection}{S.\arabic{subsection}}
\setcounter{section}{0}\renewcommand{\thesection}{S.\arabic{section}}

\section{Proof of Lemma 1} \label{supp:lem}

From (2), it is immediate that $\fx(\x)$ and
    $\fyxrone(y,\x)$ are identifiable. To further show the
    identifiability of $\bb$ and $g$, we prove by contradiction.
We assume that there exists $\bb_1$, $g_1(\u)$ and $\bb_2$, $g_2(\u)$ such that
    \bse
    \int \fyxrone(y,\x)[1+\exp\{-h(y;\bb_1)-g_1(\u)\}]dy = \int \fyxrone(y,\x)[1+\exp\{-h(y;\bb_2)-g_2(\u)\}]dy,
    \ese
    which is equivalent to
    \bse
    \int \fyxrone(y,\x) [\exp\{-h(y;\bb_1)-g_1(\u)\}-\exp\{-h(y;\bb_2)-g_2(\u)\}] dy=0.
    \ese
    According to the condition, we have
    \bse
    g_1(\u)-g_2(\u)=h(y;\bb_2)-h(y;\bb_1).
    \ese
Because the left handside is a function of $\u$ while the right
handside is a function of $y$, so we have
$g_1(\u)-g_2(\u)=h(y;\bb_2)-h(y;\bb_1)=c$, where $c$ is a
constant. Under the condition of Lemma 1, $c=0$ and
$\bb_1=\bb_2$ and $g_1(\u)=g_2(\u)$.
This completes the proof. 
\qed

\section{Proof of the Main Theorem 1}\label{sec:prooflemmaseff}

\bse
&&E\{\S^*\eff(\x,r,ry)\mid\u\}\\
&=&E\left[
\frac{\a^*(\u)
E\{e^{-h(Y)}\mid\x,1\}
-E\{e^{-h(Y)}\h'_\bb(Y;\bb) \mid\x,1\}
}{d^*(\x)}\{1-r \pi^{-1}(Y,\u;\bb,g^*)\}\mid\u\right]\\
&=&E\left(
\frac{\a^*(\u)
E\{e^{-h(Y)}\mid\x,1\}
-E\{e^{-h(Y)}\h'_\bb(Y;\bb) \mid\x,1\}
}
{d^*(\x)
}[1-E\{r \pi^{-1}(Y,\u;\bb,g^*)\mid\x\}]\mid\u\right)\\
&=&E\left(
\frac{\a^*(\u)
E\{e^{-h(Y)}\mid\x,1\}
-E\{e^{-h(Y)}\h'_\bb(Y;\bb) \mid\x,1\}
}
{d^*(\x)
}\left[1-\frac{E\{\pi^{-1}(Y,\u;\bb,g^*)\mid\x,1\}}{
E\{\pi^{-1}(Y,\u;\bb,g)\mid\x,1\}}\right]\mid\u\right)\\
&=&E\left(
\frac{\a^*(\u)
E\{e^{-h(Y)}\mid\x,1\}
-E\{e^{-h(Y)}\h'_\bb(Y;\bb) \mid\x,1\}
}
{d^*(\x) }\left[1-\frac{1+e^{-g^*(\u)}E\{e^{-h(Y)}\mid\x,1\}}{1+
e^{-g(\u)}E\{e^{-h(Y)}\mid\x,1\}}\right]\mid\u\right)\\
&=&E\left[
\frac{\a^*(\u)
E\{e^{-h(Y)}\mid\x,1\}
-E\{e^{-h(Y)}\h'_\bb(Y;\bb) \mid\x,1\}
}
{d^*(\x) }\frac{E\{e^{-h(Y)}\mid\x,1\}}{1+
e^{-g(\u)}E\{e^{-h(Y)}\mid\x,1\}}\mid\u\right]\\
&&\times\{e^{-g(\u)}-e^{-g^*(\u)}\}\\
&=&E\left[
\frac{\a^*(\u)
E\{e^{-h(Y)}\mid\x,1\}
-E\{e^{-h(Y)}\h'_\bb(Y;\bb) \mid\x,1\}
}
{d^*(\x)
}w(\x)E\{e^{-h(Y)}\mid\x,1\}\mid\u\right]\\
&&\times\{e^{-g(\u)}-e^{-g^*(\u)}\}\\
&=&E\left[
\frac{\a^*(\u)
E\{e^{-h(Y)}\mid\x,1\}
-E\{e^{-h(Y)}\h'_\bb(Y;\bb) \mid\x,1\}
}
{d^*(\x) }E\{e^{-h(Y)}\mid\x,1\}\mid\u,1\right]\\
&&\times\{e^{-g(\u)}-e^{-g^*(\u)}\}
E\{w(\x)\mid\u\}\\
&=&\left\{
\a^*(\u)
E\left(
[E\{e^{-h(Y)}\mid\x,1\}]^2/d^*(\x) \mid\u,1\right)\right.\\
&&\left.-E\left[E\{e^{-h(Y)}\h'_\bb(Y;\bb) \mid\x,1\}E\{e^{-h(Y)}\mid\x,1\}/d^*(\x)
\mid\u,1\right]\right\}\\
&&\times\{e^{-g(\u)}-e^{-g^*(\u)}\}
E\{w(\x)\mid\u\}\\
&=&\0.
\ese
\qed

\section{More Details of Estimator $\wt\bb^*$} \label{supp:detailswtbb*}

In constructing the estimator $\wt\bb^*$, we estimate the conditional expectations using the following non-parametric estimators:
  \bse
  &&\wh E\{e^{-h(Y_i)}\mid\x_i,1\}= \frac{\sumJ r_j  K_{h}(\x_j-\x_i) e^{-h(y_j)} }{\sumJ r_jK_{h} (\x_j-\x_i)},\\
  &&\wh E\{e^{-2h(Y_i)}\mid\x_i,1\}= \frac{\sumJ r_j  K_{h} (\x_j-\x_i)e^{-2h(y_j)} }{\sumJ r_jK_{h} (\x_j-\x_i)},\\
  &&\wh E\{e^{-h(Y_i)}\h'_\bb(Y_i;\bb) \mid\x_i,1\}=\frac{\sumJ r_j  K_{h} (\x_j-\x_i)e^{-h(y_j)}\h'_\bb(y_j;\bb) }{\sumJ r_jK_{h} (\x_j-\x_i)},\\
  &&\wh E\left([\wh E\{e^{-h(Y_i)}\mid\X_i,1\}]^2/\wh d (\X_i)\mid\u_i,1\right)
  =\frac{\sumJ r_j  K_{h} (\u_j-\u_i)( [\wh E\{e^{-h(Y_j)}\mid\x_j,1\}]^2/\wh d (\x_j))}{\sumJ r_jK_{h} (\u_j-\u_i)},\\
  &&\wh E\left[\wh E\{e^{-h(Y_i)}\h'_\bb(Y_i;\bb) \mid\X_i,1\} \wh E\{e^{-h(Y_i)}\mid\X_i,1\}/\wh d (\X_i)\mid\u_i,1\right]\\
  &&= \frac{\sumJ r_j  K_{h} (\u_j-\u_i)[ \wh E\{e^{-h(Y_j)}\h'_\bb(Y_j;\bb) \mid\x_j,1\} \wh E\{e^{-h(Y_j)}\mid\x_j,1\}/\wh d (\x_j)]}{\sumJ r_j K_{h} (\u_j-\u_i)}.
  \ese

\section{Proof of Theorem~1}\label{supp:idealbetarob}
When the true expectations are known, we have
\be \label{eq:esteq2}
\0&=&\sumI\S^*\eff(\x_i,r_i,r_i y_i)\n\\
&=&\sumI \frac{\a^*(\u_i)
E\{e^{-h(Y_i)}\mid\x_i,1\}
-E\{e^{-h(Y_i)}\h'_\bb(Y_i;\bb) \mid\x_i,1\}
}{  E\{e^{-h(Y_i)}\mid\x_i,1\}
+e^{-g^*(\u_i)} E\{e^{-2h(Y_i)}\mid\x_i,1\}
}\n\\
&&\times \{1-r_i  \pi^{-1}(y_i,\u_i;\bb,g^*)\}.
\ee
Let the resulting estimator be $\wh\bb^*$.
From (\ref{eq:esteq2}) we have
\bse
\0&=&\sumI\S^*\eff(\x_i,r_i,r_i y_i,\wh\bb^*) \\
&=&\sumI\S^*\eff(\x_i,r_i,r_i y_i,\bb) + \sumI\frac{\partial \S^*\eff(\x_i,r_i,r_i y_i,\bb^*) }{\partial \bb\trans} (\wh\bb^* - \bb),
\ese
where $\bb^*$ is on a line connecting $\bb$ and $\wh\bb^*$.
Thus,
\bse
N^{1/2}(\wh\bb^* - \bb)=  -\left\{N^{-1}\sumI\frac{\partial\S^*\eff(\x_i,r_i,r_i y_i,\bb^*) }{\partial \bb\trans}\right\}^{-1} N^{-1/2}\sumI\S^*\eff(\x_i,r_i,r_i y_i,\bb).
\ese
It is easy to verify that
\bse
N^{-1}\sumI\frac{\partial\S^*\eff(\x_i,r_i,r_i y_i,\bb^*)
}{\partial \bb\trans}
=E\left\{\frac{\partial\S^*\eff(\x_i,r_i,r_i y_i,\bb)
}{\partial \bb\trans}\right\}+o_p(1).
\ese
Thus, Theorem 1 is shown.\qed

\section{Proof of Theorem~2}\label{supp:parabetarob}
When the expectations are estimated parametrically, we write $\wh\S\eff^*(\x_i,r_i,r_i
y_i)$ more explicitly as  $\S\eff^*(\x_i,r_i,r_i y_i,\bb,
\wh\ba)$, where $\wh\ba$ is a standard parametric estimator based on
the full data. Since the influence function associated with $\wh\ba$ is
$\bphi_\ba$, we have $N^{1/2}(\wh\ba-\ba)=N^{-1/2}\sumI
r_i\bphi_\ba(\x_i,y_i,\ba)+o_p(1)$.
Thus,
$\wc\bb^*$ satisfies
\bse
\0&=&\sumI\S^*\eff(\x_i,r_i,r_i y_i,\wc\bb^*,\wh\ba) \\
&=&\sumI\S^*\eff(\x_i,r_i,r_i y_i,\bb) + \sumI\frac{\partial
  \S^*\eff(\x_i,r_i,r_i y_i,\bb^*) }{\partial \bb\trans} (\wc\bb^* -
\bb) + \sumI\frac{\partial
  \S^*\eff(\x_i,r_i,r_i y_i,\bb^*,\ba^*) }{\partial \ba\trans} (\wh\ba -
\ba)\\
&=&\sumI\S^*\eff(\x_i,r_i,r_i y_i,\bb) + \sumI\frac{\partial
  \S^*\eff(\x_i,r_i,r_i y_i,\bb^*) }{\partial \bb\trans} (\wc\bb^* -
\bb) \\
&&+ \sumI\frac{\partial
  \S^*\eff(\x_i,r_i,r_i y_i,\bb^*,\ba^*) }{\partial \ba\trans}
N^{-1}\sumJ r_j\bphi_\ba(\x_j,y_j,\ba)+o_p(N^{1/2})\\
&=&\sumI\S^*\eff(\x_i,r_i,r_i y_i,\bb) + \sumI\frac{\partial
  \S^*\eff(\x_i,r_i,r_i y_i,\bb^*) }{\partial \bb\trans} (\wc\bb^* -
\bb) \\
&&+ E\left\{\frac{\partial
  \S^*\eff(\X_i,R_i,R_i Y_i,\bb^*,\ba^*) }{\partial \ba\trans} \right\}
\sumI r_i\bphi_\ba(\x_i,y_i,\ba)+o_p(N^{1/2})
\ese
where $({\bb^*}\trans,{\ba^*}\trans)\trans$ is on a line connecting
$(\bb\trans,\ba\trans)\trans$ and $(\wc\bb^{*\rm T},\wc\ba^{*\rm T})\trans$.
Thus,
\bse
N^{1/2}(\wc\bb^* - \bb)&=& -N^{-1/2}\sumI
\left[E\left\{\frac{\partial\S^*\eff(\x_i,r_i,r_i y_i,\bb)
  }{\partial \bb\trans}\right\}\right]^{-1} \\
&&\times
\left[
\S^*\eff(\x_i,r_i,r_i y_i,\bb)+
E\left\{\frac{\partial
  \S^*\eff(\X_i,R_i,R_i Y_i,\bb,\ba) }{\partial \ba\trans} \right\}
r_i\bphi_\ba(\x_i,y_i,\ba)\right]+o_p(1).
\ese
Thus, Theorem 2 is shown.\qed

\section{Proof of Theorem~3}\label{supp:betarob}

For simplicity we will write $\wt\bb^*$ as $\wt\bb$ in this proof. Also, we first introduce some notations and relations that will be used in this proof. We define
\bse
\Delta_{1i}&\equiv& E\{e^{-h(Y_i)}\mid\x_i,1\},\\
\Delta_{2i}&\equiv&
E\{e^{-2h(Y_i)}\mid\x_i,1\},\\
\Delta_{3i}&\equiv&
E\{e^{-h(Y_i)}\h'_\bb(Y_i;\bb)\mid\x_i,1\},\\
A_i &\equiv&[\a^*(\u_i)
 \Delta_{1i}
-\Delta_{3i}]\{1-r_i  \pi^{-1}(y_i,\u_i;\bb,g^*)\},\\
 d^*(\x_i)&\equiv& \Delta_{1i}
+e^{-g^*(\u_i)} \Delta_{2i},\\
C_i&\equiv&E[\{\Delta_{3i}\Delta_{1i}/ d^* (\X_i) \}\mid \u_i,1]\Delta_{1i}
-E[\{\Delta^2_{1i}/ d^* (\X_i) \}\mid \u_i,1]\Delta_{3i},\\
D_i&\equiv& E[\{\Delta^2_{1i}/ d^* (\X_i) \}\mid \u_i,1],\\
T_{j,k}&\equiv&\Delta_{3k}\Delta_{1k}
-\Delta^2_{1k}\h'_\bb(y_j;\bb),\\
V_{i,k}&\equiv& d^* (\x_k)
E(R_i\mid\u_i) f_{\U}(\u_i) E(R_i\mid\x_i) f_{\X}(\x_i).
\ese
We note the relation
\bse
A_i=\frac{\{1-r_i  \pi^{-1}(y_i,\u_i;\bb,g^*)\}C_i}{D_i}.
\ese
For any $a(\x_i)$,
\bse
&&E \left[ \frac{ \{1-R_i
  \pi^{-1}(Y_i,\U_i;\bb,g^*)\}  K_h({\u_j-\U_i}) a(\X_i) }{ E(R_i\mid\U_i) f_{\U}(\U_i)  }\right]\n\\
&=& E \left\{\frac{ a(\X_j) }{ E(R_j\mid\u_j)} \mid \u_j \right\}  - E \left\{
  \pi^{-1}(Y_j,\U_j;\bb,g^*)a(\X_j) \mid \u_j,1\right\} +
O_p(h^m)\n\\
&=& \frac{E\{w(\X_j) \mid \u_j\}}{E(R_j\mid\u_j)} E \left\{ a(\X_j) w^{-1}(\X_j)  \mid \u_j,1 \right\}  - E \left\{
  \pi^{-1}(Y_j,\U_j;\bb,g^*)a(\X_j) \mid \u_j,1\right\}+O_p(h^m) \n\\
&=& E \left\{ a(\X_j) w^{-1}(\X_j) \mid \u_j,1 \right\}   - E \left\{ a(\X_j)
  w^{*-1}(\X_j) \mid \u_j,1\right\}+ O_p(h^m) \n\\
&=& E \left[ a(\X_j) \{w^{-1}(\X_j) -w^{*-1}(\X_j)\} \mid \u_j,1 \right] + O_p(h^m),
\ese
where recall that
$w(\x_i)= [E\{\pi^{-1}(Y_i,\U_i;\bb,g) \mid \x_i,1\}]^{-1}$ and $w^*(\x_i)= [E\{\pi^{-1}(Y_i,\U_i;\bb,g^*) \mid \x_i,1\}]^{-1}$.
The third equality uses the property that $w(\x_j)=E(R_j\mid\x_j)$, hence $E\{w(\x_j) \mid
  \u_j\}=E(R_j\mid\u_j)$.
Similarly for any $a(\x_i)$,
\bse
&&E \left[ \frac{ \{1-R_i
  \pi^{-1}(Y_i,\U_i;\bb,g^*)\}  K_h({\x_j-\X_i}) a(\X_i) }{ E(R_i\mid\X_i) f_{\X}(\X_i)  }\right]\n\\
&=& E \left\{\frac{ a(\x_j) }{ E(R_j\mid\x_j)} \mid \x_j \right\}  - E \left\{
  \pi^{-1}(Y_j,\U_j;\bb,g^*)a(\x_j) \mid \x_j,1\right\} + O_p(h^m)\n\\
&=& \frac{ a(\x_j) }{ E(R_j\mid\x_j)}  - a(\x_j) E \left\{
  \pi^{-1}(Y_j,\U_j;\bb,g^*) \mid \x_j,1\right\} + O_p(h^m)\n\\
&=&  a(\x_j) \{w^{-1}(\x_j) -w^{*-1}(\x_j)\}  + O_p(h^m).
\ese
By definition of $w^{-1}(\x_i)$ we have
$
w^{-1}(\x_i)= 1+e^{-g(\u_i)} \Delta_{1i},
$
and similarly, $w^{*-1}(\x_i)=1+e^{-g^*(\u_i)} \Delta_{1i}$. These
further lead to
\bse
E \left[\frac{ \Delta_{1i} \{w^{-1}(\X_i)
-w^{*-1}(\X_i)\}}{d^*(\X_i) D_i}  \mid \u_i,1 \right]
&=& E \left[\frac{ \Delta_{1i} \{e^{-g(\u_i)} \Delta_{1i}
-e^{-g^*(\u_i)} \Delta_{1i}\}}{d^*(\X_i) D_i}  \mid \u_i,1 \right]\n\\
&=& \frac{\{e^{-g(\u_i)}
-e^{-g^*(\u_i)} \}}{D_i} E \left[\frac{ \Delta_{1i}^2 }{d^*(\X_i) }  \mid \u_i,1 \right]\n\\
&=&e^{-g(\u_i)}
-e^{-g^*(\u_i)},\\
 E \left[\frac{ \Delta_{3i} \{w^{-1}(\X_i)
-w^{*-1}(\X_i)\}}{d^*(\X_i)}  \mid \u_i,1 \right]
&=&  E \left\{\frac{ \Delta_{3i} \{e^{-g(\u_i)}\Delta_{1i}-e^{-g^*(\u_i)}\Delta_{1i}\} }{d^*(\X_i)}  \mid \u_i,1 \right\}\n\\
&=& \{e^{-g(\u_i)}-e^{-g^*(\u_i)}\} E \left\{\frac{ \Delta_{1i}\Delta_{3i} }{d^*(\X_i)}  \mid \u_i,1 \right\},\\
E \left[ \frac{ \{w^{-1}(\X_i)
-w^{*-1}(\X_i)\}    C_i }{d^*(\x_i)  } \mid \u_i,1\right]
&=&E \left[ \frac{ \{e^{-g(\u_i)}-e^{-g^*(\u_i)}\}  \Delta_{1i}  C_i }{d^*(\x_i)   } \mid \u_i,1\right]\n\\
&=& \{e^{-g(\u_i)}-e^{-g^*(\u_i)}\} D_iE \left[ \frac{ \Delta_{1i} \{\a^*(\u_i)
 \Delta_{1i}- \Delta_{3i}\} }{d^*(\x_i)  } \mid \u_i,1\right]\n\\
&=&\0.
\ese

Firstly, we have
\bse
\0&=&\sumI\wh\S^*\eff(\x_i,r_i,r_i y_i,\wt\bb) \\
&=&\sumI\wh\S^*\eff(\x_i,r_i,r_i y_i,\bb) + \sumI\frac{\partial \wh\S^*\eff(\x_i,r_i,r_i y_i,\bb^*) }{\partial \bb\trans} (\wt\bb - \bb),
\ese
where $\bb^*$ is on a line connecting $\bb$ and $\wt\bb$.
Thus,
\bse
N^{1/2}(\wt\bb - \bb)=  \left\{N^{-1}\sumI\frac{\partial\wh\S^*\eff(\x_i,r_i,r_i y_i,\bb^*) }{\partial \bb\trans}\right\}^{-1} N^{-1/2}\sumI\wh\S^*\eff(\x_i,r_i,r_i y_i,\bb).
\ese
It is easy to verify that
\be \label{eq:sumseffhat}
N^{-1}\sumI\frac{\partial\wh\S^*\eff(\x_i,r_i,r_i y_i,\bb^*)
}{\partial \bb\trans}
=E\left\{\frac{\partial\S^*\eff(\x_i,r_i,r_i y_i,\bb)
}{\partial \bb\trans}\right\}+o_p(1).
\ee
By Taylor series expansion, we  get
\be \label{eq:main}
&&\frac{1}{\sqrt{N}}\sumI\wh\S^*\eff(\x_i,r_i,r_i y_i,\bb)\n\\
&=& \frac{1}{\sqrt{N}}\sumI \frac{\wh A_i}{\wh d^*(\x_i)}\n\\
&=&
\frac{1}{\sqrt{N}}\sumI \frac{A_i}{d^*(\x_i)}
+\frac{1}{\sqrt{N}}\sumI \frac{\wh A_i}{d^*(\x_i)}
-\frac{1}{\sqrt{N}}
\sumI \frac{A_i \wh d^*(\x_i)}{d^{*2} (\x_i)} +O_p(N^{1/2} h^{2m} + N^{-1/2}h^{-p})\n\\
&=&
\frac{1}{\sqrt{N}}\sumI \frac{A_i}{d^*(\x_i)}
+T_0
-T_1 +O_p(N^{1/2} h^{2m} + N^{-1/2}h^{-p}).
\ee
where
\bse
T_0&=&\frac{1}{\sqrt{N}}\sumI \frac{\wh A_i}{d^*(\x_i)},\\
T_1&=&\frac{1}{\sqrt{N}}\sumI \frac{A_i \wh d^*(\x_i)}{d^{*2}(\x_i)}.
\ese
Under the condition that $N^{1/2} h^{2m}\to0$ and
$N^{1/2}h^p\to\infty$, the last term in (\ref{eq:main}) is $o_p(1)$.
We rewrite $T_1$ as
\be \label{eq:main3tmp}
T_1&=&\frac{1}{\sqrt{N}}\sumI \frac{A_i \wh d^*(\x_i)}{d^{*2}(\x_i)}\n\\
&=& \frac{1}{\sqrt{N}}\sumI \frac{A_i }{d^{*2}(\x_i)} [\wh E\{e^{-h(Y_i)}\mid\x_i,1\}
+e^{-g^* (\u_i)}\wh E\{e^{-2h(Y_i)}\mid\x_i,1\}].
\ee
Consider
\be \label{eq:main31}
&&{N^{-1/2}}\sumI \frac{A_i }{d^{*2}(\x_i)} \wh E\{e^{-h(Y_i)}\mid\x_i,1\}\n\\
&=& N^{-3/2} \sumI \sumJ \frac{A_i }{d^{*2}(\x_i)}  \frac{r_j
  K_{h}(\x_j-\x_i) e^{-h(y_j)} }{{N^{-1}}\sumJ r_j K_{h}
  (\x_j-\x_i)}\n\\
&=& N^{-3/2} \sumI \sumJ \frac{A_i }{d^{*2}(\x_i)}  \frac{r_j
  K_{h}(\x_j-\x_i) e^{-h(y_j)} }{E(R_i\mid\x_i) f_{\X}(\x_i)}\n\\
&&- N^{-3/2} \sumI \sumJ \frac{A_i }{d^{*2}(\x_i)}  \frac{r_j
  K_{h}(\x_j-\x_i) e^{-h(y_j)} }{\{E(R_i\mid\x_i) f_{\X}(\x_i)\}^2} \{N^{-1}\sumK r_k K_{h}
  (\x_k-\x_i)-E(R_i\mid\x_i) f_{\X}(\x_i)\}\n\\
  && + O_p(N^{1/2}h^{2m}+N^{-1/2}h^{-p})\n\\
&=& 2N^{-3/2} \sumI \sumJ \frac{A_i }{d^{*2}(\x_i)}  \frac{r_j
  K_{h}(\x_j-\x_i) e^{-h(y_j)} }{E(R_i\mid\x_i) f_{\X}(\x_i)}\n\\
&&- N^{-5/2} \sumI \sumJ \sumK \frac{A_i }{d^{*2}(\x_i)}  \frac{r_j
  K_{h}(\x_j-\x_i) e^{-h(y_j)} r_k K_{h}
  (\x_k-\x_i) }{\{E(R_i\mid\x_i) f_{\X}(\x_i)\}^2}\n\\
&&+O_p(N^{1/2} h^{2m} + N^{-1/2}h^{-p}).
\ee
Similarly
\be \label{eq:main32}
&&{N^{-1/2}}\sumI \frac{A_i e^{-g^*(\u_i)} }{d^{*2}(\x_i)} \wh E\{e^{-2h(Y_i)}\mid\x_i,1\}\n\\
&=& 2N^{-3/2} \sumI \sumJ \frac{A_i e^{-g^*(\u_i)}}{d^{*2}(\x_i)}  \frac{r_j
  K_{h}(\x_j-\x_i) e^{-2h(y_j)} }{E(R_i\mid\x_i) f_{\X}(\x_i)}\n\\
&&- N^{-5/2} \sumI \sumJ \sumK \frac{A_i e^{-g^*(\u_i)}}{d^{*2}(\x_i)}  \frac{r_j
  K_{h}(\x_j-\x_i) e^{-2h(y_j)} r_k K_{h}
  (\x_k-\x_i) }{\{E(R_i\mid\x_i) f_{\X}(\x_i)\}^2}\n\\
  && + O_p(N^{1/2}h^{2m}+N^{-1/2}h^{-p}).
\ee
Therefore, inserting (\ref{eq:main31}) and (\ref{eq:main32}) into  (\ref{eq:main3tmp}),
we get
\be \label{eq:main3}
T_1&=&\frac{1}{\sqrt{N}}\sumI \frac{A_i \wh d^*(\x_i)}{d^{*2}(\x_i)}\n\\
&=& 2N^{-3/2} \sumI \sumJ \frac{A_i }{d^{*2}(\x_i)}  \frac{r_j
  K_{h}(\x_j-\x_i)}{E(R_i\mid\x_i) f_{\X}(\x_i)}\{e^{-h(y_j)}+e^{-g^*(\u_i)}e^{-2h(y_j)}\}\n\\
&&- N^{-5/2} \sumI \sumJ \sumK \frac{A_i }{d^{*2}(\x_i)}  \frac{r_j
  K_{h}(\x_j-\x_i)  r_k K_{h}
  (\x_k-\x_i) }{\{E(R_i\mid\x_i) f_{\X}(\x_i)\}^2}\{e^{-h(y_j)}+e^{-g^*(\u_i)}e^{-2h(y_j)}\}\n\\
  && + O_p(N^{1/2}h^{2m}+N^{-1/2}h^{-p})\n\\
&=& 2T_{11}-T_{12}+ O_p(N^{1/2} h^{2m} + N^{-1/2}h^{-p}),
\ee
where
\be
T_{11}&=& N^{-3/2} \sumI \sumJ \frac{A_i }{d^{*2}(\x_i)}  \frac{r_j
  K_{h}(\x_j-\x_i)}{E(R_i\mid\x_i)
  f_{\X}(\x_i)}\{e^{-h(y_j)}+e^{-g^*(\u_i)}e^{-2h(y_j)}\}\label{eq:t11}\\
T_{12}&=&N^{-5/2} \sumI \sumJ \sumK \frac{A_i }{d^{*2}(\x_i)}  \frac{r_j
  K_{h}(\x_j-\x_i)  r_k K_{h}
  (\x_k-\x_i) }{\{E(R_i\mid\x_i) f_{\X}(\x_i)\}^2}\n\\
  &&\times \{e^{-h(y_j)}+e^{-g^*(\u_i)}e^{-2h(y_j)}\}.\label{eq:t12}
\ee
Moving to the second term in (\ref{eq:main}).
We have
\be \label{eq:main2}
T_0&=&\frac{1}{\sqrt{N}}\sumI \frac{\wh A_i}{d^*(\x_i)}\n\\
&=& \frac{1}{\sqrt{N}}\sumI  \frac{\{1-r_i  \pi^{-1}(y_i,\u_i;\bb,g^*)\}}{d^*(\x_i)}[\wh\a^*(\u_i)
\wh E\{e^{-h(Y_i)}\mid\x_i,1\}
-\wh E\{e^{-h(Y_i)}\h'_\bb(Y_i;\bb) \mid\x_i,1\}].\n \\
&=& \frac{1}{\sqrt{N}}\sumI  \frac{\{1-r_i  \pi^{-1}(y_i,\u_i;\bb,g^*)\}}{d^*(\x_i)}\left(\frac{\wh E[\{\wh \Delta_{3i}\wh \Delta_{1i}/ \wh d^* (\X_i) \}\mid \u_i,1]}{\wh E[\{\wh \Delta^2_{1i}/ \wh d^* (\X_i) \}\mid \u_i,1]}
\wh \Delta_{1i}
-\wh \Delta_{3i}\right)\n\\
&=& \frac{1}{\sqrt{N}}\sumI \frac{\{1-r_i
  \pi^{-1}(Y_i,\u_i;\bb,g^*)\}}{d^*(\x_i)}\frac{\wh C_i}{\wh
    D_i}\n\\
&=&\frac{1}{\sqrt{N}}\sumI \frac{\{1-r_i
  \pi^{-1}(Y_i,\u_i;\bb,g^*)\}}{d^*(\x_i)}\frac{C_i}{
    D_i}+T_2-T_3 +O_p(N^{1/2}h^{2m}+N^{-1/2}h^{-p})\n\\
&=&\frac{1}{\sqrt{N}}\sumI \frac{A_i}{d^*(\x_i)}+T_2-T_3 +O_p(N^{1/2}h^{2m}+N^{-1/2}h^{-p}),
\ee
where
\bse
T_2&=& \frac{1}{\sqrt{N}}\sumI \frac{\{1-r_i
  \pi^{-1}(Y_i,\u_i;\bb,g^*)\}}{d^*(\x_i)}\frac{\wh
    C_i}{D_i} ,\\
T_3&=&\frac{1}{\sqrt{N}}\sumI \frac{\{1-r_i
  \pi^{-1}(Y_i,\u_i;\bb,g^*)\}}{d^*(\x_i)}\frac{C_i \wh D_i}{D_i^2}.
\ese
Now $N^{-1}\sumJ r_j K_{h}
  (\u_j-\u_i) - E(R_i\mid\u_i) f_{\U}(\u_i)=
  O_p\{h^m+(Nh^q)^{-1/2}\}$,
$\wh \Delta^2_{1j} - \Delta^2_{1j}=
O_p\{h^m+(Nh^p)^{-1/2}\}$,
and $\wh d^*(\x_j)- d^*(\x_j) =O_p\{h^m+(Nh^p)^{-1/2}\}$.
Then
\be \label{eq:t3}
T_3&=&\frac{1}{\sqrt{N}}\sumI \frac{\{1-r_i  \pi^{-1}(y_i,\u_i;\bb,g^*)\}C_i}{d^*(\x_i) D_i^2}\wh D_i \n \\
&=& \frac{1}{\sqrt{N}}\sumI \sumJ \frac{\{1-r_i  \pi^{-1}(y_i,\u_i;\bb,g^*)\}C_i}{d^*(\x_i) D_i^2} \frac{ r_j  K_{h} (\u_j-\u_i)\{\wh \Delta^2_{1j}/\wh d^* (\x_j) \}}{\sumJ r_j K_{h} (\u_j-\u_i)} \n\\
&=&N^{-3/2} \sumI \sumJ \frac{\{1-r_i  \pi^{-1}(y_i,\u_i;\bb,g^*)\}C_i r_j  K_{h} (\u_j-\u_i)}{d^*(\x_i) D_i^2}
 \left[\frac{\Delta^2_{1j}}{d^* (\x_j) E(R_i\mid\u_i) f_{\U}(\u_i)}
 \right.\n\\
&&\left.+ \frac{ \wh \Delta^2_{1j}}{d^* (\x_j) E(R_i\mid\u_i) f_{\U}(\u_i)} - \frac{ \Delta^2_{1j}\wh d^* (\x_j) \{N^{-1}\sumK r_k K_{h} (\u_k-\u_i)\}}{\{d^* (\x_j) E(R_i\mid\u_i) f_{\U}(\u_i)\}^2} \right]\n\\
&&+ O_p\{N^{1/2}h^{2m}+N^{-1/2}h^{-p}\}\n\\
&=&T_{31}+T_{32}-T_{33}+ O_p(N^{1/2}h^{2m}+N^{-1/2}h^{-p}),
\ee
where
\be
T_{31}&=&
N^{-3/2} \sumI \sumJ \frac{\{1-r_i  \pi^{-1}(y_i,\u_i;\bb,g^*)\}C_i r_j
  K_{h} (\u_j-\u_i)}{d^*(\x_i) D_i^2}
\frac{[E\{e^{-h(y_j)}\mid\x_j,1\}]^2}{d^* (\x_j) E(R_i\mid\u_i)
  f_{\U}(\u_i)} \label{eq:t31}\\
T_{32}&=&N^{-3/2} \sumI \sumJ \frac{\{1-r_i  \pi^{-1}(y_i,\u_i;\bb,g^*)\}C_i r_j  K_{h} (\u_j-\u_i)}{d^*(\x_i) D_i^2 d^* (\x_j) E(R_i\mid\u_i) f_{\U}(\u_i)} \left\{ \frac{\sumK r_k  K_{h}(\x_k-\x_j) e^{-h(y_k)} }{\sumK r_k K_{h} (\x_k-\x_j)} \right\}^2\n \\
T_{33}&=&N^{-5/2} \sumI \sumJ \sumK \frac{\{1-r_i  \pi^{-1}(y_i,\u_i;\bb,g^*)\}C_i r_j  K_{h} (\u_j-\u_i)r_k K_{h} (\u_k-\u_i)  \Delta^2_{1j}}{d^*(\x_i) D_i^2 \{d^* (\x_j) E(R_i\mid\u_i) f_{\U}(\u_i)\}^2}\n \\
&&\times \left [ \wh E\{e^{-h(y_j)}\mid\x_j,1\}
+e^{-g^* (\u_j)}\wh E\{e^{-2h(y_j)}\mid\x_j,1\} \right].\n
\ee
Now
\be
T_{33}&=&N^{-5/2} \sumI \sumJ \sumK \frac{\{1-r_i  \pi^{-1}(y_i,\u_i;\bb,g^*)\}C_i r_j  K_{h} (\u_j-\u_i)r_k K_{h} (\u_k-\u_i)  \Delta^2_{1j}}{d^*(\x_i) D_i^2 \{d^* (\x_j) E(R_i\mid\u_i) f_{\U}(\u_i)\}^2}\n \\
&&\times \left \{ \frac{\sumL r_l  K_{h}(\x_l-\x_j) e^{-h(y_l)} }{\sumL r_l K_{h} (\x_l-\x_j)}
+e^{-g^* (\u_j)} \frac{\sumL r_l  K_{h}(\x_l-\x_j) e^{-2h(y_l)} }{\sumL r_l K_{h} (\x_l-\x_j)} \right\}\n\\
&=&N^{-7/2} \sumI \sumJ \sumK \sumL \frac{\{1-r_i  \pi^{-1}(y_i,\u_i;\bb,g^*)\}C_i r_j  K_{h} (\u_j-\u_i)r_k K_{h} (\u_k-\u_i)  \Delta^2_{1j}}{d^*(\x_i) D_i^2 \{d^* (\x_j) E(R_i\mid\u_i) f_{\U}(\u_i)\}^2}\n \\
&&\times \left \{ \frac{r_l  K_{h}(\x_l-\x_j) e^{-h(y_l)}+e^{-g^* (\u_j)}r_l  K_{h}(\x_l-\x_j) e^{-2h(y_l)} }{N^{-1}\sumL r_l K_{h} (\x_l-\x_j)} \right\}\n\\
&=&N^{-7/2} \sumI \sumJ \sumK \sumL \frac{\{1-r_i  \pi^{-1}(y_i,\u_i;\bb,g^*)\}C_i r_j  K_{h} (\u_j-\u_i)r_k K_{h} (\u_k-\u_i)  \Delta^2_{1j}}{d^*(\x_i) D_i^2 \{d^* (\x_j) E(R_i\mid\u_i) f_{\U}(\u_i)\}^2}r_l \n \\
&&\times K_{h}(\x_l-\x_j) e^{-h(y_l)}\{1+e^{-g^* (\u_j)}e^{-h(y_l)}\}\left[\frac{2}{E(R_j\mid\x_j) f_{\X}(\x_j)}- \frac{N^{-1}\sumM r_m K_{h} (\x_m-\x_j)}{\{E(R_j\mid\x_j) f_{\X}(\x_j)\}^2}\right]\n \\
&&+O_p(N^{1/2}h^{2m}+N^{-1/2}h^{-p})\n\\
&=& 2T_{331} - T_{332} +O_p(N^{1/2} h^{2m} + N^{-1/2}h^{-p}),\n
\ee
where
\be
T_{331}&=&N^{-7/2} \sumI \sumJ \sumK \sumL \frac{\{1-r_i  \pi^{-1}(y_i,\u_i;\bb,g^*)\}C_i r_j  K_{h} (\u_j-\u_i)r_k K_{h} (\u_k-\u_i)  \Delta^2_{1j}}{d^*(\x_i) D_i^2 \{d^* (\x_j) E(R_i\mid\u_i) f_{\U}(\u_i)\}^2} \n\\
&&\times [ r_l  K_{h}(\x_l-\x_j) \{e^{-h(y_l)}+e^{-g^* (\u_j)}e^{-2h(y_l)}\}]\left[\frac{1}{E(R_j\mid\x_j) f_{\X}(\x_j)}\right],\label{eq:t331}\\
T_{332}&=&N^{-9/2} \sumI \sumJ \sumK \sumL \sumM \frac{\{1-r_i  \pi^{-1}(y_i,\u_i;\bb,g^*)\}C_i r_j  K_{h} (\u_j-\u_i)r_k K_{h} (\u_k-\u_i)  \Delta^2_{1j}}{d^*(\x_i) D_i^2 \{d^* (\x_j) E(R_i\mid\u_i) f_{\U}(\u_i)\}^2} \n\\
&&\times [ r_l  K_{h}(\x_l-\x_j) \{e^{-h(y_l)}+e^{-g^* (\u_j)}e^{-2h(y_l)}\}]\left[\frac{ r_m K_{h} (\x_m-\x_j)}{\{E(R_j\mid\x_j) f_{\X}(\x_j)\}^2}\right].\label{eq:t332}
\ee
Let $\wh \Delta_{4j}\equiv N^{-1}\sumK r_k  e^{-h(y_k)} K_{h}
(\x_k-\x_j), \wh \Delta_{5j}\equiv N^{-1}\sumK r_k K_{h} (\x_k-\x_j),
\Delta_{4j}\equiv E\{R_j e^{-h(Y_j)}\mid \x_j\}f_{\X}(\x_j) = \Delta_{1j} E(R_j \mid \x_j)f_{\X}(\x_j) $ and
$\Delta_{5j}\equiv E(R_j\mid\x_j) f_{\X}(\x_j)$.
We have  $\wh \Delta_{4j}- \Delta_{4j}= O_p\{h^m+(Nh^p)^{-1/2}\}$ and
$\wh \Delta_{5j}- \Delta_{5j}= O_p\{h^m+(Nh^p)^{-1/2}\}$.
Then
\be
T_{32}&=&N^{-3/2} \sumI \sumJ \frac{\{1-r_i  \pi^{-1}(y_i,\u_i;\bb,g^*)\}C_i r_j  K_{h} (\u_j-\u_i)}{d^*(\x_i) D_i^2 d^* (\x_j) E(R_i\mid\u_i) f_{\U}(\u_i)} \left\{ \frac{N^{-1}\sumK r_k  K_{h}(\x_k-\x_j) e^{-h(y_k)} }{N^{-1}\sumK r_k K_{h} (\x_k-\x_j)} \right\}^2\n \\
&=&N^{-3/2} \sumI \sumJ \frac{\{1-r_i  \pi^{-1}(y_i,\u_i;\bb,g^*)\}C_i r_j  K_{h} (\u_j-\u_i)}{d^*(\x_i) D_i^2 d^* (\x_j) E(R_i\mid\u_i) f_{\U}(\u_i)} \left( \frac{\wh \Delta_{4j}}{\wh \Delta_{5j}} \right)^2\n \\
&=&N^{-3/2} \sumI \sumJ \frac{\{1-r_i  \pi^{-1}(y_i,\u_i;\bb,g^*)\}C_i r_j  K_{h} (\u_j-\u_i)}{d^*(\x_i) D_i^2 d^* (\x_j) E(R_i\mid\u_i) f_{\U}(\u_i)} \left(\frac{ \Delta_{4j}^2}{\Delta_{5j}^2}  + \frac{2\Delta_{4j} \wh \Delta_{4j}}{\Delta_{5j}^2} - \frac{2\Delta_{4j}^2 \wh \Delta_{5j}}{\Delta_{5j}^3} \right)\n \\
&&+O_p(N^{1/2}h^{2m}+N^{-1/2}h^{-p})\n\\
&=&N^{-3/2} \sumI \sumJ \frac{\{1-r_i  \pi^{-1}(y_i,\u_i;\bb,g^*)\}C_i r_j  K_{h} (\u_j-\u_i)}{d^*(\x_i) D_i^2 d^* (\x_j) E(R_i\mid\u_i) f_{\U}(\u_i)} \frac{\{\Delta_{1j} E(R_j\mid \x_j)\}^2}{\{E(R_j\mid \x_j)\}^2}\n \\
&& +2N^{-5/2} \sumI \sumJ \sumK \frac{\{1-r_i  \pi^{-1}(y_i,\u_i;\bb,g^*)\}C_i r_j  K_{h} (\u_j-\u_i)r_k  e^{-h(y_k)} K_{h} (\x_k-\x_j)}{d^*(\x_i) D_i^2 d^* (\x_j) E(R_i\mid\u_i) f_{\U}(\u_i)}\n \\
&&\times \frac{\Delta_{1j} E(R_j\mid \x_j)}{\{E(R_j\mid \x_j)\}^2 f_{\X}(\x_j)}\n\\
&&- 2N^{-5/2} \sumI \sumJ \sumK \frac{\{1-r_i  \pi^{-1}(y_i,\u_i;\bb,g^*)\}C_i r_j  K_{h} (\u_j-\u_i)r_k K_{h} (\x_k-\x_j)}{d^*(\x_i) D_i^2 d^* (\x_j) E(R_i\mid\u_i) f_{\U}(\u_i)}\n \\
&&\times \frac{\{\Delta_{1j} E(R_j\mid \x_j)\}^2}{\{E(R_j\mid \x_j)\}^3 f_{\X}(\x_j)}+O_p(N^{1/2}h^{2m}+N^{-1/2}h^{-p})\n\\
&=& T_{321}+2T_{322}-2T_{323}+O_p(N^{1/2} h^{2m} + N^{-1/2}h^{-p}),\n
\ee
where
\be
T_{321}&=&N^{-3/2} \sumI \sumJ \frac{\{1-r_i
  \pi^{-1}(Y_i,\u_i;\bb,g^*)\}C_i r_j  K_{h} (\u_j-\u_i)}{d^*(\x_i) D_i^2
  d^* (\x_j) E(R_i\mid\u_i) f_{\U}(\u_i)} \Delta_{1j}^2,\label{eq:t321}\\
T_{322}&=&N^{-5/2} \sumI \sumJ \sumK \frac{\{1-r_i  \pi^{-1}(y_i,\u_i;\bb,g^*)\}C_i r_j  K_{h} (\u_j-\u_i)r_k  e^{-h(y_k)} K_{h} (\x_k-\x_j)}{d^*(\x_i) D_i^2 d^* (\x_j) E(R_i\mid\u_i) f_{\U}(\u_i)}\n \\
&&\times \frac{\Delta_{1j}}{E(R_j\mid \x_j) f_{\X}(\x_j)}, \label{eq:t322}\\
T_{323}&=&N^{-5/2} \sumI \sumJ \sumK \frac{\{1-r_i  \pi^{-1}(y_i,\u_i;\bb,g^*)\}C_i r_j  K_{h} (\u_j-\u_i)r_k K_{h} (\x_k-\x_j)}{d^*(\x_i) D_i^2 d^* (\x_j) E(R_i\mid\u_i) f_{\U}(\u_i)}\n \\
&&\times \frac{\Delta_{1j}^2}{E(R_j\mid \x_j) f_{\X}(\x_j)}. \label{eq:t323}
\ee
Combining the expansion of $T_{32}$ and $T_{33}$ into (\ref{eq:t3}),
we obtain the expansion of $T_3$.\\
Consider $T_2$,
\be\label{eq:t2}
T_2&=& \frac{1}{\sqrt{N}}\sumI \frac{\{1-r_i
  \pi^{-1}(Y_i,\u_i;\bb,g^*)\}}{d^*(\x_i) D_i}\wh
    C_i \n\\
&=& \frac{1}{\sqrt{N}}\sumI \frac{\{1-r_i
  \pi^{-1}(Y_i,\u_i;\bb,g^*)\}}{d^*(\x_i) D_i}\left(\wh E[\{\wh \Delta_{3i}\wh \Delta_{1i}/ \wh d^* (\X_i) \}\mid \u_i,1]\wh \Delta_{1i}\right.\n\\
&&\left.-\wh E[\{\wh \Delta^2_{1i}/ \wh d^* (\X_i) \}\mid \u_i,1]\wh \Delta_{3i}\right) \n\\
&=& \frac{1}{\sqrt{N}}\sumI \sumJ \frac{\{1-r_i
  \pi^{-1}(Y_i,\u_i;\bb,g^*)\} r_j K_h({\x_j-\x_i}) e^{-h(y_j)}}{d^*(\x_i) D_i}\n\\
&& \times \left(\frac{\wh E[\{\wh \Delta_{3i}\wh \Delta_{1i}/ \wh d^* (\X_i) \}\mid \u_i,1]
-\wh E[\{\wh \Delta^2_{1i}/ \wh d^* (\X_i) \}\mid \u_i,1]\h'_\bb(y_j;\bb)}{\sumJ r_j K_h({\x_j-\x_i})} \right) \n\\
&=& N^{-5/2} \sumI \sumJ \sumK \frac{\{1-r_i
  \pi^{-1}(Y_i,\u_i;\bb,g^*)\} r_j K_h({\x_j-\x_i}) e^{-h(y_j)} r_k K_h({\u_k-\u_i})}{d^*(\x_i) D_i}\n\\
&& \times \left[\frac{\wh \Delta_{3k}\wh \Delta_{1k}
-\wh \Delta^2_{1k}\h'_\bb(y_j;\bb)}{\wh d^* (\x_k) \{N^{-1}\sumK r_k K_h({\u_k-\u_i})\} \{N^{-1}\sumJ r_j K_h({\x_j-\x_i})\}} \right] \n\\
&=& N^{-5/2} \sumI \sumJ \sumK \frac{\{1-r_i
  \pi^{-1}(Y_i,\u_i;\bb,g^*)\} r_j K_h({\x_j-\x_i}) e^{-h(y_j)} r_k K_h({\u_k-\u_i})}{d^*(\x_i) D_i} \frac{\wh T_{j,k}}{\wh V_{i,k}}\n\\
&=&T_{21}+T_{22}-T_{23}
+O_p(N^{1/2}h^{2m}+N^{-1/2}h^{-p}),
\ee
where
\be
T_{21}&=& N^{-5/2} \sumI \sumJ \sumK \frac{\{1-r_i
  \pi^{-1}(Y_i,\u_i;\bb,g^*)\} r_j K_h({\x_j-\x_i}) e^{-h(y_j)} r_k K_h({\u_k-\u_i})}{d^*(\x_i) D_i}\n\\
  &&\times \frac{T_{j,k}}{V_{i,k}},\label{eq:t21} \\
T_{22}&=&N^{-5/2} \sumI \sumJ \sumK \frac{\{1-r_i
  \pi^{-1}(Y_i,\u_i;\bb,g^*)\} r_j K_h({\x_j-\x_i}) e^{-h(y_j)} r_k K_h({\u_k-\u_i})}{d^*(\x_i) D_i} \frac{\wh
    T_{j,k}}{V_{i,k}},\n\\
T_{23}&=&N^{-5/2} \sumI \sumJ \sumK \frac{\{1-r_i
  \pi^{-1}(Y_i,\u_i;\bb,g^*)\} r_j K_h({\x_j-\x_i}) e^{-h(y_j)} r_k K_h({\u_k-\u_i})}{d^*(\x_i) D_i}\frac{T_{j,k}\wh V_{i,k}}{V_{i,k}^2}.\n
\ee
We have
\be
T_{22}&=& N^{-5/2} \sumI \sumJ \sumK \frac{\{1-r_i
  \pi^{-1}(Y_i,\u_i;\bb,g^*)\} r_j K_h({\x_j-\x_i}) e^{-h(y_j)} r_k K_h({\u_k-\u_i})}{d^*(\x_i) D_i}\frac{\wh T_{j,k}}{V_{i,k}}\n\\
&=& N^{-5/2} \sumI \sumJ \sumK \frac{\{1-r_i
  \pi^{-1}(Y_i,\u_i;\bb,g^*)\} r_j K_h({\x_j-\x_i}) e^{-h(y_j)} r_k K_h({\u_k-\u_i})}{d^*(\x_i) D_i V_{i,k}}\n\\
&&\times \{\wh \Delta_{3k}\wh \Delta_{1k}
-\wh \Delta^2_{1k}\h'_\bb(y_j;\bb)\}\n\\
&=& N^{-5/2} \sumI \sumJ \sumK \frac{\{1-r_i
  \pi^{-1}(Y_i,\u_i;\bb,g^*)\} r_j K_h({\x_j-\x_i}) e^{-h(y_j)} r_k K_h({\u_k-\u_i})}{d^*(\x_i) D_i V_{i,k}}\wh \Delta_{3k}\wh \Delta_{1k}\n\\
&&-N^{-5/2} \sumI \sumJ \sumK \frac{\{1-r_i
  \pi^{-1}(Y_i,\u_i;\bb,g^*)\} r_j K_h({\x_j-\x_i}) e^{-h(y_j)} r_k K_h({\u_k-\u_i})}{d^*(\x_i) D_i V_{i,k}}\n\\
  &&\times \wh \Delta^2_{1k}\h'_\bb(y_j;\bb)\n\\
&=&T_{221}-T_{222},\n
\ee
where
\bse
T_{221}
&=& N^{-5/2} \sumI \sumJ \sumK \frac{\{1-r_i
  \pi^{-1}(Y_i,\u_i;\bb,g^*)\} r_j K_h({\x_j-\x_i}) e^{-h(y_j)} r_k K_h({\u_k-\u_i})}{d^*(\x_i) D_i V_{i,k}}\n\\
&&\times \left[\left\{\frac{\sumL r_l  K_{h}(\x_l-\x_k) e^{-h(y_l)} }{\sumL r_l K_{h} (\x_l-\x_k)}\right\}\left\{\frac{\sumM r_m K_{h}(\x_m-\x_k) e^{-h(y_m)}\h'_\bb(y_m;\bb) }{\sumM r_m K_{h} (\x_m-\x_k)}\right\}\right],\n\\
T_{222}&=&N^{-5/2} \sumI \sumJ \sumK \frac{\{1-r_i
  \pi^{-1}(Y_i,\u_i;\bb,g^*)\} r_j K_h({\x_j-\x_i}) e^{-h(y_j)} r_k K_h({\u_k-\u_i})}{d^*(\x_i) D_i V_{i,k}}\n\\
&&\times\left[\left\{\frac{\sumL r_l  K_{h}(\x_l-\x_k) e^{-h(y_l)} }{\sumL r_l K_{h} (\x_l-\x_k)}\right\}^2 \h'_\bb(y_j;\bb)\right].
\ese
Now
\be
T_{221}&=& N^{-5/2} \sumI \sumJ \sumK \frac{\{1-r_i
  \pi^{-1}(Y_i,\u_i;\bb,g^*)\} r_j K_h({\x_j-\x_i}) e^{-h(y_j)} r_k K_h({\u_k-\u_i})}{d^*(\x_i) D_i V_{i,k}}\n\\
&&\times \left[\left\{\frac{\sumL r_l  K_{h}(\x_l-\x_k) e^{-h(y_l)} }{\sumL r_l K_{h} (\x_l-\x_k)}\right\}\left\{\frac{\sumM r_m K_{h}(\x_m-\x_k) e^{-h(y_m)}\h'_\bb(y_m;\bb) }{\sumM r_m K_{h} (\x_m-\x_k)}\right\}\right]\n\\
&=& N^{-9/2} \sumI \sumJ \sumK \sumL \sumM \frac{\{1-r_i
  \pi^{-1}(Y_i,\u_i;\bb,g^*)\} r_j K_h({\x_j-\x_i}) e^{-h(y_j)} r_k K_h({\u_k-\u_i})}{d^*(\x_i) D_i V_{i,k}}\n\\
&&\times \left\{r_l  K_{h}(\x_l-\x_k) e^{-h(y_l)}r_m K_{h}(\x_m-\x_k) e^{-h(y_m)}\h'_\bb(y_m;\bb)\right\}\n\\
&&\times \left[\frac{2}{\{E(R_k\mid\x_k) f_{\X}(\x_k)\}^2}-\frac{N^{-2}\sumO r_o K_{h} (\x_o-\x_k) \sumS r_s K_{h} (\x_s-\x_k)}{\{E(R_k\mid\x_k) f_{\X}(\x_k)\}^4}\right]\n\\
&&+ O_p(N^{1/2}h^{2m}+N^{-1/2}h^{-p})\n\\
&=&2T_{2211}-T_{2212} +O_p(N^{1/2} h^{2m} + N^{-1/2}h^{-p}),\n
\ee
where
\be
T_{2211}&=&N^{-9/2} \sumI \sumJ \sumK \sumL \sumM \frac{\{1-r_i
  \pi^{-1}(Y_i,\u_i;\bb,g^*)\} r_j K_h({\x_j-\x_i}) e^{-h(y_j)} r_k K_h({\u_k-\u_i})}{d^*(\x_i) D_i V_{i,k}}\n\\
&&\times \left\{r_l  K_{h}(\x_l-\x_k) e^{-h(y_l)}r_m K_{h}(\x_m-\x_k) e^{-h(y_m)}\h'_\bb(y_m;\bb)\right\}\frac{1}{\{E(R_k\mid\x_k)
    f_{\X}(\x_k)\}^2},\label{eq:t2211}\\
T_{2212}&=&N^{-9/2} \sumI \sumJ \sumK \sumL \sumM \frac{\{1-r_i
  \pi^{-1}(Y_i,\u_i;\bb,g^*)\} r_j K_h({\x_j-\x_i}) e^{-h(y_j)} r_k K_h({\u_k-\u_i})}{d^*(\x_i) D_i V_{i,k}}\n\\
&&\times \left\{r_l  K_{h}(\x_l-\x_k) e^{-h(y_l)}r_m K_{h}(\x_m-\x_k) e^{-h(y_m)}\h'_\bb(y_m;\bb)\right\}\n\\
&&\times \frac{N^{-2}\sumO r_o K_{h} (\x_o-\x_k) \sumS r_s K_{h} (\x_s-\x_k)}{\{E(R_k\mid\x_k) f_{\X}(\x_k)\}^4}. \label{eq:t2212}
\ee
and
\be \label{eq:t222}
T_{222}&=& N^{-5/2} \sumI \sumJ \sumK \frac{\{1-r_i
  \pi^{-1}(Y_i,\u_i;\bb,g^*)\} r_j K_h({\x_j-\x_i}) e^{-h(y_j)} r_k K_h({\u_k-\u_i})}{d^*(\x_i) D_i V_{i,k}}\n\\
&&\times \left[\left\{\frac{\sumL r_l  K_{h}(\x_l-\x_k) e^{-h(y_l)} }{\sumL r_l K_{h} (\x_l-\x_k)}\right\}^2 \h'_\bb(y_j;\bb)\right]\n\\
&=& N^{-5/2} \sumI \sumJ \sumK \frac{\{1-r_i
  \pi^{-1}(Y_i,\u_i;\bb,g^*)\} r_j K_h({\x_j-\x_i}) e^{-h(y_j)} r_k K_h({\u_k-\u_i})\h'_\bb(y_j;\bb)}{d^*(\x_i) D_i V_{i,k}}\n\\
&&\times \frac{\{\Delta_{1k} E(R_k\mid \x_k)\}^2}{\{E(R_k \mid \x_k)\}^2}\n\\
&&+ 2N^{-7/2} \sumI \sumJ \sumK \sumL \frac{\{1-r_i
  \pi^{-1}(Y_i,\u_i;\bb,g^*)\} r_j K_h({\x_j-\x_i}) e^{-h(y_j)} r_k K_h({\u_k-\u_i})}{d^*(\x_i) D_i V_{i,k}}\n\\
&&\times \frac{ \h'_\bb(y_j;\bb)r_l  K_{h}(\x_l-\x_k) e^{-h(y_l)} \Delta_{1k} E(R_k\mid \x_k) }{\{E(R_k \mid \x_k)\}^2   f_{\X}(\x_k)}\n\\
&&- 2 N^{-7/2} \sumI \sumJ \sumK \sumL \frac{\{1-r_i
  \pi^{-1}(Y_i,\u_i;\bb,g^*)\} r_j K_h({\x_j-\x_i}) e^{-h(y_j)} r_k K_h({\u_k-\u_i})}{d^*(\x_i) D_i V_{i,k}}\n\\
&&\times \frac{\h'_\bb(y_j;\bb)r_l K_{h}(\x_l-\x_k) \{\Delta_{1k} E(R_k\mid \x_k)\}^2 }{\{E(R_k \mid \x_k)\}^3 f_{\X}(\x_k)}+ O_p(N^{1/2}h^{2m}+N^{-1/2}h^{-p})\n\\
&=&T_{2221}+ 2T_{2222}-2T_{2223}+O_p(N^{1/2} h^{2m} + N^{-1/2}h^{-p}),\n
\ee
where
\be
T_{2221}
&=&N^{-5/2} \sumI \sumJ \sumK \frac{\{1-r_i
  \pi^{-1}(Y_i,\u_i;\bb,g^*)\} r_j K_h({\x_j-\x_i}) e^{-h(y_j)} r_k K_h({\u_k-\u_i})\h'_\bb(y_j;\bb)}{d^*(\x_i) D_i V_{i,k}}\n\\
  &&\times \Delta_{1k}^2,\label{eq:t2221}\\
T_{2222}&=&N^{-7/2} \sumI \sumJ \sumK \sumL \frac{\{1-r_i
  \pi^{-1}(Y_i,\u_i;\bb,g^*)\} r_j K_h({\x_j-\x_i}) e^{-h(y_j)} r_k K_h({\u_k-\u_i})}{d^*(\x_i) D_i V_{i,k}}\n\\
&&\times \frac{\h'_\bb(y_j;\bb) r_l  K_{h}(\x_l-\x_k) e^{-h(y_l)} \Delta_{1k}  }{E(R_k \mid \x_k) f_{\X}(\x_k)}, \label{eq:t2222}\\
T_{2223}&=&  N^{-7/2} \sumI \sumJ \sumK \sumL \frac{\{1-r_i
  \pi^{-1}(Y_i,\u_i;\bb,g^*)\} r_j K_h({\x_j-\x_i}) e^{-h(y_j)} r_k K_h({\u_k-\u_i})}{d^*(\x_i) D_i V_{i,k}}\n\\
&&\times \frac{\h'_\bb(y_j;\bb)r_l K_{h}(\x_l-\x_k) \Delta_{1k} ^2
}{E(R_k \mid \x_k) f_{\X}(\x_k)}. \label{eq:t2223}
\ee

Combining the expansion of $T_{221}$ and $T_{222}$ leads to an
expansion of $T_{22}$.\\
Further,
\be \label{eq:t23}
T_{23}&=& N^{-5/2} \sumI \sumJ \sumK \frac{\{1-r_i
  \pi^{-1}(Y_i,\u_i;\bb,g^*)\} r_j K_h({\x_j-\x_i}) e^{-h(y_j)} r_k K_h({\u_k-\u_i})}{d^*(\x_i) D_i}\frac{ T_{j,k} \wh V_{i,k}}{V_{i,k}^2}\n\\
&=& N^{-9/2} \sumI \sumJ \sumK \sumL \sumM \frac{\{1-r_i
  \pi^{-1}(Y_i,\u_i;\bb,g^*)\} r_j K_h({\x_j-\x_i}) e^{-h(y_j)} r_k K_h({\u_k-\u_i})}{d^*(\x_i) D_i V_{i,k}^2}\n \\
&& \times \{ T_{j,k} r_l K_h({\u_l-\u_i}) r_m K_h({\x_m-\x_i})\} [\wh E\{e^{-h(Y_k)}\mid\x_k,1\}
+e^{-g^* (\u_k)}\wh E\{e^{-2h(Y_k)}\mid\x_k,1\}]\n\\
&=& N^{-9/2} \sumI \sumJ \sumK \sumL \sumM \frac{\{1-r_i
  \pi^{-1}(Y_i,\u_i;\bb,g^*)\} r_j K_h({\x_j-\x_i}) e^{-h(y_j)} r_k K_h({\u_k-\u_i}) }{d^*(\x_i) D_i V_{i,k}^2}\n \\
&& \times \{T_{j,k} r_l K_h({\u_l-\u_i}) r_m K_h({\x_m-\x_i})\} \left[\frac {\sumO r_o K_h({\x_o-\x_k}) e^{-h(y_o)} \{1+e^{-g^* (\u_k)} e^{-h(y_o)}\} }{\sumO r_o K_h({\x_o-\x_k})}\right]\n\\
&=& N^{-11/2} \sumI \sumJ \sumK \sumL \sumM \sumO \frac{\{1-r_i
  \pi^{-1}(Y_i,\u_i;\bb,g^*)\} r_j K_h({\x_j-\x_i}) e^{-h(y_j)} r_k K_h({\u_k-\u_i})}{d^*(\x_i) D_i V_{i,k}^2}\n \\
&& \times \left[T_{j,k}r_l K_h({\u_l-\u_i}) r_m K_h({\x_m-\x_i}) r_o K_h({\x_o-\x_k}) e^{-h(y_o)} \{1+e^{-g^* (\u_k)} e^{-h(y_o)}\}\right] \n\\
&& \times \left [ \frac{2}{\{E(R_k\mid\x_k) f_{\X}(\x_k)\}}-\frac{N^{-1} \sumS r_s K_h({\x_s-\x_k})}{\{E(R_k\mid\x_k) f_{\X}(\x_k)\}^2} \right ]+ O_p(N^{1/2}h^{2m}+N^{-1/2}h^{-p})\n\\
&=&2T_{231}-T_{232}+O_p(N^{1/2} h^{2m} + N^{-1/2}h^{-p}),\n
\ee
where
\be
T_{231}&=&N^{-11/2} \sumI \sumJ \sumK \sumL \sumM \sumO \frac{\{1-r_i
  \pi^{-1}(Y_i,\u_i;\bb,g^*)\} r_j K_h({\x_j-\x_i}) e^{-h(y_j)} r_k K_h({\u_k-\u_i}) }{d^*(\x_i) D_i V_{i,k}^2}\n \\
&& \times \left[T_{j,k}r_l K_h({\u_l-\u_i}) r_m K_h({\x_m-\x_i}) r_o K_h({\x_o-\x_k}) e^{-h(y_o)} \{1+e^{-g^* (\u_k)} e^{-h(y_o)}\}\right]\n\\
&& \times \left [ \frac{1}{\{E(R_k\mid\x_k)
    f_{\X}(\x_k)\}} \right ],\label{eq:t231}\\
T_{232}&=&N^{-11/2} \sumI \sumJ \sumK \sumL \sumM \sumO \frac{\{1-r_i
  \pi^{-1}(Y_i,\u_i;\bb,g^*)\} r_j K_h({\x_j-\x_i}) e^{-h(y_j)} r_k K_h({\u_k-\u_i})}{d^*(\x_i) D_i V_{i,k}^2}\n \\
&& \times \left[T_{j,k} r_l K_h({\u_l-\u_i}) r_m K_h({\x_m-\x_i}) r_o K_h({\x_o-\x_k}) e^{-h(y_o)} \{1+e^{-g^* (\u_k)} e^{-h(y_o)}\}\right] \n\\
&& \times \left [ \frac{N^{-1} \sumS r_s K_h({\x_s-\x_k})}{\{E(R_k\mid\x_k) f_{\X}(\x_k)\}^2} \right ]. \label{eq:t232}
\ee

Combining the expansion of $T_{22}$ and $T_{23}$ into (\ref{eq:t2})
leads to an expansion
of $T_2$.
Further combining
the results of $T_2$, $T_3$ into (\ref{eq:main2}), together with the
expansion in (\ref{eq:main3}) yields the final leading order expansion
of the estimated efficient score in (\ref{eq:main}).\\
Putting all the terms together we have the following,
\be \label{eq:final}
&&\frac{1}{\sqrt{N}}\sumI\wh\S^*\eff(\x_i,r_i,r_i y_i,\bb)\n\\
&=& \frac{1}{\sqrt{N}}\sumI \frac{A_i}{d^*(\x_i)} +T_0-T_1+O_p(N^{1/2}h^{2m}+N^{-1/2}h^{-p})\n\\
&=& \frac{1}{\sqrt{N}}\sumI \frac{A_i}{d^*(\x_i)} +\frac{1}{\sqrt{N}}\sumI \frac{A_i}{d^*(\x_i)}+T_2-T_3-T_1+O_p(N^{1/2}h^{2m}+N^{-1/2}h^{-p})\n\\
&=& \frac{2}{\sqrt{N}}\sumI \frac{A_i}{d^*(\x_i)} +(T_{21}+T_{22}-T_{23})-(T_{31}+T_{32}-T_{33})-(2T_{11}-T_{12})\n\\
&&+O_p(N^{1/2}h^{2m}+N^{-1/2}h^{-p})\n\\
&=& \frac{2}{\sqrt{N}}\sumI \frac{A_i}{d^*(\x_i)} +T_{21}+(T_{221}-T_{222})-(2T_{231}-T_{232})-T_{31}-(T_{321}+2T_{322}-2T_{323})\n\\
&&+(2T_{331}-T_{332})-2T_{11}+T_{12}+O_p(N^{1/2}h^{2m}+N^{-1/2}h^{-p}).\n\\
&=& \frac{2}{\sqrt{N}}\sumI \frac{A_i}{d^*(\x_i)} +T_{21}+2T_{2211}-T_{2212}-T_{2221}-2T_{2222}+2T_{2223}-2T_{231}+T_{232}-T_{31}\n\\
&&-T_{321}-2T_{322}+2T_{323}+2T_{331}-T_{332}-2T_{11}+T_{12}+O_p(N^{1/2}h^{2m}+N^{-1/2}h^{-p}),
\ee
where $T_{11}, T_{12}$ are defined in (\ref{eq:t11}), (\ref{eq:t12}),
$T_{31}$ is defined in  (\ref{eq:t31}), $T_{331}, T_{332}$ are
defined in
 (\ref{eq:t331}) and (\ref{eq:t332}), $T_{321}, T_{322}, T_{323}$ are defined in
(\ref{eq:t321}), (\ref{eq:t322}), (\ref{eq:t323}),  $T_{21}$ is
defined in (\ref{eq:t21}),
$T_{2211}, T_{2212}$ are in  (\ref{eq:t2211}),  (\ref{eq:t2212}) , $T_{2221},
T_{2222}, T_{2223}$ are in (\ref{eq:t2221}), (\ref{eq:t2222}), (\ref{eq:t2223}),
 and $T_{231},
T_{232}$ are defined in (\ref{eq:t231}),  (\ref{eq:t232}).

Using U-statistics property we can rewrite the term $T_{232}$ as follows:
\be \label{ustat:t232}
&&T_{232}\n\\
&=& N^{-13/2} \sumI \sumJ \sumK \sumL \sumM \sumO \sumS \frac{\{1-r_i
  \pi^{-1}(Y_i,\u_i;\bb,g^*)\} r_j K_h({\x_j-\x_i}) e^{-h(y_j)} r_k K_h({\u_k-\u_i}) }{d^*(\x_i) D_i V_{i,k}^2}\n \\
&& \times \left[T_{j,k} r_l K_h({\u_l-\u_i}) r_m K_h({\x_m-\x_i}) r_o K_h({\x_o-\x_k})  \{e^{-h(y_o)}+e^{-g^* (\u_k)} e^{-2h(y_o)}\}\right] \n\\
&& \times \left [ \frac{ r_s K_h({\x_s-\x_k})}{\{E(R_k\mid\x_k) f_{\X}(\x_k)\}^2} \right ]\n\\
&=& N^{-10/2} \sumI \sumJ \sumK \sumL \sumM  \frac{\{1-r_i
  \pi^{-1}(Y_i,\u_i;\bb,g^*)\} r_j K_h({\x_j-\x_i}) e^{-h(y_j)} r_k K_h({\u_k-\u_i}) T_{j,k}}{d^*(\x_i) D_i V_{i,k}^2\{E(R_k\mid\x_k) f_{\X}(\x_k)\}^2}\n \\
&& \times \left\{ r_l K_h({\u_l-\u_i}) r_m K_h({\x_m-\x_i}) \right\} \n\\
&& \times N^{-3/2} \sumO \sumS  \left [r_o K_h({\x_o-\x_k})  \{e^{-h(y_o)}+e^{-g^* (\u_k)} e^{-2h(y_o)}\} r_s K_h({\x_s-\x_k}) \right ]\n\\
&=& N^{-10/2} \sumI \sumJ \sumK \sumL \sumM  \frac{\{1-r_i
  \pi^{-1}(Y_i,\u_i;\bb,g^*)\} r_j K_h({\x_j-\x_i}) e^{-h(y_j)} r_k K_h({\u_k-\u_i}) T_{j,k}}{d^*(\x_i) D_i V_{i,k}^2\{E(R_k\mid\x_k) f_{\X}(\x_k)\}^2}\n \\
&& \times \left\{ r_l K_h({\u_l-\u_i}) r_m K_h({\x_m-\x_i}) \right\} \n\\
&& \times N^{-1/2} \sumO E \left [r_o K_h({\x_o-\x_k})  \{e^{-h(y_o)}+e^{-g^* (\u_k)} e^{-2h(y_o)}\} R_s K_h({\X_s-\x_k}) \mid \x_o, r_o, r_o y_o\right ]\n\\
&&+ N^{-10/2} \sumI \sumJ \sumK \sumL \sumM  \frac{\{1-r_i
  \pi^{-1}(Y_i,\u_i;\bb,g^*)\} r_j K_h({\x_j-\x_i}) e^{-h(y_j)} r_k K_h({\u_k-\u_i}) T_{j,k}}{d^*(\x_i) D_i V_{i,k}^2\{E(R_k\mid\x_k) f_{\X}(\x_k)\}^2}\n \\
&& \times \left\{ r_l K_h({\u_l-\u_i}) r_m K_h({\x_m-\x_i}) \right\} \n\\
&& \times N^{-1/2} \sumS E \left [R_o K_h({\X_o-\x_k})  \{e^{-h(Y_o)}+e^{-g^* (\u_k)} e^{-2h(Y_o)}\} r_s K_h({\x_s-\x_k}) \mid \x_s, r_s, r_s y_s\right ]\n\\
&&- N^{-10/2} \sumI \sumJ \sumK \sumL \sumM  \frac{\{1-r_i
  \pi^{-1}(Y_i,\u_i;\bb,g^*)\} r_j K_h({\x_j-\x_i}) e^{-h(y_j)} r_k K_h({\u_k-\u_i}) T_{j,k}}{d^*(\x_i) D_i V_{i,k}^2\{E(R_k\mid\x_k) f_{\X}(\x_k)\}^2}\n \\
&& \times \left\{ r_l K_h({\u_l-\u_i}) r_m K_h({\x_m-\x_i}) \right\} \n\\
&& \times N^{1/2}  E \left [R_o K_h({\X_o-\x_k})  \{e^{-h(Y_o)}+e^{-g^* (\u_k)} e^{-2h(Y_o)}\} R_s K_h({\X_s-\x_k})  \right ]+O_p(N^{-1/2})\n\\
&=& T_{232A} +  T_{232B}-  T_{232C}+O_p(N^{-1/2}).\n
\ee
Here,
\be \label{ustat:t232a}
&&T_{232A}\n\\
&=&N^{-10/2} \sumI \sumJ \sumK \sumL \sumM  \frac{\{1-r_i
  \pi^{-1}(Y_i,\u_i;\bb,g^*)\} r_j K_h({\x_j-\x_i}) e^{-h(y_j)} r_k K_h({\u_k-\u_i}) T_{j,k}}{d^*(\x_i) D_i V_{i,k}^2\{E(R_k\mid\x_k) f_{\X}(\x_k)\}^2}\n \\
&& \times \left\{ r_l K_h({\u_l-\u_i}) r_m K_h({\x_m-\x_i}) \right\} \n\\
&& \times N^{-1/2} \sumO
 E \left [r_o K_h({\x_o-\x_k})  \{e^{-h(y_o)}+e^{-g^* (\u_k)} e^{-2h(y_o)}\} R_s K_h({\X_s-\x_k}) \mid \x_o, r_o, r_o y_o\right ]\n\\
&=&N^{-10/2} \sumI \sumJ \sumK \sumL \sumM  \frac{\{1-r_i
  \pi^{-1}(Y_i,\u_i;\bb,g^*)\} r_j K_h({\x_j-\x_i}) e^{-h(y_j)} r_k K_h({\u_k-\u_i}) T_{j,k}}{d^*(\x_i) D_i V_{i,k}^2\{E(R_k\mid\x_k) f_{\X}(\x_k)\}^2}\n \\
&& \times \left\{ r_l K_h({\u_l-\u_i}) r_m K_h({\x_m-\x_i}) \right\} \n\\
&& \times N^{-1/2} \sumO r_o K_h({\x_o-\x_k}) \{e^{-h(y_o)}+e^{-g^* (\u_k)} e^{-2h(y_o)}\} E \left \{ R_s K_h({\X_s-\x_k}) \right \}\n\\
&=&N^{-10/2} \sumI \sumJ \sumK \sumL \sumM  \frac{\{1-r_i
  \pi^{-1}(Y_i,\u_i;\bb,g^*)\} r_j K_h({\x_j-\x_i}) e^{-h(y_j)} r_k K_h({\u_k-\u_i}) T_{j,k}}{d^*(\x_i) D_i V_{i,k}^2\{E(R_k\mid\x_k) f_{\X}(\x_k)\}^2}\n \\
&& \times \left\{ r_l K_h({\u_l-\u_i}) r_m K_h({\x_m-\x_i}) \right\} \n\\
&& \times N^{-1/2} \sumO r_o K_h({\x_o-\x_k}) \{e^{-h(y_o)}+e^{-g^* (\u_k)} e^{-2h(y_o)}\}  E \left(R_k \mid \x_k\right) f_{\X}(\x_k)
+O_p(N^{1/2}h^m)\n\\
&=&N^{-11/2} \sumI \sumJ \sumK \sumL \sumM \sumO \frac{\{1-r_i
  \pi^{-1}(Y_i,\u_i;\bb,g^*)\} r_j K_h({\x_j-\x_i}) e^{-h(y_j)} r_k K_h({\u_k-\u_i}) }{d^*(\x_i) D_i V_{i,k}^2E(R_k\mid\x_k) f_{\X}(\x_k)}\n \\
&& \times \left\{T_{j,k} r_l K_h({\u_l-\u_i}) r_m K_h({\x_m-\x_i}) \right\}
r_o K_h({\x_o-\x_k}) \{e^{-h(y_o)}+e^{-g^* (\u_k)} e^{-2h(y_o)}\}
+O_p(N^{1/2}h^m)\n\\
&=&T_{231}+O_p(N^{1/2}h^m),
\ee
where we used the form of $T_{231}$ in (\ref{eq:t231}) in the last
equality above.

\be \label{ustat:t232b}
&&T_{232B}\n\\
&=&N^{-10/2} \sumI \sumJ \sumK \sumL \sumM  \frac{\{1-r_i
  \pi^{-1}(Y_i,\u_i;\bb,g^*)\} r_j K_h({\x_j-\x_i}) e^{-h(y_j)} r_k K_h({\u_k-\u_i}) T_{j,k}}{d^*(\x_i) D_i V_{i,k}^2\{E(R_k\mid\x_k) f_{\X}(\x_k)\}^2}\n \\
&& \times \left\{ r_l K_h({\u_l-\u_i}) r_m K_h({\x_m-\x_i}) \right\} \n\\
&& \times N^{-1/2} \sumS E \left [R_o K_h({\X_o-\x_k})  \{e^{-h(Y_o)}+e^{-g^* (\u_k)} e^{-2h(Y_o)}\} r_s K_h({\x_s-\x_k}) \mid \x_s, r_s, r_s y_s\right ]\n\\
&=&N^{-10/2} \sumI \sumJ \sumK \sumL \sumM  \frac{\{1-r_i
  \pi^{-1}(Y_i,\u_i;\bb,g^*)\} r_j K_h({\x_j-\x_i}) e^{-h(y_j)} r_k K_h({\u_k-\u_i}) T_{j,k}}{d^*(\x_i) D_i V_{i,k}^2\{E(R_k\mid\x_k) f_{\X}(\x_k)\}^2}\n \\
&& \times \left\{ r_l K_h({\u_l-\u_i}) r_m K_h({\x_m-\x_i}) \right\} \n\\
&& \times N^{-1/2} \sumS r_s K_h({\x_s-\x_k}) E \left [R_o K_h({\X_o-\x_k})  \{e^{-h(Y_o)}+e^{-g^* (\u_k)} e^{-2h(Y_o)}\}\right ]\n\\
&=&N^{-10/2} \sumI \sumJ \sumK \sumL \sumM  \frac{\{1-r_i
  \pi^{-1}(Y_i,\u_i;\bb,g^*)\} r_j K_h({\x_j-\x_i}) e^{-h(y_j)} r_k K_h({\u_k-\u_i}) T_{j,k}}{d^*(\x_i) D_i V_{i,k}^2\{E(R_k\mid\x_k) f_{\X}(\x_k)\}^2}\n \\
&& \times \left\{ r_l K_h({\u_l-\u_i}) r_m K_h({\x_m-\x_i}) \right\} \n\\
&& \times N^{-1/2} \sumS r_s K_h({\x_s-\x_k}) \pr(R_o=1)E \left [ K_h({\X_o-\x_k})  \{e^{-h(Y_o)}+e^{-g^* (\u_k)} e^{-2h(Y_o)}\}\mid R_o=1\right ]\n\\
&=&N^{-10/2} \sumI \sumJ \sumK \sumL \sumM  \frac{\{1-r_i
  \pi^{-1}(Y_i,\u_i;\bb,g^*)\} r_j K_h({\x_j-\x_i}) e^{-h(y_j)} r_k K_h({\u_k-\u_i}) T_{j,k}}{d^*(\x_i) D_i V_{i,k}^2\{E(R_k\mid\x_k) f_{\X}(\x_k)\}^2}\n \\
&& \times \left\{ r_l K_h({\u_l-\u_i}) r_m K_h({\x_m-\x_i}) E(R_k\mid\x_k) f_{\X}(\x_k) \right\} \n\\
&& \times N^{-1/2} \sumS r_s K_h({\x_s-\x_k}) E \left [\{e^{-h(Y_k)}+e^{-g^* (\u_k)} e^{-2h(Y_k)}\}\mid \x_k, R_k=1\right] +O_p(N^{1/2} h^m)\n\\
&=&N^{-11/2} \sumI \sumJ \sumK \sumL \sumM \sumS \frac{\{1-r_i
  \pi^{-1}(Y_i,\u_i;\bb,g^*)\} r_j K_h({\x_j-\x_i}) e^{-h(y_j)} r_k K_h({\u_k-\u_i})}{d^*(\x_i) D_i V_{i,k}^2 E(R_k\mid\x_k) f_{\X}(\x_k)}\n \\
&& \times \left\{ T_{j,k} r_l K_h({\u_l-\u_i}) r_m K_h({\x_m-\x_i}) r_s K_h({\x_s-\x_k}) d^*(\x_k)\right\} +O_p(N^{1/2} h^m)\n\\
&=&N^{-11/2} \sumI \sumJ \sumK \sumL \sumM \sumO \frac{\{1-r_i
  \pi^{-1}(Y_i,\u_i;\bb,g^*)\} r_j K_h({\x_j-\x_i}) e^{-h(y_j)} r_k K_h({\u_k-\u_i})}{d^*(\x_i) D_i V_{i,k}^2 E(R_k\mid\x_k) f_{\X}(\x_k)}\n \\
&& \times \left\{ T_{j,k} r_l K_h({\u_l-\u_i}) r_m K_h({\x_m-\x_i}) r_o K_h({\x_o-\x_k}) d^*(\x_k)\right\} +O_p(N^{1/2} h^m).
\ee
In the above derivation, we used the property that $\pr(R_o=1) f_{\X
  \mid R=1}( \x_k)=E(R_k\mid\x_k) f_{\X}(\x_k)$ in the second last
equality, and we incorporated the definition of $d^*(\x)$ in the last equality.

Consider $T_{232C}$, we have
\be \label{ustat:t232c}
&&T_{232C}\n\\
&=&N^{-10/2} \sumI \sumJ \sumK \sumL \sumM  \frac{\{1-r_i
  \pi^{-1}(Y_i,\u_i;\bb,g^*)\} r_j K_h({\x_j-\x_i}) e^{-h(y_j)} r_k K_h({\u_k-\u_i}) T_{j,k}}{d^*(\x_i) D_i V_{i,k}^2\{E(R_k\mid\x_k) f_{\X}(\x_k)\}^2}\n \\
&& \times \left\{ r_l K_h({\u_l-\u_i}) r_m K_h({\x_m-\x_i}) \right\} \n\\
&& \times N^{1/2} E \left [R_o K_h({\X_o-\x_k})  \{e^{-h(Y_o)}+e^{-g^* (\u_k)} e^{-2h(Y_o)}\} R_s K_h({\X_s-\x_k}) \right ]\n\\
&=&N^{-9/2} \sumI \sumJ \sumK \sumL \sumM  \frac{\{1-r_i
  \pi^{-1}(Y_i,\u_i;\bb,g^*)\} r_j K_h({\x_j-\x_i}) e^{-h(y_j)} r_k K_h({\u_k-\u_i}) T_{j,k}}{d^*(\x_i) D_i V_{i,k}^2\{E(R_k\mid\x_k) f_{\X}(\x_k)\}^2}\n \\
&& \times \left\{ r_l K_h({\u_l-\u_i}) r_m K_h({\x_m-\x_i}) \right\} \n\\
&& \times E \left\{R_s K_h({\X_s-\x_k}) \right\} E \left [R_o K_h({\X_o-\x_k})  \{e^{-h(Y_o)}+e^{-g^* (\u_k)} e^{-2h(Y_o)}\} \right ]\n\\
&=&N^{-9/2} \sumI \sumJ \sumK \sumL \sumM  \frac{\{1-r_i
  \pi^{-1}(Y_i,\u_i;\bb,g^*)\} r_j K_h({\x_j-\x_i}) e^{-h(y_j)} r_k K_h({\u_k-\u_i}) T_{j,k}}{d^*(\x_i) D_i V_{i,k}^2\{E(R_k\mid\x_k) f_{\X}(\x_k)\}^2}\n \\
&& \times \left\{ r_l K_h({\u_l-\u_i}) r_m K_h({\x_m-\x_i}) \right\}\left\{ E(R_k\mid\x_k) f_{\X}(\x_k)\right\} \left\{ E(R_k\mid\x_k) f_{\X}(\x_k) d^*(\x_k) \right\} +O_p(N^{1/2}h^m)\n\\
&=&N^{-9/2} \sumI \sumJ \sumK \sumL \sumM  \frac{\{1-r_i
  \pi^{-1}(Y_i,\u_i;\bb,g^*)\} r_j K_h({\x_j-\x_i}) e^{-h(y_j)} r_k K_h({\u_k-\u_i}) T_{j,k}}{d^*(\x_i) D_i V_{i,k}^2}\n \\
&& \times \left\{ r_l K_h({\u_l-\u_i}) r_m K_h({\x_m-\x_i}) d^*(\x_k)\right\}
+O_p(N^{1/2}h^m).
\ee

Consider now the term $T_{2212}$ in (\ref{eq:t2212}), using
U-statistics property we can rewrite it as
\be \label{ustat:t2212}
&&T_{2212}\n\\
&=& N^{-13/2} \sumI \sumJ \sumK \sumL \sumM \sumO \sumS \frac{\{1-r_i
  \pi^{-1}(y_i,\u_i;\bb,g^*)\} r_j K_h({\x_j-\x_i}) e^{-h(y_j)} r_k K_h({\u_k-\u_i})}{d^*(\x_i) D_i V_{i,k}}\n\\
&&\times \left\{r_l  K_{h}(\x_l-\x_k) e^{-h(y_l)}r_m K_{h}(\x_m-\x_k) e^{-h(y_m)}\h'_\bb(y_m;\bb)\right\}\n\\
&&\times \left[\frac{ r_o K_{h} (\x_o-\x_k)  r_s K_{h} (\x_s-\x_k)}{\{E(R_k\mid\x_k) f_{\X}(\x_k)\}^4}\right]\n\\
&=& N^{-10/2} \sumI \sumJ \sumK \sumL \sumM   \frac{\{1-r_i
  \pi^{-1}(y_i,\u_i;\bb,g^*)\} r_j K_h({\x_j-\x_i}) e^{-h(y_j)} r_k K_h({\u_k-\u_i})}{d^*(\x_i) D_i V_{i,k} \{E(R_k\mid\x_k) f_{\X}(\x_k)\}^4}\n\\
&&\times \left\{r_l  K_{h}(\x_l-\x_k) e^{-h(y_l)}r_m K_{h}(\x_m-\x_k) e^{-h(y_m)}\h'_\bb(y_m;\bb)\right\}\n\\
&&\times   N^{-1/2} \sumO E\left\{ r_o K_{h} (\x_o-\x_k)  R_s K_{h} (\X_s-\x_k)\mid \x_o,r_o,r_o y_o\right\}\n\\
&&+ N^{-10/2} \sumI \sumJ \sumK \sumL \sumM   \frac{\{1-r_i
  \pi^{-1}(y_i,\u_i;\bb,g^*)\} r_j K_h({\x_j-\x_i}) e^{-h(y_j)} r_k K_h({\u_k-\u_i})}{d^*(\x_i) D_i V_{i,k} \{E(R_k\mid\x_k) f_{\X}(\x_k)\}^4}\n\\
&&\times \left\{r_l  K_{h}(\x_l-\x_k) e^{-h(y_l)}r_m K_{h}(\x_m-\x_k) e^{-h(y_m)}\h'_\bb(y_m;\bb)\right\}\n\\
&&\times   N^{-1/2} \sumS E\left\{ R_o K_{h} (\X_o-\x_k)  r_s K_{h} (\x_s-\x_k)\mid \x_s,r_s,r_s y_s\right\}\n\\
&&- N^{-10/2} \sumI \sumJ \sumK \sumL \sumM   \frac{\{1-r_i
  \pi^{-1}(y_i,\u_i;\bb,g^*)\} r_j K_h({\x_j-\x_i}) e^{-h(y_j)} r_k K_h({\u_k-\u_i})}{d^*(\x_i) D_i V_{i,k} \{E(R_k\mid\x_k) f_{\X}(\x_k)\}^4}\n\\
&&\times \left\{r_l  K_{h}(\x_l-\x_k) e^{-h(y_l)}r_m K_{h}(\x_m-\x_k) e^{-h(y_m)}\h'_\bb(y_m;\bb)\right\}\n\\
&&\times   N^{1/2} E\left\{ R_o K_{h} (\X_o-\x_k)  R_s K_{h} (\X_s-\x_k)\right\}
+O_p(N^{-1/2})\n\\
&=& T_{2212A} +  T_{2212B}-  T_{2212C} +O_p(N^{-1/2}).\n
\ee

Consider the term $T_{2212A}$,
\be \label{ustat:t2212a}
&&T_{2212A}\n\\
&=&N^{-10/2} \sumI \sumJ \sumK \sumL \sumM   \frac{\{1-r_i
  \pi^{-1}(y_i,\u_i;\bb,g^*)\} r_j K_h({\x_j-\x_i}) e^{-h(y_j)} r_k K_h({\u_k-\u_i})}{d^*(\x_i) D_i V_{i,k} \{E(R_k\mid\x_k) f_{\X}(\x_k)\}^4}\n\\
&&\times \left\{r_l  K_{h}(\x_l-\x_k) e^{-h(y_l)}r_m K_{h}(\x_m-\x_k) e^{-h(y_m)}\h'_\bb(y_m;\bb)\right\}\n\\
&&\times   N^{-1/2} \sumO E\left\{ r_o K_{h} (\x_o-\x_k)  R_s K_{h} (\X_s-\x_k)\mid \x_o,r_o,r_o y_o\right\}\n\\
&=& N^{-11/2} \sumI \sumJ \sumK \sumL \sumM   \frac{\{1-r_i
  \pi^{-1}(y_i,\u_i;\bb,g^*)\} r_j K_h({\x_j-\x_i}) e^{-h(y_j)} r_k K_h({\u_k-\u_i})}{d^*(\x_i) D_i V_{i,k} \{E(R_k\mid\x_k) f_{\X}(\x_k)\}^4}\n\\
&&\times \left\{r_l  K_{h}(\x_l-\x_k) e^{-h(y_l)}r_m K_{h}(\x_m-\x_k) e^{-h(y_m)}\h'_\bb(y_m;\bb)\right\}\n\\
&&\times   \sumO r_o K_{h} (\x_o-\x_k) E(R_k\mid\x_k) f_{\X}(\x_k)
+O_p(N^{1/2}h^m)\n\\
&=& N^{-11/2} \sumI \sumJ \sumK \sumL \sumM \sumO  \frac{\{1-r_i
  \pi^{-1}(y_i,\u_i;\bb,g^*)\} r_j K_h({\x_j-\x_i}) e^{-h(y_j)} r_k K_h({\u_k-\u_i})}{d^*(\x_i) D_i V_{i,k} \{E(R_k\mid\x_k) f_{\X}(\x_k)\}^3}\n\\
&&\times \left\{r_l  K_{h}(\x_l-\x_k) e^{-h(y_l)}r_m K_{h}(\x_m-\x_k) e^{-h(y_m)}\h'_\bb(y_m;\bb) r_o K_{h} (\x_o-\x_k) \right\}\n\\
&&+O_p(N^{1/2}h^m).
\ee

Similarly the term $T_{2212B}$ can be written as,
\be \label{ustat:t2212b}
&&T_{2212B}\n\\
&=&N^{-10/2} \sumI \sumJ \sumK \sumL \sumM   \frac{\{1-r_i
  \pi^{-1}(y_i,\u_i;\bb,g^*)\} r_j K_h({\x_j-\x_i}) e^{-h(y_j)} r_k K_h({\u_k-\u_i})}{d^*(\x_i) D_i V_{i,k} \{E(R_k\mid\x_k) f_{\X}(\x_k)\}^4}\n\\
&&\times \left\{r_l  K_{h}(\x_l-\x_k) e^{-h(y_l)}r_m K_{h}(\x_m-\x_k) e^{-h(y_m)}\h'_\bb(y_m;\bb)\right\}\n\\
&&\times   N^{-1/2} \sumS E\left\{ R_o K_{h} (\X_o-\x_k)  r_s K_{h} (\x_s-\x_k)\mid \x_s,r_s,r_s y_s\right\}\n\\
&=& N^{-11/2} \sumI \sumJ \sumK \sumL \sumM \sumS  \frac{\{1-r_i
  \pi^{-1}(y_i,\u_i;\bb,g^*)\} r_j K_h({\x_j-\x_i}) e^{-h(y_j)} r_k K_h({\u_k-\u_i})}{d^*(\x_i) D_i V_{i,k} \{E(R_k\mid\x_k) f_{\X}(\x_k)\}^4}\n\\
&&\times \left\{r_l  K_{h}(\x_l-\x_k) e^{-h(y_l)}r_m K_{h}(\x_m-\x_k) e^{-h(y_m)}\h'_\bb(y_m;\bb) r_s K_{h} (\x_s-\x_k) E(R_k\mid\x_k) f_{\X}(\x_k)\right\}\n\\
&&+O_p(N^{1/2}h^m).\n\\
&=& N^{-11/2} \sumI \sumJ \sumK \sumL \sumM \sumO  \frac{\{1-r_i
  \pi^{-1}(y_i,\u_i;\bb,g^*)\} r_j K_h({\x_j-\x_i}) e^{-h(y_j)} r_k K_h({\u_k-\u_i})}{d^*(\x_i) D_i V_{i,k} \{E(R_k\mid\x_k) f_{\X}(\x_k)\}^3}\n\\
&&\times \left\{r_l  K_{h}(\x_l-\x_k) e^{-h(y_l)}r_m K_{h}(\x_m-\x_k) e^{-h(y_m)}\h'_\bb(y_m;\bb) r_o K_{h} (\x_o-\x_k) \right\}+O_p(N^{1/2}h^m)\n\\
&=& T_{2212A}+O_p(N^{1/2}h^m).\n
\ee
Term $T_{2212C}$ can be written as
\be \label{ustat:t2212c}
&&T_{2212C}\n\\
&=&N^{-10/2} \sumI \sumJ \sumK \sumL \sumM   \frac{\{1-r_i
  \pi^{-1}(y_i,\u_i;\bb,g^*)\} r_j K_h({\x_j-\x_i}) e^{-h(y_j)} r_k K_h({\u_k-\u_i})}{d^*(\x_i) D_i V_{i,k} \{E(R_k\mid\x_k) f_{\X}(\x_k)\}^4}\n\\
&&\times \left\{r_l  K_{h}(\x_l-\x_k) e^{-h(y_l)}r_m K_{h}(\x_m-\x_k) e^{-h(y_m)}\h'_\bb(y_m;\bb)\right\}\n\\
&&\times   N^{1/2}  E\left\{ R_o K_{h} (\X_o-\x_k)  R_s K_{h} (\X_s-\x_k)\right\}\n\\
&=&N^{-9/2} \sumI \sumJ \sumK \sumL \sumM   \frac{\{1-r_i
  \pi^{-1}(y_i,\u_i;\bb,g^*)\} r_j K_h({\x_j-\x_i}) e^{-h(y_j)} r_k K_h({\u_k-\u_i})}{d^*(\x_i) D_i V_{i,k} \{E(R_k\mid\x_k) f_{\X}(\x_k)\}^4}\n\\
&&\times \left\{r_l  K_{h}(\x_l-\x_k) e^{-h(y_l)}r_m K_{h}(\x_m-\x_k) e^{-h(y_m)}\h'_\bb(y_m;\bb)\right\}\n\\
&&\times  \left\{ E(R_k\mid\x_k) f_{\X}(\x_k)+O_p(h^m)\right\} \left\{ E(R_k\mid\x_k) f_{\X}(\x_k)+O_p(h^m)\right\}\n\\
&=&N^{-9/2} \sumI \sumJ \sumK \sumL \sumM   \frac{\{1-r_i
  \pi^{-1}(y_i,\u_i;\bb,g^*)\} r_j K_h({\x_j-\x_i}) e^{-h(y_j)} r_k K_h({\u_k-\u_i})}{d^*(\x_i) D_i V_{i,k} \{E(R_k\mid\x_k) f_{\X}(\x_k)\}^2}\n\\
&&\times \left\{r_l  K_{h}(\x_l-\x_k) e^{-h(y_l)}r_m K_{h}(\x_m-\x_k) e^{-h(y_m)}\h'_\bb(y_m;\bb)\right\}
+O_p(N^{1/2}h^m)\n\\
&=&T_{2211}+O_p(N^{1/2}h^m),\n
\ee
where we used the expression of $T_{2211}$ given in (\ref{eq:t2211})
in the last equality.

The term $T_{2212A}$  in (\ref{ustat:t2212a}) can further be written as
\be \label{ustat:t2212a_2}
&&T_{2212A}\n\\
&=& N^{-11/2} \sumI \sumJ \sumK \sumL \sumM \sumO  \frac{\{1-r_i
  \pi^{-1}(y_i,\u_i;\bb,g^*)\} r_j K_h({\x_j-\x_i}) e^{-h(y_j)} r_k K_h({\u_k-\u_i})}{d^*(\x_i) D_i V_{i,k} \{E(R_k\mid\x_k) f_{\X}(\x_k)\}^3}\n\\
&&\times \left\{r_l  K_{h}(\x_l-\x_k) e^{-h(y_l)}r_m K_{h}(\x_m-\x_k) e^{-h(y_m)}\h'_\bb(y_m;\bb) r_o K_{h} (\x_o-\x_k) \right\}+O_p(N^{1/2}h^m)\n\\
&=& N^{-8/2} \sumI \sumJ \sumK \sumL \frac{\{1-r_i
  \pi^{-1}(y_i,\u_i;\bb,g^*)\} r_j K_h({\x_j-\x_i}) e^{-h(y_j)} r_k K_h({\u_k-\u_i})}{d^*(\x_i) D_i V_{i,k} \{E(R_k\mid\x_k) f_{\X}(\x_k)\}^3}\n\\
&&\times \left\{r_l  K_{h}(\x_l-\x_k) e^{-h(y_l)}\right\}\n\\
&& \times N^{-1/2} \sumM  E \left\{ r_m K_{h}(\x_m-\x_k) e^{-h(y_m)}\h'_\bb(y_m;\bb) R_o K_{h} (\X_o-\x_k) \mid \x_m, r_m, r_m y_m \right\}\n\\
&&+ N^{-8/2} \sumI \sumJ \sumK \sumL \frac{\{1-r_i
  \pi^{-1}(y_i,\u_i;\bb,g^*)\} r_j K_h({\x_j-\x_i}) e^{-h(y_j)} r_k K_h({\u_k-\u_i})}{d^*(\x_i) D_i V_{i,k} \{E(R_k\mid\x_k) f_{\X}(\x_k)\}^3}\n\\
&&\times \left\{r_l  K_{h}(\x_l-\x_k) e^{-h(y_l)}\right\}\n\\
&& \times N^{-1/2} \sumO  E \left\{ R_m K_{h}(\X_m-\x_k) e^{-h(Y_m)}\h'_\bb(Y_m;\bb) r_o K_{h} (\x_o-\x_k) \mid \x_o, r_o, r_o y_o \right\}\n\\
&&- N^{-8/2} \sumI \sumJ \sumK \sumL \frac{\{1-r_i
  \pi^{-1}(y_i,\u_i;\bb,g^*)\} r_j K_h({\x_j-\x_i}) e^{-h(y_j)} r_k K_h({\u_k-\u_i})}{d^*(\x_i) D_i V_{i,k} \{E(R_k\mid\x_k) f_{\X}(\x_k)\}^3}\n\\
&&\times \left\{r_l  K_{h}(\x_l-\x_k) e^{-h(y_l)}\right\} N^{1/2}   E \left\{ R_m K_{h}(\X_m-\x_k) e^{-h(Y_m)}\h'_\bb(Y_m;\bb) R_o K_{h} (\X_o-\x_k)\right\}\n\\
&&+O_p(N^{-1/2})+O_p(N^{1/2}h^m)\n\\
&=& T_{2212AA} +  T_{2212AB}-  T_{2212AC} +O_p(N^{-1/2}) +O_p(N^{1/2}h^m).\n
\ee

Consider $T_{2212AA}$,
\be \label{ustat:t2212aa}
&& T_{2212AA}\n\\
&=& N^{-8/2} \sumI \sumJ \sumK \sumL \frac{\{1-r_i
  \pi^{-1}(y_i,\u_i;\bb,g^*)\} r_j K_h({\x_j-\x_i}) e^{-h(y_j)} r_k K_h({\u_k-\u_i})}{d^*(\x_i) D_i V_{i,k} \{E(R_k\mid\x_k) f_{\X}(\x_k)\}^3}\n\\
&&\times \left\{r_l  K_{h}(\x_l-\x_k) e^{-h(y_l)}\right\}\n\\
&& \times N^{-1/2} \sumM  E \left\{ r_m K_{h}(\x_m-\x_k) e^{-h(y_m)}\h'_\bb(y_m;\bb) R_o K_{h} (\X_o-\x_k) \mid \x_m, r_m, r_m y_m \right\}\n\\
&=& N^{-9/2} \sumI \sumJ \sumK \sumL   \sumM \frac{\{1-r_i
  \pi^{-1}(y_i,\u_i;\bb,g^*)\} r_j K_h({\x_j-\x_i}) e^{-h(y_j)} r_k K_h({\u_k-\u_i})}{d^*(\x_i) D_i V_{i,k} \{E(R_k\mid\x_k) f_{\X}(\x_k)\}^3}\n\\
&&\times \left\{r_l  K_{h}(\x_l-\x_k) e^{-h(y_l)}\right\} r_m K_{h}(\x_m-\x_k) e^{-h(y_m)}\h'_\bb(y_m;\bb) E \left\{  R_o K_{h} (\X_o-\x_k)\right\}\n\\
&=& N^{-9/2} \sumI \sumJ \sumK \sumL   \sumM \frac{\{1-r_i
  \pi^{-1}(y_i,\u_i;\bb,g^*)\} r_j K_h({\x_j-\x_i}) e^{-h(y_j)} r_k K_h({\u_k-\u_i})}{d^*(\x_i) D_i V_{i,k} \{E(R_k\mid\x_k) f_{\X}(\x_k)\}^2}\n\\
&&\times \left\{r_l  K_{h}(\x_l-\x_k) e^{-h(y_l)}\right\} r_m K_{h}(\x_m-\x_k) e^{-h(y_m)}\h'_\bb(y_m;\bb) +O_p(N^{1/2}h^m)\n\\
&=&T_{2211}+O_p(N^{1/2}h^m),\n
\ee
where we used the expression of $T_{2211}$ given in (\ref{eq:t2211})
in the last equality.\\

Consider $T_{2212AB}$,
\be \label{ustat:t2212ab}
&& T_{2212AB}\n\\
&=& N^{-8/2} \sumI \sumJ \sumK \sumL \frac{\{1-r_i
  \pi^{-1}(y_i,\u_i;\bb,g^*)\} r_j K_h({\x_j-\x_i}) e^{-h(y_j)} r_k K_h({\u_k-\u_i})\left\{r_l   e^{-h(y_l)}\right\} }{d^*(\x_i) D_i V_{i,k} \{E(R_k\mid\x_k) f_{\X}(\x_k)\}^3}\n\\
&&\times K_{h}(\x_l-\x_k)  N^{-1/2} \sumO  E \left\{ R_m K_{h}(\X_m-\x_k) e^{-h(Y_m)}\h'_\bb(Y_m;\bb) r_o K_{h} (\x_o-\x_k) \mid \x_o, r_o, r_o y_o \right\}\n\\
&=& N^{-9/2} \sumI \sumJ \sumK \sumL \sumO  \frac{\{1-r_i
  \pi^{-1}(y_i,\u_i;\bb,g^*)\} r_j K_h({\x_j-\x_i}) e^{-h(y_j)} r_k K_h({\u_k-\u_i})}{d^*(\x_i) D_i V_{i,k} \{E(R_k\mid\x_k) f_{\X}(\x_k)\}^3}\n\\
&&\times \left\{r_l  K_{h}(\x_l-\x_k) e^{-h(y_l)}\right\} r_o K_{h} (\x_o-\x_k) E \left\{ R_m K_{h}(\X_m-\x_k) e^{-h(Y_m)}\h'_\bb(Y_m;\bb)  \right\}\n\\
&=& N^{-9/2} \sumI \sumJ \sumK \sumL \sumO  \frac{\{1-r_i
  \pi^{-1}(y_i,\u_i;\bb,g^*)\} r_j K_h({\x_j-\x_i}) e^{-h(y_j)} r_k K_h({\u_k-\u_i})}{d^*(\x_i) D_i V_{i,k} \{E(R_k\mid\x_k) f_{\X}(\x_k)\}^2}\n\\
&&\times \left\{r_l  K_{h}(\x_l-\x_k) e^{-h(y_l)} r_o K_{h} (\x_o-\x_k)  \Delta_{3k}\right\}+O_p(N^{1/2}h^m)\n\\
&=& N^{-9/2} \sumI \sumJ \sumK \sumL \sumM  \frac{\{1-r_i
  \pi^{-1}(y_i,\u_i;\bb,g^*)\} r_j K_h({\x_j-\x_i}) e^{-h(y_j)} r_k K_h({\u_k-\u_i})}{d^*(\x_i) D_i V_{i,k} \{E(R_k\mid\x_k) f_{\X}(\x_k)\}^2}\n\\
&&\times \left\{r_l  K_{h}(\x_l-\x_k) e^{-h(y_l)} r_m K_{h} (\x_m-\x_k)  \Delta_{3k}\right\}
+O_p(N^{1/2}h^m).
\ee

Consider $T_{2212AC}$,
\be \label{ustat:t2212ac}
&& T_{2212AC}\n\\
&=& N^{-8/2} \sumI \sumJ \sumK \sumL \frac{\{1-r_i
  \pi^{-1}(y_i,\u_i;\bb,g^*)\} r_j K_h({\x_j-\x_i}) e^{-h(y_j)} r_k K_h({\u_k-\u_i})}{d^*(\x_i) D_i V_{i,k} \{E(R_k\mid\x_k) f_{\X}(\x_k)\}^3}\n\\
&&\times \left\{r_l  K_{h}(\x_l-\x_k) e^{-h(y_l)}\right\} N^{1/2}   E \left\{ R_m K_{h}(\X_m-\x_k) e^{-h(Y_m)}\h'_\bb(Y_m;\bb) R_o K_{h} (\X_o-\x_k)\right\}\n\\
&=& N^{-8/2} \sumI \sumJ \sumK \sumL \frac{\{1-r_i
  \pi^{-1}(y_i,\u_i;\bb,g^*)\} r_j K_h({\x_j-\x_i}) e^{-h(y_j)} r_k K_h({\u_k-\u_i})}{d^*(\x_i) D_i V_{i,k} \{E(R_k\mid\x_k) f_{\X}(\x_k)\}^3}\n\\
&&\times \left\{r_l  K_{h}(\x_l-\x_k) e^{-h(y_l)}\right\} N^{1/2}   \left\{ E(R_k\mid\x_k) f_{\X}(\x_k) \Delta_{3k} +O_p(h^m) \right\} \left\{ E(R_k\mid\x_k) f_{\X}(\x_k) +O_p(h^m)\right\}\n\\
&=& N^{-7/2} \sumI \sumJ \sumK \sumL \frac{\{1-r_i
  \pi^{-1}(y_i,\u_i;\bb,g^*)\} r_j K_h({\x_j-\x_i}) e^{-h(y_j)} r_k K_h({\u_k-\u_i})}{d^*(\x_i) D_i V_{i,k} \{E(R_k\mid\x_k) f_{\X}(\x_k)\}}\n\\
&&\times \left\{r_l  K_{h}(\x_l-\x_k) e^{-h(y_l)} \Delta_{3k}\right\}
+O_p(N^{1/2}h^m).
\ee

Using U-statistics property on term $T_{231}$ we have,
\be \label{ustat:t231}
&&T_{231}\n\\
&=& N^{-11/2} \sumI \sumJ \sumK \sumL \sumM \sumO \frac{\{1-r_i
  \pi^{-1}(y_i,\u_i;\bb,g^*)\} r_j K_h({\x_j-\x_i}) e^{-h(y_j)} r_k K_h({\u_k-\u_i})}{d^*(\x_i) D_i V_{i,k}^2}\n \\
&& \times \left[  T_{j,k}r_l K_h({\u_l-\u_i}) r_m K_h({\x_m-\x_i}) r_o K_h({\x_o-\x_k})  \{e^{-h(y_o)}+e^{-g^* (\u_k)} e^{-2h(y_o)}\}\right] \n\\
&& \times \left [ \frac{1}{\{E(R_k\mid\x_k) f_{\X}(\x_k)\}}\right ]\n\\
&=& N^{-8/2} \sumI \sumJ \sumK \sumL \frac{\{1-r_i
  \pi^{-1}(y_i,\u_i;\bb,g^*)\} r_j K_h({\x_j-\x_i}) e^{-h(y_j)} r_k K_h({\u_k-\u_i})  r_l K_h({\u_l-\u_i})}{d^*(\x_i) D_i V_{i,k}^2 \{E(R_k\mid\x_k) f_{\X}(\x_k)\}}\n \\
&& \times T_{j,k}N^{-1/2} \sumM E [r_m K_h({\x_m-\x_i}) R_o K_h({\X_o-\x_k})  \{e^{-h(Y_o)}+e^{-g^* (\u_k)} e^{-2h(Y_o)}\} \mid \x_m, r_m, r_m y_m]\n\\
&&+ N^{-8/2} \sumI \sumJ \sumK \sumL \frac{\{1-r_i
  \pi^{-1}(y_i,\u_i;\bb,g^*)\} r_j K_h({\x_j-\x_i}) e^{-h(y_j)} r_k K_h({\u_k-\u_i}) r_l K_h({\u_l-\u_i})}{d^*(\x_i) D_i V_{i,k}^2 \{E(R_k\mid\x_k) f_{\X}(\x_k)\}}\n \\
&& \times T_{j,k} N^{-1/2} \sumO E [R_m K_h({\X_m-\x_i}) r_o K_h({\x_o-\x_k})  \{e^{-h(y_o)}+e^{-g^* (\u_k)} e^{-2h(y_o)}\} \mid \x_o, r_o, r_o y_o]\n\\
&&- N^{-8/2} \sumI \sumJ \sumK \sumL \frac{\{1-r_i
  \pi^{-1}(y_i,\u_i;\bb,g^*)\} r_j K_h({\x_j-\x_i}) e^{-h(y_j)} r_k K_h({\u_k-\u_i})  r_l K_h({\u_l-\u_i})}{d^*(\x_i) D_i V_{i,k}^2 \{E(R_k\mid\x_k) f_{\X}(\x_k)\}}\n \\
&& \times T_{j,k} N^{1/2} E [R_m K_h({\X_m-\x_i}) R_o K_h({\X_o-\x_k})  \{e^{-h(Y_o)}+e^{-g^* (\u_k)} e^{-2h(Y_o)}\}]+O_p(N^{-1/2})\n\\
&=& T_{231A} +  T_{231B} -  T_{231C} +O_p(N^{-1/2}).\n
\ee

Consider $T_{231A}$,
\be \label{ustat:t231a}
&&T_{231A}\n\\
&=&N^{-8/2} \sumI \sumJ \sumK \sumL \frac{\{1-r_i
  \pi^{-1}(y_i,\u_i;\bb,g^*)\} r_j K_h({\x_j-\x_i}) e^{-h(y_j)} r_k K_h({\u_k-\u_i}) r_l K_h({\u_l-\u_i})}{d^*(\x_i) D_i V_{i,k}^2 \{E(R_k\mid\x_k) f_{\X}(\x_k)\}}\n \\
&& \times T_{j,k}  N^{-1/2} \sumM E [r_m K_h({\x_m-\x_i}) R_o K_h({\X_o-\x_k})  \{e^{-h(Y_o)}+e^{-g^* (\u_k)} e^{-2h(Y_o)}\} \mid \x_m, r_m, r_m y_m]\n\\
&=&N^{-9/2} \sumI \sumJ \sumK \sumL \frac{\{1-r_i
  \pi^{-1}(y_i,\u_i;\bb,g^*)\} r_j K_h({\x_j-\x_i}) e^{-h(y_j)} r_k K_h({\u_k-\u_i}) r_l K_h({\u_l-\u_i})}{d^*(\x_i) D_i V_{i,k}^2 \{E(R_k\mid\x_k) f_{\X}(\x_k)\}}\n \\
&& \times  T_{j,k} \sumM r_m K_h({\x_m-\x_i}) E(R_k\mid\x_k) f_{\X}(\x_k) E [\{e^{-h(Y_k)}+e^{-g^* (\u_k)} e^{-2h(Y_k)}\} \mid \x_k, 1]+ O_p(N^{1/2} h^m) \n\\
&=&N^{-9/2} \sumI \sumJ \sumK \sumL \sumM  \frac{\{1-r_i
  \pi^{-1}(y_i,\u_i;\bb,g^*)\} r_j K_h({\x_j-\x_i}) e^{-h(y_j)} r_k K_h({\u_k-\u_i}) T_{j,k}}{d^*(\x_i) D_i V_{i,k}^2}\n \\
&& \times  \{r_l K_h({\u_l-\u_i}) r_m K_h({\x_m-\x_i}) d^*(\x_k)\}+ O_p(N^{1/2} h^m)\n\\
&=& T_{232C}+ O_p(N^{1/2} h^m),
\ee
where $T_{232C}$ is as defined in (\ref{ustat:t232c}).

Consider $T_{231B}$,
\be \label{ustat:t231b}
&&T_{231B}\n\\
&=& N^{-8/2} \sumI \sumJ \sumK \sumL \frac{\{1-r_i
  \pi^{-1}(y_i,\u_i;\bb,g^*)\} r_j K_h({\x_j-\x_i}) e^{-h(y_j)} r_k K_h({\u_k-\u_i}) r_l K_h({\u_l-\u_i})}{d^*(\x_i) D_i V_{i,k}^2 \{E(R_k\mid\x_k) f_{\X}(\x_k)\}}\n \\
&& \times  T_{j,k} N^{-1/2} \sumO E [R_m K_h({\X_m-\x_i}) r_o K_h({\x_o-\x_k})  \{e^{-h(y_o)}+e^{-g^* (\u_k)} e^{-2h(y_o)}\} \mid \x_o, r_o, r_o y_o]\n\\
&=& N^{-9/2} \sumI \sumJ \sumK \sumL \sumO \frac{\{1-r_i
  \pi^{-1}(y_i,\u_i;\bb,g^*)\} r_j K_h({\x_j-\x_i}) e^{-h(y_j)} r_k K_h({\u_k-\u_i}) T_{j,k}}{d^*(\x_i) D_i V_{i,k}^2 \{E(R_k\mid\x_k) f_{\X}(\x_k)\}}\n \\
&& \times E(R_i\mid\x_i) f_{\X}(\x_i)  r_l K_h({\u_l-\u_i}) r_o K_h({\x_o-\x_k})  \{e^{-h(y_o)}+e^{-g^* (\u_k)} e^{-2h(y_o)}\}  + O_p(N^{1/2} h^m)\n\\
&=& N^{-9/2} \sumI \sumJ \sumK \sumL \sumM \frac{\{1-r_i
  \pi^{-1}(y_i,\u_i;\bb,g^*)\} r_j K_h({\x_j-\x_i}) e^{-h(y_j)} r_k K_h({\u_k-\u_i}) T_{j,k}}{d^*(\x_i) D_i V_{i,k}^2 \{E(R_k\mid\x_k) f_{\X}(\x_k)\}}\n \\
&& \times E(R_i\mid\x_i) f_{\X}(\x_i)  r_l K_h({\u_l-\u_i}) r_m K_h({\x_m-\x_k})  \{e^{-h(y_m)}+e^{-g^* (\u_k)} e^{-2h(y_m)}\}  + O_p(N^{1/2} h^m).
\ee

Consider $T_{231C}$,
\be \label{ustat:t231c}
&&T_{231C}\n\\
&=& N^{-8/2} \sumI \sumJ \sumK \sumL \frac{\{1-r_i
  \pi^{-1}(y_i,\u_i;\bb,g^*)\} r_j K_h({\x_j-\x_i}) e^{-h(y_j)} r_k K_h({\u_k-\u_i}) r_l K_h({\u_l-\u_i})}{d^*(\x_i) D_i V_{i,k}^2 \{E(R_k\mid\x_k) f_{\X}(\x_k)\}}\n \\
&& \times T_{j,k} N^{1/2} E [R_m K_h({\X_m-\x_i}) R_o K_h({\X_o-\x_k})  \{e^{-h(Y_o)}+e^{-g^* (\u_k)} e^{-2h(Y_o)}\}]\n\\
&=& N^{-7/2} \sumI \sumJ \sumK \sumL \frac{\{1-r_i
  \pi^{-1}(y_i,\u_i;\bb,g^*)\} r_j K_h({\x_j-\x_i}) e^{-h(y_j)} r_k K_h({\u_k-\u_i})  r_l K_h({\u_l-\u_i})}{d^*(\x_i) D_i V_{i,k}^2 \{E(R_k\mid\x_k) f_{\X}(\x_k)\}}\n \\
&& \times T_{j,k}  \{E(R_i\mid\x_i) f_{\X}(\x_i) +O_p(h^m)\}   \{E(R_k\mid\x_k) f_{\X}(\x_k) d^*(\x_k) +O_p(h^m) \}\n\\
&=& N^{-7/2} \sumI \sumJ \sumK \sumL \frac{\{1-r_i
  \pi^{-1}(y_i,\u_i;\bb,g^*)\} r_j K_h({\x_j-\x_i}) e^{-h(y_j)} r_k K_h({\u_k-\u_i}) T_{j,k} }{d^*(\x_i) D_i V_{i,k}^2}\n \\
&& \times \{r_l K_h({\u_l-\u_i}) E(R_i\mid\x_i) f_{\X}(\x_i) d^*(\x_k) \} +O_p(N^{1/2} h^m).\n\\
\ee

Using U-stat property on term $T_{2211}$ we have the following:
\be \label{ustat:t2211}
&& T_{2211}\n\\
&=& N^{-9/2} \sumI \sumJ \sumK \sumL \sumM \frac{\{1-r_i
  \pi^{-1}(y_i,\u_i;\bb,g^*)\} r_j K_h({\x_j-\x_i}) e^{-h(y_j)} r_k K_h({\u_k-\u_i})}{d^*(\x_i) D_i V_{i,k}\{E(R_k\mid\x_k) f_{\X}(\x_k)\}^2}\n\\
&&\times \left\{r_l  K_{h}(\x_l-\x_k) e^{-h(y_l)}r_m K_{h}(\x_m-\x_k) e^{-h(y_m)}\h'_\bb(y_m;\bb)\right\}\n\\
&=& N^{-6/2} \sumI \sumJ \sumK \frac{\{1-r_i
  \pi^{-1}(y_i,\u_i;\bb,g^*)\} r_j K_h({\x_j-\x_i}) e^{-h(y_j)} r_k K_h({\u_k-\u_i})}{d^*(\x_i) D_i V_{i,k}\{E(R_k\mid\x_k) f_{\X}(\x_k)\}^2}\n\\
&& \times N^{-1/2}  \sumL E\left\{r_l  K_{h}(\x_l-\x_k) e^{-h(y_l)}R_m K_{h}(\X_m-\x_k) e^{-h(Y_m)}\h'_\bb(Y_m;\bb)\mid \x_l, r_l, r_l y_l \right\}\n\\
&&+ N^{-6/2} \sumI \sumJ \sumK \frac{\{1-r_i
  \pi^{-1}(y_i,\u_i;\bb,g^*)\} r_j K_h({\x_j-\x_i}) e^{-h(y_j)} r_k K_h({\u_k-\u_i})}{d^*(\x_i) D_i V_{i,k}\{E(R_k\mid\x_k) f_{\X}(\x_k)\}^2}\n\\
&& \times N^{-1/2}  \sumM E\left\{R_l  K_{h}(\X_l-\x_k) e^{-h(Y_l)} r_m K_{h}(\x_m-\x_k) e^{-h(y_m)}\h'_\bb(y_m;\bb)\mid \x_m, r_m, r_m y_m \right\}\n\\
&&- N^{-6/2} \sumI \sumJ \sumK \frac{\{1-r_i
  \pi^{-1}(y_i,\u_i;\bb,g^*)\} r_j K_h({\x_j-\x_i}) e^{-h(y_j)} r_k K_h({\u_k-\u_i})}{d^*(\x_i) D_i V_{i,k}\{E(R_k\mid\x_k) f_{\X}(\x_k)\}^2}\n\\
&& \times N^{1/2} E\left\{R_l  K_{h}(\X_l-\x_k) e^{-h(Y_l)} R_m K_{h}(\X_m-\x_k) e^{-h(Y_m)}\h'_\bb(Y_m;\bb) \right\}+O_p(N^{-1/2})\n\\
&=& T_{2211A}+T_{2211B}-T_{2211C}+O_p(N^{-1/2}).\n
\ee

Consider $T_{2211A}$,
\be \label{ustat:t2211a}
&&T_{2211A}\n\\
&=& N^{-6/2} \sumI \sumJ \sumK \frac{\{1-r_i
  \pi^{-1}(y_i,\u_i;\bb,g^*)\} r_j K_h({\x_j-\x_i}) e^{-h(y_j)} r_k K_h({\u_k-\u_i})}{d^*(\x_i) D_i V_{i,k}\{E(R_k\mid\x_k) f_{\X}(\x_k)\}^2}\n\\
&& \times N^{-1/2}  \sumL E\left\{r_l  K_{h}(\x_l-\x_k) e^{-h(y_l)}R_m K_{h}(\X_m-\x_k) e^{-h(Y_m)}\h'_\bb(Y_m;\bb)\mid \x_l, r_l, r_l y_l \right\}\n\\
&=& N^{-7/2} \sumI \sumJ \sumK \frac{\{1-r_i
  \pi^{-1}(y_i,\u_i;\bb,g^*)\} r_j K_h({\x_j-\x_i}) e^{-h(y_j)} r_k K_h({\u_k-\u_i})}{d^*(\x_i) D_i V_{i,k}\{E(R_k\mid\x_k) f_{\X}(\x_k)\}^2}\n\\
&& \times \sumL r_l  K_{h}(\x_l-\x_k) e^{-h(y_l)} E\left\{R_m K_{h}(\X_m-\x_k) e^{-h(Y_m)}\h'_\bb(Y_m;\bb)\right\}\n\\
&=& N^{-7/2} \sumI \sumJ \sumK \sumL \frac{\{1-r_i
  \pi^{-1}(y_i,\u_i;\bb,g^*)\} r_j K_h({\x_j-\x_i}) e^{-h(y_j)} r_k K_h({\u_k-\u_i})}{d^*(\x_i) D_i V_{i,k} E(R_k\mid\x_k) f_{\X}(\x_k)}\n\\
&&\times r_l  K_{h}(\x_l-\x_k) e^{-h(y_l)} \Delta_{3k}
+ O_p(N^{1/2}h^m)\n\\
&=& T_{2212AC},
\ee
where $T_{2212AC}$ is as defined in (\ref{ustat:t2212ac}).

Consider $T_{2211B}$,
\be \label{ustat:t2211b}
&&T_{2211B} \n\\
&=& N^{-6/2} \sumI \sumJ \sumK \frac{\{1-r_i
  \pi^{-1}(y_i,\u_i;\bb,g^*)\} r_j K_h({\x_j-\x_i}) e^{-h(y_j)} r_k K_h({\u_k-\u_i})}{d^*(\x_i) D_i V_{i,k}\{E(R_k\mid\x_k) f_{\X}(\x_k)\}^2}\n\\
&& \times N^{-1/2}  \sumM E\left\{R_l  K_{h}(\X_l-\x_k) e^{-h(Y_l)} r_m K_{h}(\x_m-\x_k) e^{-h(y_m)}\h'_\bb(y_m;\bb)\mid \x_m, r_m, r_m y_m \right\}\n\\
&=& N^{-7/2} \sumI \sumJ \sumK \sumM  \frac{\{1-r_i
  \pi^{-1}(y_i,\u_i;\bb,g^*)\} r_j K_h({\x_j-\x_i}) e^{-h(y_j)} r_k K_h({\u_k-\u_i})}{d^*(\x_i) D_i V_{i,k}\{E(R_k\mid\x_k) f_{\X}(\x_k)\}}\n\\
&& \times  r_m K_{h}(\x_m-\x_k) e^{-h(y_m)}\h'_\bb(y_m;\bb) \Delta_{1k}
+ O_p(N^{1/2}h^m)\n\\
&=& N^{-7/2} \sumI \sumJ \sumK \sumL  \frac{\{1-r_i
  \pi^{-1}(y_i,\u_i;\bb,g^*)\} r_j K_h({\x_j-\x_i}) e^{-h(y_j)} r_k K_h({\u_k-\u_i})}{d^*(\x_i) D_i V_{i,k}\{E(R_k\mid\x_k) f_{\X}(\x_k)\}}\n\\
&& \times  r_l K_{h}(\x_l-\x_k) e^{-h(y_l)}\h'_\bb(y_l;\bb) \Delta_{1k}
+ O_p(N^{1/2}h^m).
\ee

Consider $T_{2211C}$,
\be \label{ustat:t2211c}
&&T_{2211C} \n\\
&=& N^{-6/2} \sumI \sumJ \sumK \frac{\{1-r_i
  \pi^{-1}(y_i,\u_i;\bb,g^*)\} r_j K_h({\x_j-\x_i}) e^{-h(y_j)} r_k K_h({\u_k-\u_i})}{d^*(\x_i) D_i V_{i,k}\{E(R_k\mid\x_k) f_{\X}(\x_k)\}^2}\n\\
&& \times N^{1/2} E\left\{R_l  K_{h}(\X_l-\x_k) e^{-h(Y_l)} R_m K_{h}(\X_m-\x_k) e^{-h(Y_m)}\h'_\bb(Y_m;\bb) \right\}\n\\
&=& N^{-5/2} \sumI \sumJ \sumK \frac{\{1-r_i
  \pi^{-1}(y_i,\u_i;\bb,g^*)\} r_j K_h({\x_j-\x_i}) e^{-h(y_j)} r_k K_h({\u_k-\u_i})}{d^*(\x_i) D_i V_{i,k}\{E(R_k\mid\x_k) f_{\X}(\x_k)\}^2}\n\\
&& \times \{E(R_k\mid\x_k) f_{\X}(\x_k) \Delta_{1k} +O_p(h^m)\} \{E(R_k\mid\x_k) f_{\X}(\x_k) \Delta_{3k}  +O_p(h^m)\}\n\\
&=& N^{-5/2} \sumI \sumJ \sumK \frac{\{1-r_i
  \pi^{-1}(y_i,\u_i;\bb,g^*)\} r_j K_h({\x_j-\x_i}) e^{-h(y_j)} r_k K_h({\u_k-\u_i})\Delta_{1k} \Delta_{3k}}{d^*(\x_i) D_i V_{i,k}}\n\\
&&+O_p(N^{1/2}h^m).
\ee

Consider the term $T_{332}$ and applying U-statistics property,
\be \label{ustat:t332}
&&T_{332}\n\\
&=&N^{-9/2} \sumI \sumJ \sumK \sumL \sumM \frac{\{1-r_i  \pi^{-1}(y_i,\u_i;\bb,g^*)\}C_i r_j  K_{h} (\u_j-\u_i)r_k K_{h} (\u_k-\u_i)  \Delta^2_{1j}}{d^*(\x_i) D_i^2 \{d^* (\x_j) E(R_i\mid\u_i) f_{\U}(\u_i)\}^2 \{E(R_j\mid\x_j) f_{\X}(\x_j)\}^2} \n\\
&&\times  r_l  K_{h}(\x_l-\x_j) \{e^{-h(y_l)}+e^{-g^* (\u_j)}e^{-2h(y_l)}\} r_m K_{h} (\x_m-\x_j)\n\\
&=&N^{-6/2} \sumI \sumJ \sumK \frac{\{1-r_i  \pi^{-1}(y_i,\u_i;\bb,g^*)\}C_i r_j  K_{h} (\u_j-\u_i)r_k K_{h} (\u_k-\u_i)  \Delta^2_{1j}}{d^*(\x_i) D_i^2 \{d^* (\x_j) E(R_i\mid\u_i) f_{\U}(\u_i)\}^2 \{E(R_j\mid\x_j) f_{\X}(\x_j)\}^2} \n\\
&& \times N^{-1/2} \sumL E[ r_l  K_{h}(\x_l-\x_j) \{e^{-h(y_l)}+e^{-g^* (\u_j)}e^{-2h(y_l)}\} R_m K_{h} (\X_m-\x_j)\mid \x_l, r_l,r_l y_l]\n\\
&&+N^{-6/2} \sumI \sumJ \sumK \frac{\{1-r_i  \pi^{-1}(y_i,\u_i;\bb,g^*)\}C_i r_j  K_{h} (\u_j-\u_i)r_k K_{h} (\u_k-\u_i)  \Delta^2_{1j}}{d^*(\x_i) D_i^2 \{d^* (\x_j) E(R_i\mid\u_i) f_{\U}(\u_i)\}^2 \{E(R_j\mid\x_j) f_{\X}(\x_j)\}^2} \n\\
&& \times N^{-1/2} \sumM E[ R_l  K_{h}(\X_l-\x_j) \{e^{-h(Y_l)}+e^{-g^* (\u_j)}e^{-2h(Y_l)}\} r_m K_{h} (\x_m-\x_j)\mid \x_m, r_m,r_m y_m]\n\\
&&-N^{-6/2} \sumI \sumJ \sumK \frac{\{1-r_i  \pi^{-1}(y_i,\u_i;\bb,g^*)\}C_i r_j  K_{h} (\u_j-\u_i)r_k K_{h} (\u_k-\u_i)  \Delta^2_{1j}}{d^*(\x_i) D_i^2 \{d^* (\x_j) E(R_i\mid\u_i) f_{\U}(\u_i)\}^2 \{E(R_j\mid\x_j) f_{\X}(\x_j)\}^2} \n\\
&& \times N^{1/2} E[ R_l  K_{h}(\X_l-\x_j) \{e^{-h(Y_l)}+e^{-g^* (\u_j)}e^{-2h(Y_l)}\} R_m K_{h} (\X_m-\x_j)]+O_p(N^{-1/2})\n\\
&=& T_{332A}+T_{332B}-T_{332C}+O_p(N^{-1/2}).\n
\ee

Consider $T_{332A}$,
\be \label{ustat:t332a}
&&T_{332A}\n\\
&=&N^{-6/2} \sumI \sumJ \sumK \frac{\{1-r_i  \pi^{-1}(y_i,\u_i;\bb,g^*)\}C_i r_j  K_{h} (\u_j-\u_i)r_k K_{h} (\u_k-\u_i)  \Delta^2_{1j}}{d^*(\x_i) D_i^2 \{d^* (\x_j) E(R_i\mid\u_i) f_{\U}(\u_i)\}^2 \{E(R_j\mid\x_j) f_{\X}(\x_j)\}^2} \n\\
&& \times N^{-1/2} \sumL E[ r_l  K_{h}(\x_l-\x_j) \{e^{-h(y_l)}+e^{-g^* (\u_j)}e^{-2h(y_l)}\} R_m K_{h} (\X_m-\x_j)\mid \x_l, r_l,r_l y_l]\n\\
&=&N^{-7/2} \sumI \sumJ \sumK  \sumL \frac{\{1-r_i  \pi^{-1}(y_i,\u_i;\bb,g^*)\}C_i r_j  K_{h} (\u_j-\u_i)r_k K_{h} (\u_k-\u_i)  \Delta^2_{1j}}{d^*(\x_i) D_i^2 \{d^* (\x_j) E(R_i\mid\u_i) f_{\U}(\u_i)\}^2 \{E(R_j\mid\x_j) f_{\X}(\x_j)\}^2} \n\\
&& \times r_l  K_{h}(\x_l-\x_j) \{e^{-h(y_l)}+e^{-g^* (\u_j)}e^{-2h(y_l)}\} E\{R_m K_{h} (\X_m-\x_j)\}\n\\
&=&N^{-7/2} \sumI \sumJ \sumK  \sumL \frac{\{1-r_i  \pi^{-1}(y_i,\u_i;\bb,g^*)\}C_i r_j  K_{h} (\u_j-\u_i)r_k K_{h} (\u_k-\u_i)  \Delta^2_{1j}}{d^*(\x_i) D_i^2 \{d^* (\x_j) E(R_i\mid\u_i) f_{\U}(\u_i)\}^2 E(R_j\mid\x_j) f_{\X}(\x_j)} \n\\
&& \times r_l  K_{h}(\x_l-\x_j) \{e^{-h(y_l)}+e^{-g^* (\u_j)}e^{-2h(y_l)}\}
+O_p(N^{1/2}h^m) \n\\
&=&T_{331}+O_p(N^{1/2}h^m),\n
\ee
where $T_{331}$ is as defined in  (\ref{eq:t331}).

Consider $T_{332B}$,
\be \label{ustat:t332b}
&&T_{332B}\n\\
&=&N^{-6/2} \sumI \sumJ \sumK \frac{\{1-r_i  \pi^{-1}(y_i,\u_i;\bb,g^*)\}C_i r_j  K_{h} (\u_j-\u_i)r_k K_{h} (\u_k-\u_i)  \Delta^2_{1j}}{d^*(\x_i) D_i^2 \{d^* (\x_j) E(R_i\mid\u_i) f_{\U}(\u_i)\}^2 \{E(R_j\mid\x_j) f_{\X}(\x_j)\}^2} \n\\
&& \times N^{-1/2} \sumM E[ R_l  K_{h}(\X_l-\x_j) \{e^{-h(Y_l)}+e^{-g^* (\u_j)}e^{-2h(Y_l)}\} r_m K_{h} (\x_m-\x_j)\mid \x_m, r_m,r_m y_m]\n\\
&=&N^{-7/2} \sumI \sumJ \sumK \sumM \frac{\{1-r_i  \pi^{-1}(y_i,\u_i;\bb,g^*)\}C_i r_j  K_{h} (\u_j-\u_i)r_k K_{h} (\u_k-\u_i)  \Delta^2_{1j}}{d^*(\x_i) D_i^2 \{d^* (\x_j) E(R_i\mid\u_i) f_{\U}(\u_i)\}^2 \{E(R_j\mid\x_j) f_{\X}(\x_j)\}^2} \n\\
&& \times r_m K_{h} (\x_m-\x_j) E[ R_l  K_{h}(\X_l-\x_j) \{e^{-h(Y_l)}+e^{-g^* (\u_j)}e^{-2h(Y_l)}\}]\n\\
&=&N^{-7/2} \sumI \sumJ \sumK \sumM \frac{\{1-r_i  \pi^{-1}(y_i,\u_i;\bb,g^*)\}C_i r_j  K_{h} (\u_j-\u_i)r_k K_{h} (\u_k-\u_i)  \Delta^2_{1j}}{d^*(\x_i) D_i^2 \{d^* (\x_j) E(R_i\mid\u_i) f_{\U}(\u_i)\}^2 \{E(R_j\mid\x_j) f_{\X}(\x_j)\}} \n\\
&& \times r_m K_{h} (\x_m-\x_j) d^*(\x_j)
+ O_p(N^{1/2}h^m)\n\\
&=&N^{-7/2} \sumI \sumJ \sumK \sumL \frac{\{1-r_i  \pi^{-1}(y_i,\u_i;\bb,g^*)\}C_i r_j  K_{h} (\u_j-\u_i)r_k K_{h} (\u_k-\u_i)  \Delta^2_{1j}}{d^*(\x_i) D_i^2 d^* (\x_j) \{E(R_i\mid\u_i) f_{\U}(\u_i)\}^2 \{E(R_j\mid\x_j) f_{\X}(\x_j)\}} \n\\
&& \times r_l K_{h} (\x_l-\x_j)
+ O_p(N^{1/2}h^m).
\ee

Consider $T_{332C}$,
\be \label{ustat:t332c}
&&T_{332C}\n\\
&=&N^{-6/2} \sumI \sumJ \sumK \frac{\{1-r_i  \pi^{-1}(y_i,\u_i;\bb,g^*)\}C_i r_j  K_{h} (\u_j-\u_i)r_k K_{h} (\u_k-\u_i)  \Delta^2_{1j}}{d^*(\x_i) D_i^2 \{d^* (\x_j) E(R_i\mid\u_i) f_{\U}(\u_i)\}^2 \{E(R_j\mid\x_j) f_{\X}(\x_j)\}^2} \n\\
&& \times N^{1/2} E[ R_l  K_{h}(\X_l-\x_j) \{e^{-h(Y_l)}+e^{-g^* (\u_j)}e^{-2h(Y_l)}\} R_m K_{h} (\X_m-\x_j)]\n\\
&=&N^{-5/2} \sumI \sumJ \sumK \frac{\{1-r_i  \pi^{-1}(y_i,\u_i;\bb,g^*)\}C_i r_j  K_{h} (\u_j-\u_i)r_k K_{h} (\u_k-\u_i)  \Delta^2_{1j}}{d^*(\x_i) D_i^2 \{d^* (\x_j) E(R_i\mid\u_i) f_{\U}(\u_i)\}^2 \{E(R_j\mid\x_j) f_{\X}(\x_j)\}^2} \n\\
&& \times \{E(R_j\mid\x_j) f_{\X}(\x_j) d^*(\x_j) +O_p(h^m)\} \{E(R_j\mid\x_j) f_{\X}(\x_j)  +O_p(h^m)\}\n\\
&=&N^{-5/2} \sumI \sumJ \sumK \frac{\{1-r_i  \pi^{-1}(y_i,\u_i;\bb,g^*)\}C_i r_j  K_{h} (\u_j-\u_i)r_k K_{h} (\u_k-\u_i)  \Delta^2_{1j} }{d^*(\x_i) D_i^2 d^* (\x_j) \{E(R_i\mid\u_i) f_{\U}(\u_i)\}^2} \n\\
&&+O_p(N^{1/2}h^m).
\ee

Using U-statistics property and applying it to the term $T_{231A}$
from (\ref{ustat:t231a}),
\be \label{ustat:t231a2}
&&T_{231A}\n\\
&=&N^{-9/2} \sumI \sumJ \sumK \sumL \sumM  \frac{\{1-r_i
  \pi^{-1}(y_i,\u_i;\bb,g^*)\} r_j K_h({\x_j-\x_i}) e^{-h(y_j)} r_k K_h({\u_k-\u_i}) T_{j,k}}{d^*(\x_i) D_i V_{i,k}^2}\n \\
&& \times  \{r_l K_h({\u_l-\u_i}) r_m K_h({\x_m-\x_i}) d^*(\x_k)\}
+ O_p(N^{1/2} h^m)\n\\
&=&N^{-6/2} \sumI \sumJ \sumK \frac{\{1-r_i
  \pi^{-1}(y_i,\u_i;\bb,g^*)\} r_j K_h({\x_j-\x_i}) e^{-h(y_j)} r_k K_h({\u_k-\u_i}) T_{j,k}  d^*(\x_k)}{d^*(\x_i) D_i V_{i,k}^2}\n \\
&& \times N^{-1/2} \sumL E \{r_l K_h({\u_l-\u_i}) R_m K_h({\X_m-\x_i})\mid \x_l, r_l, r_l y_l\} \n\\
&&+N^{-6/2} \sumI \sumJ \sumK \frac{\{1-r_i
  \pi^{-1}(y_i,\u_i;\bb,g^*)\} r_j K_h({\x_j-\x_i}) e^{-h(y_j)} r_k K_h({\u_k-\u_i}) T_{j,k}  d^*(\x_k)}{d^*(\x_i) D_i V_{i,k}^2}\n \\
&& \times N^{-1/2} \sumM E \{R_l K_h({\U_l-\u_i}) r_m K_h({\x_m-\x_i})\mid \x_m, r_m, r_m y_m\} \n\\
&&-N^{-6/2} \sumI \sumJ \sumK \frac{\{1-r_i
  \pi^{-1}(y_i,\u_i;\bb,g^*)\} r_j K_h({\x_j-\x_i}) e^{-h(y_j)} r_k K_h({\u_k-\u_i}) T_{j,k}  d^*(\x_k)}{d^*(\x_i) D_i V_{i,k}^2}\n \\
&& \times N^{1/2}  E \{R_l K_h({\U_l-\u_i}) R_m K_h({\X_m-\x_i})\} + O_p(N^{-1/2})+ O_p(N^{1/2} h^m)\n\\
&=& T_{231AA}+T_{231AB}-T_{231AC}+O_p(N^{-1/2})+ O_p(N^{1/2} h^m).\n
\ee

Consider $T_{231AA}$,
\be \label{ustat:t231aa}
&& T_{231AA}\n\\
&=&N^{-6/2} \sumI \sumJ \sumK \frac{\{1-r_i
  \pi^{-1}(y_i,\u_i;\bb,g^*)\} r_j K_h({\x_j-\x_i}) e^{-h(y_j)} r_k K_h({\u_k-\u_i}) T_{j,k}  d^*(\x_k)}{d^*(\x_i) D_i V_{i,k}^2}\n \\
&& \times N^{-1/2} \sumL E \{r_l K_h({\u_l-\u_i}) R_m K_h({\X_m-\x_i})\mid \x_l, r_l, r_l y_l\} \n\\
&=&N^{-7/2} \sumI \sumJ \sumK \sumL  \frac{\{1-r_i
  \pi^{-1}(y_i,\u_i;\bb,g^*)\} r_j K_h({\x_j-\x_i}) e^{-h(y_j)} r_k K_h({\u_k-\u_i}) T_{j,k}  d^*(\x_k)}{d^*(\x_i) D_i V_{i,k}^2}\n \\
&& \times r_l K_h({\u_l-\u_i})  E \{R_m K_h({\X_m-\x_i})\} \n\\
&=&N^{-7/2} \sumI \sumJ \sumK \sumL  \frac{\{1-r_i
  \pi^{-1}(y_i,\u_i;\bb,g^*)\} r_j K_h({\x_j-\x_i}) e^{-h(y_j)} r_k K_h({\u_k-\u_i}) T_{j,k}  d^*(\x_k)}{d^*(\x_i) D_i V_{i,k}^2}\n \\
&& \times r_l K_h({\u_l-\u_i})  \{E(R_i\mid\x_i) f_{\X}(\x_i)\}
+ O_p(N^{1/2} h^m)\n\\
&=& T_{231C}\n
\ee
where $T_{231C}$ is defined in (\ref{ustat:t231c}).

Consider $T_{231AB}$,
\be \label{ustat:t231ab}
&& T_{231AB}\n\\
&=&N^{-6/2} \sumI \sumJ \sumK \frac{\{1-r_i
  \pi^{-1}(y_i,\u_i;\bb,g^*)\} r_j K_h({\x_j-\x_i}) e^{-h(y_j)} r_k K_h({\u_k-\u_i}) T_{j,k}  d^*(\x_k)}{d^*(\x_i) D_i V_{i,k}^2}\n \\
&& \times N^{-1/2} \sumM E \{R_l K_h({\U_l-\u_i}) r_m K_h({\x_m-\x_i})\mid \x_m, r_m, r_m y_m\} \n\\
&=&N^{-7/2} \sumI \sumJ \sumK \sumM \frac{\{1-r_i
  \pi^{-1}(y_i,\u_i;\bb,g^*)\} r_j K_h({\x_j-\x_i}) e^{-h(y_j)} r_k K_h({\u_k-\u_i}) T_{j,k}  d^*(\x_k)}{d^*(\x_i) D_i V_{i,k}^2}\n \\
&& \times r_m K_h({\x_m-\x_i}) \{E(R_i\mid\u_i) f_{\U}(\u_i)\}
+ O_p(N^{1/2} h^m)\n\\
&=&N^{-7/2} \sumI \sumJ \sumK \sumL \frac{\{1-r_i
  \pi^{-1}(y_i,\u_i;\bb,g^*)\} r_j K_h({\x_j-\x_i}) e^{-h(y_j)} r_k K_h({\u_k-\u_i}) T_{j,k}  d^*(\x_k)}{d^*(\x_i) D_i V_{i,k}^2}\n \\
&& \times r_l K_h({\x_l-\x_i}) \{E(R_i\mid\u_i) f_{\U}(\u_i)\}
+ O_p(N^{1/2} h^m).
\ee

Consider $T_{231AC}$,
\be \label{ustat:t231ac}
&& T_{231AC}\n\\
&=&N^{-6/2} \sumI \sumJ \sumK \frac{\{1-r_i
  \pi^{-1}(y_i,\u_i;\bb,g^*)\} r_j K_h({\x_j-\x_i}) e^{-h(y_j)} r_k K_h({\u_k-\u_i}) T_{j,k}  d^*(\x_k)}{d^*(\x_i) D_i V_{i,k}^2}\n \\
&& \times N^{1/2}  E \{R_l K_h({\U_l-\u_i}) R_m K_h({\X_m-\x_i})\} \n\\
&=&N^{-5/2} \sumI \sumJ \sumK \frac{\{1-r_i
  \pi^{-1}(y_i,\u_i;\bb,g^*)\} r_j K_h({\x_j-\x_i}) e^{-h(y_j)} r_k K_h({\u_k-\u_i}) T_{j,k}  d^*(\x_k)}{d^*(\x_i) D_i V_{i,k}^2}\n \\
&& \times   \{E(R_i\mid\u_i) f_{\U}(\u_i) + O_p( h^m)\} \{ E(R_i\mid\x_i) f_{\X}(\x_i) + O_p( h^m)\} \n\\
&=&N^{-5/2} \sumI \sumJ \sumK \frac{\{1-r_i
  \pi^{-1}(y_i,\u_i;\bb,g^*)\} r_j K_h({\x_j-\x_i}) e^{-h(y_j)} r_k K_h({\u_k-\u_i}) T_{j,k} }{d^*(\x_i) D_i V_{i,k}}\n \\
&&+ O_p(N^{1/2} h^m)\n\\
&=& T_{21}+ O_p(N^{1/2} h^m),
\ee
where $T_{21}$ is defined in  (\ref{eq:t21}).

Consider $T_{231B}$ in (\ref{ustat:t231b})
and applying U-statistics to the term we get the following,
\be \label{ustat:t231b2}
&&T_{231B}\n\\
&=& N^{-9/2} \sumI \sumJ \sumK \sumL \sumM \frac{\{1-r_i
  \pi^{-1}(y_i,\u_i;\bb,g^*)\} r_j K_h({\x_j-\x_i}) e^{-h(y_j)} r_k K_h({\u_k-\u_i}) T_{j,k}}{d^*(\x_i) D_i V_{i,k}^2 \{E(R_k\mid\x_k) f_{\X}(\x_k)\}}\n \\
&& \times E(R_i\mid\x_i) f_{\X}(\x_i)  r_l K_h({\u_l-\u_i}) r_m K_h({\x_m-\x_k})  \{e^{-h(y_m)}+e^{-g^* (\u_k)} e^{-2h(y_m)}\}
+ O_p(N^{1/2} h^m)\n\\
&=& N^{-6/2} \sumI \sumJ \sumK \frac{\{1-r_i
  \pi^{-1}(y_i,\u_i;\bb,g^*)\} r_j K_h({\x_j-\x_i}) e^{-h(y_j)} r_k K_h({\u_k-\u_i}) E(R_i\mid\x_i) f_{\X}(\x_i)}{d^*(\x_i) D_i V_{i,k}^2 \{E(R_k\mid\x_k) f_{\X}(\x_k)\}}\n \\
&& \times T_{j,k}  N^{-1/2} \sumL E\left[ r_l K_h({\u_l-\u_i}) R_m K_h({\X_m-\x_k})  \{e^{-h(Y_m)}+e^{-g^* (\u_k)} e^{-2h(Y_m)}\} \mid \x_l, r_l, r_l y_l\right]\n\\
&&+ N^{-6/2} \sumI \sumJ \sumK \frac{\{1-r_i
  \pi^{-1}(y_i,\u_i;\bb,g^*)\} r_j K_h({\x_j-\x_i}) e^{-h(y_j)} r_k K_h({\u_k-\u_i})  E(R_i\mid\x_i) f_{\X}(\x_i)}{d^*(\x_i) D_i V_{i,k}^2 \{E(R_k\mid\x_k) f_{\X}(\x_k)\}}\n \\
&& \times T_{j,k} N^{-1/2} \sumM E\left[ R_l K_h({\U_l-\u_i}) r_m K_h({\x_m-\x_k})  \{e^{-h(y_m)}+e^{-g^* (\u_k)} e^{-2h(y_m)}\} \mid \x_m, r_m, r_m y_m\right]\n\\
&&- N^{-6/2} \sumI \sumJ \sumK \frac{\{1-r_i
  \pi^{-1}(y_i,\u_i;\bb,g^*)\} r_j K_h({\x_j-\x_i}) e^{-h(y_j)} r_k K_h({\u_k-\u_i}) E(R_i\mid\x_i) f_{\X}(\x_i)}{d^*(\x_i) D_i V_{i,k}^2 \{E(R_k\mid\x_k) f_{\X}(\x_k)\}}\n \\
&& \times  T_{j,k} N^{1/2} E\left[ R_l K_h({\U_l-\u_i}) R_m K_h({\X_m-\x_k})  \{e^{-h(Y_m)}+e^{-g^* (\u_k)} e^{-2h(Y_m)}\} \right]\n\\
&&+ O_p(N^{-1/2})+ O_p(N^{1/2} h^m)\n\\
&=& T_{231BA}+T_{231BB}-T_{231BC}+O_p(N^{-1/2})+ O_p(N^{1/2} h^m).\n
\ee

Consider $T_{231BA}$,
\be \label{ustat:t231ba}
&&T_{231BA}\n\\
&=& N^{-6/2} \sumI \sumJ \sumK \frac{\{1-r_i
  \pi^{-1}(y_i,\u_i;\bb,g^*)\} r_j K_h({\x_j-\x_i}) e^{-h(y_j)} r_k K_h({\u_k-\u_i}) E(R_i\mid\x_i) f_{\X}(\x_i)}{d^*(\x_i) D_i V_{i,k}^2 \{E(R_k\mid\x_k) f_{\X}(\x_k)\}}\n \\
&& \times  T_{j,k}  N^{-1/2} \sumL E\left[ r_l K_h({\u_l-\u_i}) R_m K_h({\X_m-\x_k})  \{e^{-h(Y_m)}+e^{-g^* (\u_k)} e^{-2h(Y_m)}\} \mid \x_l, r_l, r_l y_l\right]\n\\
&=& N^{-7/2} \sumI \sumJ \sumK   \frac{\{1-r_i
  \pi^{-1}(y_i,\u_i;\bb,g^*)\} r_j K_h({\x_j-\x_i}) e^{-h(y_j)} r_k K_h({\u_k-\u_i})  E(R_i\mid\x_i) f_{\X}(\x_i)}{d^*(\x_i) D_i V_{i,k}^2 \{E(R_k\mid\x_k) f_{\X}(\x_k)\}}\n \\
&& \times T_{j,k} \sumL  r_l K_h({\u_l-\u_i}) E(R_k\mid\x_k) f_{\X}(\x_k) E\left[  \{e^{-h(Y_k)}+e^{-g^* (\u_k)} e^{-2h(Y_k)}\} \mid \x_k ,1 \right]
+ O_p(N^{1/2} h^m)\n\\
&=& N^{-7/2} \sumI \sumJ \sumK \sumL   \frac{\{1-r_i
  \pi^{-1}(y_i,\u_i;\bb,g^*)\} r_j K_h({\x_j-\x_i}) e^{-h(y_j)} r_k K_h({\u_k-\u_i}) T_{j,k} }{d^*(\x_i) D_i V_{i,k}^2}\n \\
&& \times  r_l K_h({\u_l-\u_i}) E(R_i\mid\x_i) f_{\X}(\x_i) d^*(\x_k)
+ O_p(N^{1/2} h^m)\n\\
&=&T_{231C}.
\ee

Consider $T_{231BB}$,
\be \label{ustat:t231bb}
&&T_{231BB}\n\\
&=&N^{-6/2} \sumI \sumJ \sumK \frac{\{1-r_i
  \pi^{-1}(y_i,\u_i;\bb,g^*)\} r_j K_h({\x_j-\x_i}) e^{-h(y_j)} r_k K_h({\u_k-\u_i})  E(R_i\mid\x_i) f_{\X}(\x_i)}{d^*(\x_i) D_i V_{i,k}^2 \{E(R_k\mid\x_k) f_{\X}(\x_k)\}}\n \\
&& \times T_{j,k} N^{-1/2} \sumM E\left[ R_l K_h({\U_l-\u_i}) r_m K_h({\x_m-\x_k})  \{e^{-h(y_m)}+e^{-g^* (\u_k)} e^{-2h(y_m)}\} \mid \x_m, r_m, r_m y_m\right]\n\\
&=& N^{-7/2} \sumI \sumJ \sumK \sumM \frac{\{1-r_i
  \pi^{-1}(y_i,\u_i;\bb,g^*)\} r_j K_h({\x_j-\x_i}) e^{-h(y_j)} r_k K_h({\u_k-\u_i}) T_{j,k} }{d^*(\x_i) D_i V_{i,k}^2 \{E(R_k\mid\x_k) f_{\X}(\x_k)\}}\n \\
&& \times E(R_i\mid\x_i) f_{\X}(\x_i) E(R_i\mid\u_i) f_{\U}(\u_i)  r_m K_h({\x_m-\x_k})  \{e^{-h(y_m)}+e^{-g^* (\u_k)} e^{-2h(y_m)}\}  + O_p(N^{1/2} h^m)\n\\
&=& N^{-7/2} \sumI \sumJ \sumK \sumL \frac{\{1-r_i
  \pi^{-1}(y_i,\u_i;\bb,g^*)\} r_j K_h({\x_j-\x_i}) e^{-h(y_j)} r_k K_h({\u_k-\u_i}) T_{j,k} }{d^*(\x_i) D_i V_{i,k}^2 \{E(R_k\mid\x_k) f_{\X}(\x_k)\}}\n \\
&& \times E(R_i\mid\x_i) f_{\X}(\x_i) E(R_i\mid\u_i) f_{\U}(\u_i)  r_l K_h({\x_l-\x_k})  \{e^{-h(y_l)}+e^{-g^* (\u_k)} e^{-2h(y_l)}\}  + O_p(N^{1/2} h^m)\n\\
&=& N^{-7/2} \sumI \sumJ \sumK \sumL \frac{\{1-r_i
  \pi^{-1}(y_i,\u_i;\bb,g^*)\} r_j K_h({\x_j-\x_i}) e^{-h(y_j)} r_k K_h({\u_k-\u_i}) T_{j,k} }{d^*(\x_i) D_i      V_{i,k}  d^* (\x_k)  E(R_k\mid\x_k) f_{\X}(\x_k)}\n \\
&& \times  r_l K_h({\x_l-\x_k})  \{e^{-h(y_l)}+e^{-g^* (\u_k)} e^{-2h(y_l)}\} + O_p(N^{1/2} h^m).
\ee

Consider $T_{231BC}$,
\be \label{ustat:t231bc}
&&T_{231BC}\n\\
&=&N^{-6/2} \sumI \sumJ \sumK \frac{\{1-r_i
  \pi^{-1}(y_i,\u_i;\bb,g^*)\} r_j K_h({\x_j-\x_i}) e^{-h(y_j)} r_k K_h({\u_k-\u_i})  E(R_i\mid\x_i) f_{\X}(\x_i)}{d^*(\x_i) D_i V_{i,k}^2 \{E(R_k\mid\x_k) f_{\X}(\x_k)\}}\n \\
&& \times T_{j,k} N^{1/2} E\left[ R_l K_h({\U_l-\u_i}) R_m K_h({\X_m-\x_k})  \{e^{-h(Y_m)}+e^{-g^* (\u_k)} e^{-2h(Y_m)}\}\right]\n\\
&=&N^{-5/2} \sumI \sumJ \sumK \frac{\{1-r_i
  \pi^{-1}(y_i,\u_i;\bb,g^*)\} r_j K_h({\x_j-\x_i}) e^{-h(y_j)} r_k K_h({\u_k-\u_i}) E(R_i\mid\x_i) f_{\X}(\x_i)}{d^*(\x_i) D_i V_{i,k}^2 \{E(R_k\mid\x_k) f_{\X}(\x_k)\}}\n \\
&& \times T_{j,k} \{E(R_i\mid\u_i) f_{\U}(\u_i)+ O_p(h^m) \} \{E(R_k\mid\x_k) f_{\X}(\x_k) d^*(\x_k) + O_p( h^m)\}\n\\
&=&N^{-5/2} \sumI \sumJ \sumK \frac{\{1-r_i
  \pi^{-1}(y_i,\u_i;\bb,g^*)\} r_j K_h({\x_j-\x_i}) e^{-h(y_j)} r_k K_h({\u_k-\u_i}) T_{j,k} }{d^*(\x_i) D_i V_{i,k} }+ O_p(N^{1/2} h^m)\n\\
&=& T_{231AC}.\n
\ee
where $T_{231AC}$ is as defined in  (\ref{ustat:t231ac}).

Applying U-statistics property to $T_{232B}$ from  (\ref{ustat:t232b}) we get,
\be \label{ustat:t232b2}
&&T_{232B}\n\\
&=&N^{-11/2} \sumI \sumJ \sumK \sumL \sumM \sumO \frac{\{1-r_i
  \pi^{-1}(Y_i,\u_i;\bb,g^*)\} r_j K_h({\x_j-\x_i}) e^{-h(y_j)} r_k K_h({\u_k-\u_i}) }{d^*(\x_i) D_i V_{i,k}^2 E(R_k\mid\x_k) f_{\X}(\x_k)}\n \\
&& \times T_{j,k} \left\{ r_l K_h({\u_l-\u_i}) r_m K_h({\x_m-\x_i}) r_o K_h({\x_o-\x_k}) d^*(\x_k)\right\}
+O_p(N^{1/2} h^m)\n\\
&=&N^{-8/2} \sumI \sumJ \sumK \sumL \frac{\{1-r_i
  \pi^{-1}(Y_i,\u_i;\bb,g^*)\} r_j K_h({\x_j-\x_i}) e^{-h(y_j)} r_k K_h({\u_k-\u_i}) T_{j,k}}{d^*(\x_i) D_i V_{i,k}^2 E(R_k\mid\x_k) f_{\X}(\x_k)}\n \\
&& \times r_l K_h({\u_l-\u_i}) d^*(\x_k)  N^{-1/2} \sumM E \left\{  r_m K_h({\x_m-\x_i}) R_o K_h({\X_o-\x_k}) \mid \x_m, r_m, r_m y_m\right\} \n\\
&&+ N^{-8/2} \sumI \sumJ \sumK \sumL \frac{\{1-r_i
  \pi^{-1}(Y_i,\u_i;\bb,g^*)\} r_j K_h({\x_j-\x_i}) e^{-h(y_j)} r_k K_h({\u_k-\u_i}) T_{j,k}}{d^*(\x_i) D_i V_{i,k}^2 E(R_k\mid\x_k) f_{\X}(\x_k)}\n \\
&& \times r_l K_h({\u_l-\u_i}) d^*(\x_k)  N^{-1/2} \sumO E \left\{ R_m K_h({\X_m-\x_i}) r_o K_h({\x_o-\x_k}) \mid \x_o, r_o, r_o y_o\right\} \n\\
&&- N^{-8/2} \sumI \sumJ \sumK \sumL \frac{\{1-r_i
  \pi^{-1}(Y_i,\u_i;\bb,g^*)\} r_j K_h({\x_j-\x_i}) e^{-h(y_j)} r_k K_h({\u_k-\u_i}) T_{j,k}}{d^*(\x_i) D_i V_{i,k}^2 E(R_k\mid\x_k) f_{\X}(\x_k)}\n \\
&& \times r_l K_h({\u_l-\u_i}) d^*(\x_k)  N^{1/2} E \left\{  R_m K_h({\X_m-\x_i}) R_o K_h({\X_o-\x_k})\right\}+ O_p(N^{-1/2})+ O_p(N^{1/2} h^m)\n\\
&=& T_{232BA}+T_{232BB}-T_{232BC}+O_p(N^{-1/2})+ O_p(N^{1/2} h^m).\n
\ee

Consider $T_{232BA}$,
\be \label{ustat:t232ba}
&& T_{232BA}\n\\
&=& N^{-8/2} \sumI \sumJ \sumK \sumL \frac{\{1-r_i
  \pi^{-1}(Y_i,\u_i;\bb,g^*)\} r_j K_h({\x_j-\x_i}) e^{-h(y_j)} r_k K_h({\u_k-\u_i}) T_{j,k}}{d^*(\x_i) D_i V_{i,k}^2 E(R_k\mid\x_k) f_{\X}(\x_k)}\n \\
&& \times r_l K_h({\u_l-\u_i}) d^*(\x_k)  N^{-1/2} \sumM E \left\{  r_m K_h({\x_m-\x_i}) R_o K_h({\X_o-\x_k}) \mid \x_m, r_m, r_m y_m\right\} \n\\
&=& N^{-9/2} \sumI \sumJ \sumK \sumL \frac{\{1-r_i
  \pi^{-1}(Y_i,\u_i;\bb,g^*)\} r_j K_h({\x_j-\x_i}) e^{-h(y_j)} r_k K_h({\u_k-\u_i}) T_{j,k}}{d^*(\x_i) D_i V_{i,k}^2 E(R_k\mid\x_k) f_{\X}(\x_k)}\n \\
&& \times r_l K_h({\u_l-\u_i}) d^*(\x_k) r_m K_h({\x_m-\x_i})   \sumM E \left\{  R_o K_h({\X_o-\x_k}) \right\} \n\\
&=& N^{-9/2} \sumI \sumJ \sumK \sumL \sumM \frac{\{1-r_i
  \pi^{-1}(Y_i,\u_i;\bb,g^*)\} r_j K_h({\x_j-\x_i}) e^{-h(y_j)} r_k K_h({\u_k-\u_i}) T_{j,k}}{d^*(\x_i) D_i V_{i,k}^2 }\n \\
&& \times r_l K_h({\u_l-\u_i})  r_m K_h({\x_m-\x_i}) d^*(\x_k)
+ O_p(N^{1/2} h^m)\n\\
&=&T_{231A},\n
\ee
where $T_{231A}$ is as defined in  (\ref{ustat:t231a}).

Consider $T_{232BB}$,
\be \label{ustat:t232bb}
&& T_{232BB}\n\\
&=& N^{-8/2} \sumI \sumJ \sumK \sumL \frac{\{1-r_i
  \pi^{-1}(Y_i,\u_i;\bb,g^*)\} r_j K_h({\x_j-\x_i}) e^{-h(y_j)} r_k K_h({\u_k-\u_i}) T_{j,k}}{d^*(\x_i) D_i V_{i,k}^2 E(R_k\mid\x_k) f_{\X}(\x_k)}\n \\
&& \times r_l K_h({\u_l-\u_i}) d^*(\x_k)  N^{-1/2} \sumO E \left\{ R_m K_h({\X_m-\x_i}) r_o K_h({\x_o-\x_k}) \mid \x_o, r_o, r_o y_o\right\} \n\\
&=& N^{-9/2} \sumI \sumJ \sumK \sumL \sumO  \frac{\{1-r_i
  \pi^{-1}(Y_i,\u_i;\bb,g^*)\} r_j K_h({\x_j-\x_i}) e^{-h(y_j)} r_k K_h({\u_k-\u_i}) T_{j,k}}{d^*(\x_i) D_i V_{i,k}^2 E(R_k\mid\x_k) f_{\X}(\x_k)}\n \\
&& \times r_l K_h({\u_l-\u_i}) d^*(\x_k)  r_o K_h({\x_o-\x_k}) E \left\{ R_m K_h({\X_m-\x_i}) \right\} \n\\
&=& N^{-9/2} \sumI \sumJ \sumK \sumL \sumM  \frac{\{1-r_i
  \pi^{-1}(Y_i,\u_i;\bb,g^*)\} r_j K_h({\x_j-\x_i}) e^{-h(y_j)} r_k K_h({\u_k-\u_i}) T_{j,k}}{d^*(\x_i) D_i V_{i,k}^2 E(R_k\mid\x_k) f_{\X}(\x_k)}\n \\
&& \times r_l K_h({\u_l-\u_i}) r_m K_h({\x_m-\x_k}) E(R_i\mid\x_i) f_{\X}(\x_i)  d^*(\x_k)
+ O_p(N^{1/2} h^m).
\ee

Consider $T_{232BC}$,
\be \label{ustat:t232bc}
&& T_{232BC}\n\\
&=&N^{-8/2} \sumI \sumJ \sumK \sumL \frac{\{1-r_i
  \pi^{-1}(Y_i,\u_i;\bb,g^*)\} r_j K_h({\x_j-\x_i}) e^{-h(y_j)} r_k K_h({\u_k-\u_i}) T_{j,k}}{d^*(\x_i) D_i V_{i,k}^2 E(R_k\mid\x_k) f_{\X}(\x_k)}\n \\
&& \times r_l K_h({\u_l-\u_i}) d^*(\x_k)  N^{1/2} E \left\{  R_m K_h({\X_m-\x_i}) R_o K_h({\X_o-\x_k})\right\} \n\\
&=&  N^{-7/2} \sumI \sumJ \sumK \sumL \frac{\{1-r_i
  \pi^{-1}(Y_i,\u_i;\bb,g^*)\} r_j K_h({\x_j-\x_i}) e^{-h(y_j)} r_k K_h({\u_k-\u_i}) T_{j,k}}{d^*(\x_i) D_i V_{i,k}^2 }\n \\
&& \times r_l K_h({\u_l-\u_i}) d^*(\x_k)   E(R_i\mid\x_i) f_{\X}(\x_i)
+ O_p(N^{1/2} h^m)\n\\
&=& T_{231BA},\n
\ee
where $T_{231BA}$ is as defined in (\ref{ustat:t231ba}).

Consider $T_{232BB}$ from  (\ref{ustat:t232bb}), applying U-statistics property leads to the following
\be \label{ustat:t232bb2}
&& T_{232BB}\n\\
&=& N^{-9/2} \sumI \sumJ \sumK \sumL \sumM  \frac{\{1-r_i
  \pi^{-1}(Y_i,\u_i;\bb,g^*)\} r_j K_h({\x_j-\x_i}) e^{-h(y_j)} r_k K_h({\u_k-\u_i}) T_{j,k}}{d^*(\x_i) D_i V_{i,k}^2 E(R_k\mid\x_k) f_{\X}(\x_k)}\n \\
&& \times r_l K_h({\u_l-\u_i}) r_m K_h({\x_m-\x_k}) E(R_i\mid\x_i) f_{\X}(\x_i)  d^*(\x_k)
+ O_p(N^{1/2} h^m)\n\\
&=& N^{-6/2} \sumI \sumJ \sumK  \frac{\{1-r_i
  \pi^{-1}(Y_i,\u_i;\bb,g^*)\} r_j K_h({\x_j-\x_i}) e^{-h(y_j)} r_k K_h({\u_k-\u_i}) T_{j,k}}{d^*(\x_i) D_i V_{i,k}^2 E(R_k\mid\x_k) f_{\X}(\x_k)}\n \\
&&  \times E(R_i\mid\x_i) f_{\X}(\x_i)  d^*(\x_k) N^{-1/2} \sumL E \{r_l K_h({\u_l-\u_i}) R_m K_h({\X_m-\x_k})\mid \x_l, r_l, r_l y_l\}  \n\\
&&+ N^{-6/2} \sumI \sumJ \sumK  \frac{\{1-r_i
  \pi^{-1}(Y_i,\u_i;\bb,g^*)\} r_j K_h({\x_j-\x_i}) e^{-h(y_j)} r_k K_h({\u_k-\u_i}) T_{j,k}}{d^*(\x_i) D_i V_{i,k}^2 E(R_k\mid\x_k) f_{\X}(\x_k)}\n \\
&&  \times E(R_i\mid\x_i) f_{\X}(\x_i)  d^*(\x_k) N^{-1/2} \sumM E \{R_l K_h({\U_l-\u_i})  r_m K_h({\x_m-\x_k})\mid \x_m, r_m, r_m y_m\}  \n\\
&&- N^{-6/2} \sumI \sumJ \sumK  \frac{\{1-r_i
  \pi^{-1}(Y_i,\u_i;\bb,g^*)\} r_j K_h({\x_j-\x_i}) e^{-h(y_j)} r_k K_h({\u_k-\u_i}) T_{j,k}}{d^*(\x_i) D_i V_{i,k}^2 E(R_k\mid\x_k) f_{\X}(\x_k)}\n \\
&&  \times E(R_i\mid\x_i) f_{\X}(\x_i)  d^*(\x_k) N^{1/2} E \{R_l K_h({\U_l-\u_i}) R_m K_h({\X_m-\x_k}) \} + O_p(N^{-1/2})+O_p(N^{1/2} h^m)\n\\
&=& T_{232BBA}+T_{232BBB}-T_{232BBC}+ O_p(N^{-1/2})+O_p(N^{1/2} h^m).\n
\ee

Consider $T_{232BBA}$,
\be \label{ustat:t232bba}
&& T_{232BBA}\n\\
&=& N^{-6/2} \sumI \sumJ \sumK  \frac{\{1-r_i
  \pi^{-1}(Y_i,\u_i;\bb,g^*)\} r_j K_h({\x_j-\x_i}) e^{-h(y_j)} r_k K_h({\u_k-\u_i}) T_{j,k}}{d^*(\x_i) D_i V_{i,k}^2 E(R_k\mid\x_k) f_{\X}(\x_k)}\n \\
&&  \times E(R_i\mid\x_i) f_{\X}(\x_i)  d^*(\x_k) N^{-1/2} \sumL E \{r_l K_h({\u_l-\u_i}) R_m K_h({\X_m-\x_k})\mid \x_l, r_l, r_l y_l\}  \n\\
&=& N^{-7/2} \sumI \sumJ \sumK \sumL \frac{\{1-r_i
  \pi^{-1}(Y_i,\u_i;\bb,g^*)\} r_j K_h({\x_j-\x_i}) e^{-h(y_j)} r_k K_h({\u_k-\u_i}) T_{j,k}}{d^*(\x_i) D_i V_{i,k}^2 }\n \\
&&  \times  r_l K_h({\u_l-\u_i}) d^*(\x_k) E(R_i\mid\x_i) f_{\X}(\x_i)
+O_p(N^{1/2} h^m)\n\\
&=& T_{231C}.\n
\ee
where $T_{231C}$ is defined in (\ref{ustat:t231c}).

Consider $T_{232BBB}$,
\be \label{ustat:t232bbb}
&& T_{232BBB}\n\\
&=& N^{-6/2} \sumI \sumJ \sumK  \frac{\{1-r_i
  \pi^{-1}(Y_i,\u_i;\bb,g^*)\} r_j K_h({\x_j-\x_i}) e^{-h(y_j)} r_k K_h({\u_k-\u_i}) T_{j,k}}{d^*(\x_i) D_i V_{i,k}^2 E(R_k\mid\x_k) f_{\X}(\x_k)}\n \\
&&  \times E(R_i\mid\x_i) f_{\X}(\x_i)  d^*(\x_k) N^{-1/2} \sumM E \{R_l K_h({\U_l-\u_i})  r_m K_h({\x_m-\x_k})\mid \x_m, r_m, r_m y_m\}  \n\\
&=&N^{-7/2} \sumI \sumJ \sumK \sumM  \frac{\{1-r_i
  \pi^{-1}(Y_i,\u_i;\bb,g^*)\} r_j K_h({\x_j-\x_i}) e^{-h(y_j)} r_k K_h({\u_k-\u_i}) T_{j,k}}{d^*(\x_i) D_i V_{i,k}^2 E(R_k\mid\x_k) f_{\X}(\x_k)}\n \\
&&  \times E(R_i\mid\x_i) f_{\X}(\x_i)  d^*(\x_k) r_m K_h({\x_m-\x_k})  E(R_i\mid\u_i) f_{\U}(\u_i)
+O_p(N^{1/2} h^m)\n\\
&=&N^{-7/2} \sumI \sumJ \sumK \sumL  \frac{\{1-r_i
  \pi^{-1}(Y_i,\u_i;\bb,g^*)\} r_j K_h({\x_j-\x_i}) e^{-h(y_j)} r_k K_h({\u_k-\u_i}) T_{j,k}}{d^*(\x_i) D_i V_{i,k}^2 E(R_k\mid\x_k) f_{\X}(\x_k)}\n \\
&&  \times E(R_i\mid\x_i) f_{\X}(\x_i)  d^*(\x_k) r_l K_h({\x_l-\x_k})  E(R_i\mid\u_i) f_{\U}(\u_i)
+O_p(N^{1/2} h^m).
\ee

Consider $T_{232BBC}$,
\be \label{ustat:t232bbc}
&& T_{232BBC}\n\\
&=&N^{-6/2} \sumI \sumJ \sumK  \frac{\{1-r_i
  \pi^{-1}(Y_i,\u_i;\bb,g^*)\} r_j K_h({\x_j-\x_i}) e^{-h(y_j)} r_k K_h({\u_k-\u_i}) T_{j,k}}{d^*(\x_i) D_i V_{i,k}^2 E(R_k\mid\x_k) f_{\X}(\x_k)}\n \\
&&  \times E(R_i\mid\x_i) f_{\X}(\x_i)  d^*(\x_k) N^{1/2} E \{R_l K_h({\U_l-\u_i}) R_m K_h({\X_m-\x_k}) \}  \n\\
&=&N^{-5/2} \sumI \sumJ \sumK  \frac{\{1-r_i
  \pi^{-1}(Y_i,\u_i;\bb,g^*)\} r_j K_h({\x_j-\x_i}) e^{-h(y_j)} r_k K_h({\u_k-\u_i}) T_{j,k}}{d^*(\x_i) D_i V_{i,k}^2 E(R_k\mid\x_k) f_{\X}(\x_k)}\n \\
&&  \times E(R_i\mid\x_i) f_{\X}(\x_i)  d^*(\x_k)  E(R_i\mid\u_i) f_{\U}(\u_i) E(R_k\mid\x_k) f_{\X}(\x_k)
+O_p(N^{1/2} h^m)\n\\
&=&N^{-5/2} \sumI \sumJ \sumK  \frac{\{1-r_i
  \pi^{-1}(Y_i,\u_i;\bb,g^*)\} r_j K_h({\x_j-\x_i}) e^{-h(y_j)} r_k K_h({\u_k-\u_i}) T_{j,k}}{d^*(\x_i) D_i V_{i,k}}\n \\
&& +O_p(N^{1/2} h^m)\n\\
&=&T_{231AC}.\n
\ee
where $T_{231AC}$ is defined in (\ref{ustat:t231ac}).

Consider $T_{2212AB}$ from (\ref{ustat:t2212ab}) and applying the U-statistics property on it results in the following,
\be \label{ustat:t2212ab2}
&& T_{2212AB}\n\\
&=& N^{-9/2} \sumI \sumJ \sumK \sumL \sumM  \frac{\{1-r_i
  \pi^{-1}(y_i,\u_i;\bb,g^*)\} r_j K_h({\x_j-\x_i}) e^{-h(y_j)} r_k K_h({\u_k-\u_i})}{d^*(\x_i) D_i V_{i,k} \{E(R_k\mid\x_k) f_{\X}(\x_k)\}^2}\n\\
&&\times \left\{r_l  K_{h}(\x_l-\x_k) e^{-h(y_l)} r_m K_{h} (\x_m-\x_k)  \Delta_{3k}\right\}
+O_p(N^{1/2}h^m)\n\\
&=& N^{-6/2} \sumI \sumJ \sumK   \frac{\{1-r_i
  \pi^{-1}(y_i,\u_i;\bb,g^*)\} r_j K_h({\x_j-\x_i}) e^{-h(y_j)} r_k K_h({\u_k-\u_i})\Delta_{3k}}{d^*(\x_i) D_i V_{i,k} \{E(R_k\mid\x_k) f_{\X}(\x_k)\}^2}\n\\
&&\times  N^{-1/2}  \sumL E \left\{r_l  K_{h}(\x_l-\x_k) e^{-h(y_l)} R_m K_{h} (\X_m-\x_k)\mid \x_l,r_l, r_l y_l \right\}\n\\
&&+ N^{-6/2} \sumI \sumJ \sumK   \frac{\{1-r_i
  \pi^{-1}(y_i,\u_i;\bb,g^*)\} r_j K_h({\x_j-\x_i}) e^{-h(y_j)} r_k K_h({\u_k-\u_i})\Delta_{3k}}{d^*(\x_i) D_i V_{i,k} \{E(R_k\mid\x_k) f_{\X}(\x_k)\}^2}\n\\
&&\times  N^{-1/2}  \sumM E \left\{R_l  K_{h}(\X_l-\x_k) e^{-h(Y_l)}r_m K_{h} (\x_m-\x_k)\mid \x_m,r_m, r_m y_m \right\}\n\\
&&- N^{-6/2} \sumI \sumJ \sumK   \frac{\{1-r_i
  \pi^{-1}(y_i,\u_i;\bb,g^*)\} r_j K_h({\x_j-\x_i}) e^{-h(y_j)} r_k K_h({\u_k-\u_i})\Delta_{3k}}{d^*(\x_i) D_i V_{i,k} \{E(R_k\mid\x_k) f_{\X}(\x_k)\}^2}\n\\
&&\times  N^{1/2}  E \left\{R_l  K_{h}(\X_l-\x_k) e^{-h(Y_l)} R_m K_{h} (\X_m-\x_k)\right\}+O_p(N^{-1/2})+ O_p(N^{1/2}h^m)\n\\
&=& T_{2212ABA}+ T_{2212ABB} -T_{2212ABC}+O_p(N^{-1/2})+ O_p(N^{1/2}h^m).\n
\ee

Consider $T_{2212ABA}$,
\be \label{ustat:t2212aba}
&&T_{2212ABA}\n\\
&=& N^{-6/2} \sumI \sumJ \sumK   \frac{\{1-r_i
  \pi^{-1}(y_i,\u_i;\bb,g^*)\} r_j K_h({\x_j-\x_i}) e^{-h(y_j)} r_k K_h({\u_k-\u_i})\Delta_{3k}}{d^*(\x_i) D_i V_{i,k} \{E(R_k\mid\x_k) f_{\X}(\x_k)\}^2}\n\\
&&\times  N^{-1/2}  \sumL E \left\{r_l  K_{h}(\x_l-\x_k) e^{-h(y_l)} R_m K_{h} (\X_m-\x_k)\mid \x_l,r_l, r_l y_l \right\}\n\\
&=& N^{-7/2} \sumI \sumJ \sumK  \sumL \frac{\{1-r_i
  \pi^{-1}(y_i,\u_i;\bb,g^*)\} r_j K_h({\x_j-\x_i}) e^{-h(y_j)} r_k K_h({\u_k-\u_i})\Delta_{3k}}{d^*(\x_i) D_i V_{i,k} E(R_k\mid\x_k) f_{\X}(\x_k)}\n\\
&&\times r_l  K_{h}(\x_l-\x_k) e^{-h(y_l)}
+O_p(N^{1/2}h^m)\n\\
&=&T_{2211A},\n
\ee
where $T_{2211A}$ is defined in (\ref{ustat:t2211a}).

Consider $T_{2212ABB}$,
\be \label{ustat:t2212abb}
&&T_{2212ABB}\n\\
&=& N^{-6/2} \sumI \sumJ \sumK   \frac{\{1-r_i
  \pi^{-1}(y_i,\u_i;\bb,g^*)\} r_j K_h({\x_j-\x_i}) e^{-h(y_j)} r_k K_h({\u_k-\u_i})\Delta_{3k}}{d^*(\x_i) D_i V_{i,k} \{E(R_k\mid\x_k) f_{\X}(\x_k)\}^2}\n\\
&&\times  N^{-1/2}  \sumM E \left\{R_l  K_{h}(\X_l-\x_k) e^{-h(Y_l)}r_m K_{h} (\x_m-\x_k)\mid \x_m,r_m, r_m y_m \right\}\n\\
&=&N^{-7/2} \sumI \sumJ \sumK \sumM    \frac{\{1-r_i
  \pi^{-1}(y_i,\u_i;\bb,g^*)\} r_j K_h({\x_j-\x_i}) e^{-h(y_j)} r_k K_h({\u_k-\u_i})}{d^*(\x_i) D_i V_{i,k} E(R_k\mid\x_k) f_{\X}(\x_k)}\n\\
&&\times  r_m K_{h} (\x_m-\x_k) \Delta_{1k} \Delta_{3k}
+O_p(N^{1/2}h^m)\n\\
&=&N^{-7/2} \sumI \sumJ \sumK \sumL    \frac{\{1-r_i
  \pi^{-1}(y_i,\u_i;\bb,g^*)\} r_j K_h({\x_j-\x_i}) e^{-h(y_j)} r_k K_h({\u_k-\u_i})}{d^*(\x_i) D_i V_{i,k} E(R_k\mid\x_k) f_{\X}(\x_k)}\n\\
&&\times  r_l K_{h} (\x_l-\x_k) \Delta_{1k} \Delta_{3k}
+O_p(N^{1/2}h^m).
\ee

Consider $T_{2212ABC}$,
\be \label{ustat:t2212abc}
&&T_{2212ABC}\n\\
&=& N^{-6/2} \sumI \sumJ \sumK   \frac{\{1-r_i
  \pi^{-1}(y_i,\u_i;\bb,g^*)\} r_j K_h({\x_j-\x_i}) e^{-h(y_j)} r_k K_h({\u_k-\u_i})\Delta_{3k}}{d^*(\x_i) D_i V_{i,k} \{E(R_k\mid\x_k) f_{\X}(\x_k)\}^2}\n\\
&&\times  N^{1/2}  E \left\{R_l  K_{h}(\X_l-\x_k) e^{-h(Y_l)} R_m K_{h} (\X_m-\x_k)\right\}\n\\
&=& N^{-5/2} \sumI \sumJ \sumK   \frac{\{1-r_i
  \pi^{-1}(y_i,\u_i;\bb,g^*)\} r_j K_h({\x_j-\x_i}) e^{-h(y_j)} r_k K_h({\u_k-\u_i})}{d^*(\x_i) D_i V_{i,k}}\n\\
&&\times  \Delta_{1k} \Delta_{3k}
+O_p(N^{1/2}h^m)\n\\
&=&T_{2211C},\n
\ee
where $T_{2211C}$ is defined in (\ref{ustat:t2211c}).

Putting all the terms together (cancelling and collecting similar terms) and rewriting (\ref{eq:final}) we get the following,
\be \label{eq:final2}
&&\frac{1}{\sqrt{N}}\sumI\wh\S^*\eff(\x_i,r_i,r_i y_i,\bb)\n\\
&=& \frac{2}{\sqrt{N}}\sumI \frac{A_i}{d^*(\x_i)} +T_{21}+2T_{2211}-T_{2212}-T_{2221}-2T_{2222}+2T_{2223}-2T_{231}+T_{232}-T_{31}\n\\
&&-T_{321}-2T_{322}+2T_{323}+2T_{331}-T_{332}-2T_{11}+T_{12}+O_p(N^{1/2}h^{2m}+N^{-1/2}h^{-p})\n\\
&=& \frac{2}{\sqrt{N}}\sumI \frac{A_i}{d^*(\x_i)} +2T_{21}+T_{2211A}+T_{2211B}+T_{2211C}-T_{2221}-2T_{2222}+2T_{2223}\n\\
&&-T_{231C}-T_{231AB}-T_{231BB}+T_{232BBB}-T_{31}-2T_{2212ABB}\n\\
&&-T_{321}-2T_{322}-T_{332B}+T_{332C}+2T_{323}+T_{331}-2T_{11}+T_{12}\n\\
&&+O_p(N^{1/2}h^{2m}+N^{-1/2}h^{-p}+N^{-1/2}+N^{1/2}h^m),
\ee
where
$T_{21}$ is from  (\ref{eq:t21}), $T_{2221}$ is from
(\ref{eq:t2221}), $T_{2222}$ is from  (\ref{eq:t2222}), $T_{2211A}$ is
from  (\ref{ustat:t2211a}), $T_{2211B}$ is from  (\ref{ustat:t2211b}),
$T_{2211C}$ is from  (\ref{ustat:t2211c}), $T_{2212ABB}$ is from  (\ref{ustat:t2212abb}),$T_{2223}$ is from
(\ref{eq:t2223}),  $T_{231AB}$ is from  (\ref{ustat:t231ab}),
$T_{231AC}$ is from  (\ref{ustat:t231ac}), $T_{231BB}$ is from
(\ref{ustat:t231bb}), $T_{232BBB}$ is from  (\ref{ustat:t232bbb}),
$T_{31}$ is from  (\ref{eq:t31}), $T_{331}$ is from  (\ref{eq:t331}),
$T_{332B}$ is from  (\ref{ustat:t332b}), $T_{332C}$ is from
(\ref{ustat:t332c}), $T_{321}$ is from  (\ref{eq:t321}), $T_{322}$ is
from  (\ref{eq:t322}), $T_{323}$ is from  (\ref{eq:t323}), $T_{11}$ is
from  (\ref{eq:t11}), and $T_{12}$ is from  (\ref{eq:t12}).

Consider the term $T_{2222}$ from  (\ref{eq:t2222}) and using U-statistic property we get,
\be \label{ustat:t2222}
&&T_{2222}\n\\
&=&N^{-7/2} \sumI \sumJ \sumK \sumL \frac{\{1-r_i
  \pi^{-1}(y_i,\u_i;\bb,g^*)\} r_j K_h({\x_j-\x_i}) e^{-h(y_j)} r_k K_h({\u_k-\u_i})\h'_\bb(y_j;\bb)}{d^*(\x_i) D_i V_{i,k}}\n\\
&&\times \frac{ r_l  K_{h}(\x_l-\x_k) e^{-h(y_l)} \Delta_{1k} }{E(R_k \mid \x_k) f_{\X}(\x_k)}\n\\
&=&N^{-4/2} \sumI  \sumK \frac{\{1-r_i
  \pi^{-1}(y_i,\u_i;\bb,g^*)\}  r_k K_h({\u_k-\u_i}) \Delta_{1k}}{d^*(\x_i) D_i V_{i,k}E(R_k \mid \x_k) f_{\X}(\x_k)}\n\\
&&\times N^{-1/2} \sumJ E\{ r_j K_h({\x_j-\x_i}) e^{-h(y_j)} \h'_\bb(y_j;\bb) R_l  K_{h}(\X_l-\x_k) e^{-h(Y_l)} \mid \x_j, r_j, r_j y_j\} \n\\
&&+ N^{-4/2} \sumI  \sumK \frac{\{1-r_i
  \pi^{-1}(y_i,\u_i;\bb,g^*)\}  r_k K_h({\u_k-\u_i}) \Delta_{1k}}{d^*(\x_i) D_i V_{i,k}E(R_k \mid \x_k) f_{\X}(\x_k)}\n\\
&&\times N^{-1/2} \sumL E\{ R_j K_h({\X_j-\x_i}) e^{-h(Y_j)} \h'_\bb(Y_j;\bb) r_l  K_{h}(\x_l-\x_k) e^{-h(y_l)} \mid \x_l, r_l, r_l y_l\} \n\\
&&- N^{-4/2} \sumI  \sumK \frac{\{1-r_i
  \pi^{-1}(y_i,\u_i;\bb,g^*)\}  r_k K_h({\u_k-\u_i}) \Delta_{1k}}{d^*(\x_i) D_i V_{i,k}E(R_k \mid \x_k) f_{\X}(\x_k)}\n\\
&&\times N^{1/2} E\{ R_j K_h({\X_j-\x_i}) e^{-h(Y_j)} \h'_\bb(Y_j;\bb) R_l  K_{h}(\X_l-\x_k) e^{-h(Y_l)}\}+O_p(N^{-1/2})\n\\
&=&T_{2222A}+T_{2222B}-T_{2222C}+O_p(N^{-1/2}).\n
\ee

Consider $T_{2222A}$,
\be \label{ustat:t2222a}
&&T_{2222A}\n\\
&=&N^{-4/2} \sumI  \sumK \frac{\{1-r_i
  \pi^{-1}(y_i,\u_i;\bb,g^*)\}  r_k K_h({\u_k-\u_i}) \Delta_{1k}}{d^*(\x_i) D_i V_{i,k}E(R_k \mid \x_k) f_{\X}(\x_k)}\n\\
&&\times N^{-1/2} \sumJ E\{ r_j K_h({\x_j-\x_i}) e^{-h(y_j)} \h'_\bb(y_j;\bb) R_l  K_{h}(\X_l-\x_k) e^{-h(Y_l)} \mid \x_j, r_j, r_j y_j\} \n\\
&=&N^{-5/2} \sumI \sumJ  \sumK \frac{\{1-r_i
  \pi^{-1}(y_i,\u_i;\bb,g^*)\} r_j K_h({\x_j-\x_i}) e^{-h(y_j)} \h'_\bb(y_j;\bb)  r_k K_h({\u_k-\u_i}) \Delta_{1k}^2}{d^*(\x_i) D_i V_{i,k}}\n\\
&&+O_p(N^{1/2}h^m)\n\\
&=& T_{2221} +O_p(N^{1/2}h^m),\n
\ee
where $T_{2221}$ is defined in (\ref{eq:t2221}).

Consider $T_{2222B}$,
\be \label{ustat:t2222b}
&&T_{2222B}\n\\
&=&N^{-4/2} \sumI  \sumK \frac{\{1-r_i
  \pi^{-1}(y_i,\u_i;\bb,g^*)\}  r_k K_h({\u_k-\u_i}) \Delta_{1k}}{d^*(\x_i) D_i V_{i,k}E(R_k \mid \x_k) f_{\X}(\x_k)}\n\\
&&\times N^{-1/2} \sumL E\{ R_j K_h({\X_j-\x_i}) e^{-h(Y_j)} \h'_\bb(Y_j;\bb) r_l  K_{h}(\x_l-\x_k) e^{-h(y_l)} \mid \x_l, r_l, r_l y_l\} \n\\
&=&N^{-5/2} \sumI \sumK \sumL  \frac{\{1-r_i
  \pi^{-1}(y_i,\u_i;\bb,g^*)\}  r_k K_h({\u_k-\u_i}) \Delta_{1k} r_l  K_{h}(\x_l-\x_k) e^{-h(y_l)} \Delta_{3i}}{d^*(\x_i) D_i d^*(\x_k) E(R_i \mid \u_i) f_{\U}(\u_i) E(R_k \mid \x_k) f_{\X}(\x_k)}\n\\
&&+O_p(N^{1/2}h^m)\n\\
&=&N^{-5/2} \sumI \sumJ \sumK  \frac{\{1-r_i
  \pi^{-1}(y_i,\u_i;\bb,g^*)\}  r_j K_h({\u_j-\u_i})  r_k  K_{h}(\x_k-\x_j) e^{-h(y_k)} \Delta_{1j} \Delta_{3i}}{d^*(\x_i) D_i d^*(\x_j) E(R_i \mid \u_i) f_{\U}(\u_i) E(R_j \mid \x_j) f_{\X}(\x_j)}\n\\
&&+O_p(N^{1/2}h^m).
\ee

Consider $T_{2222C}$,
\be \label{ustat:t2222c}
&&T_{2222C}\n\\
&=&N^{-4/2} \sumI  \sumK \frac{\{1-r_i
  \pi^{-1}(y_i,\u_i;\bb,g^*)\}  r_k K_h({\u_k-\u_i}) \Delta_{1k}}{d^*(\x_i) D_i V_{i,k}E(R_k \mid \x_k) f_{\X}(\x_k)}\n\\
&&\times N^{1/2} E\{ R_j K_h({\X_j-\x_i}) e^{-h(Y_j)} \h'_\bb(Y_j;\bb) R_l  K_{h}(\X_l-\x_k) e^{-h(Y_l)}\} \n\\
&=&N^{-3/2} \sumI  \sumK \frac{\{1-r_i
  \pi^{-1}(y_i,\u_i;\bb,g^*)\}  r_k K_h({\u_k-\u_i}) \Delta_{1k}}{d^*(\x_i) D_i V_{i,k}E(R_k \mid \x_k) f_{\X}(\x_k)}\n\\
&&\times   \Delta_{3i} E(R_i \mid \x_i) f_{\X}(\x_i)  \Delta_{1k} E(R_k \mid \x_k) f_{\X}(\x_k)
+O_p(N^{1/2}h^m)\n\\
&=&N^{-3/2} \sumI  \sumK \frac{\{1-r_i
  \pi^{-1}(y_i,\u_i;\bb,g^*)\}  r_k K_h({\u_k-\u_i}) \Delta_{1k}^2  \Delta_{3i} }{d^*(\x_i) D_i d^*(\x_k) E(R_i \mid \u_i) f_{\U}(\u_i) }
+O_p(N^{1/2}h^m)\n\\
&=&N^{-3/2} \sumI  \sumJ \frac{\{1-r_i
  \pi^{-1}(y_i,\u_i;\bb,g^*)\}  r_j K_h({\u_j-\u_i}) \Delta_{1j}^2  \Delta_{3i} }{d^*(\x_i) D_i d^*(\x_j) E(R_i \mid \u_i) f_{\U}(\u_i) }
+O_p(N^{1/2}h^m).
\ee

Consider the term $T_{2223}$ in (\ref{eq:t2223}),
\be \label{ustat:t2223}
&&T_{2223}\n\\
&=&  N^{-7/2} \sumI \sumJ \sumK \sumL \frac{\{1-r_i
  \pi^{-1}(Y_i,\u_i;\bb,g^*)\} r_j K_h({\x_j-\x_i}) e^{-h(y_j)} r_k K_h({\u_k-\u_i})\h'_\bb(y_j;\bb)}{d^*(\x_i) D_i V_{i,k}}\n\\
&&\times \frac{r_l K_{h}(\x_l-\x_k) \Delta_{1k}^2
}{E(R_k \mid \x_k)  f_{\X}(\x_k)}\n\\
&=&  N^{-4/2} \sumI  \sumK  \frac{\{1-r_i
  \pi^{-1}(Y_i,\u_i;\bb,g^*)\}  r_k K_h({\u_k-\u_i})\Delta_{1k}^2}{d^*(\x_i) D_i V_{i,k} E(R_k \mid \x_k)  f_{\X}(\x_k)}\n\\
&&\times N^{-1/2}\sumJ E\{ r_j K_h({\x_j-\x_i}) e^{-h(y_j)} \h'_\bb(y_j;\bb)  R_l K_{h}(\X_l-\x_k) \mid \x_j, r_j, r_j y_j\}
\n\\
&&+  N^{-4/2} \sumI  \sumK  \frac{\{1-r_i
  \pi^{-1}(Y_i,\u_i;\bb,g^*)\}  r_k K_h({\u_k-\u_i})\Delta_{1k}^2}{d^*(\x_i) D_i V_{i,k} E(R_k \mid \x_k)  f_{\X}(\x_k)}\n\\
&&\times N^{-1/2}\sumL E\{ R_j K_h({\X_j-\x_i}) e^{-h(Y_j)} \h'_\bb(Y_j;\bb)   r_l K_{h}(\x_l-\x_k) \mid \x_l, r_l, r_l y_l\}
\n\\
&&-  N^{-4/2} \sumI  \sumK  \frac{\{1-r_i
  \pi^{-1}(Y_i,\u_i;\bb,g^*)\}  r_k K_h({\u_k-\u_i})\Delta_{1k}^2}{d^*(\x_i) D_i V_{i,k} E(R_k \mid \x_k)  f_{\X}(\x_k)}\n\\
&&\times N^{1/2} E\{ R_j K_h({\X_j-\x_i}) e^{-h(Y_j)} \h'_\bb(Y_j;\bb)  R_l K_{h}(\X_l-\x_k) \}
+O_p(N^{-1/2})\n\\
&=&T_{2223A}+T_{2223B}-T_{2223C}+O_p(N^{-1/2}).\n
\ee

Consider $T_{2223A}$,
\be \label{ustat:t2223a}
&&T_{2223A}\n\\
&=&  N^{-4/2} \sumI  \sumK  \frac{\{1-r_i
  \pi^{-1}(Y_i,\u_i;\bb,g^*)\}  r_k K_h({\u_k-\u_i})\Delta_{1k}^2}{d^*(\x_i) D_i V_{i,k} E(R_k \mid \x_k)  f_{\X}(\x_k)}\n\\
&&\times N^{-1/2}\sumJ E\{ r_j K_h({\x_j-\x_i}) e^{-h(y_j)} \h'_\bb(y_j;\bb)  R_l K_{h}(\X_l-\x_k) \mid \x_j, r_j, r_j y_j\}
\n\\
&=&  N^{-5/2} \sumI \sumJ \sumK  \frac{\{1-r_i
  \pi^{-1}(Y_i,\u_i;\bb,g^*)\} r_j K_h({\x_j-\x_i}) e^{-h(y_j)} \h'_\bb(y_j;\bb)  r_k K_h({\u_k-\u_i})\Delta_{1k}^2}{d^*(\x_i) D_i V_{i,k}}\n\\
&&+O_p(N^{1/2}h^m)\n\\
&=&T_{2221}+O_p(N^{1/2}h^m),\n
\ee
where $T_{2221}$ is defined in  (\ref{eq:t2221}).

Consider $T_{2223B}$,
\be \label{ustat:t2223b}
&&T_{2223B}\n\\
&=&N^{-4/2} \sumI  \sumK  \frac{\{1-r_i
  \pi^{-1}(Y_i,\u_i;\bb,g^*)\}  r_k K_h({\u_k-\u_i})\Delta_{1k}^2}{d^*(\x_i) D_i V_{i,k} E(R_k \mid \x_k)  f_{\X}(\x_k)}\n\\
&&\times N^{-1/2}\sumL E\{ R_j K_h({\X_j-\x_i}) e^{-h(Y_j)} \h'_\bb(Y_j;\bb)   r_l K_{h}(\x_l-\x_k) \mid \x_l, r_l, r_l y_l\}
\n\\
&=&N^{-5/2} \sumI  \sumK \sumL \frac{\{1-r_i
  \pi^{-1}(Y_i,\u_i;\bb,g^*)\}  r_k K_h({\u_k-\u_i})\Delta_{1k}^2}{d^*(\x_i)
  D_i
d^*(\x_k)E(R_i\mid\u_i)f_\U(\u_i)
 E(R_k \mid \x_k)  f_{\X}(\x_k)}\n\\
&&\times r_l K_{h}(\x_l-\x_k) \Delta_{3i}
+O_p(N^{1/2}h^m)\n\\
&=&N^{-5/2} \sumI  \sumJ \sumK \frac{\{1-r_i
  \pi^{-1}(Y_i,\u_i;\bb,g^*)\}  r_j K_h({\u_j-\u_i})\Delta_{1j}^2 r_k K_{h}(\x_k-\x_j) \Delta_{3i}}{d^*(\x_i) D_i d^*(\x_j) E(R_i \mid \u_i)  f_{\U}(\u_i)  E(R_j \mid \x_j)  f_{\X}(\x_j)}\n\\
&&+O_p(N^{1/2}h^m).
\ee

Consider $T_{2223C}$,
\be \label{ustat:t2223c}
&&T_{2223C}\n\\
&=& N^{-4/2} \sumI  \sumK  \frac{\{1-r_i
  \pi^{-1}(Y_i,\u_i;\bb,g^*)\}  r_k K_h({\u_k-\u_i})\Delta_{1k}^2}{d^*(\x_i) D_i V_{i,k} E(R_k \mid \x_k)  f_{\X}(\x_k)}\n\\
&&\times N^{1/2} E\{ R_j K_h({\X_j-\x_i}) e^{-h(Y_j)} \h'_\bb(Y_j;\bb)  R_l K_{h}(\X_l-\x_k) \} \n\\
&=& N^{-3/2} \sumI  \sumK  \frac{\{1-r_i
  \pi^{-1}(Y_i,\u_i;\bb,g^*)\}  r_k K_h({\u_k-\u_i})\Delta_{1k}^2}{d^*(\x_i) D_i V_{i,k} E(R_k \mid \x_k)  f_{\X}(\x_k)}\n\\
&&\times \Delta_{3i}  E(R_i \mid \x_i)  f_{\X}(\x_i)  E(R_k \mid \x_k)  f_{\X}(\x_k)
+O_p(N^{1/2}h^m)\n\\
&=& N^{-3/2} \sumI  \sumK  \frac{\{1-r_i
  \pi^{-1}(Y_i,\u_i;\bb,g^*)\}  r_k K_h({\u_k-\u_i})\Delta_{1k}^2 \Delta_{3i}}{d^*(\x_i) D_i d^*(\x_k) E(R_i \mid \u_i)  f_{\U}(\u_i)}
+O_p(N^{1/2}h^m)\n\\
&=& N^{-3/2} \sumI  \sumJ  \frac{\{1-r_i
  \pi^{-1}(Y_i,\u_i;\bb,g^*)\}  r_j K_h({\u_j-\u_i})\Delta_{1j}^2 \Delta_{3i}}{d^*(\x_i) D_i d^*(\x_j) E(R_i \mid \u_i)  f_{\U}(\u_i)}
+O_p(N^{1/2}h^m)\n\\
&=& T_{2222C} +O_p(N^{1/2}h^m),\n
\ee
where $T_{2222C}$ is defined in  (\ref{ustat:t2222c}).

Consider $T_{332B}$ in (\ref{ustat:t332b}) and applying U-statistics we get,
\be \label{ustat:t332b2}
&&T_{332B}\n\\
&=&N^{-7/2} \sumI \sumJ \sumK \sumL \frac{\{1-r_i  \pi^{-1}(y_i,\u_i;\bb,g^*)\}C_i r_j  K_{h} (\u_j-\u_i)r_k K_{h} (\u_k-\u_i)  \Delta^2_{1j}}{d^*(\x_i) D_i^2 d^* (\x_j) \{E(R_i\mid\u_i) f_{\U}(\u_i)\}^2 \{E(R_j\mid\x_j) f_{\X}(\x_j)\}} \n\\
&& \times r_l K_{h} (\x_l-\x_j)
+ O_p(N^{1/2}h^m)\n\\
&=&N^{-4/2} \sumI \sumJ  \frac{\{1-r_i  \pi^{-1}(y_i,\u_i;\bb,g^*)\}C_i r_j  K_{h} (\u_j-\u_i)  \Delta^2_{1j} }{d^*(\x_i) D_i^2 d^* (\x_j) \{E(R_i\mid\u_i) f_{\U}(\u_i)\}^2 \{E(R_j\mid\x_j) f_{\X}(\x_j)\}} \n\\
&& \times N^{-1/2} \sumK  E \{ R_l K_{h} (\X_l-\x_j) r_k K_{h} (\u_k-\u_i) \mid \x_k, r_k, r_k y_k \} \n\\
&&+N^{-4/2} \sumI \sumJ  \frac{\{1-r_i  \pi^{-1}(y_i,\u_i;\bb,g^*)\}C_i r_j  K_{h} (\u_j-\u_i)  \Delta^2_{1j}  }{d^*(\x_i) D_i^2 d^* (\x_j) \{E(R_i\mid\u_i) f_{\U}(\u_i)\}^2 \{E(R_j\mid\x_j) f_{\X}(\x_j)\}} \n\\
&& \times N^{-1/2}  \sumL E\{ r_l K_{h} (\x_l-\x_j)  R_k K_{h} (\U_k-\u_i)\mid \x_l, r_l, r_l y_l \} \n\\
&&-N^{-4/2} \sumI \sumJ  \frac{\{1-r_i  \pi^{-1}(y_i,\u_i;\bb,g^*)\}C_i r_j  K_{h} (\u_j-\u_i)  \Delta^2_{1j} }{d^*(\x_i) D_i^2 d^* (\x_j) \{E(R_i\mid\u_i) f_{\U}(\u_i)\}^2 \{E(R_j\mid\x_j) f_{\X}(\x_j)\}} \n\\
&&  \times N^{1/2}  E\{ R_l K_{h} (\X_l-\x_j) R_k K_{h} (\U_k-\u_i)\}
+O_p(N^{-1/2})+ O_p(N^{1/2}h^m)\n\\
&=&T_{332BA}+T_{332BB}-T_{332BC}+O_p(N^{-1/2})+ O_p(N^{1/2}h^m).\n
\ee

Consider $T_{332BA}$,
\be \label{ustat:t332ba}
&&T_{332BA}\n\\
&=&N^{-4/2} \sumI \sumJ  \frac{\{1-r_i  \pi^{-1}(y_i,\u_i;\bb,g^*)\}C_i r_j  K_{h} (\u_j-\u_i)  \Delta^2_{1j}}{d^*(\x_i) D_i^2 d^* (\x_j) \{E(R_i\mid\u_i) f_{\U}(\u_i)\}^2 \{E(R_j\mid\x_j) f_{\X}(\x_j)\}} \n\\
&& \times N^{-1/2} \sumK  E \{ R_l K_{h} (\X_l-\x_j) r_k K_{h} (\u_k-\u_i) \mid \x_k, r_k, r_k y_k \} \n\\
&=&N^{-5/2} \sumI \sumJ  \sumK  \frac{\{1-r_i  \pi^{-1}(y_i,\u_i;\bb,g^*)\}C_i r_j  K_{h} (\u_j-\u_i)  \Delta^2_{1j} r_k K_{h} (\u_k-\u_i) }{d^*(\x_i) D_i^2 d^* (\x_j) \{E(R_i\mid\u_i) f_{\U}(\u_i)\}^2 } \n\\
&&+ O_p(N^{1/2}h^m).
\ee

Consider $T_{332BB}$,
\be \label{ustat:t332bb}
&&T_{332BB}\n\\
&=&N^{-4/2} \sumI \sumJ  \frac{\{1-r_i  \pi^{-1}(y_i,\u_i;\bb,g^*)\}C_i r_j  K_{h} (\u_j-\u_i)  \Delta^2_{1j} }{d^*(\x_i) D_i^2 d^* (\x_j) \{E(R_i\mid\u_i) f_{\U}(\u_i)\}^2 \{E(R_j\mid\x_j) f_{\X}(\x_j)\}} \n\\
&& \times N^{-1/2}  \sumL E\{ r_l K_{h} (\x_l-\x_j)  R_k K_{h} (\U_k-\u_i)\mid \x_l, r_l, r_l y_l \} \n\\
&=&N^{-5/2} \sumI \sumJ   \sumL \frac{\{1-r_i  \pi^{-1}(y_i,\u_i;\bb,g^*)\}C_i r_j  K_{h} (\u_j-\u_i)  \Delta^2_{1j}  }{d^*(\x_i) D_i^2 d^* (\x_j)E(R_i\mid\u_i) f_{\U}(\u_i) \{E(R_j\mid\x_j) f_{\X}(\x_j)\}} \n\\
&& \times r_l K_{h} (\x_l-\x_j)
+ O_p(N^{1/2}h^m)\n\\
&=&N^{-5/2} \sumI \sumJ   \sumK \frac{\{1-r_i  \pi^{-1}(y_i,\u_i;\bb,g^*)\}C_i r_j  K_{h} (\u_j-\u_i)  \Delta^2_{1j} }{d^*(\x_i) D_i^2 d^* (\x_j) E(R_i\mid\u_i) f_{\U}(\u_i) \{E(R_j\mid\x_j) f_{\X}(\x_j)\}} \n\\
&& \times r_k K_{h} (\x_k-\x_j)
+ O_p(N^{1/2}h^m)\n\\
&=&T_{323}+ O_p(N^{1/2}h^m),\n
\ee
where $T_{323}$ is as defined in (\ref{eq:t323}).

Consider $T_{332BC}$,
\be \label{ustat:t332bc}
&&T_{332BC}\n\\
&=&N^{-4/2} \sumI \sumJ  \frac{\{1-r_i  \pi^{-1}(y_i,\u_i;\bb,g^*)\}C_i r_j  K_{h} (\u_j-\u_i)  \Delta^2_{1j}  }{d^*(\x_i) D_i^2 d^* (\x_j) \{E(R_i\mid\u_i) f_{\U}(\u_i)\}^2 \{E(R_j\mid\x_j) f_{\X}(\x_j)\}} \n\\
&&  \times N^{1/2}  E\{ R_l K_{h} (\X_l-\x_j) R_k K_{h} (\U_k-\u_i)\} \n\\
&=&N^{-3/2} \sumI \sumJ  \frac{\{1-r_i  \pi^{-1}(y_i,\u_i;\bb,g^*)\}C_i r_j  K_{h} (\u_j-\u_i)  \Delta^2_{1j} }{d^*(\x_i) D_i^2 d^* (\x_j) E(R_i\mid\u_i) f_{\U}(\u_i) }
+ O_p(N^{1/2}h^m)\n\\
&=& T_{31}+ O_p(N^{1/2}h^m),\n
\ee
where $T_{31}$ is as defined in (\ref{eq:t31}).

Applying U-statistics property to $T_{2211A}$ from (\ref{ustat:t2211a}),
\be \label{ustat:t2211a2}
&&T_{2211A}\n\\
&=& N^{-7/2} \sumI \sumJ\sumK \sumL \frac{\{1-r_i
  \pi^{-1}(y_i,\u_i;\bb,g^*)\} r_j K_h({\x_j-\x_i}) e^{-h(y_j)} r_k K_h({\u_k-\u_i})}{d^*(\x_i) D_i V_{i,k} E(R_k\mid\x_k) f_{\X}(\x_k)}\n\\
&&\times r_l  K_{h}(\x_l-\x_k) e^{-h(y_l)} \Delta_{3k}
+ O_p(N^{1/2}h^m)\n\\
&=& N^{-4/2} \sumI  \sumK \frac{\{1-r_i
  \pi^{-1}(y_i,\u_i;\bb,g^*)\}  r_k K_h({\u_k-\u_i})\Delta_{3k} }{d^*(\x_i) D_iV_{i,k} E(R_k\mid\x_k) f_{\X}(\x_k) }\n\\
&&\times N^{-1/2}  \sumJ  E\{ r_j K_h({\x_j-\x_i}) e^{-h(y_j)}  R_l  K_{h}(\X_l-\x_k) e^{-h(Y_l)} \mid \x_j,r_j, r_j y_j\}\n\\
&&+ N^{-4/2} \sumI  \sumK \frac{\{1-r_i
  \pi^{-1}(y_i,\u_i;\bb,g^*)\}  r_k K_h({\u_k-\u_i})\Delta_{3k} }{d^*(\x_i) D_iV_{i,k} E(R_k\mid\x_k) f_{\X}(\x_k) }\n\\
&&\times N^{-1/2}  \sumL  E\{ R_j K_h({\X_j-\x_i}) e^{-h(Y_j)} r_l  K_{h}(\x_l-\x_k) e^{-h(y_l)} \mid \x_l,r_l, r_l y_l\}\n\\
&&- N^{-4/2} \sumI  \sumK \frac{\{1-r_i
  \pi^{-1}(y_i,\u_i;\bb,g^*)\}  r_k K_h({\u_k-\u_i})\Delta_{3k} }{d^*(\x_i) D_iV_{i,k} E(R_k\mid\x_k) f_{\X}(\x_k) }\n\\
&&\times N^{1/2}  E\{ R_j K_h({\X_j-\x_i}) e^{-h(Y_j)} R_l  K_{h}(\X_l-\x_k) e^{-h(Y_l)}\}
+ O_p(N^{1/2}h^m)+O_p(N^{-1/2})\n\\
&=&T_{2211AA}+T_{2211AB}-T_{2211AC}+ O_p(N^{1/2}h^m)+O_p(N^{-1/2}).\n
\ee

Consider $T_{2211AA}$,
\be \label{ustat:t2211aa}
&&T_{2211AA}\n\\
&=&N^{-4/2} \sumI  \sumK \frac{\{1-r_i
  \pi^{-1}(y_i,\u_i;\bb,g^*)\}  r_k K_h({\u_k-\u_i})\Delta_{3k} }{d^*(\x_i) D_iV_{i,k} E(R_k\mid\x_k) f_{\X}(\x_k) }\n\\
&&\times N^{-1/2}  \sumJ  E\{ r_j K_h({\x_j-\x_i}) e^{-h(y_j)}  R_l  K_{h}(\X_l-\x_k) e^{-h(Y_l)} \mid \x_j,r_j, r_j y_j\}\n\\
&=&N^{-5/2} \sumI \sumJ \sumK \frac{\{1-r_i
  \pi^{-1}(y_i,\u_i;\bb,g^*)\} r_j K_h({\x_j-\x_i}) e^{-h(y_j)}  r_k K_h({\u_k-\u_i})\Delta_{3k} }{d^*(\x_i) D_i V_{i,k} E(R_k\mid\x_k) f_{\X}(\x_k) }\n\\
&&\times \Delta_{1k} E(R_k\mid\x_k) f_{\X}(\x_k)
+ O_p(N^{1/2}h^m)\n\\
&=&N^{-5/2} \sumI \sumJ \sumK \frac{\{1-r_i
  \pi^{-1}(y_i,\u_i;\bb,g^*)\} r_j K_h({\x_j-\x_i}) e^{-h(y_j)}  r_k K_h({\u_k-\u_i}) \Delta_{1k} \Delta_{3k} }{d^*(\x_i) D_i V_{i,k}}\n\\
&&+ O_p(N^{1/2}h^m)\n\\
&=&T_{2211C},\n
\ee
where $T_{2211C}$ is as defined in (\ref{ustat:t2211c}).

Consider $T_{2211AB}$,
\be \label{ustat:t2211ab}
&&T_{2211AB}\n\\
&=& N^{-4/2} \sumI  \sumK \frac{\{1-r_i
  \pi^{-1}(y_i,\u_i;\bb,g^*)\}  r_k K_h({\u_k-\u_i})\Delta_{3k} }{d^*(\x_i) D_iV_{i,k} E(R_k\mid\x_k) f_{\X}(\x_k) }\n\\
&&\times N^{-1/2}  \sumL  E\{ R_j K_h({\X_j-\x_i}) e^{-h(Y_j)} r_l  K_{h}(\x_l-\x_k) e^{-h(y_l)} \mid \x_l,r_l, r_l y_l\}\n\\
&=& N^{-5/2} \sumI  \sumK \sumL \frac{\{1-r_i
  \pi^{-1}(y_i,\u_i;\bb,g^*)\}  r_k K_h({\u_k-\u_i})\Delta_{3k} r_l  K_{h}(\x_l-\x_k) e^{-h(y_l)} }{d^*(\x_i) D_i V_{i,k} E(R_k\mid\x_k) f_{\X}(\x_k) }\n\\
&&\times \Delta_{1i} E(R_i\mid\x_i) f_{\X}(\x_i)
+ O_p(N^{1/2}h^m)\n\\
&=& N^{-5/2} \sumI  \sumJ \sumK \frac{\{1-r_i
  \pi^{-1}(y_i,\u_i;\bb,g^*)\}  r_j K_h({\u_j-\u_i})   r_k  K_{h}(\x_k-\x_j) e^{-h(y_k)} \Delta_{1i} \Delta_{3j} }{d^*(\x_i) D_i d^*(\x_j)E(R_i\mid\u_i) f_{\U}(\u_i) E(R_j\mid\x_j) f_{\X}(\x_j) }\n\\
&&+ O_p(N^{1/2}h^m).
\ee

Consider $T_{2211AC}$,
\be \label{ustat:t2211ac}
&&T_{2211AC}\n\\
&=&N^{-4/2} \sumI  \sumK \frac{\{1-r_i
  \pi^{-1}(y_i,\u_i;\bb,g^*)\}  r_k K_h({\u_k-\u_i})\Delta_{3k} }{d^*(\x_i) D_iV_{i,k} E(R_k\mid\x_k) f_{\X}(\x_k) }\n\\
&&\times N^{1/2}  E\{ R_j K_h({\X_j-\x_i}) e^{-h(Y_j)} R_l  K_{h}(\X_l-\x_k) e^{-h(Y_l)}\}\n\\
&=&N^{-3/2} \sumI  \sumK \frac{\{1-r_i
  \pi^{-1}(y_i,\u_i;\bb,g^*)\}  r_k K_h({\u_k-\u_i}) \Delta_{1i} \Delta_{1k}\Delta_{3k} }{d^*(\x_i) D_i d^*(\x_k) E(R_i\mid\u_i) f_{\U}(\u_i)   }
+ O_p(N^{1/2}h^m)\n\\
&=&N^{-3/2} \sumI  \sumJ \frac{\{1-r_i
  \pi^{-1}(y_i,\u_i;\bb,g^*)\}  r_j K_h({\u_j-\u_i}) \Delta_{1i} \Delta_{1j}\Delta_{3j}}{d^*(\x_i) D_i d^*(\x_j) E(R_i\mid\u_i) f_{\U}(\u_i)   }\n\\
&&+ O_p(N^{1/2}h^m).
\ee

Applying U-statistics property to $T_{2211B}$ from  (\ref{ustat:t2211b}),
\be \label{ustat:t2211b2}
&&T_{2211B} \n\\
&=& N^{-7/2} \sumI \sumJ \sumK \sumL  \frac{\{1-r_i
  \pi^{-1}(y_i,\u_i;\bb,g^*)\} r_j K_h({\x_j-\x_i}) e^{-h(y_j)} r_k K_h({\u_k-\u_i})}{d^*(\x_i) D_i V_{i,k}\{E(R_k\mid\x_k) f_{\X}(\x_k)\}}\n\\
&& \times  r_l K_{h}(\x_l-\x_k) e^{-h(y_l)}\h'_\bb(y_l;\bb) \Delta_{1k}
+ O_p(N^{1/2}h^m)\n\\
&=& N^{-4/2} \sumI \sumK \frac{\{1-r_i
  \pi^{-1}(y_i,\u_i;\bb,g^*)\} r_k K_h({\u_k-\u_i}) \Delta_{1k} }{d^*(\x_i) D_i V_{i,k} E(R_k\mid\x_k) f_{\X}(\x_k)}\n\\
&& \times N^{-1/2} \sumJ  E\{ r_j K_h({\x_j-\x_i}) e^{-h(y_j)} R_l K_{h}(\X_l-\x_k) e^{-h(Y_l)}\h'_\bb(Y_l;\bb)\mid \x_j, r_j, r_j y_j\}\n\\
&&+ N^{-4/2} \sumI \sumK \frac{\{1-r_i
  \pi^{-1}(y_i,\u_i;\bb,g^*)\} r_k K_h({\u_k-\u_i}) \Delta_{1k} }{d^*(\x_i) D_i V_{i,k} E(R_k\mid\x_k) f_{\X}(\x_k)}\n\\
&& \times N^{-1/2} \sumL  E\{ R_j K_h({\X_j-\x_i}) e^{-h(Y_j)}   r_l K_{h}(\x_l-\x_k) e^{-h(y_l)}\h'_\bb(y_l;\bb)\mid \x_l, r_l, r_l y_l\}\n\\
&&- N^{-4/2} \sumI \sumK \frac{\{1-r_i
  \pi^{-1}(y_i,\u_i;\bb,g^*)\} r_k K_h({\u_k-\u_i}) \Delta_{1k} }{d^*(\x_i) D_i V_{i,k} E(R_k\mid\x_k) f_{\X}(\x_k)}\n\\
&& \times N^{1/2}  E\{ R_j K_h({\X_j-\x_i}) e^{-h(Y_j)}  R_l K_{h}(\X_l-\x_k) e^{-h(Y_l)}\h'_\bb(Y_l;\bb)\}+ O_p(N^{1/2}h^m)+O_p(N^{-1/2})\n\\
&=&T_{2211BA}+T_{2211BB}-T_{2211BC} + O_p(N^{1/2}h^m)+O_p(N^{-1/2}).\n
\ee

Consider $T_{2211BA}$,
\be \label{ustat:t2211ba}
&&T_{2211BA}\n\\
&=& N^{-4/2} \sumI \sumK \frac{\{1-r_i
  \pi^{-1}(y_i,\u_i;\bb,g^*)\} r_k K_h({\u_k-\u_i}) \Delta_{1k} }{d^*(\x_i) D_i V_{i,k} E(R_k\mid\x_k) f_{\X}(\x_k)}\n\\
&& \times N^{-1/2} \sumJ  E\{ r_j K_h({\x_j-\x_i}) e^{-h(y_j)} R_l K_{h}(\X_l-\x_k) e^{-h(Y_l)}\h'_\bb(Y_l;\bb)\mid \x_j, r_j, r_j y_j\}\n\\
&=& N^{-5/2} \sumI \sumJ \sumK \frac{\{1-r_i
  \pi^{-1}(y_i,\u_i;\bb,g^*)\} r_j K_h({\x_j-\x_i}) e^{-h(y_j)} r_k K_h({\u_k-\u_i}) \Delta_{1k}  \Delta_{3k} }{d^*(\x_i) D_i V_{i,k} }\n\\
&&+ O_p(N^{1/2}h^m)\n\\
&=&T_{2211C},\n
\ee
where $T_{2211C}$ is as defined in (\ref{ustat:t2211c}).

Consider $T_{2211BB}$,
\be \label{ustat:t2211bb}
&&T_{2211BB}\n\\
&=&N^{-4/2} \sumI \sumK \frac{\{1-r_i
  \pi^{-1}(y_i,\u_i;\bb,g^*)\} r_k K_h({\u_k-\u_i}) \Delta_{1k} }{d^*(\x_i) D_i V_{i,k} E(R_k\mid\x_k) f_{\X}(\x_k)}\n\\
&& \times N^{-1/2} \sumL  E\{ R_j K_h({\X_j-\x_i}) e^{-h(Y_j)}   r_l K_{h}(\x_l-\x_k) e^{-h(y_l)}\h'_\bb(y_l;\bb)\mid \x_l, r_l, r_l y_l\}\n\\
&=&N^{-5/2} \sumI \sumK \sumL \frac{\{1-r_i
  \pi^{-1}(y_i,\u_i;\bb,g^*)\} r_k K_h({\u_k-\u_i}) r_l K_{h}(\x_l-\x_k) e^{-h(y_l)}\h'_\bb(y_l;\bb)  \Delta_{1i} \Delta_{1k} }{d^*(\x_i) D_i d^*(\x_k) E(R_i\mid\u_i) f_{\U}(\u_i) E(R_k\mid\x_k) f_{\X}(\x_k)}\n\\
&&+ O_p(N^{1/2}h^m)\n\\
&=&N^{-5/2} \sumI \sumJ \sumK \frac{\{1-r_i
  \pi^{-1}(y_i,\u_i;\bb,g^*)\} r_j K_h({\u_j-\u_i}) r_k K_{h}(\x_k-\x_j) e^{-h(y_k)}\h'_\bb(y_k;\bb)  \Delta_{1i} \Delta_{1j} }{d^*(\x_i) D_i d^*(\x_j) E(R_i\mid\u_i) f_{\U}(\u_i) E(R_j\mid\x_j) f_{\X}(\x_j)}\n\\
&&+ O_p(N^{1/2}h^m).
\ee

Consider $T_{2211BC}$,
\be \label{ustat:t2211bc}
&&T_{2211BC}\n\\
&=& N^{-4/2} \sumI \sumK \frac{\{1-r_i
  \pi^{-1}(y_i,\u_i;\bb,g^*)\} r_k K_h({\u_k-\u_i}) \Delta_{1k} }{d^*(\x_i) D_i V_{i,k} E(R_k\mid\x_k) f_{\X}(\x_k)}\n\\
&& \times N^{1/2}  E\{ R_j K_h({\X_j-\x_i}) e^{-h(Y_j)}  R_l K_{h}(\X_l-\x_k) e^{-h(Y_l)}\h'_\bb(Y_l;\bb)\}\n\\
&=& N^{-3/2} \sumI \sumK \frac{\{1-r_i
  \pi^{-1}(y_i,\u_i;\bb,g^*)\} r_k K_h({\u_k-\u_i}) \Delta_{1k} }{d^*(\x_i) D_i V_{i,k} E(R_k\mid\x_k) f_{\X}(\x_k)}\n\\
&& \times \Delta_{1i} E(R_i\mid\x_i) f_{\X}(\x_i)  \Delta_{3k} E(R_k\mid\x_k) f_{\X}(\x_k)
+ O_p(N^{1/2}h^m)\n\\
&=& N^{-3/2} \sumI \sumJ \frac{\{1-r_i
  \pi^{-1}(y_i,\u_i;\bb,g^*)\} r_j K_h({\u_j-\u_i}) \Delta_{1i} \Delta_{1j} \Delta_{3j}}{d^*(\x_i) D_i d^*(\x_j) E(R_i\mid\u_i) f_{\U}(\u_i)  }
+ O_p(N^{1/2}h^m)\n\\
&=&T_{2211AC},\n
\ee
where $T_{2211AC}$ is defined in (\ref{ustat:t2211ac}).

Applying U-statistics property to $T_{2212ABB}$ from  (\ref{ustat:t2212abb}),
\be \label{ustat:t2212abb2}
&&T_{2212ABB}\n\\
&=&N^{-7/2} \sumI \sumJ \sumK \sumL    \frac{\{1-r_i
  \pi^{-1}(y_i,\u_i;\bb,g^*)\} r_j K_h({\x_j-\x_i}) e^{-h(y_j)} r_k K_h({\u_k-\u_i})}{d^*(\x_i) D_i V_{i,k} E(R_k\mid\x_k) f_{\X}(\x_k)}\n\\
&&\times  r_l K_{h} (\x_l-\x_k) \Delta_{1k} \Delta_{3k}
+O_p(N^{1/2}h^m)\n\\
&=&N^{-4/2} \sumI  \sumK    \frac{\{1-r_i
  \pi^{-1}(y_i,\u_i;\bb,g^*)\} r_k K_h({\u_k-\u_i})\Delta_{1k} \Delta_{3k} }{d^*(\x_i) D_i V_{i,k} E(R_k\mid\x_k) f_{\X}(\x_k)}\n\\
&&\times N^{-1/2} \sumJ E \left \{ r_j K_h({\x_j-\x_i}) e^{-h(y_j)} R_l K_{h} (\X_l-\x_k) \mid \x_j, r_j, r_j y_j \right\}\n\\
&&+N^{-4/2} \sumI  \sumK    \frac{\{1-r_i
  \pi^{-1}(y_i,\u_i;\bb,g^*)\} r_k K_h({\u_k-\u_i})\Delta_{1k} \Delta_{3k} }{d^*(\x_i) D_i V_{i,k} E(R_k\mid\x_k) f_{\X}(\x_k)}\n\\
&&\times N^{-1/2} \sumL E \left \{ R_j K_h({\X_j-\x_i}) e^{-h(Y_j)} r_l K_{h} (\x_l-\x_k) \mid \x_l, r_l, r_l y_l \right\}\n\\
&&-N^{-4/2} \sumI  \sumK    \frac{\{1-r_i
  \pi^{-1}(y_i,\u_i;\bb,g^*)\} r_k K_h({\u_k-\u_i})\Delta_{1k} \Delta_{3k} }{d^*(\x_i) D_i V_{i,k} E(R_k\mid\x_k) f_{\X}(\x_k)}\n\\
&&\times N^{1/2}  E \left \{ R_j K_h({\X_j-\x_i}) e^{-h(Y_j)} R_l K_{h} (\X_l-\x_k)  \right\}+ O_p(N^{1/2}h^m)+O_p(N^{-1/2})\n\\
&=&T_{2212ABBA} +T_{2212ABBB}-T_{2212ABBC}+ O_p(N^{1/2}h^m)+O_p(N^{-1/2}).\n
\ee

Consider $T_{2212ABBA}$,
\be \label{ustat:t2212abba}
&&T_{2212ABBA}\n\\
&=&N^{-4/2} \sumI  \sumK    \frac{\{1-r_i
  \pi^{-1}(y_i,\u_i;\bb,g^*)\} r_k K_h({\u_k-\u_i})\Delta_{1k} \Delta_{3k} }{d^*(\x_i) D_i V_{i,k} E(R_k\mid\x_k) f_{\X}(\x_k)}\n\\
&&\times N^{-1/2} \sumJ E \left \{ r_j K_h({\x_j-\x_i}) e^{-h(y_j)} R_l K_{h} (\X_l-\x_k) \mid \x_j, r_j, r_j y_j \right\}\n\\
&=&N^{-5/2} \sumI \sumJ \sumK    \frac{\{1-r_i
  \pi^{-1}(y_i,\u_i;\bb,g^*)\} r_j K_h({\x_j-\x_i}) e^{-h(y_j)} r_k K_h({\u_k-\u_i})\Delta_{1k} \Delta_{3k} }{d^*(\x_i) D_i V_{i,k} }\n\\
&&+ O_p(N^{1/2}h^m)\n\\
&=&T_{2211C},\n
\ee
where $T_{2211C}$ is as defined in (\ref{ustat:t2211c}).

Consider $T_{2212ABBB}$,
\be \label{ustat:t2212abbb}
&&T_{2212ABBB}\n\\
&=&N^{-4/2} \sumI  \sumK    \frac{\{1-r_i
  \pi^{-1}(y_i,\u_i;\bb,g^*)\} r_k K_h({\u_k-\u_i})\Delta_{1k} \Delta_{3k} }{d^*(\x_i) D_i V_{i,k} E(R_k\mid\x_k) f_{\X}(\x_k)}\n\\
&&\times N^{-1/2} \sumL E \left \{ R_j K_h({\X_j-\x_i}) e^{-h(Y_j)} r_l K_{h} (\x_l-\x_k) \mid \x_l, r_l, r_l y_l \right\}\n\\
&=&N^{-5/2} \sumI \sumK \sumL   \frac{\{1-r_i
  \pi^{-1}(y_i,\u_i;\bb,g^*)\} r_k K_h({\u_k-\u_i}) \Delta_{1i} \Delta_{1k} \Delta_{3k} r_l K_{h} (\x_l-\x_k)  }{d^*(\x_i) D_i d^*(\x_k)  E(R_i\mid\u_i) f_{\U}(\u_i) E(R_k\mid\x_k) f_{\X}(\x_k)}\n\\
&&+ O_p(N^{1/2}h^m)\n\\
&=&N^{-5/2} \sumI \sumJ \sumK  \frac{\{1-r_i
  \pi^{-1}(y_i,\u_i;\bb,g^*)\} r_j K_h({\u_j-\u_i})  r_k K_{h} (\x_k-\x_j) \Delta_{1i} \Delta_{1j} \Delta_{3j} }{d^*(\x_i) D_i d^*(\x_j)  E(R_i\mid\u_i) f_{\U}(\u_i) E(R_j\mid\x_j) f_{\X}(\x_j)}\n\\
&&+ O_p(N^{1/2}h^m).
\ee

Consider $T_{2212ABBC}$,
\be \label{ustat:t2212abbc}
&&T_{2212ABBC}\n\\
&=&N^{-4/2} \sumI  \sumK    \frac{\{1-r_i
  \pi^{-1}(y_i,\u_i;\bb,g^*)\} r_k K_h({\u_k-\u_i})\Delta_{1k} \Delta_{3k} }{d^*(\x_i) D_i V_{i,k} E(R_k\mid\x_k) f_{\X}(\x_k)}\n\\
&&\times N^{1/2}  E \left \{ R_j K_h({\X_j-\x_i}) e^{-h(Y_j)} R_l K_{h} (\X_l-\x_k)  \right\}\n\\
&=&N^{-3/2} \sumI  \sumK    \frac{\{1-r_i
  \pi^{-1}(y_i,\u_i;\bb,g^*)\} r_k K_h({\u_k-\u_i}) \Delta_{1i} \Delta_{1k} \Delta_{3k} }{d^*(\x_i) D_i d^*(\x_k)  E(R_i\mid\u_i) f_{\U}(\u_i)}
+ O_p(N^{1/2}h^m)\n\\
&=&N^{-3/2} \sumI  \sumJ    \frac{\{1-r_i
  \pi^{-1}(y_i,\u_i;\bb,g^*)\} r_j K_h({\u_j-\u_i}) \Delta_{1i} \Delta_{1j} \Delta_{3j} }{d^*(\x_i) D_i d^*(\x_j)  E(R_i\mid\u_i) f_{\U}(\u_i)}
+ O_p(N^{1/2}h^m)\n\\
&=& T_{2211AC},\n
\ee
where $T_{2211AC}$ is defined in (\ref{ustat:t2211ac}).

Applying U-statistics property to $T_{231BB}$ from (\ref{ustat:t231bb}),
\be \label{ustat:t231bb2}
&&T_{231BB}\n\\
&=& N^{-7/2} \sumI \sumJ \sumK \sumL \frac{\{1-r_i
  \pi^{-1}(y_i,\u_i;\bb,g^*)\} r_j K_h({\x_j-\x_i}) e^{-h(y_j)} r_k K_h({\u_k-\u_i}) T_{j,k} }{d^*(\x_i) D_i      V_{i,k}  d^* (\x_k)  E(R_k\mid\x_k) f_{\X}(\x_k)}\n \\
&& \times  r_l K_h({\x_l-\x_k})  \{e^{-h(y_l)}+e^{-g^* (\u_k)} e^{-2h(y_l)}\} + O_p(N^{1/2} h^m)\n\\
&=& N^{-4/2} \sumI \sumK \frac{\{1-r_i
  \pi^{-1}(y_i,\u_i;\bb,g^*)\} r_k K_h({\u_k-\u_i}) }{d^*(\x_i) D_i      V_{i,k}  d^* (\x_k)  E(R_k\mid\x_k) f_{\X}(\x_k) }\n \\
&& \times N^{-1/2} \sumJ E [r_j K_h({\x_j-\x_i}) e^{-h(y_j)} T_{j,k}  R_l K_h({\X_l-\x_k})  \{e^{-h(Y_l)}+e^{-g^* (\u_k)} e^{-2h(Y_l)}\} \mid x_j, r_j ,r_j y_j ] \n\\
&&+ N^{-4/2} \sumI \sumK \frac{\{1-r_i
  \pi^{-1}(y_i,\u_i;\bb,g^*)\} r_k K_h({\u_k-\u_i})  }{d^*(\x_i) D_i      V_{i,k}  d^* (\x_k)  E(R_k\mid\x_k) f_{\X}(\x_k) }\n \\
&& \times N^{-1/2} \sumL E [R_j K_h({\X_j-\x_i}) e^{-h(Y_j)}  T_{j,k} r_l K_h({\x_l-\x_k})  \{e^{-h(y_l)}+e^{-g^* (\u_k)} e^{-2h(y_l)}\} \mid x_l, r_l ,r_l y_l ] \n\\
&&- N^{-4/2} \sumI \sumK \frac{\{1-r_i
  \pi^{-1}(y_i,\u_i;\bb,g^*)\} r_k K_h({\u_k-\u_i}) }{d^*(\x_i) D_i      V_{i,k}  d^* (\x_k)  E(R_k\mid\x_k) f_{\X}(\x_k) }\n \\
&& \times N^{1/2}  E [R_j K_h({\X_j-\x_i}) e^{-h(Y_j)}T_{j,k}   R_l K_h({\X_l-\x_k})  \{e^{-h(Y_l)}+e^{-g^* (\u_k)} e^{-2h(Y_l)}\}] \n\\
&&+ O_p(N^{1/2}h^m)+O_p(N^{-1/2})\n\\
&=&T_{231BBA} +T_{231BBB}-T_{231BBC}+ O_p(N^{1/2}h^m)+O_p(N^{-1/2}).\n
\ee

Consider $T_{231BBA}$,
\be \label{ustat:t231bba}
&&T_{231BBA}\n\\
&=& N^{-4/2} \sumI \sumK \frac{\{1-r_i
  \pi^{-1}(y_i,\u_i;\bb,g^*)\} r_k K_h({\u_k-\u_i})  }{d^*(\x_i) D_i      V_{i,k}  d^* (\x_k)  E(R_k\mid\x_k) f_{\X}(\x_k) }\n \\
&& \times N^{-1/2} \sumJ E [r_j K_h({\x_j-\x_i}) e^{-h(y_j)} T_{j,k} R_l K_h({\X_l-\x_k})  \{e^{-h(Y_l)}+e^{-g^* (\u_k)} e^{-2h(Y_l)}\} \mid x_j, r_j ,r_j y_j ] \n\\
&=& N^{-5/2} \sumI \sumJ \sumK \frac{\{1-r_i
  \pi^{-1}(y_i,\u_i;\bb,g^*)\} r_j K_h({\x_j-\x_i}) e^{-h(y_j)}  r_k K_h({\u_k-\u_i}) T_{j,k} }{d^*(\x_i) D_i      V_{i,k}  d^* (\x_k)  E(R_k\mid\x_k) f_{\X}(\x_k) }\n \\
&& \times E [ R_l K_h({\X_l-\x_k})  \{e^{-h(Y_l)}+e^{-g^* (\u_k)} e^{-2h(Y_l)}\}] \n\\
&=& N^{-5/2} \sumI \sumJ  \sumK \frac{\{1-r_i
  \pi^{-1}(y_i,\u_i;\bb,g^*)\} r_j K_h({\x_j-\x_i}) e^{-h(y_j)}  r_k K_h({\u_k-\u_i}) T_{j,k}   }{d^*(\x_i) D_i V_{i,k}   }\n \\
&&+ O_p(N^{1/2}h^m)\n\\
&=&T_{21}+ O_p(N^{1/2}h^m),\n
\ee
where $T_{21}$ is as defined in (\ref{eq:t21}).

Consider $T_{231BBB}$,
\be \label{ustat:t231bbb}
&&T_{231BBB}\n\\
&=&N^{-4/2} \sumI \sumK \frac{\{1-r_i
  \pi^{-1}(y_i,\u_i;\bb,g^*)\} r_k K_h({\u_k-\u_i})  }{d^*(\x_i) D_i  V_{i,k}  d^* (\x_k)  E(R_k\mid\x_k) f_{\X}(\x_k) }\n \\
&& \times N^{-1/2} \sumL E [R_j K_h({\X_j-\x_i}) e^{-h(Y_j)}  T_{j,k} r_l K_h({\x_l-\x_k})  \{e^{-h(y_l)}+e^{-g^* (\u_k)} e^{-2h(y_l)}\} \mid x_l, r_l ,r_l y_l ] \n\\
&=&N^{-5/2} \sumI \sumK \sumL  \frac{\{1-r_i
  \pi^{-1}(y_i,\u_i;\bb,g^*)\} r_k K_h({\u_k-\u_i})  r_l K_h({\x_l-\x_k})    }{d^*(\x_i) D_i      V_{i,k}  d^* (\x_k)  E(R_k\mid\x_k) f_{\X}(\x_k) }\n \\
&& \times \{e^{-h(y_l)}+e^{-g^* (\u_k)} e^{-2h(y_l)}\} E [R_j K_h({\X_j-\x_i}) e^{-h(Y_j)}\{\Delta_{1k} \Delta_{3k} -\Delta_{1k}^2 \h'_\bb(Y_j;\bb) \}] \n\\
&=&N^{-5/2} \sumI \sumK \sumL  \frac{\{1-r_i
  \pi^{-1}(y_i,\u_i;\bb,g^*)\} r_k K_h({\u_k-\u_i})  r_l K_h({\x_l-\x_k})  }{d^*(\x_i) D_i \{d^* (\x_k)\}^2 E(R_i\mid\u_i) f_{\U}(\u_i) E(R_k\mid\x_k) f_{\X}(\x_k) }\n \\
&& \times \{e^{-h(y_l)}+e^{-g^* (\u_k)} e^{-2h(y_l)}\} \{\Delta_{1i}\Delta_{1k} \Delta_{3k} -\Delta_{1k}^2 \Delta_{3i} \}
+ O_p(N^{1/2}h^m)\n\\
&=&N^{-5/2} \sumI \sumJ \sumK  \frac{\{1-r_i
  \pi^{-1}(y_i,\u_i;\bb,g^*)\} r_j K_h({\u_j-\u_i})  r_k K_h({\x_k-\x_j})  }{d^*(\x_i) D_i \{d^* (\x_j)\}^2 E(R_i\mid\u_i) f_{\U}(\u_i) E(R_j\mid\x_j) f_{\X}(\x_j) }\n \\
&& \times  \{e^{-h(y_k)}+e^{-g^* (\u_j)} e^{-2h(y_k)}\}  \{\Delta_{1i}\Delta_{1j} \Delta_{3j} -\Delta_{1j}^2 \Delta_{3i} \}
+ O_p(N^{1/2}h^m).
\ee

Consider $T_{231BBC}$,
\be \label{ustat:t231bbc}
&&T_{231BBC}\n\\
&=&N^{-4/2} \sumI \sumK \frac{\{1-r_i
  \pi^{-1}(y_i,\u_i;\bb,g^*)\} r_k K_h({\u_k-\u_i}) }{d^*(\x_i) D_i  V_{i,k}  d^* (\x_k)  E(R_k\mid\x_k) f_{\X}(\x_k) }\n \\
&& \times N^{1/2}  E [R_j K_h({\X_j-\x_i}) e^{-h(Y_j)}T_{j,k}   R_l K_h({\X_l-\x_k})  \{e^{-h(Y_l)}+e^{-g^* (\u_k)} e^{-2h(Y_l)}\}] \n\\
&=&N^{-3/2} \sumI \sumK \frac{\{1-r_i
  \pi^{-1}(y_i,\u_i;\bb,g^*)\} r_k K_h({\u_k-\u_i}) }{d^*(\x_i) D_i V_{i,k}  d^* (\x_k)  E(R_k\mid\x_k) f_{\X}(\x_k) }\n \\
&& \times  \{\Delta_{1i}\Delta_{1k} \Delta_{3k} -\Delta_{1k}^2 \Delta_{3i} \} E(R_i\mid\x_i) f_{\X}(\x_i)  d^* (\x_k)  E(R_k\mid\x_k) f_{\X}(\x_k)
+ O_p(N^{1/2}h^m)\n\\
&=&N^{-3/2} \sumI \sumJ \frac{\{1-r_i
  \pi^{-1}(y_i,\u_i;\bb,g^*)\} r_j K_h({\u_j-\u_i}) }{d^*(\x_i) D_i d^* (\x_j) E(R_i\mid\u_i) f_{\U}(\u_i)  }
 \{\Delta_{1i}\Delta_{1j} \Delta_{3j} -\Delta_{1j}^2 \Delta_{3i} \}
+ O_p(N^{1/2}h^m)\n\\
&=& T_{2211AC} - T_{2222C},
\ee
where $T_{2211AC}$ is defined in (\ref{ustat:t2211ac}) and $T_{2222C}$ is defined in (\ref{ustat:t2222c}).

Applying U-statistic property $T_{231AB}$ from  (\ref{ustat:t231ab}) we get,
\be \label{ustat:t231ab2}
&& T_{231AB}\n\\
&=&N^{-7/2} \sumI \sumJ \sumK \sumL \frac{\{1-r_i
  \pi^{-1}(y_i,\u_i;\bb,g^*)\} r_j K_h({\x_j-\x_i}) e^{-h(y_j)} r_k K_h({\u_k-\u_i}) T_{j,k}  d^*(\x_k)}{d^*(\x_i) D_i V_{i,k}^2}\n \\
&& \times r_l K_h({\x_l-\x_i}) \{E(R_i\mid\u_i) f_{\U}(\u_i)\}
+ O_p(N^{1/2} h^m)\n\\
&=&N^{-4/2} \sumI \sumK
\frac{\{1-r_i
  \pi^{-1}(y_i,\u_i;\bb,g^*)\} r_k K_h({\u_k-\u_i}) }{d^*(\x_i) D_i  d^*(\x_k) \{E(R_i\mid\x_i) f_{\X}(\x_i)\}^2 E(R_i\mid\u_i) f_{\U}(\u_i) }\n \\
&& \times N^{-1/2} \sumJ E\left\{  r_j K_h({\x_j-\x_i}) e^{-h(y_j)} T_{j,k}  R_l K_h({\X_l-\x_i})  \mid \x_j, r_j, r_j y_j \right\}\n\\
&&+N^{-4/2} \sumI \sumK
\frac{\{1-r_i
  \pi^{-1}(y_i,\u_i;\bb,g^*)\} r_k K_h({\u_k-\u_i}) }{d^*(\x_i) D_i  d^*(\x_k) \{E(R_i\mid\x_i) f_{\X}(\x_i)\}^2 E(R_i\mid\u_i) f_{\U}(\u_i) }\n \\
&& \times N^{-1/2} \sumL E\left\{R_j K_h({\X_j-\x_i}) e^{-h(Y_j)} T_{j,k}  r_l K_h({\x_l-\x_i}) \mid \x_l, r_l, r_l y_l \right\}\n\\
&&-N^{-4/2} \sumI \sumK
\frac{\{1-r_i
  \pi^{-1}(y_i,\u_i;\bb,g^*)\} r_k K_h({\u_k-\u_i}) }{d^*(\x_i) D_i  d^*(\x_k) \{E(R_i\mid\x_i) f_{\X}(\x_i)\}^2 E(R_i\mid\u_i) f_{\U}(\u_i) }\n \\
&& \times N^{1/2}  E\left\{ R_j K_h({\X_j-\x_i}) e^{-h(Y_j)} T_{j,k} R_l K_h({\X_l-\x_i}) \right\}
+ O_p(N^{1/2}h^m)+O_p(N^{-1/2})\n\\
&=&T_{231ABA} +T_{231ABB}-T_{231ABC}+ O_p(N^{1/2}h^m)+O_p(N^{-1/2}).\n
\ee

Consider $T_{231ABA}$,
\be \label{ustat:t231aba}
&&T_{231ABA}\n\\
&=&N^{-4/2} \sumI \sumK
\frac{\{1-r_i
  \pi^{-1}(y_i,\u_i;\bb,g^*)\} r_k K_h({\u_k-\u_i}) }{d^*(\x_i) D_i  d^*(\x_k) \{E(R_i\mid\x_i) f_{\X}(\x_i)\}^2 E(R_i\mid\u_i) f_{\U}(\u_i) }\n \\
&& \times N^{-1/2} \sumJ E\left\{  r_j K_h({\x_j-\x_i}) e^{-h(y_j)} T_{j,k}  R_l K_h({\X_l-\x_i})  \mid \x_j, r_j, r_j y_j \right\}\n\\
&=&N^{-5/2} \sumI  \sumJ \sumK
\frac{\{1-r_i
  \pi^{-1}(y_i,\u_i;\bb,g^*)\}  r_j K_h({\x_j-\x_i}) e^{-h(y_j)} r_k K_h({\u_k-\u_i}) T_{j,k} }{d^*(\x_i) D_i  V_{i,k} }\n \\
&&+ O_p(N^{1/2}h^m)\n\\
&=& T_{21} + O_p(N^{1/2}h^m).\n
\ee
where $T_{21}$ is defined in  (\ref{eq:t21}).

Consider $T_{231ABB}$,
\be \label{ustat:t231abb}
&&T_{231ABB}\n\\
&=&N^{-4/2} \sumI \sumK
\frac{\{1-r_i
  \pi^{-1}(y_i,\u_i;\bb,g^*)\} r_k K_h({\u_k-\u_i}) }{d^*(\x_i) D_i  d^*(\x_k) \{E(R_i\mid\x_i) f_{\X}(\x_i)\}^2 E(R_i\mid\u_i) f_{\U}(\u_i) }\n \\
&& \times N^{-1/2} \sumL E\left\{R_j K_h({\X_j-\x_i}) e^{-h(Y_j)} T_{j,k}  r_l K_h({\x_l-\x_i}) \mid \x_l, r_l, r_l y_l \right\}\n\\
&=&N^{-5/2} \sumI \sumK \sumL
\frac{\{1-r_i
  \pi^{-1}(y_i,\u_i;\bb,g^*)\} r_k K_h({\u_k-\u_i})  r_l K_h({\x_l-\x_i}) }{d^*(\x_i) D_i  d^*(\x_k) \{E(R_i\mid\x_i) f_{\X}(\x_i)\}^2 E(R_i\mid\u_i) f_{\U}(\u_i) }\n \\
&& \times E\left\{R_j K_h({\X_j-\x_i}) e^{-h(Y_j)} T_{j,k}  \right\}\n\\
&=&N^{-5/2} \sumI \sumK \sumL
\frac{\{1-r_i
  \pi^{-1}(y_i,\u_i;\bb,g^*)\} r_k K_h({\u_k-\u_i})  r_l K_h({\x_l-\x_i}) }{d^*(\x_i) D_i  V_{i,k} }\n \\
&& \times \{\Delta_{1i}\Delta_{1k} \Delta_{3k} -\Delta_{1k}^2 \Delta_{3i} \}
+ O_p(N^{1/2}h^m)\n\\
&=&N^{-5/2} \sumI \sumJ \sumK
\frac{\{1-r_i
  \pi^{-1}(y_i,\u_i;\bb,g^*)\} r_j K_h({\u_j-\u_i})  r_k K_h({\x_k-\x_i}) }{d^*(\x_i) D_i  V_{i,j} }\n \\
&& \times \{\Delta_{1i}\Delta_{1j} \Delta_{3j} -\Delta_{1j}^2 \Delta_{3i} \}
+ O_p(N^{1/2}h^m).
\ee

Consider $T_{231ABC}$,
\be \label{ustat:t231abc}
&&T_{231ABC}\n\\
&=&N^{-4/2} \sumI \sumK
\frac{\{1-r_i
  \pi^{-1}(y_i,\u_i;\bb,g^*)\} r_k K_h({\u_k-\u_i}) }{d^*(\x_i) D_i  d^*(\x_k) \{E(R_i\mid\x_i) f_{\X}(\x_i)\}^2 E(R_i\mid\u_i) f_{\U}(\u_i) }\n \\
&& \times N^{1/2}  E\left\{ R_j K_h({\X_j-\x_i}) e^{-h(Y_j)} T_{j,k} R_l K_h({\X_l-\x_i}) \right\}\n\\
&=&N^{-3/2} \sumI \sumK
\frac{\{1-r_i
  \pi^{-1}(y_i,\u_i;\bb,g^*)\} r_k K_h({\u_k-\u_i}) }{d^*(\x_i) D_i  d^*(\x_k) E(R_i\mid\u_i) f_{\U}(\u_i) }\{\Delta_{1i}\Delta_{1k} \Delta_{3k} -\Delta_{1k}^2 \Delta_{3i} \}
+O_p(N^{1/2}h^m)\n\\
&=&N^{-3/2} \sumI \sumJ
\frac{\{1-r_i
  \pi^{-1}(y_i,\u_i;\bb,g^*)\} r_j K_h({\u_j-\u_i}) }{d^*(\x_i) D_i  d^*(\x_j) E(R_i\mid\u_i) f_{\U}(\u_i) }\{\Delta_{1i}\Delta_{1j} \Delta_{3j} -\Delta_{1j}^2 \Delta_{3i} \}
+O_p(N^{1/2}h^m)\n\\
&=& T_{2211AC} - T_{2222C},\n
\ee
where $T_{2211AC}$ is defined in (\ref{ustat:t2211ac}) and $T_{2222C}$ is defined in (\ref{ustat:t2222c}).

Consider $T_{232BBB}$ from (\ref{ustat:t232bbb}) and using U-statistics property we get,
\be \label{ustat:t232bbb2}
&& T_{232BBB}\n\\
&=&N^{-7/2} \sumI \sumJ \sumK \sumL  \frac{\{1-r_i
  \pi^{-1}(Y_i,\u_i;\bb,g^*)\} r_j K_h({\x_j-\x_i}) e^{-h(y_j)} r_k K_h({\u_k-\u_i}) T_{j,k}}{d^*(\x_i) D_i V_{i,k}^2 E(R_k\mid\x_k) f_{\X}(\x_k)}\n \\
&&  \times E(R_i\mid\x_i) f_{\X}(\x_i)  d^*(\x_k) r_l K_h({\x_l-\x_k})  E(R_i\mid\u_i) f_{\U}(\u_i)
+O_p(N^{1/2} h^m)\n\\
&=&N^{-4/2} \sumI \sumK \frac{\{1-r_i
  \pi^{-1}(Y_i,\u_i;\bb,g^*)\} r_k K_h({\u_k-\u_i}) }{d^*(\x_i) D_i V_{i,k} E(R_k\mid\x_k) f_{\X}(\x_k)}\n \\
&&  \times N^{-1/2} \sumJ  E\left\{ r_j K_h({\x_j-\x_i}) e^{-h(y_j)} T_{j,k} R_l K_h({\X_l-\x_k}) \mid \x_j, r_j, r_j y_j \right\}\n\\
&&+N^{-4/2} \sumI \sumK \frac{\{1-r_i
  \pi^{-1}(Y_i,\u_i;\bb,g^*)\} r_k K_h({\u_k-\u_i}) }{d^*(\x_i) D_i V_{i,k} E(R_k\mid\x_k) f_{\X}(\x_k)}\n \\
&&  \times N^{-1/2} \sumL  E\left\{  R_j K_h({\X_j-\x_i}) e^{-h(Y_j)} T_{j,k} r_l K_h({\x_l-\x_k}) \mid \x_l, r_l, r_l y_l \right\}\n\\
&&-N^{-4/2} \sumI \sumK \frac{\{1-r_i
  \pi^{-1}(Y_i,\u_i;\bb,g^*)\} r_k K_h({\u_k-\u_i}) }{d^*(\x_i) D_i V_{i,k} E(R_k\mid\x_k) f_{\X}(\x_k)}\n \\
&&  \times N^{1/2}   E\left\{ R_j K_h({\X_j-\x_i}) e^{-h(Y_j)} T_{j,k} R_l K_h({\X_l-\x_k}) \right\}
+ O_p(N^{1/2}h^m)+O_p(N^{-1/2})\n\\
&=&T_{232BBBA} +T_{232BBBB}-T_{232BBBC}+ O_p(N^{1/2}h^m)+O_p(N^{-1/2}).\n
\ee

Consider $T_{232BBBA}$,
\be \label{ustat:t232bbba}
&&T_{232BBBA}\n\\
&=&N^{-4/2} \sumI \sumK \frac{\{1-r_i
  \pi^{-1}(Y_i,\u_i;\bb,g^*)\} r_k K_h({\u_k-\u_i}) }{d^*(\x_i) D_i V_{i,k} E(R_k\mid\x_k) f_{\X}(\x_k)}\n \\
&&  \times N^{-1/2} \sumJ  E\left\{ r_j K_h({\x_j-\x_i}) e^{-h(y_j)} T_{j,k} R_l K_h({\X_l-\x_k}) \mid \x_j, r_j, r_j y_j \right\}\n\\
&=&N^{-5/2} \sumI \sumJ \sumK  \frac{\{1-r_i
  \pi^{-1}(Y_i,\u_i;\bb,g^*)\} r_j K_h({\x_j-\x_i}) e^{-h(y_j)} r_k K_h({\u_k-\u_i}) T_{j,k} }{d^*(\x_i) D_i V_{i,k} }\n\\
&&+ O_p(N^{1/2}h^m)\n\\
&=&T_{21}+ O_p(N^{1/2}h^m),\n
\ee
where $T_{21}$ is defined in  (\ref{eq:t21}).

Consider $T_{232BBBB}$,
\be \label{ustat:t232bbbb}
&&T_{232BBBB}\n\\
&=&N^{-4/2} \sumI \sumK \frac{\{1-r_i
  \pi^{-1}(Y_i,\u_i;\bb,g^*)\} r_k K_h({\u_k-\u_i}) }{d^*(\x_i) D_i V_{i,k} E(R_k\mid\x_k) f_{\X}(\x_k)}\n \\
&&  \times N^{-1/2} \sumL  E\left\{  R_j K_h({\X_j-\x_i}) e^{-h(Y_j)} T_{j,k} r_l K_h({\x_l-\x_k}) \mid \x_l, r_l, r_l y_l \right\}\n\\
&=&N^{-5/2} \sumI \sumK  \sumL \frac{\{1-r_i
  \pi^{-1}(Y_i,\u_i;\bb,g^*)\} r_k K_h({\u_k-\u_i}) }{d^*(\x_i) D_i V_{i,k} E(R_k\mid\x_k) f_{\X}(\x_k)}\n \\
&&  \times  r_l K_h({\x_l-\x_k})  E\left\{  R_j K_h({\X_j-\x_i}) e^{-h(Y_j)} T_{j,k} \right\}\n\\
&=&N^{-5/2} \sumI \sumK  \sumL \frac{\{1-r_i
  \pi^{-1}(Y_i,\u_i;\bb,g^*)\} r_k K_h({\u_k-\u_i}) r_l K_h({\x_l-\x_k})  }{d^*(\x_i) D_i d^*(\x_k) E(R_i\mid\u_i) f_{\U}(\u_i) E(R_k\mid\x_k) f_{\X}(\x_k)}\n \\
&&  \times   \{\Delta_{1i}\Delta_{1k} \Delta_{3k} -\Delta_{1k}^2 \Delta_{3i} \}
+ O_p(N^{1/2}h^m)\n\\
&=&N^{-5/2} \sumI \sumJ  \sumK \frac{\{1-r_i
  \pi^{-1}(Y_i,\u_i;\bb,g^*)\} r_j K_h({\u_j-\u_i}) r_k K_h({\x_k-\x_j})  }{d^*(\x_i) D_i d^*(\x_j) E(R_i\mid\u_i) f_{\U}(\u_i) E(R_j\mid\x_j) f_{\X}(\x_j)}\n \\
&&  \times   \{\Delta_{1i}\Delta_{1j} \Delta_{3j} -\Delta_{1j}^2 \Delta_{3i} \}
+ O_p(N^{1/2}h^m).
\ee

Consider $T_{232BBBC}$,
\be \label{ustat:t232bbbc}
&&T_{232BBBC}\n\\
&=&N^{-4/2} \sumI \sumK \frac{\{1-r_i
  \pi^{-1}(Y_i,\u_i;\bb,g^*)\} r_k K_h({\u_k-\u_i}) }{d^*(\x_i) D_i V_{i,k} E(R_k\mid\x_k) f_{\X}(\x_k)}\n \\
&&  \times N^{1/2}   E\left\{ R_j K_h({\X_j-\x_i}) e^{-h(Y_j)}T_{j,k} R_l K_h({\X_l-\x_k}) \right\}\n\\
&=&N^{-3/2} \sumI \sumK \frac{\{1-r_i
  \pi^{-1}(Y_i,\u_i;\bb,g^*)\} r_k K_h({\u_k-\u_i}) }{d^*(\x_i) D_i d^*(\x_k) E(R_i\mid\u_i) f_{\U}(\u_i)} \{\Delta_{1i}\Delta_{1k} \Delta_{3k} -\Delta_{1k}^2 \Delta_{3i} \}
+ O_p(N^{1/2}h^m)\n\\
&=&N^{-3/2} \sumI \sumJ
\frac{\{1-r_i
  \pi^{-1}(y_i,\u_i;\bb,g^*)\} r_j K_h({\u_j-\u_i}) }{d^*(\x_i) D_i  d^*(\x_j) E(R_i\mid\u_i) f_{\U}(\u_i) }\{\Delta_{1i}\Delta_{1j} \Delta_{3j} -\Delta_{1j}^2 \Delta_{3i} \}
+ O_p(N^{1/2}h^m)\n\\
&=& T_{2211AC} - T_{2222C},\n
\ee
where $T_{2211AC}$ is defined in (\ref{ustat:t2211ac}) and $T_{2222C}$ is defined in (\ref{ustat:t2222c}).

Consider $T_{231C}$ from  (\ref{ustat:t231c}) and applying the U-statistic property we get,
\be \label{ustat:t231c2}
&&T_{231C}\n\\
&=& N^{-7/2} \sumI \sumJ \sumK \sumL \frac{\{1-r_i
  \pi^{-1}(y_i,\u_i;\bb,g^*)\} r_j K_h({\x_j-\x_i}) e^{-h(y_j)} r_k K_h({\u_k-\u_i}) T_{j,k} }{d^*(\x_i) D_i V_{i,k}^2}\n \\
&& \times \{r_l K_h({\u_l-\u_i}) E(R_i\mid\x_i) f_{\X}(\x_i) d^*(\x_k) \} +O_p(N^{1/2} h^m)\n\\
&=& N^{-4/2} \sumI \sumK  \frac{\{1-r_i
  \pi^{-1}(y_i,\u_i;\bb,g^*)\}  r_k K_h({\u_k-\u_i}) }{d^*(\x_i) D_i  V_{i,k} E(R_i\mid\u_i) f_{\U}(\u_i) }\n \\
&& \times N^{-1/2} \sumJ E \left\{ r_j K_h({\x_j-\x_i}) e^{-h(y_j)} T_{j,k} R_l K_h({\U_l-\u_i})\mid \x_j, r_j, r_j y_j \right\} \n\\
&&+ N^{-4/2} \sumI \sumK  \frac{\{1-r_i
  \pi^{-1}(y_i,\u_i;\bb,g^*)\}  r_k K_h({\u_k-\u_i}) }{d^*(\x_i) D_i  V_{i,k} E(R_i\mid\u_i) f_{\U}(\u_i) }\n \\
&& \times N^{-1/2} \sumL E \left\{ R_j K_h({\X_j-\x_i}) e^{-h(Y_j)} T_{j,k} r_l K_h({\u_l-\u_i})\mid \x_l, r_l, r_l y_l \right\} \n\\
&&- N^{-4/2} \sumI \sumK  \frac{\{1-r_i
  \pi^{-1}(y_i,\u_i;\bb,g^*)\}  r_k K_h({\u_k-\u_i}) }{d^*(\x_i) D_i  V_{i,k} E(R_i\mid\u_i) f_{\U}(\u_i) }\n \\
&& \times N^{1/2} E \left\{ R_j K_h({\X_j-\x_i}) e^{-h(Y_j)} T_{j,k} R_l K_h({\U_l-\u_i})\right\}
+O_p(N^{-1/2})+O_p(N^{1/2} h^m)\n\\
&=&T_{231CA}+T_{231CB}-T_{231CC}+O_p(N^{-1/2})+O_p(N^{1/2} h^m).\n
\ee

Consider $T_{231CA}$,
\be \label{ustat:t231ca}
&&T_{231CA}\n\\
&=&N^{-4/2} \sumI \sumK  \frac{\{1-r_i
  \pi^{-1}(y_i,\u_i;\bb,g^*)\}  r_k K_h({\u_k-\u_i}) }{d^*(\x_i) D_i  V_{i,k} E(R_i\mid\u_i) f_{\U}(\u_i) }\n \\
&& \times N^{-1/2} \sumJ E \left\{ r_j K_h({\x_j-\x_i}) e^{-h(y_j)} T_{j,k} R_l K_h({\U_l-\u_i})\mid \x_j, r_j, r_j y_j \right\} \n\\
&=&N^{-5/2} \sumI \sumJ \sumK \frac{\{1-r_i
  \pi^{-1}(y_i,\u_i;\bb,g^*)\} r_j K_h({\x_j-\x_i}) e^{-h(y_j)} r_k K_h({\u_k-\u_i}) T_{j,k}  }{d^*(\x_i) D_i V_{i,k}   }\n \\
&&+O_p(N^{1/2} h^m)\n\\
&=& T_{21}+O_p(N^{1/2} h^m),\n
\ee
where $T_{21}$ is defined in  (\ref{eq:t21}).

Consider $T_{231CB}$,
\be \label{ustat:t231cb}
&&T_{231CB}\n\\
&=&N^{-4/2} \sumI \sumK  \frac{\{1-r_i
  \pi^{-1}(y_i,\u_i;\bb,g^*)\}  r_k K_h({\u_k-\u_i}) }{d^*(\x_i) D_i  V_{i,k} E(R_i\mid\u_i) f_{\U}(\u_i) }\n \\
&& \times N^{-1/2} \sumL E \left\{ R_j K_h({\X_j-\x_i}) e^{-h(Y_j)} T_{j,k} r_l K_h({\u_l-\u_i})\mid \x_l, r_l, r_l y_l \right\} \n\\
&=&N^{-5/2} \sumI \sumK  \sumL \frac{\{1-r_i
  \pi^{-1}(y_i,\u_i;\bb,g^*)\}  r_k K_h({\u_k-\u_i}) r_l K_h({\u_l-\u_i}) }{d^*(\x_i) D_i  V_{i,k} E(R_i\mid\u_i) f_{\U}(\u_i) }\n \\
&& \times  E \left\{ R_j K_h({\X_j-\x_i}) e^{-h(Y_j)} T_{j,k} \right\} \n\\
&=&N^{-5/2} \sumI \sumK  \sumL  \frac{\{1-r_i
  \pi^{-1}(y_i,\u_i;\bb,g^*)\}  r_k K_h({\u_k-\u_i}) r_l K_h({\u_l-\u_i}) }{d^*(\x_i) D_i d^*(\x_k) \{E(R_i\mid\u_i) f_{\U}(\u_i)\}^2 }\n \\
&& \times
\{\Delta_{1i}\Delta_{1k} \Delta_{3k} -\Delta_{1k}^2 \Delta_{3i} \}
+ O_p(N^{1/2}h^m)\n\\
&=&N^{-5/2} \sumI \sumJ \sumK   \frac{\{1-r_i
  \pi^{-1}(y_i,\u_i;\bb,g^*)\}  r_j K_h({\u_j-\u_i}) r_k K_h({\u_k-\u_i}) }{d^*(\x_i) D_i d^*(\x_j) \{E(R_i\mid\u_i) f_{\U}(\u_i)\}^2 }\n \\
&& \times
\{\Delta_{1i}\Delta_{1j} \Delta_{3j} -\Delta_{1j}^2 \Delta_{3i} \}
+ O_p(N^{1/2}h^m).
\ee

Consider $T_{231CC}$,
\be \label{ustat:t231cc}
&&T_{231CC}\n\\
&=& N^{-4/2} \sumI \sumK  \frac{\{1-r_i
  \pi^{-1}(y_i,\u_i;\bb,g^*)\}  r_k K_h({\u_k-\u_i}) }{d^*(\x_i) D_i  V_{i,k} E(R_i\mid\u_i) f_{\U}(\u_i) }\n \\
&& \times N^{1/2} E \left\{ R_j K_h({\X_j-\x_i}) e^{-h(Y_j)} T_{j,k} R_l K_h({\U_l-\u_i})\right\} \n\\
&=& N^{-3/2} \sumI \sumK  \frac{\{1-r_i
  \pi^{-1}(y_i,\u_i;\bb,g^*)\}  r_k K_h({\u_k-\u_i}) \{\Delta_{1i}\Delta_{1k} \Delta_{3k} -\Delta_{1k}^2 \Delta_{3i} \}   }{d^*(\x_i) D_i  d^*(\x_k) E(R_i\mid\u_i) f_{\U}(\u_i) }
+ O_p(N^{1/2}h^m)\n\\
&=&T_{231BBC},\n
\ee
where $T_{231BBC}$ is defined in (\ref{ustat:t231bbc}).

Applying U-statistics property to $T_{331}$ from  (\ref{eq:t331}),
\be \label{ustat:t331}
&&T_{331}\n\\
&=&N^{-7/2} \sumI \sumJ \sumK \sumL \frac{\{1-r_i  \pi^{-1}(y_i,\u_i;\bb,g^*)\}C_i r_j  K_{h} (\u_j-\u_i)r_k K_{h} (\u_k-\u_i)  \Delta^2_{1j}}{d^*(\x_i) D_i^2 \{d^* (\x_j) E(R_i\mid\u_i) f_{\U}(\u_i)\}^2 E(R_j\mid\x_j) f_{\X}(\x_j)} \n\\
&&\times [ r_l  K_{h}(\x_l-\x_j) \{e^{-h(y_l)}+e^{-g^* (\u_j)}e^{-2h(y_l)}\}] \n\\
&=&N^{-4/2} \sumI \sumJ\frac{\{1-r_i  \pi^{-1}(y_i,\u_i;\bb,g^*)\}C_i r_j  K_{h} (\u_j-\u_i)  \Delta^2_{1j}}{d^*(\x_i) D_i^2 \{d^* (\x_j) E(R_i\mid\u_i) f_{\U}(\u_i)\}^2 E(R_j\mid\x_j) f_{\X}(\x_j)} \n\\
&&\times N^{-1/2} \sumK E[ r_k K_{h} (\u_k-\u_i) R_l  K_{h}(\X_l-\x_j) \{e^{-h(Y_l)}+e^{-g^* (\u_j)}e^{-2h(Y_l)}\} \mid \x_k, r_k, r_k y_k] \n\\
&&+N^{-4/2} \sumI \sumJ\frac{\{1-r_i  \pi^{-1}(y_i,\u_i;\bb,g^*)\}C_i r_j  K_{h} (\u_j-\u_i)  \Delta^2_{1j}}{d^*(\x_i) D_i^2 \{d^* (\x_j) E(R_i\mid\u_i) f_{\U}(\u_i)\}^2 E(R_j\mid\x_j) f_{\X}(\x_j)} \n\\
&&\times N^{-1/2} \sumL E[  R_k K_{h} (\U_k-\u_i) r_l  K_{h}(\x_l-\x_j) \{e^{-h(y_l)}+e^{-g^* (\u_j)}e^{-2h(y_l)}\} \mid \x_l, r_l, r_l y_l] \n\\
&&-N^{-4/2} \sumI \sumJ\frac{\{1-r_i  \pi^{-1}(y_i,\u_i;\bb,g^*)\}C_i r_j  K_{h} (\u_j-\u_i)  \Delta^2_{1j}}{d^*(\x_i) D_i^2 \{d^* (\x_j) E(R_i\mid\u_i) f_{\U}(\u_i)\}^2 E(R_j\mid\x_j) f_{\X}(\x_j)} \n\\
&&\times N^{1/2}E[ R_k K_{h} (\U_k-\u_i) R_l  K_{h}(\X_l-\x_j) \{e^{-h(Y_l)}+e^{-g^* (\u_j)}e^{-2h(Y_l)}\} ] \n\\
&&+O_p(N^{-1/2})+ O_p(N^{1/2}h^m)\n\\
&=& T_{331A}+T_{331B}-T_{331C}+O_p(N^{-1/2})+ O_p(N^{1/2}h^m).\n
\ee

Consider $T_{331A}$,
\be \label{ustat:t331a}
&& T_{331A}\n\\
&=&N^{-4/2} \sumI \sumJ\frac{\{1-r_i  \pi^{-1}(y_i,\u_i;\bb,g^*)\}C_i r_j  K_{h} (\u_j-\u_i)  \Delta^2_{1j}}{d^*(\x_i) D_i^2 \{d^* (\x_j) E(R_i\mid\u_i) f_{\U}(\u_i)\}^2 E(R_j\mid\x_j) f_{\X}(\x_j)} \n\\
&&\times N^{-1/2} \sumK E[ r_k K_{h} (\u_k-\u_i) R_l  K_{h}(\X_l-\x_j) \{e^{-h(Y_l)}+e^{-g^* (\u_j)}e^{-2h(Y_l)}\} \mid \x_k, r_k, r_k y_k] \n\\
&=&N^{-5/2} \sumI \sumJ  \sumK \frac{\{1-r_i  \pi^{-1}(y_i,\u_i;\bb,g^*)\}C_i r_j  K_{h} (\u_j-\u_i)  \Delta^2_{1j}}{d^*(\x_i) D_i^2 \{d^* (\x_j) E(R_i\mid\u_i) f_{\U}(\u_i)\}^2 E(R_j\mid\x_j) f_{\X}(\x_j)} \n\\
&&\times r_k K_{h} (\u_k-\u_i)  E[ R_l  K_{h}(\X_l-\x_j) \{e^{-h(Y_l)}+e^{-g^* (\u_j)}e^{-2h(Y_l)}\}] \n\\
&=&N^{-5/2} \sumI \sumJ  \sumK \frac{\{1-r_i  \pi^{-1}(y_i,\u_i;\bb,g^*)\}C_i r_j  K_{h} (\u_j-\u_i)r_k K_{h} (\u_k-\u_i)  \Delta^2_{1j}}{d^*(\x_i) D_i^2 d^* (\x_j) \{E(R_i\mid\u_i) f_{\U}(\u_i)\}^2}
+O_p(N^{1/2}h^m)\n\\
&=&T_{332BA},\n
\ee
where $T_{332BA}$ is defined in (\ref{ustat:t332ba}).

Consider $T_{331B}$,
\be \label{ustat:t331b}
&& T_{331B}\n\\
&=&N^{-4/2} \sumI \sumJ\frac{\{1-r_i  \pi^{-1}(y_i,\u_i;\bb,g^*)\}C_i r_j  K_{h} (\u_j-\u_i)  \Delta^2_{1j}}{d^*(\x_i) D_i^2 \{d^* (\x_j) E(R_i\mid\u_i) f_{\U}(\u_i)\}^2 E(R_j\mid\x_j) f_{\X}(\x_j)} \n\\
&&\times N^{-1/2} \sumL E[  R_k K_{h} (\U_k-\u_i) r_l  K_{h}(\x_l-\x_j) \{e^{-h(y_l)}+e^{-g^* (\u_j)}e^{-2h(y_l)}\} \mid \x_l, r_l, r_l y_l] \n\\
&=&N^{-5/2} \sumI \sumJ \sumL  \frac{\{1-r_i  \pi^{-1}(y_i,\u_i;\bb,g^*)\}C_i r_j  K_{h} (\u_j-\u_i)  \Delta^2_{1j}}{d^*(\x_i) D_i^2 \{d^* (\x_j)\}^2 E(R_i\mid\u_i) f_{\U}(\u_i) E(R_j\mid\x_j) f_{\X}(\x_j)} \n\\
&&\times r_l  K_{h}(\x_l-\x_j) \{e^{-h(y_l)}+e^{-g^* (\u_j)}e^{-2h(y_l)}\}
+O_p(N^{1/2}h^m)\n\\
&=&N^{-5/2} \sumI \sumJ \sumK  \frac{\{1-r_i  \pi^{-1}(y_i,\u_i;\bb,g^*)\}C_i r_j  K_{h} (\u_j-\u_i)  \Delta^2_{1j}}{d^*(\x_i) D_i^2 \{d^* (\x_j)\}^2 E(R_i\mid\u_i) f_{\U}(\u_i) E(R_j\mid\x_j) f_{\X}(\x_j)} \n\\
&&\times r_k  K_{h}(\x_k-\x_j) \{e^{-h(y_k)}+e^{-g^* (\u_j)}e^{-2h(y_k)}\}
+O_p(N^{1/2}h^m).
\ee

Consider $T_{331C}$,
\be \label{ustat:t331c}
&& T_{331C}\n\\
&=&N^{-4/2} \sumI \sumJ\frac{\{1-r_i  \pi^{-1}(y_i,\u_i;\bb,g^*)\}C_i r_j  K_{h} (\u_j-\u_i)  \Delta^2_{1j}}{d^*(\x_i) D_i^2 \{d^* (\x_j) E(R_i\mid\u_i) f_{\U}(\u_i)\}^2 E(R_j\mid\x_j) f_{\X}(\x_j)} \n\\
&&\times N^{1/2}E[ R_k K_{h} (\U_k-\u_i) R_l  K_{h}(\X_l-\x_j) \{e^{-h(Y_l)}+e^{-g^* (\u_j)}e^{-2h(Y_l)}\} ] \n\\
&=&N^{-3/2} \sumI \sumJ\frac{\{1-r_i  \pi^{-1}(y_i,\u_i;\bb,g^*)\}C_i r_j  K_{h} (\u_j-\u_i)  \Delta^2_{1j}}{d^*(\x_i) D_i^2 d^* (\x_j) E(R_i\mid\u_i) f_{\U}(\u_i) }
+O_p(N^{1/2}h^m)\n\\
&=&T_{31}+O_p(N^{1/2}h^m),\n
\ee
where $T_{31}$ is as defined in (\ref{eq:t31}).

Putting all the terms together (cancelling and collecting similar terms) and rewriting (\ref{eq:final2}) we get the following,
\be \label{eq:final3}
&&\frac{1}{\sqrt{N}}\sumI\wh\S^*\eff(\x_i,r_i,r_i y_i,\bb)\n\\
&=& \frac{2}{\sqrt{N}}\sumI \frac{A_i}{d^*(\x_i)} +2T_{21}+T_{2211A}+T_{2211B}+T_{2211C}-T_{2221}-2T_{2222}+2T_{2223}\n\\
&&-T_{231C}-T_{231AB}-T_{231BB}+T_{232BBB}-T_{31}-2T_{2212ABB}\n\\
&&-T_{321}-2T_{322}-T_{332B}+T_{332C}+2T_{323}+T_{331}-2T_{11}+T_{12}\n\\
&&+O_p(N^{1/2}h^{2m}+ N^{-1/2}h^{-p}+N^{-1/2}+N^{1/2}h^m)\n\\
&=& \frac{2}{\sqrt{N}}\sumI \frac{A_i}{d^*(\x_i)}
+T_{2211AB}+T_{2211BB}+T_{2211C}+2T_{2211AC}-2T_{2212ABBB}
\n\\
&&-T_{2221}-2T_{2222B}-2T_{2222C} +2T_{2223B}-T_{231ABB}-T_{231BBB}\n\\
&&-T_{231CB}+T_{232BBBB}-T_{31}-2T_{322}+T_{323}-T_{321}\n\\
&&+T_{332C}+T_{331B}-2T_{11}+T_{12}+O_p(N^{1/2}h^{2m}+ N^{-1/2}h^{-p}+N^{-1/2}+N^{1/2}h^m),
\ee
where
$T_{2211AB}$ is defined in  (\ref{ustat:t2211ab}),
$T_{2211BB}$ is defined in  (\ref{ustat:t2211bb}),
$T_{2211C}$ is defined in  (\ref{ustat:t2211c}),
$T_{2211AC}$ is defined in  (\ref{ustat:t2211ac}),
$T_{2212ABBB}$ is defined in  (\ref{ustat:t2212abbb}),
$T_{2221}$ is defined in  (\ref{eq:t2221}),
$T_{2222B}$ is defined in  (\ref{ustat:t2222b}),
$T_{2222C}$ is defined in  (\ref{ustat:t2222c}),
$T_{2223B}$ is defined in  (\ref{ustat:t2223b}),
$T_{231ABB}$ is defined in  (\ref{ustat:t231abb}),
$T_{231BBB}$ is defined in  (\ref{ustat:t231bbb}),
$T_{231CB}$ is defined in  (\ref{ustat:t231cb}),
$T_{232BBBB}$ is defined in  (\ref{ustat:t232bbbb}),
$T_{31}$ is defined in  (\ref{eq:t31}),
$T_{322}$ is defined in  (\ref{eq:t322}),
$T_{323}$ is defined in  (\ref{eq:t323}),
$T_{321}$ is defined in  (\ref{eq:t321}),
$T_{332C}$ is defined in  (\ref{ustat:t332c}),
$T_{331B}$ is defined in  (\ref{ustat:t331b}),
$T_{11}$ is defined in  (\ref{eq:t11}), and
$T_{12}$ is defined in  (\ref{eq:t12}).

Consider $T_{2211AB}$ as defined in  (\ref{ustat:t2211ab}), applying the U-statistics property to it we get,
\be \label{ustat:t2211ab2}
&&T_{2211AB}\n\\
&=& N^{-5/2} \sumI  \sumJ \sumK \frac{\{1-r_i
  \pi^{-1}(y_i,\u_i;\bb,g^*)\}  r_j K_h({\u_j-\u_i})   r_k  K_{h}(\x_k-\x_j) e^{-h(y_k)} \Delta_{1i} \Delta_{3j} }{d^*(\x_i) D_i d^*(\x_j)E(R_i\mid\u_i) f_{\U}(\u_i) E(R_j\mid\x_j) f_{\X}(\x_j) }\n\\
&&+ O_p(N^{1/2}h^m)\n\\
&=& N^{-2/2} \sumJ \frac{r_j \Delta_{3j} }{ d^*(\x_j) E(R_j\mid\x_j) f_{\X}(\x_j) }\n\\
&& \times N^{-1/2} \sumI E \left[ \frac{\{1-r_i
  \pi^{-1}(y_i,\u_i;\bb,g^*)\}  K_h({\u_j-\u_i})   R_k  K_{h}(\X_k-\x_j) e^{-h(Y_k)} \Delta_{1i} }{d^*(\x_i) D_i E(R_i\mid\u_i) f_{\U}(\u_i)  } \mid \x_i, r_i, r_i y_i\right]\n\\
&&+ N^{-2/2} \sumJ \frac{r_j \Delta_{3j} }{ d^*(\x_j) E(R_j\mid\x_j) f_{\X}(\x_j) }\n\\
&& \times N^{-1/2} \sumK E \left[ \frac{ \{1-R_i
  \pi^{-1}(Y_i,\U_i;\bb,g^*)\}  K_h({\u_j-\U_i})  r_k  K_{h}(\x_k-\x_j) e^{-h(y_k)} \Delta_{1i} }{d^*(\X_i) D_i E(R_i\mid\U_i) f_{\U}(\U_i)  } \mid \x_k, r_k, r_k y_k\right]\n\\
&&- N^{-2/2} \sumJ \frac{r_j \Delta_{3j} }{ d^*(\x_j) E(R_j\mid\x_j) f_{\X}(\x_j) }\n\\
&& \times N^{1/2}E \left[ \frac{\{1-R_i
  \pi^{-1}(Y_i,\U_i;\bb,g^*)\}  K_h({\u_j-\U_i})   R_k  K_{h}(\X_k-\x_j) e^{-h(Y_k)} \Delta_{1i} }{d^*(\X_i) D_i E(R_i\mid\U_i) f_{\U}(\U_i)  } \right]\n\\
&&+ O_p(N^{1/2}h^m)+ O_p(N^{-1/2}) \n\\
&=&T_{2211ABA}+T_{2211ABB}-T_{2211ABC}+ O_p(N^{1/2}h^m)+ O_p(N^{-1/2}).\n
\ee

Consider $T_{2211ABA}$,
\be \label{ustat:t2211aba}
&&T_{2211ABA}\n\\
&=& N^{-2/2} \sumJ \frac{r_j \Delta_{3j} }{ d^*(\x_j) E(R_j\mid\x_j) f_{\X}(\x_j) }\n\\
&& \times N^{-1/2} \sumI E \left[ \frac{\{1-r_i
  \pi^{-1}(y_i,\u_i;\bb,g^*)\}  K_h({\u_j-\u_i})   R_k  K_{h}(\X_k-\x_j) e^{-h(Y_k)} \Delta_{1i} }{d^*(\x_i) D_i E(R_i\mid\u_i) f_{\U}(\u_i)  } \mid \x_i, r_i, r_i y_i\right]\n\\
&=& N^{-3/2}  \sumI \sumJ \frac{r_j \Delta_{3j} }{ d^*(\x_j) E(R_j\mid\x_j) f_{\X}(\x_j) }\n\\
&& \times \frac{\{1-r_i
  \pi^{-1}(y_i,\u_i;\bb,g^*)\}  K_h({\u_j-\u_i})    \Delta_{1i} }{d^*(\x_i) D_i E(R_i\mid\u_i) f_{\U}(\u_i)  }   E \left\{R_k  K_{h}(\X_k-\x_j) e^{-h(Y_k)} \right\}\n\\
&=& N^{-3/2}  \sumI \sumJ \frac{\{1-r_i
  \pi^{-1}(y_i,\u_i;\bb,g^*)\}  r_j K_h({\u_j-\u_i})    \Delta_{1i}  \Delta_{1j} \Delta_{3j} }{ d^*(\x_i) D_i  d^*(\x_j)  E(R_i\mid\u_i) f_{\U}(\u_i)  }
+ O_p(N^{1/2}h^m)\n\\
&=& T_{2211AC},\n
\ee
where $T_{2211AC}$ is as defined in (\ref{ustat:t2211ac}).

Consider $T_{2211ABB}$,
\be \label{ustat:t2211abb}
&&T_{2211ABB}\n\\
&=&N^{-2/2} \sumJ \frac{r_j \Delta_{3j} }{ d^*(\x_j) E(R_j\mid\x_j) f_{\X}(\x_j) }\n\\
&& \times N^{-1/2} \sumK E \left[ \frac{ \{1-R_i
  \pi^{-1}(Y_i,\U_i;\bb,g^*)\}  K_h({\u_j-\U_i})  r_k  K_{h}(\x_k-\x_j) e^{-h(y_k)} \Delta_{1i} }{d^*(\X_i) D_i E(R_i\mid\U_i) f_{\U}(\U_i)  } \mid \x_k, r_k, r_k y_k\right]\n\\
&=&N^{-3/2} \sumJ  \sumK \frac{r_j r_k  K_{h}(\x_k-\x_j) e^{-h(y_k)} \Delta_{3j} }{ d^*(\x_j) E(R_j\mid\x_j) f_{\X}(\x_j) }E \left[ \frac{ \{1-R_i
  \pi^{-1}(Y_i,\U_i;\bb,g^*)\}  K_h({\u_j-\U_i}) \Delta_{1i} }{d^*(\X_i) D_i E(R_i\mid\U_i) f_{\U}(\U_i)  } \right]\n\\
&=&N^{-3/2} \sumJ  \sumK \frac{r_j r_k  K_{h}(\x_k-\x_j) e^{-h(y_k)} \Delta_{3j} }{ d^*(\x_j) E(R_j\mid\x_j) f_{\X}(\x_j)}
   E \left[\frac{ \Delta_{1j} \{w^{-1}(\X_j)
-w^{*-1}(\X_j)\}}{d^*(\X_j) D_j}  \mid \u_j,1 \right]\n\\
&&+ O_p(N^{1/2}h^m)\n\\
&=&N^{-3/2} \sumJ  \sumK \frac{r_j r_k  K_{h}(\x_k-\x_j) e^{-h(y_k)} \Delta_{3j} \{e^{-g(\u_j)}
-e^{-g^*(\u_j)} \} }{ d^*(\x_j) E(R_j\mid\x_j) f_{\X}(\x_j)} + O_p(N^{1/2}h^m)\n\\
&=&N^{-3/2} \sumI  \sumJ \frac{r_i r_j  K_{h}(\x_j-\x_i) e^{-h(y_j)}
  \Delta_{3i}\{e^{-g(\u_i)}
-e^{-g^*(\u_i)} \} }{ d^*(\x_i) E(R_i\mid\x_i) f_{\X}(\x_i) } + O_p(N^{1/2}h^m).
\ee

Consider $T_{2211ABC}$,
\be \label{ustat:t2211abc}
&&T_{2211ABC}\n\\
&=& N^{-2/2} \sumJ \frac{r_j \Delta_{3j} }{ d^*(\x_j) E(R_j\mid\x_j) f_{\X}(\x_j) }\n\\
&& \times N^{1/2}E \left[ \frac{\{1-R_i
  \pi^{-1}(Y_i,\U_i;\bb,g^*)\}  K_h({\u_j-\U_i})   R_k  K_{h}(\X_k-\x_j) e^{-h(Y_k)} \Delta_{1i} }{d^*(\X_i) D_i E(R_i\mid\U_i) f_{\U}(\U_i)  } \right]\n\\
&=& N^{-1/2} \sumJ \frac{r_j \Delta_{3j} }{ d^*(\x_j) E(R_j\mid\x_j) f_{\X}(\x_j) }\n\\
&& \times E\{ R_k  K_{h}(\X_k-\x_j) e^{-h(Y_k)}\} E \left[ \frac{\{1-R_i
  \pi^{-1}(Y_i,\U_i;\bb,g^*)\}  K_h({\u_j-\U_i})   \Delta_{1i} }{d^*(\X_i) D_i E(R_i\mid\U_i) f_{\U}(\U_i)  } \right]\n\\
&=& N^{-1/2} \sumJ \frac{r_j  \Delta_{1j} \Delta_{3j} }{ d^*(\x_j) } E \left[\frac{ \Delta_{1j} \{w^{-1}(\X_j)
-w^{*-1}(\X_j)\}}{d^*(\X_j) D_j}  \mid \u_j,1 \right]
+ O_p(N^{1/2}h^m)\n\\
&=& N^{-1/2} \sumJ \frac{r_j  \Delta_{1j} \Delta_{3j} \{e^{-g(\u_j)}
-e^{-g^*(\u_j)} \}}{ d^*(\x_j) }
+ O_p(N^{1/2}h^m)\n\\
&=& N^{-1/2} \sumI \frac{r_i  \Delta_{1i} \Delta_{3i}\{e^{-g(\u_i)}
-e^{-g^*(\u_i)} \} }{ d^*(\x_i)   }
+ O_p(N^{1/2}h^m).
\ee

Consider $T_{2211BB}$ as defined in  (\ref{ustat:t2211bb}), applying the U-statistics property to it we get,
\be \label{ustat:t2211bb2}
&&T_{2211BB}\n\\
&=&N^{-5/2} \sumI \sumJ \sumK \frac{\{1-r_i
  \pi^{-1}(y_i,\u_i;\bb,g^*)\} r_j K_h({\u_j-\u_i}) r_k K_{h}(\x_k-\x_j) e^{-h(y_k)}\h'_\bb(y_k;\bb)  \Delta_{1i} \Delta_{1j} }{d^*(\x_i) D_i d^*(\x_j) E(R_i\mid\u_i) f_{\U}(\u_i) E(R_j\mid\x_j) f_{\X}(\x_j)}\n\\
&&+ O_p(N^{1/2}h^m)\n\\
&=&N^{-2/2} \sumJ \frac{ r_j \Delta_{1j} }{ d^*(\x_j) E(R_j\mid\x_j) f_{\X}(\x_j)}\n\\
&& \times N^{-1/2} \sumI E\left[\frac{\{1-r_i
  \pi^{-1}(y_i,\u_i;\bb,g^*)\} K_h({\u_j-\u_i}) R_k K_{h}(\X_k-\x_j) e^{-h(Y_k)}\h'_\bb(Y_k;\bb)  \Delta_{1i}  }{d^*(\x_i) D_i E(R_i\mid\u_i) f_{\U}(\u_i) } \right]\n\\
&&+N^{-2/2} \sumJ \frac{ r_j \Delta_{1j} }{ d^*(\x_j) E(R_j\mid\x_j) f_{\X}(\x_j)}\n\\
&& \times N^{-1/2} \sumK E\left[\frac{\{1-R_i
  \pi^{-1}(Y_i,\U_i;\bb,g^*)\} K_h({\u_j-\U_i}) r_k K_{h}(\x_k-\x_j) e^{-h(y_k)}\h'_\bb(y_k;\bb)  \Delta_{1i}  }{d^*(\X_i) D_i E(R_i\mid\U_i) f_{\U}(\U_i) } \right]\n\\
&&-N^{-2/2} \sumJ \frac{ r_j \Delta_{1j} }{ d^*(\x_j) E(R_j\mid\x_j) f_{\X}(\x_j)}\n\\
&& \times N^{1/2}  E\left[\frac{\{1-R_i
  \pi^{-1}(Y_i,\U_i;\bb,g^*)\} K_h({\u_j-\U_i}) R_k K_{h}(\X_k-\x_j) e^{-h(Y_k)}\h'_\bb(Y_k;\bb)  \Delta_{1i}  }{d^*(\X_i) D_i E(R_i\mid\U_i) f_{\U}(\U_i) } \right]\n\\
&&+ O_p(N^{1/2}h^m)+O_p(N^{-1/2})\n\\
&=&T_{2211BBA}+T_{2211BBB}-T_{2211BBC}+ O_p(N^{1/2}h^m)+O_p(N^{-1/2}).\n
\ee

Consider $T_{2211BBA}$,
\be \label{ustat:t2211bba}
&&T_{2211BBA}\n\\
&=&N^{-2/2} \sumJ \frac{ r_j \Delta_{1j} }{ d^*(\x_j) E(R_j\mid\x_j) f_{\X}(\x_j)}\n\\
&& \times N^{-1/2} \sumI E\left[\frac{\{1-r_i
  \pi^{-1}(y_i,\u_i;\bb,g^*)\} K_h({\u_j-\u_i}) R_k K_{h}(\X_k-\x_j) e^{-h(Y_k)}\h'_\bb(Y_k;\bb)  \Delta_{1i}  }{d^*(\x_i) D_i E(R_i\mid\u_i) f_{\U}(\u_i) } \right]\n\\
&=&N^{-3/2} \sumI \sumJ \frac{ \{1-r_i
  \pi^{-1}(y_i,\u_i;\bb,g^*)\} r_j K_h({\u_j-\u_i}) \Delta_{1i} \Delta_{1j} }{ d^*(\x_i) D_i E(R_i\mid\u_i) f_{\U}(\u_i)  d^*(\x_j) E(R_j\mid\x_j) f_{\X}(\x_j)}\n\\
&& \times  E\left\{ R_k K_{h}(\X_k-\x_j) e^{-h(Y_k)}\h'_\bb(Y_k;\bb)   \right\}\n\\
&=&N^{-3/2} \sumI \sumJ \frac{ \{1-r_i
  \pi^{-1}(y_i,\u_i;\bb,g^*)\} r_j K_h({\u_j-\u_i}) \Delta_{1i} \Delta_{1j} \Delta_{3j} }{ d^*(\x_i) D_i d^*(\x_j) E(R_i\mid\u_i) f_{\U}(\u_i)   }
+O_p(N^{1/2}h^m)\n\\
&=&T_{2211AC},\n
\ee
where $T_{2211AC}$ is as defined in  (\ref{ustat:t2211ac}).

Consider $T_{2211BBB}$,
\be \label{ustat:t2211bbb}
&&T_{2211BBB}\n\\
&=&N^{-2/2} \sumJ \frac{ r_j \Delta_{1j} }{ d^*(\x_j) E(R_j\mid\x_j) f_{\X}(\x_j)}\n\\
&& \times N^{-1/2} \sumK E\left[\frac{\{1-R_i
  \pi^{-1}(Y_i,\U_i;\bb,g^*)\} K_h({\u_j-\U_i}) r_k K_{h}(\x_k-\x_j) e^{-h(y_k)}\h'_\bb(y_k;\bb)  \Delta_{1i}  }{d^*(\X_i) D_i E(R_i\mid\U_i) f_{\U}(\U_i) } \right]\n\\
&=&N^{-3/2} \sumJ \sumK \frac{ r_j r_k K_{h}(\x_k-\x_j) e^{-h(y_k)}\h'_\bb(y_k;\bb) \Delta_{1j} }{ d^*(\x_j) E(R_j\mid\x_j) f_{\X}(\x_j)} E\left[\frac{\{1-R_i
  \pi^{-1}(Y_i,\U_i;\bb,g^*)\} K_h({\u_j-\U_i})   \Delta_{1i}  }{d^*(\X_i) D_i E(R_i\mid\U_i) f_{\U}(\U_i) } \right]\n\\
&=&N^{-3/2} \sumJ \sumK \frac{ r_j r_k K_{h}(\x_k-\x_j) e^{-h(y_k)}\h'_\bb(y_k;\bb) \Delta_{1j} }{ d^*(\x_j) E(R_j\mid\x_j) f_{\X}(\x_j)} E \left[\frac{ \Delta_{1j} \{w^{-1}(\X_j)
-w^{*-1}(\X_j)\}}{d^*(\X_j) D_j}  \mid \u_j,1 \right] \n\\
&&+O_p(N^{1/2}h^m)\n\\
&=&N^{-3/2} \sumJ \sumK \frac{ r_j r_k K_{h}(\x_k-\x_j) e^{-h(y_k)}\h'_\bb(y_k;\bb) \Delta_{1j}\{e^{-g(\u_j)}-e^{-g^*(\u_j)}\} }{ d^*(\x_j) E(R_j\mid\x_j) f_{\X}(\x_j)} +O_p(N^{1/2}h^m)\n\\
&=&N^{-3/2} \sumI \sumJ \frac{ r_i r_j K_{h}(\x_j-\x_i) e^{-h(y_j)}\h'_\bb(y_j;\bb) \Delta_{1i}\{e^{-g(\u_i)}-e^{-g^*(\u_i)}\}  }{ d^*(\x_i) E(R_i\mid\x_i) f_{\X}(\x_i)}+O_p(N^{1/2}h^m).
\ee

Consider $T_{2211BBC}$,
\be \label{ustat:t2211bbc}
&&T_{2211BBC}\n\\
&=&N^{-2/2} \sumJ \frac{ r_j \Delta_{1j} }{ d^*(\x_j) E(R_j\mid\x_j) f_{\X}(\x_j)}\n\\
&& \times N^{1/2}  E\left[\frac{\{1-R_i
  \pi^{-1}(Y_i,\U_i;\bb,g^*)\} K_h({\u_j-\U_i}) R_k K_{h}(\X_k-\x_j) e^{-h(Y_k)}\h'_\bb(Y_k;\bb)  \Delta_{1i}  }{d^*(\X_i) D_i E(R_i\mid\U_i) f_{\U}(\U_i) } \right]\n\\
&=&N^{-1/2} \sumJ \frac{ r_j \Delta_{1j} }{ d^*(\x_j) E(R_j\mid\x_j) f_{\X}(\x_j)}\n\\
&& \times E\{R_k K_{h}(\X_k-\x_j) e^{-h(Y_k)}\h'_\bb(Y_k;\bb)\}  E\left[\frac{\{1-R_i
  \pi^{-1}(Y_i,\U_i;\bb,g^*)\} K_h({\u_j-\U_i})   \Delta_{1i}  }{d^*(\X_i) D_i E(R_i\mid\U_i) f_{\U}(\U_i) } \right]\n\\
&=&N^{-1/2} \sumJ \frac{ r_j \Delta_{1j}  \Delta_{3j}}{ d^*(\x_j)} E \left[\frac{ \Delta_{1j} \{w^{-1}(\X_j)
-w^{*-1}(\X_j)\}}{d^*(\X_j) D_j}  \mid \u_j,1 \right]
+O_p(N^{1/2}h^m)\n\\
&=&N^{-1/2} \sumJ \frac{ r_j \Delta_{1j}  \Delta_{3j}\{e^{-g(\u_j)}-e^{-g^*(\u_j)}\}}{ d^*(\x_j)}
+O_p(N^{1/2}h^m)\n\\
&=&T_{2211ABC},\n
\ee
where $T_{2211ABC}$ is as defined in (\ref{ustat:t2211abc}).

Consider $T_{2211C}$ as defined in  (\ref{ustat:t2211c}), applying the U-statistics property to it we get,
\be \label{ustat:t2211c2}
&&T_{2211C}\n\\
&=&N^{-5/2} \sumI \sumJ \sumK \frac{\{1-r_i
  \pi^{-1}(y_i,\u_i;\bb,g^*)\} r_j K_h({\x_j-\x_i}) e^{-h(y_j)} r_k K_h({\u_k-\u_i})\Delta_{1k} \Delta_{3k}}{d^*(\x_i) D_i V_{i,k}}\n\\
&&+ O_p(N^{1/2}h^m)\n\\
&=&N^{-2/2} \sumI \frac{\{1-r_i
  \pi^{-1}(y_i,\u_i;\bb,g^*)\}}{d^*(\x_i) D_i E(R_i\mid\u_i) f_{\U}(\u_i) E(R_i\mid\x_i) f_{\X}(\x_i)}   \n\\
&& \times N^{-1/2}  \sumJ E\left\{ \frac{r_j e^{-h(y_j)} K_h({\x_j-\x_i})   R_k K_h({\U_k-\u_i})\Delta_{1k} \Delta_{3k}}{d^* (\X_k)} \mid \x_j, r_j, r_j y_j\right\}\n\\
&&+N^{-2/2} \sumI \frac{\{1-r_i
  \pi^{-1}(y_i,\u_i;\bb,g^*)\}}{d^*(\x_i) D_i E(R_i\mid\u_i) f_{\U}(\u_i) E(R_i\mid\x_i) f_{\X}(\x_i)}   \n\\
&& \times N^{-1/2}  \sumK E\left\{ \frac{R_j e^{-h(Y_j)} K_h({\X_j-\x_i})  r_k K_h({\u_k-\u_i})\Delta_{1k} \Delta_{3k}}{d^* (\x_k)} \mid \x_k, r_k, r_k y_k \right\}\n\\
&&-N^{-2/2} \sumI \frac{\{1-r_i
  \pi^{-1}(y_i,\u_i;\bb,g^*)\}}{d^*(\x_i) D_i E(R_i\mid\u_i) f_{\U}(\u_i) E(R_i\mid\x_i) f_{\X}(\x_i)}   \n\\
&& \times N^{1/2}  E\left\{ \frac{R_j e^{-h(Y_j)} K_h({\X_j-\x_i})  R_k K_h({\U_k-\u_i})\Delta_{1k} \Delta_{3k}}{d^* (\X_k)} \right\}+ O_p(N^{1/2}h^m)+O_p(N^{-1/2})\n\\
&=&T_{2211CA}+T_{2211CB}-T_{2211CC}+ O_p(N^{1/2}h^m)+O_p(N^{-1/2}).\n
\ee

Consider $T_{2211CA}$,
\be \label{ustat:t2211ca}
&&T_{2211CA}\n\\
&=&N^{-2/2} \sumI \frac{\{1-r_i
  \pi^{-1}(y_i,\u_i;\bb,g^*)\}}{d^*(\x_i) D_i E(R_i\mid\u_i) f_{\U}(\u_i) E(R_i\mid\x_i) f_{\X}(\x_i)}   \n\\
&& \times N^{-1/2}  \sumJ E\left\{ \frac{r_j e^{-h(y_j)} K_h({\x_j-\x_i})   R_k K_h({\U_k-\u_i})\Delta_{1k} \Delta_{3k}}{d^* (\X_k)} \mid \x_j, r_j, r_j y_j\right\}\n\\
&=&N^{-3/2} \sumI  \sumJ \frac{\{1-r_i
  \pi^{-1}(y_i,\u_i;\bb,g^*)\}r_j e^{-h(y_j)} K_h({\x_j-\x_i})  }{d^*(\x_i) D_i E(R_i\mid\u_i) f_{\U}(\u_i) E(R_i\mid\x_i) f_{\X}(\x_i)}    E\left\{ \frac{ R_k K_h({\U_k-\u_i})\Delta_{1k} \Delta_{3k}}{d^* (\X_k)}  \right\}\n\\
&=&N^{-3/2} \sumI  \sumJ \frac{\{1-r_i
  \pi^{-1}(y_i,\u_i;\bb,g^*)\}r_j e^{-h(y_j)} K_h({\x_j-\x_i})  }{d^*(\x_i) D_i E(R_i\mid\x_i) f_{\X}(\x_i)}    E\left[\{ \Delta_{1i} \Delta_{3i}/d^* (\X_i)\} \mid \u_i,1  \right]\n\\
&&+ O_p(N^{1/2}h^m).
\ee

Consider $T_{2211CB}$,
\be \label{ustat:t2211cb}
&&T_{2211CB}\n\\
&=&N^{-2/2} \sumI \frac{\{1-r_i
  \pi^{-1}(y_i,\u_i;\bb,g^*)\}}{d^*(\x_i) D_i E(R_i\mid\u_i) f_{\U}(\u_i) E(R_i\mid\x_i) f_{\X}(\x_i)}   \n\\
&& \times N^{-1/2}  \sumK E\left\{ \frac{R_j e^{-h(Y_j)} K_h({\X_j-\x_i})  r_k K_h({\u_k-\u_i})\Delta_{1k} \Delta_{3k}}{d^* (\x_k)} \mid \x_k, r_k, r_k y_k \right\}\n\\
&=&N^{-3/2} \sumI \sumK \frac{\{1-r_i
  \pi^{-1}(y_i,\u_i;\bb,g^*)\} r_k K_h({\u_k-\u_i})\Delta_{1k} \Delta_{3k}}{d^*(\x_i) D_i d^* (\x_k) E(R_i\mid\u_i) f_{\U}(\u_i) E(R_i\mid\x_i) f_{\X}(\x_i)} E\left\{R_j e^{-h(Y_j)} K_h({\X_j-\x_i})   \right\}\n\\
&=&N^{-3/2} \sumI \sumK \frac{\{1-r_i
  \pi^{-1}(y_i,\u_i;\bb,g^*)\} r_k K_h({\u_k-\u_i})  \Delta_{1i}\Delta_{1k} \Delta_{3k}}{d^*(\x_i) D_i d^* (\x_k) E(R_i\mid\u_i) f_{\U}(\u_i) }
+ O_p(N^{1/2}h^m)\n\\
&=&N^{-3/2} \sumI \sumJ \frac{\{1-r_i
  \pi^{-1}(y_i,\u_i;\bb,g^*)\} r_j K_h({\u_j-\u_i})  \Delta_{1i}\Delta_{1j} \Delta_{3j}}{d^*(\x_i) D_i d^* (\x_j) E(R_i\mid\u_i) f_{\U}(\u_i) }
+ O_p(N^{1/2}h^m)\n\\
&=&T_{2211AC},\n
\ee
where $T_{2211AC}$ is as defined in (\ref{ustat:t2211ac}).

Consider $T_{2211CC}$,
\be \label{ustat:t2211cc}
&&T_{2211CC}\n\\
&=&N^{-2/2} \sumI \frac{\{1-r_i
  \pi^{-1}(y_i,\u_i;\bb,g^*)\}}{d^*(\x_i) D_i E(R_i\mid\u_i) f_{\U}(\u_i) E(R_i\mid\x_i) f_{\X}(\x_i)}   \n\\
&& \times N^{1/2}  E\left\{ \frac{R_j e^{-h(Y_j)} K_h({\X_j-\x_i})  R_k K_h({\U_k-\u_i})\Delta_{1k} \Delta_{3k}}{d^* (\X_k)} \right\}\n\\
&=&N^{-1/2} \sumI \frac{\{1-r_i
  \pi^{-1}(y_i,\u_i;\bb,g^*)\}}{d^*(\x_i) D_i E(R_i\mid\u_i) f_{\U}(\u_i) E(R_i\mid\x_i) f_{\X}(\x_i)}   \n\\
&& \times \Delta_{1i} E(R_i\mid\x_i) f_{\X}(\x_i) E\left[\{ \Delta_{1i} \Delta_{3i}/d^* (\X_i)\} \mid \u_i,1  \right]E(R_i\mid\u_i) f_{\U}(\u_i)
+ O_p(N^{1/2}h^m)\n\\
&=&N^{-1/2} \sumI \frac{\{1-r_i
  \pi^{-1}(y_i,\u_i;\bb,g^*)   \} \Delta_{1i}}{d^*(\x_i) D_i}   E\left[\{ \Delta_{1i} \Delta_{3i}/d^* (\X_i)\} \mid \u_i,1  \right] \n\\
&&+ O_p(N^{1/2}h^m).
\ee

Consider $T_{2212ABBB}$ as defined in  (\ref{ustat:t2212abbb}), applying the U-statistics property to it we get,
\be \label{ustat:t2212abbb2}
&&T_{2212ABBB}\n\\
&=&N^{-5/2} \sumI \sumJ \sumK  \frac{\{1-r_i
  \pi^{-1}(y_i,\u_i;\bb,g^*)\} r_j K_h({\u_j-\u_i})  r_k K_{h} (\x_k-\x_j) \Delta_{1i} \Delta_{1j} \Delta_{3j} }{d^*(\x_i) D_i d^*(\x_j)  E(R_i\mid\u_i) f_{\U}(\u_i) E(R_j\mid\x_j) f_{\X}(\x_j)}\n\\
&&+ O_p(N^{1/2}h^m)\n\\
&=&N^{-2/2}  \sumJ   \frac{ r_j \Delta_{1j} \Delta_{3j} }{d^*(\x_j) E(R_j\mid\x_j) f_{\X}(\x_j)}\n\\
&&\times N^{-1/2} \sumI E\left[  \frac{\{1-r_i
  \pi^{-1}(y_i,\u_i;\bb,g^*)\} K_h({\u_j-\u_i})  R_k K_{h} (\X_k-\x_j) \Delta_{1i}  }{d^*(\x_i) D_i E(R_i\mid\u_i) f_{\U}(\u_i)} \mid \x_i, r_i, r_i y_i \right] \n\\
&&+N^{-2/2}  \sumJ   \frac{ r_j \Delta_{1j} \Delta_{3j} }{d^*(\x_j) E(R_j\mid\x_j) f_{\X}(\x_j)}\n\\
&&\times N^{-1/2} \sumK E\left[  \frac{\{1-R_i
  \pi^{-1}(Y_i,\U_i;\bb,g^*)\} K_h({\u_j-\U_i})   r_k K_{h} (\x_k-\x_j) \Delta_{1i}  }{d^*(\X_i) D_i E(R_i\mid\U_i) f_{\U}(\U_i)} \mid \x_k, r_k, r_k y_k \right] \n\\
&&-N^{-2/2}  \sumJ   \frac{ r_j \Delta_{1j} \Delta_{3j} }{d^*(\x_j) E(R_j\mid\x_j) f_{\X}(\x_j)}\n\\
&&\times N^{1/2} E\left[  \frac{\{1-R_i
  \pi^{-1}(Y_i,\U_i;\bb,g^*)\} K_h({\u_j-\U_i})  R_k K_{h} (\X_k-\x_j) \Delta_{1i}  }{d^*(\X_i) D_i E(R_i\mid\U_i) f_{\U}(\U_i)}  \right] \n\\
&&+ O_p(N^{1/2}h^m)+O_p(N^{-1/2})\n\\
&=&T_{2212ABBBA}+T_{2212ABBBB}-T_{2212ABBBC}+ O_p(N^{1/2}h^m)+O_p(N^{-1/2}).\n
\ee

Consider $T_{2212ABBBA}$,
\be \label{ustat:t2212abbba}
&&T_{2212ABBBA}\n\\
&=&N^{-2/2}  \sumJ   \frac{ r_j \Delta_{1j} \Delta_{3j} }{d^*(\x_j) E(R_j\mid\x_j) f_{\X}(\x_j)}\n\\
&&\times N^{-1/2} \sumI E\left[  \frac{\{1-r_i
  \pi^{-1}(y_i,\u_i;\bb,g^*)\} K_h({\u_j-\u_i})  R_k K_{h} (\X_k-\x_j) \Delta_{1i}  }{d^*(\x_i) D_i E(R_i\mid\u_i) f_{\U}(\u_i)} \mid \x_i, r_i, r_i y_i \right] \n\\
&=&N^{-3/2}  \sumI \sumJ   \frac{ \{1-r_i
  \pi^{-1}(y_i,\u_i;\bb,g^*)\} r_j K_h({\u_j-\u_i})   \Delta_{1i} \Delta_{1j}  \Delta_{3j} }{ d^*(\x_i) D_i d^*(\x_j) E(R_i\mid\u_i) f_{\U}(\u_i)}
+O_p(N^{1/2}h^m)\n\\
&=&T_{2211AC},\n
\ee
where $T_{2211AC}$ is as defined in (\ref{ustat:t2211ac}).

Consider $T_{2212ABBBB}$,
\be \label{ustat:t2212abbbb}
&&T_{2212ABBBB}\n\\
&=&N^{-2/2}  \sumJ   \frac{ r_j \Delta_{1j} \Delta_{3j} }{d^*(\x_j) E(R_j\mid\x_j) f_{\X}(\x_j)}\n\\
&&\times N^{-1/2} \sumK E\left[  \frac{\{1-R_i
  \pi^{-1}(Y_i,\U_i;\bb,g^*)\} K_h({\u_j-\U_i})   r_k K_{h} (\x_k-\x_j) \Delta_{1i}  }{d^*(\X_i) D_i E(R_i\mid\U_i) f_{\U}(\U_i)} \mid \x_k, r_k, r_k y_k \right] \n\\
&=&N^{-3/2}  \sumJ \sumK  \frac{ r_j r_k K_{h} (\x_k-\x_j)  \Delta_{1j} \Delta_{3j} }{d^*(\x_j) E(R_j\mid\x_j) f_{\X}(\x_j)} E\left[  \frac{\{1-R_i
  \pi^{-1}(Y_i,\U_i;\bb,g^*)\} K_h({\u_j-\U_i})   \Delta_{1i}  }{d^*(\X_i) D_i E(R_i\mid\U_i) f_{\U}(\U_i)} \right] \n\\
&=&N^{-3/2}  \sumJ \sumK  \frac{ r_j r_k K_{h} (\x_k-\x_j)  \Delta_{1j} \Delta_{3j} }{d^*(\x_j) E(R_j\mid\x_j) f_{\X}(\x_j)} E \left[\frac{ \Delta_{1j} \{w^{-1}(\X_j)
-w^{*-1}(\X_j)\}}{d^*(\X_j) D_j}  \mid \u_j,1 \right]+O_p(N^{1/2}h^m)\n\\
&=&N^{-3/2}  \sumJ \sumK  \frac{ r_j r_k K_{h} (\x_k-\x_j)  \Delta_{1j} \Delta_{3j} \{e^{-g(\u_j}-e^{-g^*(\u_j}\}}{d^*(\x_j) E(R_j\mid\x_j) f_{\X}(\x_j)} +O_p(N^{1/2}h^m)\n\\
&=&N^{-3/2}  \sumI \sumJ  \frac{ r_i r_j K_{h} (\x_j-\x_i)  \Delta_{1i} \Delta_{3i} \{e^{-g(\u_i}-e^{-g^*(\u_i}\} }{d^*(\x_i) E(R_i\mid\x_i) f_{\X}(\x_i)}+O_p(N^{1/2}h^m).
\ee

Consider $T_{2212ABBBC}$,
\be \label{ustat:t2212abbbc}
&&T_{2212ABBBC}\n\\
&=&N^{-2/2}  \sumJ   \frac{ r_j \Delta_{1j} \Delta_{3j} }{d^*(\x_j) E(R_j\mid\x_j) f_{\X}(\x_j)}\n\\
&&\times N^{1/2} E\left[  \frac{\{1-R_i
  \pi^{-1}(Y_i,\U_i;\bb,g^*)\} K_h({\u_j-\U_i})  R_k K_{h} (\X_k-\x_j) \Delta_{1i}  }{d^*(\X_i) D_i E(R_i\mid\U_i) f_{\U}(\U_i)}  \right] \n\\
&=&N^{-1/2}  \sumJ   \frac{ r_j \Delta_{1j} \Delta_{3j} }{d^*(\x_j) } E \left[\frac{ \Delta_{1j} \{w^{-1}(\X_j)
-w^{*-1}(\X_j)\}}{d^*(\X_j) D_j}  \mid \u_j,1 \right]
+O_p(N^{1/2}h^m)\n\\
&=&N^{-1/2}  \sumJ   \frac{ r_j \Delta_{1j} \Delta_{3j}\{e^{-g(\u_j)}-e^{-g^*(\u_j)}\} }{d^*(\x_j) }
+O_p(N^{1/2}h^m)\n\\
&=&T_{2211ABC},\n
\ee
where $T_{2211ABC}$ is as defined in (\ref{ustat:t2211abc}).

Consider $T_{2221}$ as defined in  (\ref{eq:t2221}), applying the U-statistics property to it we get,
\be \label{ustat:t2221}
&&T_{2221}\n\\
&=&N^{-5/2} \sumI \sumJ \sumK \frac{\{1-r_i
  \pi^{-1}(Y_i,\u_i;\bb,g^*)\} r_j K_h({\x_j-\x_i}) e^{-h(y_j)} r_k K_h({\u_k-\u_i})\h'_\bb(y_j;\bb) \Delta_{1k}^2}{d^*(\x_i) D_i V_{i,k}}\n\\
&=&N^{-2/2} \sumI  \frac{\{1-r_i
  \pi^{-1}(Y_i,\u_i;\bb,g^*)\}}{d^*(\x_i) D_i
E(R_i\mid\u_i) f_{\U}(\u_i) E(R_i\mid\x_i) f_{\X}(\x_i)}\n\\
&& \times N^{-1/2} \sumJ E\left\{ \frac{ r_j K_h({\x_j-\x_i}) e^{-h(y_j)} \h'_\bb(y_j;\bb) R_k K_h({\U_k-\u_i}) \Delta_{1k}^2}{d^* (\X_k)}\mid \x_j,r_j,r_j y_j \right\}\n\\
&&+N^{-2/2} \sumI  \frac{\{1-r_i
  \pi^{-1}(Y_i,\u_i;\bb,g^*)\}}{d^*(\x_i) D_i
E(R_i\mid\u_i) f_{\U}(\u_i) E(R_i\mid\x_i) f_{\X}(\x_i)}\n\\
&& \times N^{-1/2} \sumK E\left\{ \frac{ R_j K_h({\X_j-\x_i}) e^{-h(Y_j)} \h'_\bb(Y_j;\bb) r_k K_h({\u_k-\u_i}) \Delta_{1k}^2}{d^* (\x_k)}\mid \x_k,r_k,r_k y_k \right\}\n\\
&&-N^{-2/2} \sumI  \frac{\{1-r_i
  \pi^{-1}(Y_i,\u_i;\bb,g^*)\}}{d^*(\x_i) D_i
E(R_i\mid\u_i) f_{\U}(\u_i) E(R_i\mid\x_i) f_{\X}(\x_i)}\n\\
&& \times N^{1/2}  E\left\{ \frac{ R_j K_h({\X_j-\x_i}) e^{-h(Y_j)} \h'_\bb(Y_j;\bb) R_k K_h({\U_k-\u_i}) \Delta_{1k}^2}{d^* (\X_k)} \right\} +O_p(N^{-1/2}) \n\\
&=&T_{2221A}+T_{2221B}-T_{2221C}+O_p(N^{-1/2}).\n
\ee

Consider $T_{2221A}$,
\be \label{ustat:t2221a}
&&T_{2221A}\n\\
&=&N^{-2/2} \sumI  \frac{\{1-r_i
  \pi^{-1}(Y_i,\u_i;\bb,g^*)\}}{d^*(\x_i) D_i
E(R_i\mid\u_i) f_{\U}(\u_i) E(R_i\mid\x_i) f_{\X}(\x_i)}\n\\
&& \times N^{-1/2} \sumJ E\left\{ \frac{ r_j K_h({\x_j-\x_i}) e^{-h(y_j)} \h'_\bb(y_j;\bb) R_k K_h({\U_k-\u_i}) \Delta_{1k}^2}{d^* (\X_k)}\mid \x_j,r_j,r_j y_j \right\}\n\\
&=&N^{-3/2} \sumI \sumJ \frac{\{1-r_i
  \pi^{-1}(Y_i,\u_i;\bb,g^*)\} r_j K_h({\x_j-\x_i}) e^{-h(y_j)} \h'_\bb(y_j;\bb)}{d^*(\x_i) D_i
 E(R_i\mid\x_i) f_{\X}(\x_i)}  E\left[\{ \Delta_{1i}^2 /d^* (\X_i)\} \mid \u_i,1  \right]\n\\
&&+O_p(N^{1/2}h^m)\n\\
&=&N^{-3/2} \sumI \sumJ \frac{\{1-r_i
  \pi^{-1}(Y_i,\u_i;\bb,g^*)\} r_j K_h({\x_j-\x_i}) e^{-h(y_j)} \h'_\bb(y_j;\bb)}{d^*(\x_i)
 E(R_i\mid\x_i) f_{\X}(\x_i)} +O_p(N^{1/2}h^m).
\ee

Consider $T_{2221B}$,
\be \label{ustat:t2221b}
&&T_{2221B}\n\\
&=&N^{-2/2} \sumI  \frac{\{1-r_i
  \pi^{-1}(Y_i,\u_i;\bb,g^*)\}}{d^*(\x_i) D_i
E(R_i\mid\u_i) f_{\U}(\u_i) E(R_i\mid\x_i) f_{\X}(\x_i)}\n\\
&& \times N^{-1/2} \sumK E\left\{ \frac{ R_j K_h({\X_j-\x_i}) e^{-h(Y_j)} \h'_\bb(Y_j;\bb) r_k K_h({\u_k-\u_i}) \Delta_{1k}^2}{d^* (\x_k)}\mid \x_k,r_k,r_k y_k \right\}\n\\
&=&N^{-3/2} \sumI \sumK \frac{\{1-r_i
  \pi^{-1}(Y_i,\u_i;\bb,g^*)\}r_k K_h({\u_k-\u_i}) \Delta_{1k}^2}{d^*(\x_i) D_i d^* (\x_k)
E(R_i\mid\u_i) f_{\U}(\u_i) E(R_i\mid\x_i) f_{\X}(\x_i)} \n\\
&&\times E\left\{ R_j K_h({\X_j-\x_i}) e^{-h(Y_j)} \h'_\bb(Y_j;\bb)  \right\}\n\\
&=&N^{-3/2} \sumI \sumK \frac{\{1-r_i
  \pi^{-1}(Y_i,\u_i;\bb,g^*)\}r_k K_h({\u_k-\u_i}) \Delta_{3i} \Delta_{1k}^2 }{d^*(\x_i) D_i d^* (\x_k)
E(R_i\mid\u_i) f_{\U}(\u_i) }
+O_p(N^{1/2}h^m)\n\\
&=&N^{-3/2} \sumI \sumJ \frac{\{1-r_i
  \pi^{-1}(Y_i,\u_i;\bb,g^*)\}r_j K_h({\u_j-\u_i}) \Delta_{3i} \Delta_{1j}^2 }{d^*(\x_i) D_i d^* (\x_j)
E(R_i\mid\u_i) f_{\U}(\u_i) }
+O_p(N^{1/2}h^m)\n\\
&=&T_{2222C},\n
\ee
where $T_{2222C}$ is as defined in (\ref{ustat:t2222c}).

Consider $T_{2221C}$,
\be \label{ustat:t2221c}
&&T_{2221C}\n\\
&=&N^{-2/2} \sumI  \frac{\{1-r_i
  \pi^{-1}(Y_i,\u_i;\bb,g^*)\}}{d^*(\x_i) D_i
E(R_i\mid\u_i) f_{\U}(\u_i) E(R_i\mid\x_i) f_{\X}(\x_i)}\n\\
&& \times N^{1/2}  E\left\{ \frac{ R_j K_h({\X_j-\x_i}) e^{-h(Y_j)} \h'_\bb(Y_j;\bb) R_k K_h({\U_k-\u_i}) \Delta_{1k}^2}{d^* (\X_k)} \right\}\n\\
&=&N^{-1/2} \sumI  \frac{\{1-r_i
  \pi^{-1}(Y_i,\u_i;\bb,g^*)\} \Delta_{3i} }{d^*(\x_i) D_i} E\left[\{ \Delta_{1i}^2 /d^* (\X_i)\} \mid \u_i,1  \right]+O_p(N^{1/2}h^m)\n\\
&=&N^{-1/2} \sumI  \frac{\{1-r_i
  \pi^{-1}(Y_i,\u_i;\bb,g^*)\} \Delta_{3i} }{d^*(\x_i) } +O_p(N^{1/2}h^m).
\ee

Consider $T_{2222B}$ as defined in  (\ref{ustat:t2222b}), applying the U-statistics property to it we get,
\be \label{ustat:t2222b2}
&&T_{2222B}\n\\
&=&N^{-5/2} \sumI \sumJ \sumK  \frac{\{1-r_i
  \pi^{-1}(y_i,\u_i;\bb,g^*)\}  r_j K_h({\u_j-\u_i})  r_k  K_{h}(\x_k-\x_j) e^{-h(y_k)} \Delta_{1j} \Delta_{3i}}{d^*(\x_i) D_i d^*(\x_j) E(R_i \mid \u_i) f_{\U}(\u_i) E(R_j \mid \x_j) f_{\X}(\x_j)}\n\\
&&+O_p(N^{1/2}h^m)\n\\
&=&N^{-2/2} \sumJ\frac{  r_j   \Delta_{1j} }{ d^*(\x_j)  E(R_j \mid \x_j) f_{\X}(\x_j)}\n\\
&& \times N^{-1/2} \sumI E \left[ \frac{\{1-r_i
  \pi^{-1}(y_i,\u_i;\bb,g^*)\}  K_h({\u_j-\u_i})   R_k  K_{h}(\X_k-\x_j) e^{-h(Y_k)} \Delta_{3i}}{d^*(\x_i) D_i  E(R_i \mid \u_i) f_{\U}(\u_i)} \mid \x_i, r_i, r_i y_i\right]\n\\
&&+N^{-2/2} \sumJ\frac{  r_j   \Delta_{1j} }{ d^*(\x_j)  E(R_j \mid \x_j) f_{\X}(\x_j)}\n\\
&& \times N^{-1/2} \sumK E \left[ \frac{\{1-R_i
  \pi^{-1}(Y_i,\U_i;\bb,g^*)\}  K_h({\u_j-\U_i}) r_k  K_{h}(\x_k-\x_j) e^{-h(y_k)} \Delta_{3i}}{d^*(\X_i) D_i  E(R_i \mid \U_i) f_{\U}(\U_i)} \mid \x_k, r_k, r_k y_k\right]\n\\
&&-N^{-2/2} \sumJ\frac{  r_j   \Delta_{1j} }{ d^*(\x_j)  E(R_j \mid \x_j) f_{\X}(\x_j)}\n\\
&& \times N^{1/2} E \left[ \frac{\{1-R_i
  \pi^{-1}(Y_i,\U_i;\bb,g^*)\}  K_h({\u_j-\U_i})  R_k  K_{h}(\X_k-\x_j) e^{-h(Y_k)} \Delta_{3i}}{d^*(\X_i) D_i  E(R_i \mid \U_i) f_{\U}(\U_i)} \right]\n\\
&&+O_p(N^{1/2}h^m)+O_p(N^{-1/2}) \n\\
&=&T_{2222BA}+T_{2222BB}-T_{2222BC}+O_p(N^{1/2}h^m)+O_p(N^{-1/2}).\n
\ee

Consider $T_{2222BA}$,
\be \label{ustat:t2222ba}
&&T_{2222BA}\n\\
&=&N^{-2/2} \sumJ\frac{  r_j   \Delta_{1j} }{ d^*(\x_j)  E(R_j \mid \x_j) f_{\X}(\x_j)}\n\\
&& \times N^{-1/2} \sumI E \left[ \frac{\{1-r_i
  \pi^{-1}(y_i,\u_i;\bb,g^*)\}  K_h({\u_j-\u_i})   R_k  K_{h}(\X_k-\x_j) e^{-h(Y_k)} \Delta_{3i}}{d^*(\x_i) D_i  E(R_i \mid \u_i) f_{\U}(\u_i)} \mid \x_i, r_i, r_i y_i\right]\n\\
&=&N^{-3/2} \sumI \sumJ\frac{\{1-r_i
  \pi^{-1}(y_i,\u_i;\bb,g^*)\}  r_j K_h({\u_j-\u_i})   \Delta_{1j} \Delta_{3i} }{d^*(\x_i) D_i d^*(\x_j) E(R_i \mid \u_i) f_{\U}(\u_i)   E(R_j \mid \x_j) f_{\X}(\x_j)} E \left\{ R_k  K_{h}(\X_k-\x_j) e^{-h(Y_k)} \right\}\n\\
&=&N^{-3/2} \sumI \sumJ\frac{\{1-r_i
  \pi^{-1}(y_i,\u_i;\bb,g^*)\}  r_j K_h({\u_j-\u_i})   \Delta_{1j}^2 \Delta_{3i} }{d^*(\x_i) D_i d^*(\x_j) E(R_i \mid \u_i) f_{\U}(\u_i)}
+O_p(N^{1/2}h^m)\n\\
&=&T_{2222C},\n
\ee
where $T_{2222C}$ is as defined in (\ref{ustat:t2222c}).

Consider $T_{2222BB}$,
\be \label{ustat:t2222bb}
&&T_{2222BB}\n\\
&=&N^{-2/2} \sumJ\frac{  r_j   \Delta_{1j} }{ d^*(\x_j)  E(R_j \mid \x_j) f_{\X}(\x_j)}\n\\
&& \times N^{-1/2} \sumK E \left[ \frac{\{1-R_i
  \pi^{-1}(Y_i,\U_i;\bb,g^*)\}  K_h({\u_j-\U_i}) r_k  K_{h}(\x_k-\x_j) e^{-h(y_k)} \Delta_{3i}}{d^*(\X_i) D_i  E(R_i \mid \U_i) f_{\U}(\U_i)} \mid \x_k, r_k, r_k y_k\right]\n\\
&=&N^{-3/2} \sumJ \sumK \frac{  r_j  r_k  K_{h}(\x_k-\x_j) e^{-h(y_k)} \Delta_{1j} }{ d^*(\x_j)  E(R_j \mid \x_j) f_{\X}(\x_j)}E \left[ \frac{\{1-R_i
  \pi^{-1}(Y_i,\U_i;\bb,g^*)\}  K_h({\u_j-\U_i})  \Delta_{3i}}{d^*(\X_i) D_i  E(R_i \mid \U_i) f_{\U}(\U_i)} \right]\n\\
&=&N^{-3/2} \sumJ \sumK \frac{  r_j  r_k  K_{h}(\x_k-\x_j) e^{-h(y_k)} \Delta_{1j} }{ d^*(\x_j)  E(R_j \mid \x_j) f_{\X}(\x_j)}  E \left[\frac{ \Delta_{3j} \{w^{-1}(\X_j)
-w^{*-1}(\X_j)\}}{d^*(\X_j) D_j}  \mid \u_j,1 \right]+O_p(N^{1/2}h^m)\n\\
&=&N^{-3/2} \sumJ \sumK \frac{  r_j  r_k  K_{h}(\x_k-\x_j) e^{-h(y_k)} \Delta_{1j} }{ d^*(\x_j) D_j  E(R_j \mid \x_j) f_{\X}(\x_j)}  \{e^{-g(\u_j)}-e^{-g^*(\u_j)}\} E \left\{\frac{ \Delta_{1j}\Delta_{3j} }{d^*(\X_j)}  \mid \u_j,1 \right\}+O_p(N^{1/2}h^m)\n\\
&=&N^{-3/2} \sumI \sumJ \frac{  r_i  r_j  K_{h}(\x_j-\x_i) e^{-h(y_j)} \Delta_{1i} }{ d^*(\x_i)D_i  E(R_i \mid \x_i) f_{\X}(\x_i)} \{e^{-g(\u_i)}-e^{-g^*(\u_i)}\} E \left\{\frac{ \Delta_{1i}\Delta_{3i} }{d^*(\X_i)}  \mid \u_i,1 \right\}\n\\
&&+O_p(N^{1/2}h^m).
\ee

Consider $T_{2222BC}$,
\be \label{ustat:t2222bc}
&&T_{2222BC}\n\\
&=&N^{-2/2} \sumJ\frac{  r_j   \Delta_{1j} }{ d^*(\x_j)  E(R_j \mid \x_j) f_{\X}(\x_j)}\n\\
&& \times N^{1/2} E \left[ \frac{\{1-R_i
  \pi^{-1}(Y_i,\U_i;\bb,g^*)\}  K_h({\u_j-\U_i})  R_k  K_{h}(\X_k-\x_j) e^{-h(Y_k)} \Delta_{3i}}{d^*(\x_i) D_i  E(R_i \mid \U_i) f_{\U}(\U_i)} \right]\n\\
&=&N^{-1/2} \sumJ\frac{  r_j   \Delta_{1j}^2 }{ d^*(\x_j)  } E \left[\frac{ \Delta_{3j} \{w^{-1}(\X_j)
-w^{*-1}(\X_j)\}}{d^*(\x_j) D_j}  \mid \u_j,1 \right]
+O_p(N^{1/2}h^m)\n\\
&=&N^{-1/2} \sumJ\frac{  r_j   \Delta_{1j}^2 }{ d^*(\x_j) D_j  } \{e^{-g(\u_j)}-e^{-g^*(\u_j)}\} E \left\{\frac{ \Delta_{1j}\Delta_{3j} }{d^*(\X_j)}  \mid \u_j,1 \right\}
+O_p(N^{1/2}h^m)\n\\
&=&N^{-1/2} \sumI \frac{  r_i \a^*(\u_i) \Delta_{1i}^2 \{e^{-g(\u_i)}-e^{-g^*(\u_i)}\} }{ d^*(\x_i)  }
+O_p(N^{1/2}h^m).
\ee

Consider $T_{2223B}$ as defined in  (\ref{ustat:t2223b}), applying the U-statistics property to it we get,
\be \label{ustat:t2223b2}
&&T_{2223B}\n\\
&=&N^{-5/2} \sumI \sumJ \sumK \frac{\{1-r_i
  \pi^{-1}(Y_i,\u_i;\bb,g^*)\}  r_j K_h({\u_j-\u_i})\Delta_{1j}^2 r_k K_{h}(\x_k-\x_j) \Delta_{3i}}{d^*(\x_i) D_i d^*(\x_j) E(R_i \mid \u_i)  f_{\U}(\u_i)  E(R_j \mid \x_j)  f_{\X}(\x_j)}\n\\
&&+O_p(N^{1/2}h^m)\n\\
&=&N^{-2/2}  \sumJ \frac{  r_j \Delta_{1j}^2 }{ d^*(\x_j)  E(R_j \mid \x_j)  f_{\X}(\x_j)}\n\\
&&\times N^{-1/2} \sumI E \left[ \frac{\{1-r_i
  \pi^{-1}(Y_i,\u_i;\bb,g^*)\}   K_h({\u_j-\u_i}) \Delta_{3i} R_k K_{h}(\X_k-\x_j) }{d^*(\x_i) D_i  E(R_i \mid \u_i)  f_{\U}(\u_i)} \mid \x_i, r_i, r_i y_i \right]\n\\
&&+N^{-2/2}  \sumJ \frac{  r_j \Delta_{1j}^2 }{ d^*(\x_j)  E(R_j \mid \x_j)  f_{\X}(\x_j)}\n\\
&&\times N^{-1/2} \sumK E \left[ \frac{\{1-R_i
  \pi^{-1}(Y_i,\U_i;\bb,g^*)\}   K_h({\u_j-\U_i}) \Delta_{3i}  r_k K_{h}(\x_k-\x_j) }{d^*(\X_i) D_i  E(R_i \mid \U_i)  f_{\U}(\U_i)} \mid \x_k, r_k, r_k y_k  \right]\n\\
&&-N^{-2/2}  \sumJ \frac{  r_j \Delta_{1j}^2 }{ d^*(\x_j)  E(R_j \mid \x_j)  f_{\X}(\x_j)}\n\\
&&\times N^{1/2} E \left[ \frac{\{1-R_i
  \pi^{-1}(Y_i,\U_i;\bb,g^*)\}   K_h({\u_j-\U_i}) \Delta_{3i} R_k K_{h}(\X_k-\x_j) }{d^*(\X_i) D_i  E(R_i \mid \U_i)  f_{\U}(\U_i)}  \right]\n\\
&&+O_p(N^{1/2}h^m)+O_p(N^{-1/2}) \n\\
&=&T_{2223BA}+T_{2223BB}-T_{2223BC}+O_p(N^{1/2}h^m+O_p(N^{-1/2}).\n
\ee

Consider $T_{2223BA}$,
\be \label{ustat:t2223ba}
&&T_{2223BA}\n\\
&=&N^{-2/2}  \sumJ \frac{  r_j \Delta_{1j}^2 }{ d^*(\x_j)  E(R_j \mid \x_j)  f_{\X}(\x_j)}\n\\
&&\times N^{-1/2} \sumI E \left[ \frac{\{1-r_i
  \pi^{-1}(Y_i,\u_i;\bb,g^*)\}   K_h({\u_j-\u_i}) \Delta_{3i} R_k K_{h}(\X_k-\x_j) }{d^*(\x_i) D_i  E(R_i \mid \u_i)  f_{\U}(\u_i)} \mid \x_i, r_i, r_i y_i \right]\n\\
&=&N^{-3/2} \sumI \sumJ \frac{ \{1-r_i
  \pi^{-1}(Y_i,\u_i;\bb,g^*)\} r_j K_h({\u_j-\u_i}) \Delta_{1j}^2  \Delta_{3i} }{d^*(\x_i) D_i  d^*(\x_j) E(R_i \mid \u_i)  f_{\U}(\u_i)}
+O_p(N^{1/2}h^m)\n\\
&=&T_{2222C},\n
\ee
where $T_{2222C}$ is as defined in (\ref{ustat:t2222c}).

Consider $T_{2223BB}$,
\be \label{ustat:t2223bb}
&&T_{2223BB}\n\\
&=&N^{-2/2}  \sumJ \frac{  r_j \Delta_{1j}^2 }{ d^*(\x_j)  E(R_j \mid \x_j)  f_{\X}(\x_j)}\n\\
&&\times N^{-1/2} \sumK E \left[ \frac{\{1-R_i
  \pi^{-1}(Y_i,\U_i;\bb,g^*)\}   K_h({\u_j-\U_i}) \Delta_{3i}  r_k K_{h}(\x_k-\x_j) }{d^*(\X_i) D_i  E(R_i \mid \U_i)  f_{\U}(\U_i)} \mid \x_k, r_k, r_k y_k  \right]\n\\
&=&N^{-3/2}  \sumJ \sumK \frac{  r_j r_k K_{h}(\x_k-\x_j) \Delta_{1j}^2 }{ d^*(\x_j)  E(R_j \mid \x_j)  f_{\X}(\x_j)} E \left[ \frac{\{1-R_i
  \pi^{-1}(Y_i,\U_i;\bb,g^*)\}   K_h({\u_j-\U_i}) \Delta_{3i} }{d^*(\X_i) D_i  E(R_i \mid \U_i)  f_{\U}(\U_i)} \right]\n\\
&=&N^{-3/2}  \sumJ \sumK \frac{  r_j r_k K_{h}(\x_k-\x_j) \Delta_{1j}^2 }{ d^*(\x_j)  E(R_j \mid \x_j)  f_{\X}(\x_j)} E \left[\frac{ \Delta_{3j} \{w^{-1}(\X_j)
-w^{*-1}(\X_j)\}}{d^*(\X_j) D_j}  \mid \u_j,1 \right]+O_p(N^{1/2}h^m)\n\\
&=&N^{-3/2}  \sumJ \sumK \frac{  r_j r_k K_{h}(\x_k-\x_j) \Delta_{1j}^2 \{e^{-g(\u_j)}-e^{-g^*(\u_j)}\} }{ d^*(\x_j) D_j E(R_j \mid \x_j)  f_{\X}(\x_j)}  E \left\{\frac{ \Delta_{1j}\Delta_{3j} }{d^*(\X_j)}  \mid \u_j,1 \right\}+O_p(N^{1/2}h^m)\n\\
&=&N^{-3/2}  \sumI \sumJ \frac{  r_i r_j K_{h}(\x_j-\x_i) \Delta_{1i}^2 \{e^{-g(\u_i)}-e^{-g^*(\u_i)}\} }{ d^*(\x_i) D_i  E(R_i \mid \x_i)  f_{\X}(\x_i)}  E \left\{\frac{ \Delta_{1i}\Delta_{3i} }{d^*(\X_i)}  \mid \u_i,1 \right\}+O_p(N^{1/2}h^m).
\ee

Consider $T_{2223BC}$,
\be \label{ustat:t2223bc}
&&T_{2223BC}\n\\
&=&N^{-2/2}  \sumJ \frac{  r_j \Delta_{1j}^2 }{ d^*(\x_j)  E(R_j \mid \x_j)  f_{\X}(\x_j)}\n\\
&&\times N^{1/2} E \left[ \frac{\{1-R_i
  \pi^{-1}(Y_i,\U_i;\bb,g^*)\}   K_h({\u_j-\U_i}) \Delta_{3i} R_k K_{h}(\X_k-\x_j) }{d^*(\X_i) D_i  E(R_i \mid \U_i)  f_{\U}(\U_i)}  \right]\n\\
&=&N^{-1/2}  \sumJ \frac{  r_j \Delta_{1j}^2 }{ d^*(\x_j)}E \left[\frac{ \Delta_{3j} \{w^{-1}(\X_j)
-w^{*-1}(\X_j)\}}{d^*(\X_j) D_j}  \mid \u_j,1 \right]
+O_p(N^{1/2}h^m)\n\\
&=&N^{-3/2}  \sumJ \sumK \frac{  r_j  \Delta_{1j}^2 }{ d^*(\x_j) D_j } \{e^{-g(\u_j)}-e^{-g^*(\u_j)}\} E \left\{\frac{ \Delta_{1j}\Delta_{3j} }{d^*(\X_j)}  \mid \u_j,1 \right\}+O_p(N^{1/2}h^m)\n\\
&=&N^{-3/2}  \sumI \sumJ \frac{  r_i \Delta_{1i}^2 }{ d^*(\x_i) D_i } \{e^{-g(\u_i)}-e^{-g^*(\u_i)}\} E \left\{\frac{ \Delta_{1i}\Delta_{3i} }{d^*(\X_i)}  \mid \u_i,1 \right\}+O_p(N^{1/2}h^m)\n\\
&=&T_{2222BC},\n
\ee
where $T_{2222BC}$ is as defined in (\ref{ustat:t2222bc}).

Consider $T_{231ABB}$ as defined in  (\ref{ustat:t231abb}), applying the U-statistics property to it we get,
\be \label{ustat:t231abb2}
&&T_{231ABB}\n\\
&=&N^{-5/2} \sumI \sumJ \sumK
\frac{\{1-r_i
  \pi^{-1}(y_i,\u_i;\bb,g^*)\} r_j K_h({\u_j-\u_i})  r_k K_h({\x_k-\x_i}) }{d^*(\x_i) D_i  V_{i,j} }\n \\
&& \times \{\Delta_{1i}\Delta_{1j} \Delta_{3j} -\Delta_{1j}^2 \Delta_{3i} \} +O_p(N^{1/2}h^m)\n\\
&=&N^{-2/2} \sumI
\frac{\{1-r_i
  \pi^{-1}(y_i,\u_i;\bb,g^*)\}}{d^*(\x_i) D_i
E(R_i\mid\u_i) f_{\U}(\u_i) E(R_i\mid\x_i) f_{\X}(\x_i) }\n \\
&& \times N^{-1/2} \sumJ E \left[\frac{r_j K_h({\u_j-\u_i})  R_k K_h({\X_k-\x_i}) \{\Delta_{1i}\Delta_{1j} \Delta_{3j} -\Delta_{1j}^2 \Delta_{3i} \} }{d^* (\x_j) } \mid \x_j, r_j, r_j y_j\right] \n\\
&&+N^{-2/2} \sumI
\frac{\{1-r_i
  \pi^{-1}(y_i,\u_i;\bb,g^*)\}}{d^*(\x_i) D_i
E(R_i\mid\u_i) f_{\U}(\u_i) E(R_i\mid\x_i) f_{\X}(\x_i) }\n \\
&& \times N^{-1/2} \sumK E \left[\frac{R_j K_h({\U_j-\u_i})  r_k K_h({\x_k-\x_i}) \{\Delta_{1i}\Delta_{1j} \Delta_{3j} -\Delta_{1j}^2 \Delta_{3i} \} }{d^* (\X_j) } \mid \x_k, r_k, r_k y_k\right] \n\\
&&-N^{-2/2} \sumI
\frac{\{1-r_i
  \pi^{-1}(y_i,\u_i;\bb,g^*)\}}{d^*(\x_i) D_i
E(R_i\mid\u_i) f_{\U}(\u_i) E(R_i\mid\x_i) f_{\X}(\x_i) }\n \\
&& \times N^{1/2}  E \left[\frac{R_j K_h({\U_j-\u_i})  R_k K_h({\X_k-\x_i}) \{\Delta_{1i}\Delta_{1j} \Delta_{3j} -\Delta_{1j}^2 \Delta_{3i} \} }{d^* (\X_j) } \right]\n\\ &&+O_p(N^{1/2}h^m)+O_p(N^{-1/2}) \n\\
&=&T_{231ABBA}+T_{231ABBB}-T_{231ABBC}+O_p(N^{-1/2}).\n
\ee

Consider $T_{231ABBA}$,
\be \label{ustat:t231abba}
&&T_{231ABBA}\n\\
&=&N^{-2/2} \sumI
\frac{\{1-r_i
  \pi^{-1}(y_i,\u_i;\bb,g^*)\}}{d^*(\x_i) D_i
E(R_i\mid\u_i) f_{\U}(\u_i) E(R_i\mid\x_i) f_{\X}(\x_i) }\n \\
&& \times N^{-1/2} \sumJ E \left[\frac{r_j K_h({\u_j-\u_i})  R_k K_h({\X_k-\x_i}) \{\Delta_{1i}\Delta_{1j} \Delta_{3j} -\Delta_{1j}^2 \Delta_{3i} \} }{d^* (\x_j) } \mid \x_j, r_j, r_j y_j\right] \n\\
&=&N^{-3/2} \sumI \sumJ
\frac{\{1-r_i \pi^{-1}(y_i,\u_i;\bb,g^*)\}r_j K_h({\u_j-\u_i}) \{\Delta_{1i}\Delta_{1j} \Delta_{3j} -\Delta_{1j}^2 \Delta_{3i} \} }{d^*(\x_i) D_i d^* (\x_j)
E(R_i\mid\u_i) f_{\U}(\u_i) }+O_p(N^{1/2}h^m)\n\\
&=&T_{2211AC}-T_{2222C},\n
\ee
where $T_{2211AC}$ is as defined in (\ref{ustat:t2211ac}) and
where $T_{2222C}$ is as defined in (\ref{ustat:t2222c}).

Consider $T_{231ABBB}$,
\be \label{ustat:t231abbb}
&&T_{231ABBB}\n\\
&=&N^{-2/2} \sumI
\frac{\{1-r_i
  \pi^{-1}(y_i,\u_i;\bb,g^*)\}}{d^*(\x_i) D_i
E(R_i\mid\u_i) f_{\U}(\u_i) E(R_i\mid\x_i) f_{\X}(\x_i) }\n \\
&& \times N^{-1/2} \sumK E \left[\frac{R_j K_h({\U_j-\u_i})  r_k K_h({\x_k-\x_i}) \{\Delta_{1i}\Delta_{1j} \Delta_{3j} -\Delta_{1j}^2 \Delta_{3i} \} }{d^* (\X_j) } \mid \x_k, r_k, r_k y_k\right] \n\\
&=&N^{-3/2} \sumI  \sumK
\frac{\{1-r_i \pi^{-1}(y_i,\u_i;\bb,g^*)\} r_k K_h({\x_k-\x_i})}{d^*(\x_i) D_i
 E(R_i\mid\x_i) f_{\X}(\x_i) } E \left[\frac{\{\Delta_{1i}\Delta_{1i} \Delta_{3i} -\Delta_{1i}^2 \Delta_{3i} \} }{d^* (\X_i) } \mid \u_i, 1\right] \n\\
&&+O_p(N^{1/2}h^m)\n\\
&=& O_p(N^{1/2}h^m).\n
\ee

Consider $T_{231ABBC}$,
\be \label{ustat:t231abbc}
&&T_{231ABBC}\n\\
&=&N^{-2/2} \sumI
\frac{\{1-r_i
  \pi^{-1}(y_i,\u_i;\bb,g^*)\}}{d^*(\x_i) D_i
E(R_i\mid\u_i) f_{\U}(\u_i) E(R_i\mid\x_i) f_{\X}(\x_i) }\n \\
&& \times N^{1/2}  E \left[\frac{R_j K_h({\U_j-\u_i})  R_k K_h({\X_k-\x_i}) \{\Delta_{1i}\Delta_{1j} \Delta_{3j} -\Delta_{1j}^2 \Delta_{3i} \} }{d^* (\X_j) } \right] \n\\
&=&N^{-1/2} \sumI
\frac{\{1-r_i
  \pi^{-1}(y_i,\u_i;\bb,g^*)\}}{d^*(\x_i) D_i
 } E \left[\frac{\{\Delta_{1i}\Delta_{1i} \Delta_{3i} -\Delta_{1i}^2 \Delta_{3i} \} }{d^* (\X_i) } \mid \u_i, 1\right]+O_p(N^{1/2}h^m)\n\\
&=&O_p(N^{1/2}h^m).\n
\ee

Consider $T_{231BBB}$ as defined in  (\ref{ustat:t231bbb}), applying the U-statistics property to it we get,
\be \label{ustat:t231bbb2}
&&T_{231BBB}\n\\
&=&N^{-5/2} \sumI \sumJ \sumK  \frac{\{1-r_i
  \pi^{-1}(y_i,\u_i;\bb,g^*)\} r_j K_h({\u_j-\u_i})  r_k K_h({\x_k-\x_j})   }{d^*(\x_i) D_i \{d^* (\x_j)\}^2 E(R_i\mid\u_i) f_{\U}(\u_i) E(R_j\mid\x_j) f_{\X}(\x_j) }\n \\
&& \times \{e^{-h(y_k)}+e^{-g^* (\u_j)} e^{-2h(y_k)}\}  \{\Delta_{1i}\Delta_{1j} \Delta_{3j} -\Delta_{1j}^2 \Delta_{3i} \}
+O_p(N^{1/2}h^m)\n\\
&=&N^{-3/2} \sumI \sumJ  \frac{ \{1-r_i
  \pi^{-1}(y_i,\u_i;\bb,g^*)\} r_j K_h({\u_j-\u_i})  \{\Delta_{1i}\Delta_{1j} \Delta_{3j} -\Delta_{1j}^2 \Delta_{3i} \}    }{d^*(\x_i) D_i   \{d^* (\x_j)\}^2 E(R_i\mid\u_i) f_{\U}(\u_i)   E(R_j\mid\x_j) f_{\X}(\x_j) }  \n \\
&&\times  E \left[  R_k K_h({\X_k-\x_j})  \{e^{-h(Y_k)}+e^{-g^* (\u_j)} e^{-2h(Y_k)}\} \mid \x_i, r_r, r_i y_i \right] \n\\
&&+ N^{-3/2}  \sumJ \sumK  \frac{ r_j r_k K_h({\x_k-\x_j})   \{e^{-h(y_k)}+e^{-g^* (\u_j)} e^{-2h(y_k)}\}   }{\{d^* (\x_j)\}^2  E(R_j\mid\x_j) f_{\X}(\x_j) }\n \\
&& \times E\left[\frac{\{1-R_i
  \pi^{-1}(Y_i,\U_i;\bb,g^*)\} K_h({\u_j-\U_i})  \{\Delta_{1i}\Delta_{1j} \Delta_{3j} -\Delta_{1j}^2 \Delta_{3i} \}  }{d^*(\X_i) D_i  E(R_i\mid\U_i) f_{\U}(\U_i)  } \mid \x_k, r_k, r_k y_k \right]\n\\
&&- N^{-1/2}  \sumJ  \frac{ r_j   }{\{d^* (\x_j)\}^2  E(R_j\mid\x_j) f_{\X}(\x_j) }\n \\
&& \times E \left[  R_k K_h({\X_k-\x_j})  \{e^{-h(Y_k)}+e^{-g^* (\u_j)} e^{-2h(Y_k)}\}  \right]\n\\
&& \times E\left[\frac{\{1-R_i
  \pi^{-1}(Y_i,\U_i;\bb,g^*)\} K_h({\u_j-\U_i})  \{\Delta_{1i}\Delta_{1j} \Delta_{3j} -\Delta_{1j}^2 \Delta_{3i} \}  }{d^*(\X_i) D_i  E(R_i\mid\U_i) f_{\U}(\U_i)  }\right]\n\\
&&+O_p(N^{1/2}h^m)+O_p(N^{-1/2}) \n\\
&=&T_{231BBBA}+T_{231BBBB}-T_{231BBBC}+O_p(N^{1/2}h^m)+O_p(N^{-1/2}).\n
\ee

Consider $T_{231BBBA}$,
\be \label{ustat:t231bbba}
&&T_{231BBBA}\n\\
&=&N^{-3/2} \sumI \sumJ  \frac{ \{1-r_i
  \pi^{-1}(y_i,\u_i;\bb,g^*)\} r_j K_h({\u_j-\u_i})  \{\Delta_{1i}\Delta_{1j} \Delta_{3j} -\Delta_{1j}^2 \Delta_{3i} \}    }{d^*(\x_i) D_i   \{d^* (\x_j)\}^2 E(R_i\mid\u_i) f_{\U}(\u_i)   E(R_j\mid\x_j) f_{\X}(\x_j) }  \n \\
&&\times  E \left[  R_k K_h({\X_k-\x_j})  \{e^{-h(Y_k)}+e^{-g^* (\u_j)} e^{-2h(Y_k)}\} \mid \x_i, r_r, r_i y_i \right] \n\\
&=&N^{-3/2} \sumI \sumJ  \frac{ \{1-r_i
  \pi^{-1}(y_i,\u_i;\bb,g^*)\} r_j K_h({\u_j-\u_i})  \{\Delta_{1i}\Delta_{1j} \Delta_{3j} -\Delta_{1j}^2 \Delta_{3i} \}    }{d^*(\x_i) D_i   d^* (\x_j) E(R_i\mid\u_i) f_{\U}(\u_i) }  \n \\
&&+O_p(N^{1/2}h^m)\n\\
&=&T_{2211AC}-T_{2222C},\n
\ee
where $T_{2211AC}$ is as defined in (\ref{ustat:t2211ac}) and
where $T_{2222C}$ is as defined in (\ref{ustat:t2222c}).

Consider $T_{231BBBB}$,
\be \label{ustat:t231bbbb}
&&T_{231BBBB}\n\\
&=&N^{-3/2}  \sumJ \sumK  \frac{ r_j r_k K_h({\x_k-\x_j})   \{e^{-h(y_k)}+e^{-g^* (\u_j)} e^{-2h(y_k)}\}   }{\{d^* (\x_j)\}^2  E(R_j\mid\x_j) f_{\X}(\x_j) }\n \\
&& \times E\left[\frac{\{1-R_i
  \pi^{-1}(Y_i,\U_i;\bb,g^*)\} K_h({\u_j-\U_i})  \{\Delta_{1i}\Delta_{1j} \Delta_{3j} -\Delta_{1j}^2 \Delta_{3i} \}  }{d^*(\X_i) D_i  E(R_i\mid\U_i) f_{\U}(\U_i)  } \mid \x_k, r_k, r_k y_k \right]\n\\
&=&N^{-3/2}  \sumJ \sumK  \frac{ r_j r_k K_h({\x_k-\x_j})   \{e^{-h(y_k)}+e^{-g^* (\u_j)} e^{-2h(y_k)}\}   }{\{d^* (\x_j)\}^2  E(R_j\mid\x_j) f_{\X}(\x_j) }\n \\
&& \times  E \left[\frac{ \{\Delta_{1j}\Delta_{1j} \Delta_{3j} -\Delta_{1j}^2 \Delta_{3j} \} \{w^{-1}(\X_j)
-w^{*-1}(\X_j)\}}{d^*(\X_j) D_j}  \mid \u_j,1 \right]
+O_p(N^{1/2}h^m)\n\\
&=&O_p(N^{1/2}h^m).\n
\ee

Consider $T_{231BBBC}$,
\be \label{ustat:t231bbbc}
&&T_{231BBBC}\n\\
&=&N^{-1/2}  \sumJ  \frac{ r_j   }{\{d^* (\x_j)\}^2  E(R_j\mid\x_j) f_{\X}(\x_j) }\n \\
&& \times E \left[  R_k K_h({\X_k-\x_j})  \{e^{-h(Y_k)}+e^{-g^* (\u_j)} e^{-2h(Y_k)}\}  \right]\n\\
&& \times E\left[\frac{\{1-R_i
  \pi^{-1}(Y_i,\U_i;\bb,g^*)\} K_h({\u_j-\U_i})  \{\Delta_{1i}\Delta_{1j} \Delta_{3j} -\Delta_{1j}^2 \Delta_{3i} \}  }{d^*(\X_i) D_i  E(R_i\mid\U_i) f_{\U}(\U_i)  }\right]\n\\
&=&N^{-1/2}  \sumJ  \frac{ r_j   }{d^* (\x_j)} E \left[\frac{ \{\Delta_{1j}\Delta_{1j} \Delta_{3j} -\Delta_{1j}^2 \Delta_{3j} \} \{w^{-1}(\X_j)
-w^{*-1}(\X_j)\}}{d^*(\X_j) D_j}  \mid \u_j,1 \right]\n\\
&&+ O_p(N^{1/2}h^m)\n\\
&=&O_p(N^{1/2}h^m).\n
\ee

Consider $T_{231CB}$ as defined in  (\ref{ustat:t231cb}), applying the U-statistics property to it we get,
\be \label{ustat:t231cb2}
&&T_{231CB}\n\\
&=&N^{-5/2} \sumI \sumJ \sumK   \frac{\{1-r_i
  \pi^{-1}(y_i,\u_i;\bb,g^*)\}  r_j K_h({\u_j-\u_i}) r_k K_h({\u_k-\u_i}) }{d^*(\x_i) D_i d^*(\x_j) \{E(R_i\mid\u_i) f_{\U}(\u_i)\}^2 }\n \\
&& \times
\{\Delta_{1i}\Delta_{1j} \Delta_{3j} -\Delta_{1j}^2 \Delta_{3i} \}
+ O_p(N^{1/2}h^m)\n\\
&=&N^{-2/2} \sumI   \frac{\{1-r_i
  \pi^{-1}(y_i,\u_i;\bb,g^*)\} }{d^*(\x_i) D_i \{E(R_i\mid\u_i) f_{\U}(\u_i)\}^2 }\n \\
&& \times N^{-1/2} \sumJ E \left[\frac{ r_j K_h({\u_j-\u_i}) R_k K_h({\U_k-\u_i})\{\Delta_{1i}\Delta_{1j} \Delta_{3j} -\Delta_{1j}^2 \Delta_{3i} \} }{ d^*(\x_j) } \mid \x_j, r_j, r_j y_j\right]
 \n\\
&&+N^{-2/2} \sumI   \frac{\{1-r_i
  \pi^{-1}(y_i,\u_i;\bb,g^*)\} }{d^*(\x_i) D_i \{E(R_i\mid\u_i) f_{\U}(\u_i)\}^2 }\n \\
&& \times N^{-1/2} \sumK E \left[\frac{ R_j K_h({\U_j-\u_i}) r_k K_h({\u_k-\u_i})\{\Delta_{1i}\Delta_{1j} \Delta_{3j} -\Delta_{1j}^2 \Delta_{3i} \} }{ d^*(\X_j) } \mid \x_k, r_k, r_k y_k\right]
 \n\\
&&-N^{-2/2} \sumI   \frac{\{1-r_i
  \pi^{-1}(y_i,\u_i;\bb,g^*)\} }{d^*(\x_i) D_i \{E(R_i\mid\u_i) f_{\U}(\u_i)\}^2 }\n \\
&& \times N^{1/2} E \left[\frac{ R_j K_h({\U_j-\u_i}) R_k K_h({\U_k-\u_i})\{\Delta_{1i}\Delta_{1j} \Delta_{3j} -\Delta_{1j}^2 \Delta_{3i} \} }{ d^*(\X_j) } \right]
 \n\\
&&+O_p(N^{1/2}h^m)+O_p(N^{-1/2}) \n\\
&=&T_{231CBA}+T_{231CBB}-T_{231CBC}+O_p(N^{1/2}h^m)+O_p(N^{-1/2}).\n
\ee

Consider $T_{231CBA}$,
\be \label{ustat:t231cba}
&&T_{231CBA}\n\\
&=&N^{-2/2} \sumI   \frac{\{1-r_i
  \pi^{-1}(y_i,\u_i;\bb,g^*)\} }{d^*(\x_i) D_i \{E(R_i\mid\u_i) f_{\U}(\u_i)\}^2 }\n \\
&& \times N^{-1/2} \sumJ E \left[\frac{ r_j K_h({\u_j-\u_i}) R_k K_h({\U_k-\u_i})\{\Delta_{1i}\Delta_{1j} \Delta_{3j} -\Delta_{1j}^2 \Delta_{3i} \} }{ d^*(\x_j) } \mid \x_j, r_j, r_j y_j\right]
 \n\\
&=&N^{-3/2} \sumI \sumJ   \frac{\{1-r_i
  \pi^{-1}(y_i,\u_i;\bb,g^*)\} r_j K_h({\u_j-\u_i}) \{\Delta_{1i}\Delta_{1j} \Delta_{3j} -\Delta_{1j}^2 \Delta_{3i} \} }{d^*(\x_i) D_i d^*(\x_j) E(R_i\mid\u_i) f_{\U}(\u_i)}+O_p(N^{1/2}h^m)\n\\
&=&T_{2211AC}-T_{2222C},\n
\ee
where $T_{2211AC}$ is as defined in (\ref{ustat:t2211ac}) and
where $T_{2222C}$ is as defined in (\ref{ustat:t2222c}).

Consider $T_{231CBB}$,
\be \label{ustat:t231cbb}
&&T_{231CBB}\n\\
&=&N^{-2/2} \sumI   \frac{\{1-r_i
  \pi^{-1}(y_i,\u_i;\bb,g^*)\} }{d^*(\x_i) D_i \{E(R_i\mid\u_i) f_{\U}(\u_i)\}^2 }\n \\
&& \times N^{-1/2} \sumK E \left[\frac{ R_j K_h({\U_j-\u_i}) r_k K_h({\u_k-\u_i})\{\Delta_{1i}\Delta_{1j} \Delta_{3j} -\Delta_{1j}^2 \Delta_{3i} \} }{ d^*(\X_j) } \mid \x_k, r_k, r_k y_k\right]
 \n\\
&=&N^{-3/2} \sumI  \sumK   \frac{\{1-r_i
  \pi^{-1}(y_i,\u_i;\bb,g^*)\} r_k K_h({\u_k-\u_i}) }{d^*(\x_i) D_i E(R_i\mid\u_i) f_{\U}(\u_i) } E\left[\frac{  \{\Delta_{1i}\Delta_{1i} \Delta_{3i} -\Delta_{1i}^2 \Delta_{3i} \}  }{d^*(\X_i) } \mid \u_i, 1 \right]
 \n\\
&&+O_p(N^{1/2}h^m)\n\\
&=&O_p(N^{1/2}h^m).\n
\ee

Consider $T_{231CBC}$,
\be \label{ustat:t231cbc}
&&T_{231CBC}\n\\
&=&N^{-2/2} \sumI   \frac{\{1-r_i
  \pi^{-1}(y_i,\u_i;\bb,g^*)\} }{d^*(\x_i) D_i \{E(R_i\mid\u_i) f_{\U}(\u_i)\}^2 }\n \\
&& \times N^{1/2} E \left[\frac{ R_j K_h({\U_j-\u_i}) R_k K_h({\U_k-\u_i})\{\Delta_{1i}\Delta_{1j} \Delta_{3j} -\Delta_{1j}^2 \Delta_{3i} \} }{ d^*(\X_j) } \right]
 \n\\
&=&N^{-1/2} \sumI   \frac{\{1-r_i
  \pi^{-1}(y_i,\u_i;\bb,g^*)\} }{d^*(\x_i) D_i } E\left[\frac{  \{\Delta_{1i}\Delta_{1i} \Delta_{3i} -\Delta_{1i}^2 \Delta_{3i} \}  }{d^*(\X_i) } \mid \u_i, 1 \right]
+O_p(N^{1/2}h^m)\n\\
&=&O_p(N^{1/2}h^m).\n
\ee

Consider $T_{232BBBB}$ as defined in  (\ref{ustat:t232bbbb}), applying the U-statistics property to it we get,
\be \label{ustat:t232bbbb2}
&&T_{232BBBB}\n\\
&=&N^{-5/2} \sumI \sumJ  \sumK \frac{\{1-r_i
  \pi^{-1}(Y_i,\u_i;\bb,g^*)\} r_j K_h({\u_j-\u_i}) r_k K_h({\x_k-\x_j})  }{d^*(\x_i) D_i d^*(\x_j) E(R_i\mid\u_i) f_{\U}(\u_i) E(R_j\mid\x_j) f_{\X}(\x_j)}\n \\
&&  \times   \{\Delta_{1i}\Delta_{1j} \Delta_{3j} -\Delta_{1j}^2 \Delta_{3i} \} + O_p(N^{1/2}h^m)\n\\
&=&N^{-3/2}  \sumI \sumJ  \frac{ r_j  }{ d^*(\x_j) E(R_j\mid\x_j) f_{\X}(\x_j)}\n \\
&&  \times E \left[\frac{\{1-r_i
  \pi^{-1}(Y_i,\u_i;\bb,g^*)\}  K_h({\u_j-\u_i}) R_k K_h({\X_k-\x_j})  }{d^*(\x_i) D_i  E(R_i\mid\u_i) f_{\U}(\u_i) } \{\Delta_{1i}\Delta_{1j} \Delta_{3j} -\Delta_{1j}^2 \Delta_{3i} \} \mid \x_i, r_i, r_i y_i \right]\n\\
&&+N^{-3/2}  \sumJ \sumK  \frac{ r_j  }{ d^*(\x_j) E(R_j\mid\x_j) f_{\X}(\x_j)}\n \\
&&  \times E \left[\frac{\{1-R_i
  \pi^{-1}(Y_i,\U_i;\bb,g^*)\}  K_h({\u_j-\U_i}) r_k K_h({\x_k-\x_j})  }{d^*(\X_i) D_i  E(R_i\mid\U_i) f_{\U}(\U_i) } \{\Delta_{1i}\Delta_{1j} \Delta_{3j} -\Delta_{1j}^2 \Delta_{3i} \} \mid \x_k, r_k, r_k y_k \right]\n\\
&&-N^{-1/2}  \sumJ  \frac{ r_j  }{ d^*(\x_j) E(R_j\mid\x_j) f_{\X}(\x_j)}\n \\
&&  \times E \left[\frac{\{1-R_i
  \pi^{-1}(Y_i,\U_i;\bb,g^*)\}   K_h({\u_j-\U_i}) R_k K_h({\X_k-\x_j})  }{d^*(\X_i) D_i  E(R_i\mid\U_i) f_{\U}(\U_i) } \{\Delta_{1i}\Delta_{1j} \Delta_{3j} -\Delta_{1j}^2 \Delta_{3i} \} \right]\n\\
&&+O_p(N^{1/2}h^m)+O_p(N^{-1/2}) \n\\
&=&T_{232BBBBA}+T_{232BBBBB}-T_{232BBBBC}+O_p(N^{1/2}h^m)+O_p(N^{-1/2}).\n
\ee

Consider $T_{232BBBBA}$,
\be \label{ustat:t232bbbba}
&&T_{232BBBBA}\n\\
&=&N^{-3/2}  \sumI \sumJ  \frac{ r_j  }{ d^*(\x_j) E(R_j\mid\x_j) f_{\X}(\x_j)}\n \\
&&  \times E \left[\frac{\{1-r_i
  \pi^{-1}(Y_i,\u_i;\bb,g^*)\}  K_h({\u_j-\u_i}) R_k K_h({\X_k-\x_j})  }{d^*(\x_i) D_i  E(R_i\mid\u_i) f_{\U}(\u_i) } \{\Delta_{1i}\Delta_{1j} \Delta_{3j} -\Delta_{1j}^2 \Delta_{3i} \} \mid \x_i, r_i, r_i y_i \right]\n\\
&=&N^{-3/2}  \sumI \sumJ  \frac{ \{1-r_i
  \pi^{-1}(Y_i,\u_i;\bb,g^*)\}  r_j K_h({\u_j-\u_i}) \{\Delta_{1i}\Delta_{1j} \Delta_{3j} -\Delta_{1j}^2 \Delta_{3i} \} }{ d^*(\x_i) D_i  d^*(\x_j) E(R_i\mid\u_i) f_{\U}(\u_i)}+O_p(N^{1/2}h^m)\n\\
&=&T_{2211AC}-T_{2222C},\n
\ee
where $T_{2211AC}$ is as defined in (\ref{ustat:t2211ac}) and
where $T_{2222C}$ is as defined in (\ref{ustat:t2222c}).

Consider $T_{232BBBBB}$,
\be \label{ustat:t232bbbbb}
&&T_{232BBBBB}\n\\
&=&N^{-3/2}  \sumJ \sumK  \frac{ r_j  }{ d^*(\x_j) E(R_j\mid\x_j) f_{\X}(\x_j)}\n \\
&&  \times E \left[\frac{\{1-R_i
  \pi^{-1}(Y_i,\U_i;\bb,g^*)\}  K_h({\u_j-\U_i}) r_k K_h({\x_k-\x_j})  }{d^*(\X_i) D_i  E(R_i\mid\U_i) f_{\U}(\U_i) } \{\Delta_{1i}\Delta_{1j} \Delta_{3j} -\Delta_{1j}^2 \Delta_{3i} \} \mid \x_k, r_k, r_k y_k \right]\n\\
&=&N^{-3/2}  \sumJ \sumK  \frac{ r_j r_k K_h({\x_k-\x_j})  }{ d^*(\x_j) E(R_j\mid\x_j) f_{\X}(\x_j)}\n \\
&&  \times E \left[\frac{ \{\Delta_{1j}\Delta_{1j} \Delta_{3j} -\Delta_{1j}^2 \Delta_{3j} \} \{w^{-1}(\X_j)
-w^{*-1}(\X_j)\}}{d^*(\X_j) D_j}  \mid \u_j,1 \right]
+O_p(N^{1/2}h^m)\n\\
&=&O_p(N^{1/2}h^m).\n
\ee

Consider $T_{232BBBBC}$,
\be \label{ustat:t232bbbbc}
&&T_{232BBBBC}\n\\
&=&N^{-1/2}  \sumJ  \frac{ r_j  }{ d^*(\x_j) E(R_j\mid\x_j) f_{\X}(\x_j)}\n \\
&&  \times E \left[\frac{\{1-R_i
  \pi^{-1}(Y_i,\U_i;\bb,g^*)\}   K_h({\u_j-\U_i}) R_k K_h({\X_k-\x_j})  }{d^*(\X_i) D_i  E(R_i\mid\U_i) f_{\U}(\U_i) } \{\Delta_{1i}\Delta_{1j} \Delta_{3j} -\Delta_{1j}^2 \Delta_{3i} \} \right]\n\\
&=&N^{-1/2}  \sumJ  \frac{ r_j  }{ d^*(\x_j)} E \left[\frac{ \{\Delta_{1j}\Delta_{1j} \Delta_{3j} -\Delta_{1j}^2 \Delta_{3j} \} \{w^{-1}(\X_j)
-w^{*-1}(\X_j)\}}{d^*(\X_j) D_j}  \mid \u_j,1 \right]
+O_p(N^{1/2}h^m)\n\\
&=&O_p(N^{1/2}h^m).\n
\ee

Consider $T_{322}$ as defined in  (\ref{eq:t322}), applying the U-statistics property to it we get,
\be \label{ustat:t322}
&&T_{322}\n\\
&=&N^{-5/2} \sumI \sumJ \sumK \frac{\{1-r_i  \pi^{-1}(y_i,\u_i;\bb,g^*)\}C_i r_j  K_{h} (\u_j-\u_i)r_k  e^{-h(y_k)} K_{h} (\x_k-\x_j) \Delta_{1j}}{d^*(\x_i) D_i^2 d^* (\x_j) E(R_i\mid\u_i) f_{\U}(\u_i) E(R_j\mid \x_j) f_{\X}(\x_j)}\n\\
&=&N^{-2/2} \sumJ \frac{ r_j \Delta_{1j}}{ d^* (\x_j) E(R_j\mid \x_j) f_{\X}(\x_j)}\n\\
&& \times N^{-1/2} \sumI E \left[ \frac{\{1-r_i  \pi^{-1}(y_i,\u_i;\bb,g^*)\}C_i  K_{h} (\u_j-\u_i)R_k  e^{-h(Y_k)} K_{h} (\X_k-\x_j)}{d^*(\x_i) D_i^2 E(R_i\mid\u_i) f_{\U}(\u_i)} \mid \x_i, r_i, r_i y_i \right]\n\\
&&+N^{-2/2} \sumJ \frac{ r_j \Delta_{1j}}{ d^* (\x_j) E(R_j\mid \x_j) f_{\X}(\x_j)}\n\\
&& \times N^{-1/2} \sumK E \left[ \frac{\{1-R_i  \pi^{-1}(Y_i,\U_i;\bb,g^*)\} C_i  K_{h} (\u_j-\U_i)r_k  e^{-h(y_k)} K_{h} (\x_k-\x_j) }{d^*(\X_i) D_i^2E(R_i\mid\U_i) f_{\U}(\U_i)} \mid \x_k, r_k, r_k y_k \right]\n\\
&&-N^{-2/2} \sumJ \frac{ r_j \Delta_{1j}}{ d^* (\x_j) E(R_j\mid \x_j) f_{\X}(\x_j)}\n\\
&& \times N^{1/2} E \left[ \frac{\{1-R_i  \pi^{-1}(Y_i,\U_i;\bb,g^*)\} C_i  K_{h} (\u_j-\U_i)R_k  e^{-h(Y_k)} K_{h} (\X_k-\x_j) }{d^*(\X_i) D_i^2 E(R_i\mid\U_i) f_{\U}(\U_i)} \right] +O_p(N^{-1/2})\n\\
&=& T_{322A}+T_{322B}-T_{322C}+O_p(N^{-1/2}).\n
\ee

Consider $T_{322A}$,
\be \label{ustat:t322a}
&&T_{322A}\n\\
&=&N^{-2/2} \sumJ \frac{ r_j \Delta_{1j}}{ d^* (\x_j) E(R_j\mid \x_j) f_{\X}(\x_j)}\n\\
&& \times N^{-1/2} \sumI E \left[ \frac{\{1-r_i  \pi^{-1}(y_i,\u_i;\bb,g^*)\}C_i  K_{h} (\u_j-\u_i)R_k  e^{-h(Y_k)} K_{h} (\X_k-\x_j)}{d^*(\x_i) D_i^2 E(R_i\mid\u_i) f_{\U}(\u_i)} \mid \x_i, r_i, r_i y_i \right]\n\\
&=&N^{-3/2} \sumI \sumJ \frac{ \{1-r_i  \pi^{-1}(y_i,\u_i;\bb,g^*)\}C_i  r_j K_{h} (\u_j-\u_i) \Delta_{1j}^2}{ d^*(\x_i) D_i^2 E(R_i\mid\u_i) f_{\U}(\u_i)  d^* (\x_j) }
+O_p(N^{1/2}h^m)\n\\
&=&T_{31}+O_p(N^{1/2}h^m).\n
\ee
where $T_{31}$ is as defined in (\ref{eq:t31}).

Consider $T_{322B}$,
\be \label{ustat:t322b}
&&T_{322B}\n\\
&=&N^{-2/2} \sumJ \frac{ r_j \Delta_{1j}}{ d^* (\x_j) E(R_j\mid \x_j) f_{\X}(\x_j)}\n\\
&& \times N^{-1/2} \sumK E \left[ \frac{\{1-R_i  \pi^{-1}(Y_i,\U_i;\bb,g^*)\} C_i  K_{h} (\u_j-\U_i)r_k  e^{-h(y_k)} K_{h} (\x_k-\x_j) }{d^*(\X_i) D_i^2E(R_i\mid\U_i) f_{\U}(\U_i)} \mid \x_k, r_k, r_k y_k \right]\n\\
&=&N^{-3/2} \sumJ \sumK \frac{ r_j \Delta_{1j} r_k  e^{-h(y_k)} K_{h} (\x_k-\x_j) }{ d^* (\x_j) E(R_j\mid \x_j) f_{\X}(\x_j)}
  E \left[ \frac{\{1-R_i  \pi^{-1}(Y_i,\U_i;\bb,g^*)\} C_i  K_{h} (\u_j-\U_i) }{d^*(\X_i) D_i^2E(R_i\mid\U_i) f_{\U}(\U_i)}  \right]\n\\
&=&N^{-3/2} \sumJ \sumK \frac{ r_j \Delta_{1j} r_k  e^{-h(y_k)} K_{h} (\x_k-\x_j) }{ d^* (\x_j) E(R_j\mid \x_j) f_{\X}(\x_j)} E\left[\frac{ C_j \{w^{-1}(\X_j)
-w^{*-1}(\X_j)\}}{d^*(\X_j) D_j^2 } \mid \u_j, 1 \right]+O_p(N^{1/2}h^m)\n\\
&=&O_p(N^{1/2}h^m).\n
\ee

Consider $T_{322C}$,
\be \label{ustat:t322c}
&&T_{322C}\n\\
&=&N^{-2/2} \sumJ \frac{ r_j \Delta_{1j}}{ d^* (\x_j) E(R_j\mid \x_j) f_{\X}(\x_j)}\n\\
&& \times N^{1/2} E \left[ \frac{\{1-R_i  \pi^{-1}(Y_i,\U_i;\bb,g^*)\} C_i  K_{h} (\u_j-\U_i)R_k  e^{-h(Y_k)} K_{h} (\X_k-\x_j) }{d^*(\X_i) D_i^2 E(R_i\mid\U_i) f_{\U}(\U_i)} \right]\n\\
&=&N^{-1/2} \sumJ \frac{ r_j \Delta_{1j}^2}{ d^* (\x_j) } E\left[\frac{\{w^{-1}(\X_j)
-w^{*-1}(\X_j)\}  C_j }{d^*(\X_j) D_j^2 } \mid \u_j, 1 \right]
+O_p(N^{1/2}h^m)\n\\
&=&O_p(N^{1/2}h^m).\n
\ee

Consider $T_{323}$ as defined in  (\ref{eq:t323}), applying the U-statistics property to it we get,
\be \label{ustat:t323}
&&T_{323}\n\\
&=&N^{-5/2} \sumI \sumJ \sumK \frac{\{1-r_i  \pi^{-1}(y_i,\u_i;\bb,g^*)\}C_i r_j  K_{h} (\u_j-\u_i)r_k K_{h} (\x_k-\x_j) \Delta_{1j}^2}{d^*(\x_i) D_i^2 d^* (\x_j) E(R_i\mid\u_i) f_{\U}(\u_i) E(R_j\mid \x_j) f_{\X}(\x_j)}\n \\
&=&N^{-2/2} \sumJ  \frac{ r_j  \Delta_{1j}^2}{d^* (\x_j) E(R_j\mid \x_j) f_{\X}(\x_j)}\n \\
&&\times N^{-1/2} \sumI E \left[\frac{\{1-r_i  \pi^{-1}(y_i,\u_i;\bb,g^*)\}C_i   K_{h} (\u_j-\u_i)R_k K_{h} (\X_k-\x_j) }{d^*(\x_i) D_i^2 E(R_i\mid\u_i) f_{\U}(\u_i)} \mid \x_i, r_i, r_i y_i\right]\n \\
&&+N^{-2/2} \sumJ  \frac{ r_j  \Delta_{1j}^2}{d^* (\x_j) E(R_j\mid \x_j) f_{\X}(\x_j)}\n \\
&&\times N^{-1/2} \sumK E \left[\frac{\{1-R_i  \pi^{-1}(Y_i,\U_i;\bb,g^*)\}C_i   K_{h} (\u_j-\U_i)r_k K_{h} (\x_k-\x_j) }{d^*(\X_i) D_i^2  E(R_i\mid\U_i) f_{\U}(\U_i)} \mid \x_k, r_k, r_k y_k\right]\n \\
&&-N^{-2/2} \sumJ  \frac{ r_j  \Delta_{1j}^2}{d^* (\x_j) E(R_j\mid \x_j) f_{\X}(\x_j)}\n \\
&&\times N^{1/2} E \left[\frac{\{1-R_i  \pi^{-1}(Y_i,\U_i;\bb,g^*)\}C_i   K_{h} (\u_j-\U_i)R_k K_{h} (\X_k-\x_j) }{d^*(\X_i) D_i^2 E(R_i\mid\U_i) f_{\U}(\U_i)} \right]+O_p(N^{-1/2})\n\\
&=& T_{323A}+T_{323B}-T_{323C}+O_p(N^{-1/2}).\n
\ee

Consider $T_{323A}$,
\be \label{ustat:t323a}
&&T_{323A}\n\\
&=&N^{-2/2} \sumJ  \frac{ r_j  \Delta_{1j}^2}{d^* (\x_j) E(R_j\mid \x_j) f_{\X}(\x_j)}\n \\
&&\times N^{-1/2} \sumI E \left[\frac{\{1-r_i  \pi^{-1}(y_i,\u_i;\bb,g^*)\}C_i   K_{h} (\u_j-\u_i)R_k K_{h} (\X_k-\x_j) }{d^*(\x_i) D_i^2 E(R_i\mid\u_i) f_{\U}(\u_i)} \mid \x_i, r_i, r_i y_i\right]\n \\
&=&N^{-3/2} \sumI \sumJ  \frac{ \{1-r_i  \pi^{-1}(y_i,\u_i;\bb,g^*)\}C_i  r_j  K_{h} (\u_j-\u_i) \Delta_{1j}^2}{ d^*(\x_i) D_i^2 d^* (\x_j) E(R_i\mid\u_i) f_{\U}(\u_i)}
+O_p(N^{1/2}h^m)\n\\
&=&T_{31}+O_p(N^{1/2}h^m),\n
\ee
where $T_{31}$ is as defined in (\ref{eq:t31}).

Consider $T_{323B}$,
\be \label{ustat:t323b}
&&T_{323B}\n\\
&=&N^{-2/2} \sumJ  \frac{ r_j  \Delta_{1j}^2}{d^* (\x_j) E(R_j\mid \x_j) f_{\X}(\x_j)}\n \\
&&\times N^{-1/2} \sumK E \left[\frac{\{1-R_i  \pi^{-1}(Y_i,\U_i;\bb,g^*)\}C_i   K_{h} (\u_j-\U_i)r_k K_{h} (\x_k-\x_j) }{d^*(\X_i) D_i^2  E(R_i\mid\U_i) f_{\U}(\U_i)} \mid \x_k, r_k, r_k y_k\right]\n \\
&=&N^{-3/2} \sumJ \sumK \frac{ r_j r_k K_{h} (\x_k-\x_j)  \Delta_{1j}^2}{d^* (\x_j) E(R_j\mid \x_j) f_{\X}(\x_j)} E \left[\frac{\{1-R_i  \pi^{-1}(Y_i,\U_i;\bb,g^*)\}C_i   K_{h} (\u_j-\U_i) }{d^*(\X_i) D_i^2  E(R_i\mid\U_i) f_{\U}(\U_i)} \right]\n \\
&=&N^{-3/2} \sumJ \sumK \frac{ r_j r_k K_{h} (\x_k-\x_j)  \Delta_{1j}^2}{d^* (\x_j) E(R_j\mid \x_j) f_{\X}(\x_j)} E\left[\frac{\{w^{-1}(\X_j)
-w^{*-1}(\X_j)\}  C_j }{d^*(\X_j) D_j^2 } \mid \u_j, 1 \right]
+O_p(N^{1/2}h^m)\n\\
&=&O_p(N^{1/2}h^m).\n
\ee

Consider $T_{323C}$,
\be \label{ustat:t323c}
&&T_{323C}\n\\
&=&N^{-2/2} \sumJ  \frac{ r_j  \Delta_{1j}^2}{d^* (\x_j) E(R_j\mid \x_j) f_{\X}(\x_j)}\n \\
&&\times N^{1/2} E \left[\frac{\{1-R_i  \pi^{-1}(Y_i,\U_i;\bb,g^*)\}C_i   K_{h} (\u_j-\U_i)R_k K_{h} (\X_k-\x_j) }{d^*(\X_i) D_i^2 E(R_i\mid\U_i) f_{\U}(\U_i)} \right]\n \\
&=&N^{-1/2} \sumJ  \frac{ r_j  \Delta_{1j}^2}{d^* (\x_j) }  E\left[\frac{\{w^{-1}(\X_j)
-w^{*-1}(\X_j)\} C_j }{d^*(\X_j) D_j^2 } \mid \u_j, 1 \right]
+O_p(N^{1/2}h^m)\n\\
&=&O_p(N^{1/2}h^m).\n
\ee

Consider $T_{332C}$ as defined in  (\ref{ustat:t332c}), applying the U-statistics property to it we get,
\be \label{ustat:t332c2}
&&T_{332C}\n\\
&=&N^{-5/2} \sumI \sumJ \sumK \frac{\{1-r_i  \pi^{-1}(y_i,\u_i;\bb,g^*)\}C_i r_j  K_{h} (\u_j-\u_i)r_k K_{h} (\u_k-\u_i)  \Delta^2_{1j} }{d^*(\x_i) D_i^2 d^* (\x_j) \{E(R_i\mid\u_i) f_{\U}(\u_i)\}^2} \n\\
&&+O_p(N^{1/2}h^m)\n\\
&=&N^{-2/2} \sumI  \frac{\{1-r_i  \pi^{-1}(y_i,\u_i;\bb,g^*)\}C_i  }{d^*(\x_i) D_i^2 \{E(R_i\mid\u_i) f_{\U}(\u_i)\}^2} \n\\
&&\times N^{-1/2} \sumJ E\left\{ \frac{ r_j  K_{h} (\u_j-\u_i)R_k K_{h} (\U_k-\u_i)  \Delta^2_{1j} }{ d^* (\x_j)} \mid \x_j, r_j, r_j y_j\right\}\n\\
&&+N^{-2/2} \sumI  \frac{\{1-r_i  \pi^{-1}(y_i,\u_i;\bb,g^*)\}C_i  }{d^*(\x_i) D_i^2 \{E(R_i\mid\u_i) f_{\U}(\u_i)\}^2} \n\\
&&\times N^{-1/2} \sumK E\left\{ \frac{ R_j  K_{h} (\U_j-\u_i)r_k K_{h} (\u_k-\u_i)  \Delta^2_{1j} }{ d^* (\X_j)} \mid \x_k, r_k, r_k y_k \right\}\n\\
&&-N^{-2/2} \sumI  \frac{\{1-r_i  \pi^{-1}(y_i,\u_i;\bb,g^*)\}C_i  }{d^*(\x_i) D_i^2 \{E(R_i\mid\u_i) f_{\U}(\u_i)\}^2} N^{1/2}  E\left\{ \frac{ R_j  K_{h} (\U_j-\u_i)R_k K_{h} (\U_k-\u_i)  \Delta^2_{1j} }{ d^* (\X_j)} \right\}\n\\
&&+O_p(N^{1/2}h^m)+O_p(N^{-1/2})\n\\
&=& T_{332CA}+T_{332CB}-T_{332CC}+O_p(N^{1/2}h^m)+O_p(N^{-1/2}).\n
\ee

Consider $T_{332CA}$,
\be \label{ustat:t332ca}
&&T_{332CA}\n\\
&=&N^{-2/2} \sumI  \frac{\{1-r_i  \pi^{-1}(y_i,\u_i;\bb,g^*)\}C_i  }{d^*(\x_i) D_i^2 \{E(R_i\mid\u_i) f_{\U}(\u_i)\}^2} \n\\
&&\times N^{-1/2} \sumJ E\left\{ \frac{ r_j  K_{h} (\u_j-\u_i)R_k K_{h} (\U_k-\u_i)  \Delta^2_{1j} }{ d^* (\x_j)} \mid \x_j, r_j, r_j y_j\right\}\n\\
&=&N^{-3/2} \sumI \sumJ \frac{\{1-r_i  \pi^{-1}(y_i,\u_i;\bb,g^*)\}C_i r_j  K_{h} (\u_j-\u_i) \Delta^2_{1j} }{d^*(\x_i) D_i^2 d^* (\x_j) E(R_i\mid\u_i) f_{\U}(\u_i)}
+O_p(N^{1/2}h^m)\n\\
&=&T_{31}+O_p(N^{1/2}h^m),\n
\ee
where $T_{31}$ is as defined in (\ref{eq:t31}).

Consider $T_{332CB}$,
\be \label{ustat:t332cb}
&&T_{332CB}\n\\
&=&N^{-2/2} \sumI  \frac{\{1-r_i  \pi^{-1}(y_i,\u_i;\bb,g^*)\}C_i  }{d^*(\x_i) D_i^2 \{E(R_i\mid\u_i) f_{\U}(\u_i)\}^2} \n\\
&&\times N^{-1/2} \sumK E\left\{ \frac{ R_j  K_{h} (\U_j-\u_i)r_k K_{h} (\u_k-\u_i)  \Delta^2_{1j} }{ d^* (\X_j)} \mid \x_k, r_k, r_k y_k \right\}\n\\
&=&N^{-3/2} \sumI  \sumJ \frac{\{1-r_i  \pi^{-1}(y_i,\u_i;\bb,g^*)\}C_i r_j K_{h} (\u_j-\u_i)  }{d^*(\x_i) D_i^2 E(R_i\mid\u_i) f_{\U}(\u_i) } E\left[\{ \Delta_{1i}^2 /d^* (\X_i)\} \mid \u_i,1  \right]\n\\
&&+O_p(N^{1/2}h^m)\n\\
&=&N^{-3/2} \sumI  \sumJ \frac{\{1-r_i  \pi^{-1}(y_i,\u_i;\bb,g^*)\}C_i r_j K_{h} (\u_j-\u_i)  }{d^*(\x_i) D_i E(R_i\mid\u_i) f_{\U}(\u_i) }+O_p(N^{1/2}h^m).
\ee

Consider $T_{332CC}$,
\be \label{ustat:t332cc}
&&T_{332CC}\n\\
&=&N^{-2/2} \sumI  \frac{\{1-r_i  \pi^{-1}(y_i,\u_i;\bb,g^*)\}C_i  }{d^*(\x_i) D_i^2 \{E(R_i\mid\u_i) f_{\U}(\u_i)\}^2}  N^{1/2}  E\left\{ \frac{ R_j  K_{h} (\U_j-\u_i)R_k K_{h} (\U_k-\u_i)  \Delta^2_{1j} }{ d^* (\X_j)} \right\}\n\\
&=&N^{-1/2} \sumI  \frac{\{1-r_i  \pi^{-1}(y_i,\u_i;\bb,g^*)\}C_i  }{d^*(\x_i) D_i^2 } E\left[\{ \Delta_{1i}^2 /d^* (\X_i)\} \mid \u_i,1  \right]+O_p(N^{1/2}h^m)\n\\
&=&N^{-1/2} \sumI  \frac{\{1-r_i  \pi^{-1}(y_i,\u_i;\bb,g^*)\}C_i  }{d^*(\x_i) D_i } +O_p(N^{1/2}h^m)\n\\
&=&N^{-1/2} \sumI  \frac{A_i  }{d^*(\x_i) } +O_p(N^{1/2}h^m).\n
\ee

Consider $T_{331B}$ as defined in  (\ref{ustat:t331b}), applying the U-statistics property to it we get,
\be \label{ustat:t331b2}
&&T_{331B}\n\\
&=&N^{-5/2} \sumI \sumJ \sumK  \frac{\{1-r_i  \pi^{-1}(y_i,\u_i;\bb,g^*)\}C_i r_j  K_{h} (\u_j-\u_i)  \Delta^2_{1j}}{d^*(\x_i) D_i^2 \{d^* (\x_j)\}^2 E(R_i\mid\u_i) f_{\U}(\u_i) E(R_j\mid\x_j) f_{\X}(\x_j)} \n\\
&&\times r_k  K_{h}(\x_k-\x_j) \{e^{-h(y_k)}+e^{-g^* (\u_j)}e^{-2h(y_k)}\}
+O_p(N^{1/2}h^m)\n\\
&=&N^{-3/2} \sumI \sumJ  \frac{ r_j \Delta^2_{1j}}{ \{d^* (\x_j)\}^2  E(R_j\mid\x_j) f_{\X}(\x_j)} \n\\
&&\times E \left[\frac{ \{1-r_i  \pi^{-1}(y_i,\u_i;\bb,g^*)\}C_i K_{h} (\u_j-\u_i)  R_k  K_{h}(\X_k-\x_j) \{e^{-h(Y_k)}+e^{-g^* (\u_j)}e^{-2h(Y_k)}\}}{d^*(\x_i) D_i^2 E(R_i\mid\u_i) f_{\U}(\u_i)} \right] \n\\
&&+N^{-3/2}  \sumJ \sumK \frac{ r_j \Delta^2_{1j}}{ \{d^* (\x_j)\}^2  E(R_j\mid\x_j) f_{\X}(\x_j)} \n\\
&&\times E \left[\frac{ \{1-R_i  \pi^{-1}(y_i,\u_i;\bb,g^*)\}C_i K_{h} (\u_j-\U_i)  r_k  K_{h}(\x_k-\x_j) \{e^{-h(y_k)}+e^{-g^* (\u_j)}e^{-2h(y_k)}\}}{d^*(\X_i) D_i^2 E(R_i\mid\U_i) f_{\U}(\U_i)} \right] \n\\
&&-N^{-1/2}  \sumJ \frac{ r_j \Delta^2_{1j}}{ \{d^* (\x_j)\}^2  E(R_j\mid\x_j) f_{\X}(\x_j)} \n\\
&&\times E \left[\frac{ \{1-R_i  \pi^{-1}(y_i,\u_i;\bb,g^*)\}C_i K_{h} (\u_j-\U_i)  R_k  K_{h}(\X_k-\x_j) \{e^{-h(Y_k)}+e^{-g^* (\u_j)}e^{-2h(Y_k)}\}}{d^*(\X_i) D_i^2 E(R_i\mid\U_i) f_{\U}(\U_i)}\right] \n\\
&&+O_p(N^{1/2}h^m)+O_p(N^{-1/2})\n\\
&=& T_{331BA}+T_{331BB}-T_{331BC}+O_p(N^{1/2}h^m)+O_p(N^{-1/2}).\n
\ee

Consider $T_{331BA}$,
\be \label{ustat:t331ba}
&&T_{331BA}\n\\
&=&N^{-3/2} \sumI \sumJ  \frac{ r_j \Delta^2_{1j}}{ \{d^* (\x_j)\}^2  E(R_j\mid\x_j) f_{\X}(\x_j)} \n\\
&&\times E \left[\frac{ \{1-r_i  \pi^{-1}(y_i,\u_i;\bb,g^*)\}C_i K_{h} (\u_j-\u_i)  R_k  K_{h}(\X_k-\x_j) \{e^{-h(Y_k)}+e^{-g^* (\u_j)}e^{-2h(Y_k)}\}}{d^*(\x_i) D_i^2 E(R_i\mid\u_i) f_{\U}(\u_i)} \right] \n\\
&=&N^{-3/2} \sumI \sumJ  \frac{  \{1-r_i  \pi^{-1}(y_i,\u_i;\bb,g^*)\}C_i  r_j K_{h} (\u_j-\u_i)  \Delta^2_{1j}}{ d^*(\x_i) D_i^2 \{d^* (\x_j)\}^2 E(R_i\mid\u_i) f_{\U}(\u_i) E(R_j\mid\x_j) f_{\X}(\x_j)} \n\\
&&\times d^* (\x_j) E(R_j\mid\x_j) f_{\X}(\x_j)
+O_p(N^{1/2}h^m)\n\\
&=&T_{31}+O_p(N^{1/2}h^m),\n
\ee
where $T_{31}$ is as defined in (\ref{eq:t31}).

Consider $T_{331BB}$,
\be \label{ustat:t331bb}
&&T_{331BB}\n\\
&=&N^{-3/2}  \sumJ \sumK \frac{ r_j \Delta^2_{1j}}{ \{d^* (\x_j)\}^2  E(R_j\mid\x_j) f_{\X}(\x_j)} \n\\
&&\times E \left[\frac{ \{1-R_i  \pi^{-1}(y_i,\u_i;\bb,g^*)\}C_i K_{h} (\u_j-\U_i)  r_k  K_{h}(\x_k-\x_j) \{e^{-h(y_k)}+e^{-g^* (\u_j)}e^{-2h(y_k)}\}}{d^*(\X_i) D_i^2 E(R_i\mid\U_i) f_{\U}(\U_i)} \right] \n\\
&=&N^{-3/2}  \sumJ \sumK \frac{ r_j r_k  K_{h}(\x_k-\x_j) \{e^{-h(y_k)}+e^{-g^* (\u_j)}e^{-2h(y_k)}\} \Delta^2_{1j}}{ \{d^* (\x_j)\}^2  E(R_j\mid\x_j) f_{\X}(\x_j)}\n\\
&&\times E\left[\frac{\{w^{-1}(\X_j)
-w^{*-1}(\X_j)\} C_j }{d^*(\X_j) D_j^2 } \mid \u_j, 1 \right]+O_p(N^{1/2}h^m)\n\\
&=&O_p(N^{1/2}h^m).\n
\ee

Consider $T_{331BC}$,
\be \label{ustat:t331bc}
&&T_{331BC}\n\\
&=&N^{-1/2}  \sumJ \frac{ r_j \Delta^2_{1j}}{ \{d^* (\x_j)\}^2  E(R_j\mid\x_j) f_{\X}(\x_j)} \n\\
&&\times E \left[\frac{ \{1-R_i  \pi^{-1}(y_i,\u_i;\bb,g^*)\}C_i K_{h} (\u_j-\U_i)  R_k  K_{h}(\X_k-\x_j) \{e^{-h(Y_k)}+e^{-g^* (\u_j)}e^{-2h(Y_k)}\}}{d^*(\X_i) D_i^2 E(R_i\mid\U_i) f_{\U}(\U_i)}\right] \n\\
&=&N^{-3/2}  \sumJ  \frac{ r_j \Delta^2_{1j}}{ d^* (\x_j) } E\left[\frac{\{w^{-1}(\X_j)
-w^{*-1}(\X_j)\}  C_j }{d^*(\X_j) D_j^2 } \mid \u_j, 1 \right]
+O_p(N^{1/2}h^m)\n\\
&=&O_p(N^{1/2}h^m).\n
\ee

Consider $T_{12}$ as defined in  (\ref{eq:t12}), applying the U-statistics property to it we get,
\be \label{ustat:t12}
&&T_{12}\n\\
&=&N^{-5/2} \sumI \sumJ \sumK \frac{A_i }{d^{*2}(\x_i)}  \frac{r_j
  K_{h}(\x_j-\x_i)  r_k K_{h}
  (\x_k-\x_i) }{\{E(R_i\mid\x_i) f_{\X}(\x_i)\}^2}\{e^{-h(y_j)}+e^{-g^*(\u_i)}e^{-2h(y_j)}\}\n\\
&=&N^{-2/2} \sumI  \frac{A_i }{d^{*2}(\x_i)\{E(R_i\mid\x_i) f_{\X}(\x_i)\}^2} \n\\
&&\times N^{-1/2}  \sumJ E[ r_j
  K_{h}(\x_j-\x_i)  R_k K_{h}
  (\X_k-\x_i)  \{e^{-h(y_j)}+e^{-g^*(\u_i)}e^{-2h(y_j)}\}\mid \x_j, r_j, r_j y_j]\n\\
&&+N^{-2/2} \sumI  \frac{A_i }{d^{*2}(\x_i)\{E(R_i\mid\x_i) f_{\X}(\x_i)\}^2} \n\\
&&\times N^{-1/2}  \sumK E[ R_j
  K_{h}(\X_j-\x_i)   r_k K_{h}
  (\x_k-\x_i) \{e^{-h(Y_j)}+e^{-g^*(\u_i)}e^{-2h(Y_j)}\}\mid \x_k, r_k, r_k y_k]\n\\
&&-N^{-2/2} \sumI  \frac{A_i }{d^{*2}(\x_i)\{E(R_i\mid\x_i) f_{\X}(\x_i)\}^2} \n\\
&&\times N^{1/2}   E[ R_j
  K_{h}(\X_j-\x_i)  R_k K_{h}
  (\X_k-\x_i) \{e^{-h(Y_j)}+e^{-g^*(\u_i)}e^{-2h(Y_j)}\}]
+O_p(N^{-1/2})\n\\
&=& T_{12A}+T_{12B}-T_{12C}+O_p(N^{-1/2}).\n
\ee

Consider $T_{12A}$,
\be \label{ustat:t12a}
&&T_{12A}\n\\
&=&N^{-2/2} \sumI  \frac{A_i }{d^{*2}(\x_i)\{E(R_i\mid\x_i) f_{\X}(\x_i)\}^2} \n\\
&&\times N^{-1/2}  \sumJ E[ r_j
  K_{h}(\x_j-\x_i)  R_k K_{h}
  (\X_k-\x_i)  \{e^{-h(y_j)}+e^{-g^*(\u_i)}e^{-2h(y_j)}\}\mid \x_j, r_j, r_j y_j]\n\\
&=&N^{-3/2} \sumI  \sumJ  \frac{A_i r_j
  K_{h}(\x_j-\x_i)\{e^{-h(y_j)}+e^{-g^*(\u_i)}e^{-2h(y_j)}\} }{d^{*2}(\x_i)E(R_i\mid\x_i) f_{\X}(\x_i)}
+O_p(N^{1/2}h^m)\n\\
&=&T_{11}+O_p(N^{1/2}h^m),\n
\ee
where $T_{11}$ is as defined in (\ref{eq:t11}).

Consider $T_{12B}$,
\be \label{ustat:t12b}
&&T_{12B}\n\\
&=&N^{-2/2} \sumI  \frac{A_i }{d^{*2}(\x_i)\{E(R_i\mid\x_i) f_{\X}(\x_i)\}^2} \n\\
&&\times N^{-1/2}  \sumK E[ R_j
  K_{h}(\X_j-\x_i)   r_k K_{h}
  (\x_k-\x_i) \{e^{-h(Y_j)}+e^{-g^*(\u_i)}e^{-2h(Y_j)}\}\mid \x_k, r_k, r_k y_k]\n\\
&=&N^{-3/2} \sumI \sumK \frac{A_i r_k K_{h}
  (\x_k-\x_i) }{d^{*2}(\x_i)\{E(R_i\mid\x_i) f_{\X}(\x_i)\}^2} d^*(\x_i) E(R_i\mid\x_i) f_{\X}(\x_i)
+O_p(N^{1/2}h^m)\n\\
&=&N^{-3/2} \sumI \sumK \frac{A_i r_k K_{h}
  (\x_k-\x_i) }{d^*(\x_i) E(R_i\mid\x_i) f_{\X}(\x_i)}
+O_p(N^{1/2}h^m)\n\\
&=&N^{-3/2} \sumI \sumJ \frac{A_i r_j K_{h}
  (\x_j-\x_i) }{d^*(\x_i) E(R_i\mid\x_i) f_{\X}(\x_i)}
+O_p(N^{1/2}h^m).
\ee

Consider $T_{12C}$,
\be \label{ustat:t12c}
&&T_{12C}\n\\
&=&N^{-2/2} \sumI  \frac{A_i }{d^{*2}(\x_i)\{E(R_i\mid\x_i) f_{\X}(\x_i)\}^2} \n\\
&&\times N^{1/2}   E[ R_j
  K_{h}(\X_j-\x_i)  R_k K_{h}
  (\X_k-\x_i) \{e^{-h(Y_j)}+e^{-g^*(\u_i)}e^{-2h(Y_j)}\}]\n\\
&=&N^{-1/2} \sumI  \frac{A_i }{d^*(\x_i)}+O_p(N^{1/2}h^m).\n
\ee

Putting all the terms together (cancelling and collecting similar terms) and rewriting (\ref{eq:final3}) we get the following (terms with 2 or 1 summations),
\be \label{eq:final4}
&&\frac{1}{\sqrt{N}}\sumI\wh\S^*\eff(\x_i,r_i,r_i y_i,\bb)\n\\
&=& \frac{2}{\sqrt{N}}\sumI \frac{A_i}{d^*(\x_i)}
+T_{2211AB}+T_{2211BB}+T_{2211C}+2T_{2211AC}-2T_{2212ABBB}
\n\\
&&-T_{2221}-2T_{2222B}-2T_{2222C} +2T_{2223B}-T_{231ABB}-T_{231BBB}\n\\
&&-T_{231CB}+T_{232BBBB}-T_{31}-2T_{322}+T_{323}-T_{321}\n\\
&&+T_{332C}+T_{331B}-2T_{11}+T_{12}+O_p(N^{1/2}h^{2m}+
N^{-1/2}h^{-p}+N^{-1/2}+N^{1/2}h^m)\n\\
&=& T_{2211ABB}+T_{2211BBB} +T_{2211CA}-T_{2211CC}+T_{2211AC}-2T_{2212ABBBB}
-T_{2221A}\n\\
&&+T_{2221C}-2T_{2222BB}-T_{2222C}+2T_{2223BB}
-T_{321}+T_{332CB}-T_{11}+T_{12B}\n\\
&&+O_p(N^{1/2}h^{2m}+ N^{-1/2}h^{-p}+N^{-1/2}+N^{1/2}h^m),
\ee
where
$T_{2211ABB}$ is defined in  (\ref{ustat:t2211abb}),
$T_{2211BBB}$ is defined in  (\ref{ustat:t2211bbb}),
$T_{2211CA}$ is defined in  (\ref{ustat:t2211ca}),
$T_{2211CC}$ is defined in  (\ref{ustat:t2211cc}),
$T_{2211AC}$ is defined in  (\ref{ustat:t2211ac}),
$T_{2212ABBBB}$ is defined in  (\ref{ustat:t2212abbbb}),
$T_{2221A}$ is defined in  (\ref{ustat:t2221a}),
$T_{2221C}$ is defined in  (\ref{ustat:t2221c}),
$T_{2222BB}$ is defined in  (\ref{ustat:t2222bb}),
$T_{2222C}$ is defined in  (\ref{ustat:t2222c}),
$T_{2223BB}$ is defined in  (\ref{ustat:t2223bb}),
$T_{321}$ is defined in  (\ref{eq:t321}),
$T_{332CB}$ is defined in  (\ref{ustat:t332cb}),
$T_{11}$ is defined in  (\ref{eq:t11}), and
$T_{12B}$ is defined in  (\ref{ustat:t12b}).

Consider $T_{2211ABB}$ as defined in  (\ref{ustat:t2211abb}), applying the U-statistics property to it we get,
\be \label{ustat:t2211abb2}
&&T_{2211ABB}\n\\
&=&N^{-3/2} \sumI  \sumJ \frac{r_i r_j  K_{h}(\x_j-\x_i) e^{-h(y_j)}
  \Delta_{3i}\{e^{-g(\u_i)}
-e^{-g^*(\u_i)} \} }{ d^*(\x_i) E(R_i\mid\x_i) f_{\X}(\x_i) } + O_p(N^{1/2}h^m)\n\\
&=&N^{-1/2} \sumI E \left[\frac{r_i R_j  K_{h}(\X_j-\x_i) e^{-h(Y_j)} \Delta_{3i} }{ d^*(\x_i) E(R_i\mid\x_i) f_{\X}(\x_i) } \{e^{-g(\u_i)}
-e^{-g^*(\u_i)} \} \mid \x_i, r_i, r_i y_i \right] \n\\
&&+N^{-1/2} \sumJ E \left[\frac{R_i r_j  K_{h}(\x_j-\X_i) e^{-h(y_j)} \Delta_{3i} }{ d^*(\X_i) E(R_i\mid\X_i) f_{\X}(\X_i) } \{e^{-g(\U_i)}
-e^{-g^*(\U_i)} \} \mid \x_j, r_j, r_j y_j \right] \n\\
&&-N^{1/2}  E \left[\frac{R_i R_j  K_{h}(\X_j-\X_i) e^{-h(Y_j)} \Delta_{3i} }{ d^*(\X_i) E(R_i\mid\X_i) f_{\X}(\X_i) } \{e^{-g(\U_i)}
-e^{-g^*(\U_i)} \} \right] + O_p(N^{1/2}h^m)+O_p(N^{-1/2})\n\\
&=& T_{2211ABBA}+T_{2211ABBB}-T_{2211ABBC}+ O_p(N^{1/2}h^m)+O_p(N^{-1/2}).\n
\ee

Consider $T_{2211ABBA}$,
\be \label{ustat:t2211abba}
&&T_{2211ABBA}\n\\
&=&N^{-1/2} \sumI E \left[\frac{r_i R_j  K_{h}(\X_j-\x_i) e^{-h(Y_j)} \Delta_{3i} }{ d^*(\x_i) E(R_i\mid\x_i) f_{\X}(\x_i) } \{e^{-g(\u_i)}
-e^{-g^*(\u_i)} \} \mid \x_i, r_i, r_i y_i \right] \n\\
&=&N^{-1/2} \sumI \frac{r_i  \Delta_{3i} \{e^{-g(\u_i)}
-e^{-g^*(\u_i)} \}}{ d^*(\x_i) E(R_i\mid\x_i) f_{\X}(\x_i) }  E \left\{ R_j  K_{h}(\X_j-\x_i) e^{-h(Y_j)}  \right\} \n\\
&=&N^{-1/2} \sumI \frac{r_i \Delta_{1i} \Delta_{3i} \{e^{-g(\u_i)}
-e^{-g^*(\u_i)} \}}{ d^*(\x_i) }
+ O_p(N^{1/2}h^m)\n\\
&=&T_{2211ABC},\n
\ee
where $T_{2211ABC}$ is defined in (\ref{ustat:t2211abc}).

Consider $T_{2211ABBB}$,
\be \label{ustat:t2211abbb}
&&T_{2211ABBB}\n\\
&=&N^{-1/2} \sumJ E \left[\frac{R_i r_j  K_{h}(\x_j-\X_i) e^{-h(y_j)} \Delta_{3i} }{ d^*(\X_i) E(R_i\mid\X_i) f_{\X}(\X_i) } \{e^{-g(\U_i)}
-e^{-g^*(\U_i)} \} \mid \x_j, r_j, r_j y_j \right] \n\\
&=&N^{-1/2} \sumJ r_j   e^{-h(y_j)}   E \left[\frac{ \Delta_{3j} }{ d^*(\x_j) } \{e^{-g(\u_j)}
-e^{-g^*(\u_j)} \}  \mid \x_j,1 \right] \n\\
&=&N^{-1/2} \sumJ   \frac{ r_j   e^{-h(y_j)}\Delta_{3j} }{ d^*(\x_j) } \{e^{-g(\u_j)}
-e^{-g^*(\u_j)} \}
+ O_p(N^{1/2}h^m)\n\\
&=&N^{-1/2} \sumI   \frac{ r_i   e^{-h(y_i)}\Delta_{3i} }{ d^*(\x_i) } \{e^{-g(\u_i)}
-e^{-g^*(\u_i)} \}
+ O_p(N^{1/2}h^m).
\ee

Consider $T_{2211ABBC}$,
\be \label{ustat:t2211abbc}
&&T_{2211ABBC}\n\\
&=&N^{1/2}  E \left[\frac{R_i R_j  K_{h}(\X_j-\X_i) e^{-h(Y_j)} \Delta_{3i} }{ d^*(\X_i) E(R_i\mid\X_i) f_{\X}(\X_i) } \{e^{-g(\U_i)}
-e^{-g^*(\U_i)} \} \right] \n\\
&=&N^{1/2} E \left\{ E \left(\frac{r_i R_j  K_{h}(\X_j-\x_i) e^{-h(Y_j)} \Delta_{3i} }{ d^*(\x_i) E(R_i\mid\x_i) f_{\X}(\x_i) }  \{e^{-g(\u_i)}
-e^{-g^*(\u_i)} \} \mid \x_i, r_i, r_i y_i \right)  \right\}\n\\
&=&N^{1/2} E \left(\frac{R_i \Delta_{1i}\Delta_{3i} }{ d^*(\X_i) }\{e^{-g(\U_i)}
-e^{-g^*(\U_i)} \}  \right)
+ O_p(N^{1/2}h^m).
\ee

Consider $T_{2211BBB}$ as defined in  (\ref{ustat:t2211bbb}), applying the U-statistics property to it we get,
\be \label{ustat:t2211bbb2}
&&T_{2211BBB}\n\\
&=&N^{-3/2} \sumI \sumJ \frac{ r_i r_j K_{h}(\x_j-\x_i) e^{-h(y_j)}\h'_\bb(y_j;\bb) \Delta_{1i}\{e^{-g(\u_i)}-e^{-g^*(\u_i)}\}  }{ d^*(\x_i) E(R_i\mid\x_i) f_{\X}(\x_i)} \n\\
&&+O_p(N^{1/2}h^m)\n\\
&=&N^{-1/2} \sumI \frac{ r_i\Delta_{1i} \{e^{-g(\u_i)}-e^{-g^*(\u_i)}\} }{ d^*(\x_i) E(R_i\mid\x_i) f_{\X}(\x_i)}  E\left\{R_j K_{h}(\X_j-\x_i) e^{-h(Y_j)}\h'_\bb(Y_j;\bb) \mid \x_i, r_i, r_i y_i \right\}\n\\
&&+N^{-1/2} \sumJ  r_j  e^{-h(y_j)}\h'_\bb(y_j;\bb)  E\left[\frac{  R_i K_{h}(\x_j-\X_i)\Delta_{1i} }{ d^*(\X_i) E(R_i\mid\X_i) f_{\X}(\X_i)}\{e^{-g(\U_i)}-e^{-g^*(\U_i)}\} \mid \x_j, r_j, r_j y_j \right]\n\\
&&-N^{1/2} E\left[\frac{ R_i R_j K_{h}(\X_j-\X_i) e^{-h(Y_j)}\h'_\bb(Y_j;\bb) \Delta_{1i} }{ d^*(\X_i) E(R_i\mid\X_i) f_{\X}(\X_i)} \{e^{-g(\U_i)}-e^{-g^*(\U_i)}\} \right]\n\\
&&+O_p(N^{1/2}h^m)+O_p(N^{-1/2})\n\\
&=& T_{2211BBBA}+T_{2211BBBB}-T_{2211BBBC}+ O_p(N^{1/2}h^m)+O_p(N^{-1/2}).\n
\ee

Consider $T_{2211BBBA}$,
\be \label{ustat:t2211bbba}
&&T_{2211BBBA}\n\\
&=&N^{-1/2} \sumI \frac{ r_i\Delta_{1i}\{e^{-g(\u_i)}-e^{-g^*(\u_i)}\} }{ d^*(\x_i) E(R_i\mid\x_i) f_{\X}(\x_i)}  E\left\{R_j K_{h}(\X_j-\x_i) e^{-h(Y_j)}\h'_\bb(Y_j;\bb) \mid \x_i, r_i, r_i y_i \right\}\n\\
&=&N^{-1/2} \sumI \frac{ r_i\Delta_{1i} \Delta_{3i} }{ d^*(\x_i) } \{e^{-g(\u_i)}-e^{-g^*(\u_i)}\}
+ O_p(N^{1/2}h^m)\n\\
&=&T_{2211ABC},\n
\ee
where $T_{2211ABC}$ is defined in  (\ref{ustat:t2211abc}).

Consider $T_{2211BBBB}$,
\be \label{ustat:t2211bbbb}
&&T_{2211BBBB}\n\\
&=&N^{-1/2} \sumJ  r_j  e^{-h(y_j)}\h'_\bb(y_j;\bb) E\left[\frac{  R_i K_{h}(\x_j-\X_i)\Delta_{1i} }{ d^*(\X_i) E(R_i\mid\X_i) f_{\X}(\X_i)}\{e^{-g(\U_i)}-e^{-g^*(\U_i)}\} \mid \x_j, r_j, r_j y_j \right]\n\\
&=&N^{-1/2} \sumJ  r_j  e^{-h(y_j)}\h'_\bb(y_j;\bb)  E\left[\frac{  \Delta_{1j} }{ d^*(\x_j) } \{e^{-g(\u_j)}-e^{-g^*(\u_j)}\} \mid \x_j, 1 \right]
+ O_p(N^{1/2}h^m)\n\\
&=&N^{-1/2} \sumJ  \frac{ r_j  e^{-h(y_j)}\h'_\bb(y_j;\bb)  \Delta_{1j} }{ d^*(\x_j) } \{e^{-g(\u_j)}-e^{-g^*(\u_j)}\}
+ O_p(N^{1/2}h^m)\n\\
&=&N^{-1/2} \sumI  \frac{ r_i  e^{-h(y_i)}\h'_\bb(y_i;\bb)  \Delta_{1i} }{ d^*(\x_i) } \{e^{-g(\u_i)}-e^{-g^*(\u_i)}\}+ O_p(N^{1/2}h^m).
\ee

Consider $T_{2211BBBC}$,
\be \label{ustat:t2211bbbc}
&&T_{2211BBBC}\n\\
&=&N^{1/2} E\left[\frac{ R_i R_j K_{h}(\X_j-\X_i) e^{-h(Y_j)}\h'_\bb(Y_j;\bb) \Delta_{1i} }{ d^*(\X_i) E(R_i\mid\X_i) f_{\X}(\X_i)} \{e^{-g(\U_i)}-e^{-g^*(\U_i)}\} \right]\n\\
&=&N^{1/2} E \left\{ E\left(\frac{ r_i R_j K_{h}(\X_j-\x_i) e^{-h(Y_j)}\h'_\bb(Y_j;\bb) \Delta_{1i} }{ d^*(\x_i) E(R_i\mid\x_i) f_{\X}(\x_i)} \{e^{-g(\u_i)}-e^{-g^*(\u_i)}\} \mid \x_i, r_i, r_i y_i \right)  \right\}\n\\
&=&N^{1/2} E\left(\frac{ R_i  \Delta_{1i} \Delta_{3i}}{ d^*(\X_i) } \{e^{-g(\U_i)}-e^{-g^*(\U_i)}\}\right)+ O_p(N^{1/2}h^m)\n\\
&=&T_{2211ABBC},\n
\ee
where $T_{2211ABBC}$ is defined in  (\ref{ustat:t2211abbc}).

Consider $T_{2211CA}$ as defined in  (\ref{ustat:t2211ca}), applying the U-statistics property to it we get,
\be \label{ustat:t2211ca2}
&&T_{2211CA}\n\\
&=&N^{-3/2} \sumI  \sumJ \frac{\{1-r_i
  \pi^{-1}(y_i,\u_i;\bb,g^*)\}r_j e^{-h(y_j)} K_h({\x_j-\x_i})  }{d^*(\x_i) D_i E(R_i\mid\x_i) f_{\X}(\x_i)}    E\left[\{ \Delta_{1i} \Delta_{3i}/d^* (\X_i)\} \mid \u_i,1  \right]\n\\
&&+ O_p(N^{1/2}h^m)\n\\
&=&N^{-1/2} \sumI \frac{\{1-r_i
  \pi^{-1}(y_i,\u_i;\bb,g^*)\}  }{d^*(\x_i) D_i E(R_i\mid\x_i) f_{\X}(\x_i)}    E\left[\{ \Delta_{1i} \Delta_{3i}/d^* (\X_i)\} \mid \u_i,1  \right]\n\\
&& \times E\{ R_j e^{-h(Y_j)} K_h({\X_j-\x_i})\mid \x_i, r_i, r_i y_i  \}\n\\
&&+N^{-1/2} \sumJ r_j e^{-h(y_j)}\n\\
&& \times E \left(\frac{\{1-R_i
  \pi^{-1}(Y_i,\U_i;\bb,g^*)\} K_h({\x_j-\X_i})  }{d^*(\X_i) D_i E(R_i\mid\X_i) f_{\X}(\X_i)}    E\left[\{ \Delta_{1i} \Delta_{3i}/d^* (\X_i)\} \mid \U_i,1  \right] \mid \x_j, r_j, r_j y_j \right)\n\\
&&-N^{1/2} E \left(\frac{\{1-R_i
  \pi^{-1}(Y_i,\U_i;\bb,g^*)\}R_j e^{-h(Y_j)} K_h({\X_j-\X_i})  }{d^*(\X_i) D_i E(R_i\mid\X_i) f_{\X}(\X_i)}    E\left[\{ \Delta_{1i} \Delta_{3i}/d^* (\X_i)\} \mid \U_i,1  \right] \right)\n\\
&&+ O_p(N^{1/2}h^m)+O_p(N^{-1/2})\n\\
&=& T_{2211CAA}+T_{2211CAB}-T_{2211CAC}+ O_p(N^{1/2}h^m)+O_p(N^{-1/2}).\n
\ee

Consider $T_{2211CAA}$,
\be \label{ustat:t2211caa}
&&T_{2211CAA}\n\\
&=&N^{-1/2} \sumI \frac{\{1-r_i
  \pi^{-1}(y_i,\u_i;\bb,g^*)\}  }{d^*(\x_i) D_i E(R_i\mid\x_i) f_{\X}(\x_i)}    E\left[\{ \Delta_{1i} \Delta_{3i}/d^* (\X_i)\} \mid \u_i,1  \right]\n\\
&& \times E\{ R_j e^{-h(Y_j)} K_h({\X_j-\x_i})\mid \x_i, r_i, r_i y_i  \}\n\\
&=&N^{-1/2} \sumI \frac{\{1-r_i
  \pi^{-1}(y_i,\u_i;\bb,g^*)\} \Delta_{1i} }{d^*(\x_i) D_i }    E\left[\{ \Delta_{1i} \Delta_{3i}/d^* (\X_i)\} \mid \u_i,1  \right]
+ O_p(N^{1/2}h^m)\n\\
&=&T_{2211CC},\n
\ee
where $T_{2211CC}$ is defined in  (\ref{ustat:t2211cc}).

Consider $T_{2211CAB}$,
\be \label{ustat:t2211cab}
&&T_{2211CAB}\n\\
&=&N^{-1/2} \sumJ r_j e^{-h(y_j)}\n\\
&& \times E \left(\frac{\{1-R_i
  \pi^{-1}(Y_i,\U_i;\bb,g^*)\} K_h({\x_j-\X_i})  }{d^*(\X_i) D_i E(R_i\mid\X_i) f_{\X}(\X_i)}    E\left[\{ \Delta_{1i} \Delta_{3i}/d^* (\X_i)\} \mid \U_i,1  \right] \mid \x_j, r_j, r_j y_j \right)\n\\
&=&N^{-1/2} \sumJ \frac{  r_j e^{-h(y_j)} \{w^{-1}(\x_j) -w^{*-1}(\x_j)\} }{d^*(\x_j) D_j }    E\left[\{ \Delta_{1j} \Delta_{3j}/d^* (\X_j)\} \mid \u_j,1  \right]+ O_p(N^{1/2}h^m)\n\\
&=&N^{-1/2} \sumI   \frac{\{w^{-1}(\x_i) -w^{*-1}(\x_i)\} r_i e^{-h(y_i)} }{d^*(\x_i) D_i }    E\left[\{ \Delta_{1i} \Delta_{3i}/d^* (\X_i)\} \mid \u_i,1  \right] + O_p(N^{1/2}h^m).
\ee

Consider $T_{2211CAC}$,
\be \label{ustat:t2211cac}
&&T_{2211CAC}\n\\
&=&N^{1/2} E \left(\frac{\{1-R_i
  \pi^{-1}(Y_i,\U_i;\bb,g^*)\}R_j e^{-h(Y_j)} K_h({\X_j-\X_i})  }{d^*(\X_i) D_i E(R_i\mid\X_i) f_{\X}(\X_i)}    E\left[\{ \Delta_{1i} \Delta_{3i}/d^* (\X_i)\} \mid \U_i,1  \right] \right)\n\\
&=&N^{1/2} E \left(\frac{\{1-r_i
  \pi^{-1}(y_i,\u_i;\bb,g^*)\}  }{d^*(\X_i) D_i E(R_i\mid\x_i) f_{\X}(\x_i)}    E\left[\{ \Delta_{1i} \Delta_{3i}/d^* (\X_i)\} \mid \u_i,1  \right] E\{ R_j e^{-h(Y_j)} K_h({\X_j-\x_i})\} \right)\n\\
&=&N^{1/2} E \left(\frac{\{1-R_i
  \pi^{-1}(Y_i,\U_i;\bb,g^*)\} \Delta_{1i}  }{d^*(\X_i) D_i }    E\left[\{ \Delta_{1i} \Delta_{3i}/d^* (\X_i)\} \mid \U_i,1  \right] \right)+ O_p(N^{1/2}h^m).
\ee

Consider $T_{2211AC}$ as defined in  (\ref{ustat:t2211ac}), applying the U-statistics property to it we get,
\be \label{ustat:t2211ac2}
&&T_{2211AC}\n\\
&=&N^{-3/2} \sumI  \sumJ \frac{\{1-r_i
  \pi^{-1}(y_i,\u_i;\bb,g^*)\}  r_j K_h({\u_j-\u_i}) \Delta_{1i} \Delta_{1j}\Delta_{3j}}{d^*(\x_i) D_i d^*(\x_j) E(R_i\mid\u_i) f_{\U}(\u_i)   }
+ O_p(N^{1/2}h^m)\n\\
&=&N^{-1/2} \sumI  E\left[ \frac{\{1-r_i
  \pi^{-1}(y_i,\u_i;\bb,g^*)\}  R_j K_h({\U_j-\u_i}) \Delta_{1i} \Delta_{1j}\Delta_{3j}}{d^*(\x_i) D_i d^*(\X_j) E(R_i\mid\u_i) f_{\U}(\u_i)} \mid \x_i, r_i, r_i y_i \right]\n\\
&&+N^{-1/2} \sumJ  E\left[ \frac{\{1-R_i
  \pi^{-1}(Y_i,\U_i;\bb,g^*)\}  r_j K_h({\u_j-\U_i}) \Delta_{1i} \Delta_{1j}\Delta_{3j}}{d^*(\X_i) D_i d^*(\x_j) E(R_i\mid\U_i) f_{\U}(\U_i)} \mid \x_j, r_j, r_j y_j \right]\n\\
&&-N^{1/2} E\left[ \frac{\{1-R_i
  \pi^{-1}(Y_i,\U_i;\bb,g^*)\} R_j K_h({\U_j-\U_i}) \Delta_{1i} \Delta_{1j}\Delta_{3j}}{d^*(\X_i) D_i d^*(\X_j) E(R_i\mid\U_i) f_{\U}(\U_i)} \right]\n\\
&&+ O_p(N^{1/2}h^m)+O_p(N^{-1/2})\n\\
&=& T_{2211ACA}+T_{2211ACB}-T_{2211ACC}+ O_p(N^{1/2}h^m)+O_p(N^{-1/2}).\n
\ee

Consider $T_{2211ACA}$,
\be \label{ustat:t2211aca}
&&T_{2211ACA}\n\\
&=&N^{-1/2} \sumI  E\left[ \frac{\{1-r_i
  \pi^{-1}(y_i,\u_i;\bb,g^*)\}  R_j K_h({\U_j-\u_i}) \Delta_{1i} \Delta_{1j}\Delta_{3j}}{d^*(\x_i) D_i d^*(\X_j) E(R_i\mid\u_i) f_{\U}(\u_i)} \mid \x_i, r_i, r_i y_i \right]\n\\
&=&N^{-1/2} \sumI \frac{\{1-r_i
  \pi^{-1}(y_i,\u_i;\bb,g^*)\} \Delta_{1i}}{d^*(\x_i) D_i E(R_i\mid\u_i) f_{\U}(\u_i)}  E\left\{ \frac{ R_j K_h({\U_j-\u_i}) \Delta_{1j}\Delta_{3j}}{d^*(\X_j)} \right\}\n\\
&=&N^{-1/2} \sumI \frac{\{1-r_i
  \pi^{-1}(y_i,\u_i;\bb,g^*)   \} \Delta_{1i}}{d^*(\x_i) D_i}   E\left[\{ \Delta_{1i} \Delta_{3i}/d^* (\X_i)\} \mid \u_i,1  \right]
+ O_p(N^{1/2}h^m)\n\\
&=&T_{2211CC},\n
\ee
where $T_{2211CC}$ is defined in  (\ref{ustat:t2211cc}).

Consider $T_{2211ACB}$,
\be \label{ustat:t2211acb}
&&T_{2211ACB}\n\\
&=&N^{-1/2} \sumJ  E\left[ \frac{\{1-R_i
  \pi^{-1}(Y_i,\U_i;\bb,g^*)\}  r_j K_h({\u_j-\U_i}) \Delta_{1i} \Delta_{1j}\Delta_{3j}}{d^*(\X_i) D_i d^*(\x_j) E(R_i\mid\U_i) f_{\U}(\U_i)} \mid \x_j, r_j, r_j y_j \right]\n\\
&=&N^{-1/2} \sumJ \frac{  r_j \Delta_{1j}\Delta_{3j}}{d^*(\x_j)  }  E\left[ \frac{\{1-R_i
  \pi^{-1}(Y_i,\U_i;\bb,g^*)\}   K_h({\u_j-\U_i}) \Delta_{1i} }{d^*(\X_i) D_i  E(R_i\mid\U_i) f_{\U}(\U_i)}\right]\n\\
&=&N^{-1/2} \sumJ \frac{  r_j \Delta_{1j}\Delta_{3j}}{d^*(\x_j)  }  E \left[ \frac{ \{w^{-1}(\X_j)
-w^{*-1}(\X_j)\}     \Delta_{1j} }{d^*(\X_j)  D_j } \mid \u_j,1\right]
+ O_p(N^{1/2}h^m)\n\\
&=&N^{-1/2} \sumJ \frac{  r_j \Delta_{1j}\Delta_{3j}}{d^*(\x_j)  }  \{e^{-g(\u_j)}-e^{-g^*(\u_j)}\}
+ O_p(N^{1/2}h^m)\n\\
&=&N^{-1/2} \sumI \frac{  r_i \Delta_{1i}\Delta_{3i}}{d^*(\x_i)  }  \{e^{-g(\u_i)}-e^{-g^*(\u_i)}\}
+ O_p(N^{1/2}h^m)\n\\
&=&T_{2211ABC},\n
\ee
where $T_{2211ABC}$ is defined in  (\ref{ustat:t2211abc}).

Consider $T_{2211ACC}$,
\be \label{ustat:t2211acc}
&&T_{2211ACC}\n\\
&=&N^{1/2} E\left[ \frac{\{1-R_i
  \pi^{-1}(Y_i,\U_i;\bb,g^*)\} R_j K_h({\U_j-\U_i}) \Delta_{1i} \Delta_{1j}\Delta_{3j}}{d^*(\X_i) D_i d^*(\X_j) E(R_i\mid\U_i) f_{\U}(\U_i)} \right]\n\\
&=&N^{1/2} E \left( E\left[ \frac{\{1-r_i
  \pi^{-1}(y_i,\u_i;\bb,g^*)\} R_j K_h({\U_j-\u_i}) \Delta_{1i} \Delta_{1j}\Delta_{3j}}{d^*(\x_i) D_i d^*(\X_j) E(R_i\mid\u_i) f_{\U}(\u_i)} \mid \x_i, r_i, r_i y_i \right] \right)\n\\
&=&N^{1/2} E \left( \frac{\{1-R_i
  \pi^{-1}(Y_i,\U_i;\bb,g^*)   \} \Delta_{1i}}{d^*(\X_i) D_i}   E\left[\{ \Delta_{1i} \Delta_{3i}/d^* (\X_i)\} \mid \U_i,1  \right] \right)+ O_p(N^{1/2}h^m)\n\\
&=& T_{2211CAC},\n
\ee
where $T_{2211CAC}$ is defined in  (\ref{ustat:t2211cac}).

Consider $T_{2212ABBBB}$ as defined in  (\ref{ustat:t2212abbbb}), applying the U-statistics property to it we get,
\be \label{ustat:t2212abbbb2}
&&T_{2212ABBBB}\n\\
&=&N^{-3/2}  \sumI \sumJ  \frac{ r_i r_j K_{h} (\x_j-\x_i)  \Delta_{1i} \Delta_{3i} \{e^{-g(\u_i)}-e^{-g^*(\u_i)}\} }{d^*(\x_i) E(R_i\mid\x_i) f_{\X}(\x_i)}
+O_p(N^{1/2}h^m)\n\\
&=&N^{-1/2}  \sumI E\left[  \frac{ r_i R_j K_{h} (\X_j-\x_i)  \Delta_{1i} \Delta_{3i} \{e^{-g(\u_i)}-e^{-g^*(\u_i)}\} }{d^*(\x_i) E(R_i\mid\x_i) f_{\X}(\x_i)} \mid \x_i, r_i, r_i y_i \right]\n\\
&&+ N^{-1/2}  \sumJ E\left[  \frac{ R_i r_j K_{h} (\x_j-\X_i)  \Delta_{1i} \Delta_{3i}  \{e^{-g(\U_i)}-e^{-g^*(\U_i)}\}}{d^*(\X_i) E(R_i\mid\X_i) f_{\X}(\X_i)} \mid \x_j, r_j, r_j y_j \right]\n\\
&&- N^{1/2}   E\left[  \frac{ R_i R_j K_{h} (\X_j-\X_i)  \Delta_{1i} \Delta_{3i} \{e^{-g(\U_i)}-e^{-g^*(\U_i)}\}}{d^*(\X_i) E(R_i\mid\X_i) f_{\X}(\X_i)} \right]+ O_p(N^{1/2}h^m)+O_p(N^{-1/2})\n\\
&=& T_{2212ABBBBA}+T_{2212ABBBBB}-T_{2212ABBBBC}+ O_p(N^{1/2}h^m)+O_p(N^{-1/2}).\n
\ee

Consider $T_{2212ABBBBA}$,
\be \label{ustat:t2212abbbba}
&&T_{2212ABBBBA}\n\\
&=&N^{-1/2}  \sumI E\left[  \frac{ r_i R_j K_{h} (\X_j-\x_i)  \Delta_{1i} \Delta_{3i} \{e^{-g(\u_i)}-e^{-g^*(\u_i)}\} }{d^*(\x_i) E(R_i\mid\x_i) f_{\X}(\x_i)} \mid \x_i, r_i, r_i y_i \right]\n\\
&=&N^{-1/2}  \sumI \frac{ r_i  \Delta_{1i} \Delta_{3i} }{d^*(\x_i)} \{e^{-g(\u_i)}-e^{-g^*(\u_i)}\}
+ O_p(N^{1/2}h^m)\n\\
&=&T_{2211ABC},\n
\ee
where $T_{2211ABC}$ is defined in  (\ref{ustat:t2211abc}).

Consider $T_{2212ABBBBB}$,
\be \label{ustat:t2212abbbbb}
&&T_{2212ABBBBB}\n\\
&=&N^{-1/2}  \sumJ E\left[  \frac{ R_i r_j K_{h} (\x_j-\X_i)  \Delta_{1i} \Delta_{3i}  \{e^{-g(\U_i)}-e^{-g^*(\U_i)}\}}{d^*(\X_i) E(R_i\mid\X_i) f_{\X}(\X_i)} \mid \x_j, r_j, r_j y_j \right]\n\\
&=&N^{-1/2}  \sumJ r_j E\left[  \frac{ \Delta_{1j} \Delta_{3j} }{d^*(\x_j) } \{e^{-g(\u_j)}-e^{-g^*(\u_j)}\}  \mid \x_j, 1\right]+ O_p(N^{1/2}h^m)\n\\
&=&N^{-1/2}  \sumJ  \frac{ r_j \Delta_{1j} \Delta_{3j} }{d^*(\x_j) } \{e^{-g(\u_j)}-e^{-g^*(\u_j)}\}
+ O_p(N^{1/2}h^m)\n\\
&=&N^{-1/2}  \sumI \frac{ r_i  \Delta_{1i} \Delta_{3i} }{d^*(\x_i)} \{e^{-g(\u_i)}-e^{-g^*(\u_i)}\}
+ O_p(N^{1/2}h^m)\n\\
&=&T_{2211ABC},\n
\ee
where $T_{2211ABC}$ is defined in  (\ref{ustat:t2211abc}).

Consider $T_{2212ABBBBC}$,
\be \label{ustat:t2212abbbbc}
&&T_{2212ABBBBC}\n\\
&=&N^{1/2}   E\left[  \frac{ R_i R_j K_{h} (\X_j-\X_i)  \Delta_{1i} \Delta_{3i} \{e^{-g(\U_i)}-e^{-g^*(\U_i)}\}}{d^*(\X_i) E(R_i\mid\X_i) f_{\X}(\X_i)} \right]\n\\
&=&N^{1/2}   E\left(   E \left[\frac{ r_i R_j K_{h} (\X_j-\x_i)  \Delta_{1i} \Delta_{3i} }{d^*(\x_i) E(R_i\mid\x_i) f_{\X}(\x_i)} \{e^{-g(\u_i)}-e^{-g^*(\u_i)}\} \mid \x_i, r_i, r_i y_i\right] \right) \n\\
&=&N^{1/2}   E\left[  \frac{ R_i  \Delta_{1i} \Delta_{3i} }{d^*(\X_i) }\{e^{-g(\U_i)}-e^{-g^*(\U_i)}\} \right]
+ O_p(N^{1/2}h^m)\n\\
&=&T_{2211ABBC},\n
\ee
where $T_{2211ABBC}$ is defined in  (\ref{ustat:t2211abbc}).

Consider $T_{2221A}$ as defined in  (\ref{ustat:t2221a}), applying the U-statistics property to it we get,
\be \label{ustat:t2221a2}
&&T_{2221A}\n\\
&=&N^{-3/2} \sumI \sumJ \frac{\{1-r_i
  \pi^{-1}(Y_i,\u_i;\bb,g^*)\} r_j K_h({\x_j-\x_i}) e^{-h(y_j)} \h'_\bb(y_j;\bb)}{d^*(\x_i)
 E(R_i\mid\x_i) f_{\X}(\x_i)}+O_p(N^{1/2}h^m)\n\\
&=&N^{-1/2} \sumI \frac{\{1-r_i
  \pi^{-1}(Y_i,\u_i;\bb,g^*)\} }{d^*(\x_i)
 E(R_i\mid\x_i) f_{\X}(\x_i)} E \{ R_j K_h({\X_j-\x_i}) e^{-h(Y_j)} \h'_\bb(Y_j;\bb) \mid \x_i, r_, r_i y_i\}\n\\
&&+N^{-1/2} \sumJ r_j e^{-h(y_j)} \h'_\bb(y_j;\bb) E\left[ \frac{\{1-R_i
  \pi^{-1}(Y_i,\U_i;\bb,g^*)\} K_h({\x_j-\X_i}) }{d^*(\X_i)
 E(R_i\mid\X_i) f_{\X}(\X_i)}  \mid \x_j, r_j, r_j y_j \right]\n\\
&&-N^{1/2} E\left[\frac{\{1-R_i
  \pi^{-1}(Y_i,\U_i;\bb,g^*)\} R_j K_h({\X_j-\X_i}) e^{-h(Y_j)} \h'_\bb(Y_j;\bb)}{d^*(\X_i)
 E(R_i\mid\X_i) f_{\X}(\X_i)} \right]+ O_p(N^{1/2}h^m)\n\\
 &&+O_p(N^{-1/2})\n\\
&=& T_{2221AA}+T_{2221AB}-T_{2221AC}+ O_p(N^{1/2}h^m)+O_p(N^{-1/2}).\n
\ee

Consider $T_{2221AA}$,
\be \label{ustat:t2221aa}
&&T_{2221AA}\n\\
&=&N^{-1/2} \sumI \frac{\{1-r_i
  \pi^{-1}(Y_i,\u_i;\bb,g^*)\} }{d^*(\x_i)
 E(R_i\mid\x_i) f_{\X}(\x_i)} E \{ R_j K_h({\X_j-\x_i}) e^{-h(Y_j)} \h'_\bb(Y_j;\bb) \mid \x_i, r_, r_i y_i\}\n\\
&=&N^{-1/2} \sumI \frac{\{1-r_i
  \pi^{-1}(Y_i,\u_i;\bb,g^*)\}\Delta_{3i} }{d^*(\x_i) }
+ O_p(N^{1/2}h^m)\n\\
&=&T_{2221C},\n
\ee
where $T_{2221C}$ is defined in  (\ref{ustat:t2221c}).

Consider $T_{2221AB}$,
\be \label{ustat:t2221ab}
&&T_{2221AB}\n\\
&=&N^{-1/2} \sumJ r_j e^{-h(y_j)} \h'_\bb(y_j;\bb) E\left[ \frac{\{1-R_i
  \pi^{-1}(Y_i,\U_i;\bb,g^*)\} K_h({\x_j-\X_i}) }{d^*(\X_i)
 E(R_i\mid\X_i) f_{\X}(\X_i)}\mid \x_j, r_j, r_j y_j \right]\n\\
&=&N^{-1/2} \sumJ r_j e^{-h(y_j)} \h'_\bb(y_j;\bb) E\left[ \frac{\{1-R_i
  \pi^{-1}(Y_i,\U_i;\bb,g^*)\} K_h({\x_j-\X_i}) }{d^*(\X_i)
 E(R_i\mid\X_i) f_{\X}(\X_i)}  \right]\n\\
&=&N^{-1/2} \sumJ \frac{\{w^{-1}(\x_j) -w^{*-1}(\x_j)\} r_j e^{-h(y_j)} \h'_\bb(y_j;\bb)}{d^*(\x_j)} + O_p(N^{1/2}h^m)\n\\
&=&N^{-1/2} \sumI \frac{\{w^{-1}(\x_i) -w^{*-1}(\x_i)\} r_i e^{-h(y_i)} \h'_\bb(y_i;\bb)}{d^*(\x_i)} + O_p(N^{1/2}h^m).
\ee

Consider $T_{2221AC}$,
\be \label{ustat:t2221ac}
&&T_{2221AC}\n\\
&=&N^{1/2} E\left[\frac{\{1-R_i
  \pi^{-1}(Y_i,\U_i;\bb,g^*)\} R_j K_h({\X_j-\X_i}) e^{-h(Y_j)} \h'_\bb(Y_j;\bb)}{d^*(\X_i)
 E(R_i\mid\X_i) f_{\X}(\X_i)}  \right]\n\\
&=&N^{1/2} E\left[\frac{\{1-R_i
  \pi^{-1}(Y_i,\U_i;\bb,g^*)\} }{d^*(\X_i)
 E(R_i\mid\X_i) f_{\X}(\X_i)}   E\left\{ R_j K_h({\X_j-\x_i}) e^{-h(Y_j)} \h'_\bb(Y_j;\bb)\right\} \right]\n\\
&=&N^{1/2} E\left[ \frac{\{1-R_i
  \pi^{-1}(Y_i,\U_i;\bb,g^*)\} \Delta_{3i} }{d^*(\X_i)
 }   \right]
+ O_p(N^{1/2}h^m).
\ee

Consider $T_{2222BB}$ as defined in  (\ref{ustat:t2222bb}), applying the U-statistics property to it we get,
\be \label{ustat:t2222bb2}
&&T_{2222BB}\n\\
&=&N^{-3/2} \sumI \sumJ \frac{  r_i  r_j  K_{h}(\x_j-\x_i) e^{-h(y_j)} \Delta_{1i} }{ d^*(\x_i)D_i  E(R_i \mid \x_i) f_{\X}(\x_i)} \{e^{-g(\u_i)}-e^{-g^*(\u_i)}\} E \left\{\frac{ \Delta_{1i}\Delta_{3i} }{d^*(\X_i)}  \mid \u_i,1 \right\}\n\\
&&+O_p(N^{1/2}h^m)\n\\
&=&N^{-1/2} \sumI E \left[\frac{  r_i  R_j  K_{h}(\X_j-\x_i) e^{-h(Y_j)} \Delta_{1i} }{ d^*(\x_i) D_i E(R_i \mid \x_i) f_{\X}(\x_i)}  \{e^{-g(\u_i)}-e^{-g^*(\u_i)}\} E \left\{\frac{ \Delta_{1i}\Delta_{3i} }{d^*(\X_i)}  \mid \u_i,1 \right\} \right]\n\\
&&+N^{-1/2} \sumJ  E \left[\frac{  R_i  r_j  K_{h}(\x_j-\X_i) e^{-h(y_j)} \Delta_{1i} }{ d^*(\X_i) D_i E(R_i \mid \X_i) f_{\X}(\X_i)}  \{e^{-g(\U_i)}-e^{-g^*(\U_i)}\} E \left\{\frac{ \Delta_{1i}\Delta_{3i} }{d^*(\X_i)}  \mid \U_i,1 \right\} \right]\n\\
&&-N^{1/2}  E \left[\frac{  R_i  R_j  K_{h}(\X_j-\X_i) e^{-h(Y_j)} \Delta_{1i} }{ d^*(\X_i) D_i E(R_i \mid \X_i) f_{\X}(\X_i)}  \{e^{-g(\U_i)}-e^{-g^*(\U_i)}\} E \left\{\frac{ \Delta_{1i}\Delta_{3i} }{d^*(\X_i)}  \mid \U_i,1 \right\} \right] \n\\
&&+ O_p(N^{1/2}h^m)+O_p(N^{-1/2})\n\\
&=& T_{2222BBA}+T_{2222BBB}-T_{2222BBC}+ O_p(N^{1/2}h^m)+O_p(N^{-1/2}).\n
\ee

Consider $T_{2222BBA}$,
\be \label{ustat:t2222bba}
&&T_{2222BBA}\n\\
&=&N^{-1/2} \sumI E \left[\frac{  r_i  R_j  K_{h}(\X_j-\x_i) e^{-h(Y_j)} \Delta_{1i} }{ d^*(\x_i) D_i E(R_i \mid \x_i) f_{\X}(\x_i)}  \{e^{-g(\u_i)}-e^{-g^*(\u_i)}\} E \left\{\frac{ \Delta_{1i}\Delta_{3i} }{d^*(\X_i)}  \mid \u_i,1 \right\} \right]\n\\
&=&N^{-1/2} \sumI \frac{  r_i   \Delta_{1i}^2 }{ d^*(\x_i) D_i }  \{e^{-g(\u_i)}-e^{-g^*(\u_i)}\} E \left\{\frac{ \Delta_{1i}\Delta_{3i} }{d^*(\X_i)}  \mid \u_i,1 \right\}
+ O_p(N^{1/2}h^m)\n\\
&=&T_{2222BC},\n
\ee
where $T_{2222BC}$ is defined in  (\ref{ustat:t2222bc}).

Consider $T_{2222BBB}$,
\be \label{ustat:t2222bbb}
&&T_{2222BBB}\n\\
&=&N^{-1/2} \sumJ  E \left[\frac{  R_i  r_j  K_{h}(\x_j-\X_i) e^{-h(y_j)} \Delta_{1i} }{ d^*(\X_i) D_i E(R_i \mid \X_i) f_{\X}(\X_i)}  \{e^{-g(\U_i)}-e^{-g^*(\U_i)}\} E \left\{\frac{ \Delta_{1i}\Delta_{3i} }{d^*(\X_i)}  \mid \U_i,1 \right\} \right]\n\\
&=&N^{-1/2} \sumJ r_j e^{-h(y_j)} E \left[\frac{ \Delta_{1j} }{ d^*(\x_j) D_j }\{e^{-g(\u_j)}-e^{-g^*(\u_j)}\} E \left\{\frac{ \Delta_{1j}\Delta_{3j} }{d^*(\X_j)}  \mid \u_j,1 \right\}  \mid \x_j, 1\right]+ O_p(N^{1/2}h^m)\n\\
&=&N^{-1/2} \sumI \frac{r_i e^{-h(y_i)} \Delta_{1i} }{ d^*(\x_i)D_i } \{e^{-g(\u_i)}-e^{-g^*(\u_i)}\} E \left\{\frac{ \Delta_{1i}\Delta_{3i} }{d^*(\X_i)}  \mid \u_i,1 \right\}
+ O_p(N^{1/2}h^m)\n\\
&=&N^{-1/2} \sumI \frac{r_i e^{-h(y_i)} \Delta_{1i} \a^*(\u_i)}{ d^*(\x_i) } \{e^{-g(\u_i)}-e^{-g^*(\u_i)}\} + O_p(N^{1/2}h^m)\n\\.
\ee

Consider $T_{2222BBC}$,
\be \label{ustat:t2222bbc}
&&T_{2222BBC}\n\\
&=&N^{1/2}  E \left[\frac{  R_i  R_j  K_{h}(\X_j-\X_i) e^{-h(Y_j)} \Delta_{1i} }{ d^*(\X_i) D_i E(R_i \mid \X_i) f_{\X}(\X_i)}  \{e^{-g(\U_i)}-e^{-g^*(\U_i)}\} E \left\{\frac{ \Delta_{1i}\Delta_{3i} }{d^*(\X_i)}  \mid \U_i,1 \right\} \right] \n\\
&=&N^{1/2} E\left[ \frac{  R_i  \Delta_{1i} \{e^{-g(\U_i)}-e^{-g^*(\U_i)}\}}{ d^*(\X_i) D_i E(R_i \mid \X_i) f_{\X}(\X_i)} E \left\{\frac{ \Delta_{1i}\Delta_{3i} }{d^*(\X_i)}  \mid \U_i,1 \right\}  E \left\{ R_j  K_{h}(\X_j-\x_i) e^{-h(Y_j)}\right\} \right] \n\\
&=&N^{1/2} E\left[\frac{  R_i  \Delta_{1i}^2 \{e^{-g(\U_i)}-e^{-g^*(\U_i)}\}}{ d^*(\X_i) D_i} E \left\{\frac{ \Delta_{1i}\Delta_{3i} }{d^*(\X_i)}  \mid \U_i,1 \right\}  \right]
+ O_p(N^{1/2}h^m).
\ee

Consider $T_{2222C}$ as defined in  (\ref{ustat:t2222c}), applying the U-statistics property to it we get,
\be \label{ustat:t2222c2}
&&T_{2222C}\n\\
&=&N^{-3/2} \sumI  \sumJ \frac{\{1-r_i
  \pi^{-1}(y_i,\u_i;\bb,g^*)\}  r_j K_h({\u_j-\u_i}) \Delta_{1j}^2  \Delta_{3i} }{d^*(\x_i) D_i d^*(\x_j) E(R_i \mid \u_i) f_{\U}(\u_i) }
+O_p(N^{1/2}h^m)\n\\
&=&N^{-1/2} \sumI E \left[\frac{\{1-r_i
  \pi^{-1}(y_i,\u_i;\bb,g^*)\}  R_j K_h({\U_j-\u_i}) \Delta_{1j}^2  \Delta_{3i} }{d^*(\x_i) D_i d^*(\X_j) E(R_i \mid \u_i) f_{\U}(\u_i) } \mid \x_i, r_i, r_i y_i \right]\n\\
&&+N^{-1/2} \sumJ E \left[\frac{\{1-R_i
  \pi^{-1}(Y_i,\U_i;\bb,g^*)\}  r_j K_h({\u_j-\U_i}) \Delta_{1j}^2  \Delta_{3i} }{d^*(\X_i) D_i d^*(\x_j) E(R_i \mid \U_i) f_{\U}(\U_i) } \mid \x_j, r_j, r_j y_j \right] \n\\
&&-N^{1/2} E \left[\frac{\{1-R_i
  \pi^{-1}(Y_i,\U_i;\bb,g^*)\}  R_j K_h({\U_j-\U_i}) \Delta_{1j}^2  \Delta_{3i} }{d^*(\X_i) D_i d^*(\X_j) E(R_i \mid \U_i) f_{\U}(\U_i) } \right]  \n\\
&&+ O_p(N^{1/2}h^m)+O_p(N^{-1/2})\n\\
&=& T_{2222CA}+T_{2222CB}-T_{2222CC}+ O_p(N^{1/2}h^m)+O_p(N^{-1/2}).\n
\ee

Consider $T_{2222CA}$,
\be \label{ustat:t2222ca}
&&T_{2222CA}\n\\
&=&N^{-1/2} \sumI E \left[\frac{\{1-r_i
  \pi^{-1}(y_i,\u_i;\bb,g^*)\}  R_j K_h({\U_j-\u_i}) \Delta_{1j}^2  \Delta_{3i} }{d^*(\x_i) D_i d^*(\X_j) E(R_i \mid \u_i) f_{\U}(\u_i) } \mid \x_i, r_i, r_i y_i \right]\n\\
&=&N^{-1/2} \sumI \frac{\{1-r_i
  \pi^{-1}(y_i,\u_i;\bb,g^*)\} \Delta_{3i} }{d^*(\x_i) D_i} E \left[ \{\Delta_{1i}^2 /d^*(\X_i)\} \mid \u_i,1  \right]
+ O_p(N^{1/2}h^m)\n\\
&=&N^{-1/2} \sumI \frac{\{1-r_i
  \pi^{-1}(y_i,\u_i;\bb,g^*)\} \Delta_{3i} }{d^*(\x_i) }
+ O_p(N^{1/2}h^m)\n\\
&=&T_{2221C},\n
\ee
where $T_{2221C}$ is defined in  (\ref{ustat:t2221c}).

Consider $T_{2222CB}$,
\be \label{ustat:t2222cb}
&&T_{2222CB}\n\\
&=&N^{-1/2} \sumJ E \left[\frac{\{1-R_i
  \pi^{-1}(Y_i,\U_i;\bb,g^*)\}  r_j K_h({\u_j-\U_i}) \Delta_{1j}^2  \Delta_{3i} }{d^*(\X_i) D_i d^*(\x_j) E(R_i \mid \U_i) f_{\U}(\U_i) } \mid \x_j, r_j, r_j y_j \right] \n\\
&=&N^{-1/2} \sumJ \frac{  r_j \Delta_{1j}^2 }{ d^*(\x_j)} E \left[\frac{\{1-R_i
  \pi^{-1}(Y_i,\U_i;\bb,g^*)\}  K_h({\u_j-\U_i})  \Delta_{3i} }{d^*(\X_i) D_i E(R_i \mid \U_i) f_{\U}(\U_i) } \right] \n\\
&=&N^{-1/2} \sumJ \frac{  r_j \Delta_{1j}^2 }{ d^*(\x_j)} E \left[\frac{ \{w^{-1}(\X_j)
-w^{*-1}(\X_j)\} \Delta_{3j} }{d^*(\X_j) D_j } \mid \u_j, 1 \right]
+ O_p(N^{1/2}h^m)\n\\
&=&N^{-1/2} \sumJ \frac{  r_j \Delta_{1j}^2\{e^{-g(\u_j)}-e^{-g^*(\u_j)}\}  }{ d^*(\x_j) D_j} E \left\{\frac{ \Delta_{1j}\Delta_{3j} }{d^*(\X_j)}  \mid \u_j,1 \right\}
+ O_p(N^{1/2}h^m)\n\\
&=&T_{2222BC},\n
\ee
where $T_{2222BC}$ is defined in  (\ref{ustat:t2222bc}).

Consider $T_{2222CC}$,
\be \label{ustat:t2222cc}
&&T_{2222CC}\n\\
&=&N^{1/2} E \left[\frac{\{1-R_i
  \pi^{-1}(Y_i,\U_i;\bb,g^*)\}  R_j K_h({\U_j-\U_i}) \Delta_{1j}^2  \Delta_{3i} }{d^*(\X_i) D_i d^*(\X_j) E(R_i \mid \U_i) f_{\U}(\U_i) } \right]  \n\\
&=&N^{1/2} E \left[\frac{\{1-R_i
  \pi^{-1}(Y_i,\U_i;\bb,g^*)\}  \Delta_{3i} }{d^*(\X_i) D_i  E(R_i \mid \U_i) f_{\U}(\U_i) } E\left\{\frac{R_j K_h({\U_j-\u_i}) \Delta_{1j}^2   }{d^*(\X_j)  } \right\}\right]  \n\\
&=&N^{1/2} E \left[\frac{\{1-R_i
  \pi^{-1}(Y_i,\U_i;\bb,g^*)\}  \Delta_{3i} }{d^*(\X_i)}  \right]
+ O_p(N^{1/2}h^m)\n\\
&=&T_{2221AC},\n
\ee
where $T_{2221AC}$ is defined in  (\ref{ustat:t2221ac}).

Consider $T_{2223BB}$ as defined in  (\ref{ustat:t2223bb}), applying the U-statistics property to it we get,
\be \label{ustat:t2223bb2}
&&T_{2223BB}\n\\
&=&N^{-3/2}  \sumI \sumJ \frac{  r_i r_j K_{h}(\x_j-\x_i) \Delta_{1i}^2 \{e^{-g(\u_i)}-e^{-g^*(\u_i)}\}}{ d^*(\x_i) D_i  E(R_i \mid \x_i)  f_{\X}(\x_i)}  E \left\{\frac{ \Delta_{1i}\Delta_{3i} }{d^*(\X_i)}  \mid \u_i,1 \right\}
+O_p(N^{1/2}h^m)\n\\
&=&N^{-1/2} \sumI E \left[\frac{  r_i R_j K_{h}(\X_j-\x_i) \Delta_{1i}^2\{e^{-g(\u_i)}-e^{-g^*(\u_i)}\} }{ d^*(\x_i) D_i  E(R_i \mid \x_i)  f_{\X}(\x_i)} E \left\{\frac{ \Delta_{1i}\Delta_{3i} }{d^*(\X_i)}  \mid \u_i,1 \right\} \mid \x_i, r_i, r_i y_i \right]\n\\
&&+N^{-1/2} \sumJ E \left[\frac{  R_i r_j K_{h}(\x_j-\X_i) \Delta_{1i}^2 \{e^{-g(\U_i)}-e^{-g^*(\U_i)}\} }{ d^*(\X_i) D_i E(R_i \mid \X_i)  f_{\X}(\X_i)} E \left\{\frac{ \Delta_{1i}\Delta_{3i} }{d^*(\X_i)}  \mid \U_i,1 \right\} \mid \x_j, r_j, r_j y_j \right] \n\\
&&-N^{1/2} E \left[\frac{  R_i R_j K_{h}(\X_j-\X_i) \Delta_{1i}^2\{e^{-g(\U_i)}-e^{-g^*(\U_i)}\} }{ d^*(\X_i) D_i E(R_i \mid \X_i)  f_{\X}(\X_i)} E \left\{\frac{ \Delta_{1i}\Delta_{3i} }{d^*(\X_i)}  \mid \U_i,1 \right\}  \right]  \n\\
&&+ O_p(N^{1/2}h^m)+O_p(N^{-1/2})\n\\
&=& T_{2223BBA}+T_{2223BBB}-T_{2223BBC}+ O_p(N^{1/2}h^m)+O_p(N^{-1/2}).\n
\ee

Consider $T_{2223BBA}$,
\be \label{ustat:t2223bba}
&&T_{2223BBA}\n\\
&=&N^{-1/2} \sumI E \left[\frac{  r_i R_j K_{h}(\X_j-\x_i) \Delta_{1i}^2\{e^{-g(\u_i)}-e^{-g^*(\u_i)}\} }{ d^*(\x_i) D_i  E(R_i \mid \x_i)  f_{\X}(\x_i)} E \left\{\frac{ \Delta_{1i}\Delta_{3i} }{d^*(\X_i)}  \mid \u_i,1 \right\} \mid \x_i, r_i, r_i y_i \right]\n\\
&=&N^{-1/2} \sumI \frac{  r_i  \Delta_{1i}^2 \{e^{-g(\u_i)}-e^{-g^*(\u_i)}\}}{ d^*(\x_i)D_i} E \left\{\frac{ \Delta_{1i}\Delta_{3i} }{d^*(\X_i)}  \mid \u_i,1 \right\}
+ O_p(N^{1/2}h^m)\n\\
&=&T_{2222BC}\n
\ee
where $T_{2222BC}$ is defined in  (\ref{ustat:t2222bc}).

Consider $T_{2223BBB}$,
\be \label{ustat:t2223bbb}
&&T_{2223BBB}\n\\
&=&N^{-1/2} \sumJ E \left[\frac{  R_i r_j K_{h}(\x_j-\X_i) \Delta_{1i}^2 \{e^{-g(\U_i)}-e^{-g^*(\U_i)}\} }{ d^*(\X_i) D_i E(R_i \mid \X_i)  f_{\X}(\X_i)} E \left\{\frac{ \Delta_{1i}\Delta_{3i} }{d^*(\X_i)}  \mid \U_i,1 \right\} \mid \x_j, r_j, r_j y_j \right] \n\\
&=&N^{-1/2} \sumJ r_j  E \left[\frac{\Delta_{1j}^2 \{e^{-g(\u_j)}-e^{-g^*(\u_j)}\}}{ d^*(\x_j) D_j } E \left\{\frac{ \Delta_{1j}\Delta_{3j} }{d^*(\X_j)}  \mid \u_j,1 \right\}\mid \x_j, 1 \right]
+ O_p(N^{1/2}h^m)\n\\
&=&N^{-1/2} \sumJ  \frac{ r_j \Delta_{1j}^2 \{e^{-g(\u_j)}-e^{-g^*(\u_j)}\}}{ d^*(\x_j) D_j } E \left\{\frac{ \Delta_{1j}\Delta_{3j} }{d^*(\X_j)}  \mid \u_j,1 \right\}
+ O_p(N^{1/2}h^m)\n\\
&=&T_{2222BC},\n
\ee
where $T_{2222BC}$ is defined in  (\ref{ustat:t2222bc}).

Consider $T_{2223BBC}$,
\be \label{ustat:t2223bbc}
&&T_{2223BBC}\n\\
&=&N^{1/2} E \left[\frac{  R_i R_j K_{h}(\X_j-\X_i) \Delta_{1i}^2\{e^{-g(\U_i)}-e^{-g^*(\U_i)}\} }{ d^*(\X_i) D_i E(R_i \mid \X_i)  f_{\X}(\X_i)} E \left\{\frac{ \Delta_{1i}\Delta_{3i} }{d^*(\X_i)}  \mid \U_i,1 \right\}  \right]  \n\\
&=&N^{1/2} E \left[\frac{  R_i \Delta_{1i}^2\{e^{-g(\U_i)}-e^{-g^*(\U_i)}\} }{ d^*(\X_i) D_i E(R_i \mid \X_i)  f_{\X}(\X_i)} E \left\{\frac{ \Delta_{1i}\Delta_{3i} }{d^*(\X_i)}  \mid \U_i,1 \right\} E\{R_j K_{h}(\X_j-\x_i)\} \right]  \n\\
&=&N^{1/2} E \left[\frac{  R_i \Delta_{1i}^2\{e^{-g(\U_i)}-e^{-g^*(\U_i)}\} }{ d^*(\X_i) D_i } E \left\{\frac{ \Delta_{1i}\Delta_{3i} }{d^*(\X_i)}  \mid \U_i,1 \right\}  \right]
+ O_p(N^{1/2}h^m)\n\\
&=&T_{2222BBC},\n
\ee
where $T_{2222BBC}$ is defined in  (\ref{ustat:t2222bbc}).

Consider $T_{321}$ as defined in  (\ref{eq:t321}), applying the U-statistics property to it we get,
\be \label{ustat:t321}
&&T_{321}\n\\
&=&N^{-3/2} \sumI \sumJ \frac{\{1-r_i
  \pi^{-1}(Y_i,\u_i;\bb,g^*)\}C_i r_j  K_{h} (\u_j-\u_i) \Delta_{1j}^2 }{d^*(\x_i) D_i^2
  d^* (\x_j) E(R_i\mid\u_i) f_{\U}(\u_i)} \n\\
&=&N^{-1/2} \sumI E\left[ \frac{\{1-r_i
  \pi^{-1}(Y_i,\u_i;\bb,g^*)\}C_i R_j  K_{h} (\U_j-\u_i) \Delta_{1j}^2 }{d^*(\x_i) D_i^2
  d^* (\X_j) E(R_i\mid\u_i) f_{\U}(\u_i)} \mid \x_i, r_i, r_i y_i\right] \n\\
&&+N^{-1/2} \sumJ E\left[ \frac{\{1-R_i
  \pi^{-1}(Y_i,\U_i;\bb,g^*)\}C_i r_j  K_{h} (\u_j-\U_i) \Delta_{1j}^2 }{d^*(\X_i) D_i^2
  d^* (\x_j) E(R_i\mid\U_i) f_{\U}(\U_i)} \mid \x_j, r_j, r_j y_j\right] \n\\
&&-N^{1/2} E\left[ \frac{\{1-R_i
  \pi^{-1}(Y_i,\U_i;\bb,g^*)\}C_i R_j  K_{h} (\U_j-\U_i) \Delta_{1j}^2 }{d^*(\X_i) D_i^2
  d^* (\X_j) E(R_i\mid\U_i) f_{\U}(\U_i)}\right] + O_p(N^{1/2}h^m)+O_p(N^{-1/2})\n\\
&=& T_{321A}+T_{321B}-T_{321C}+ O_p(N^{1/2}h^m)+O_p(N^{-1/2}).\n
\ee

Consider $T_{321A}$,
\be \label{ustat:t321a}
&&T_{321A}\n\\
&=&N^{-1/2} \sumI E\left[ \frac{\{1-r_i
  \pi^{-1}(Y_i,\u_i;\bb,g^*)\}C_i R_j  K_{h} (\U_j-\u_i) \Delta_{1j}^2 }{d^*(\x_i) D_i^2
  d^* (\X_j) E(R_i\mid\u_i) f_{\U}(\u_i)} \mid \x_i, r_i, r_i y_i\right] \n\\
&=&N^{-1/2} \sumI \frac{\{1-r_i
  \pi^{-1}(Y_i,\u_i;\bb,g^*)\}C_i  }{d^*(\x_i) D_i^2 }  E\left[\{\Delta_{1i}^2/ d^* (\X_i)\} \mid \u_i, 1\right]
+ O_p(N^{1/2}h^m)\n\\
&=&N^{-1/2} \sumI \frac{\{1-r_i
  \pi^{-1}(Y_i,\u_i;\bb,g^*)\}C_i  }{d^*(\x_i) D_i}
+ O_p(N^{1/2}h^m)\n\\
&=&N^{-1/2} \sumI \frac{A_i  }{d^*(\x_i)}
+ O_p(N^{1/2}h^m).\n
\ee

Consider $T_{321B}$,
\be \label{ustat:t321b}
&&T_{321B}\n\\
&=&N^{-1/2} \sumJ E\left[ \frac{\{1-R_i
  \pi^{-1}(Y_i,\U_i;\bb,g^*)\}C_i r_j  K_{h} (\u_j-\U_i) \Delta_{1j}^2 }{d^*(\X_i) D_i^2
  d^* (\x_j) E(R_i\mid\U_i) f_{\U}(\U_i)} \mid \x_j, r_j, r_j y_j\right] \n\\
&=&N^{-1/2} \sumJ\frac{ r_j \Delta_{1j}^2}{d^* (\x_j)} E\left[ \frac{\{1-R_i
  \pi^{-1}(Y_i,\U_i;\bb,g^*)\}C_i K_{h} (\u_j-\U_i)  }{d^*(\X_i) D_i^2
   E(R_i\mid\U_i) f_{\U}(\U_i)} \right] \n\\
&=&N^{-1/2} \sumJ\frac{ r_j \Delta_{1j}^2}{d^* (\x_j)} E\left[ \frac{\{w^{-1}(\X_j)
-w^{*-1}(\X_j)\} C_j   }{d^*(\X_j) D_j^2} \mid \u_j, 1\right]
+ O_p(N^{1/2}h^m)\n\\
&=& O_p(N^{1/2}h^m).\n
\ee

Consider $T_{321C}$,
\be \label{ustat:t321c}
&&T_{321C}\n\\
&=&N^{1/2} E\left[ \frac{\{1-R_i
  \pi^{-1}(Y_i,\U_i;\bb,g^*)\}C_i R_j  K_{h} (\U_j-\U_i) \Delta_{1j}^2 }{d^*(\X_i) D_i^2
  d^* (\X_j) E(R_i\mid\U_i) f_{\U}(\U_i)}\right] \n\\
&=&N^{1/2} E\left[ \frac{\{1-R_i
  \pi^{-1}(Y_i,\U_i;\bb,g^*)\}C_i }{d^*(\X_i) D_i^2
   E(R_i\mid\U_i) f_{\U}(\U_i)} E\left\{ \frac{R_j  K_{h} (\U_j-\u_i) \Delta_{1j}^2}{d^* (\X_j)}\right\}\right] \n\\
&=&N^{1/2} E\left( \frac{\{1-R_i
  \pi^{-1}(Y_i,\U_i;\bb,g^*)\}C_i }{d^*(\X_i) D_i^2
  } E\left[\{ \Delta_{1i}^2/d^* (\X_i)\} \mid  \U_i, 1\right]\right) + O_p(N^{1/2}h^m)\n\\
&=&N^{1/2} E\left\{\frac{A_i }{d^*(\X_i)} \right\} + O_p(N^{1/2}h^m)\n\\
&=&N^{1/2} E\{\S^*\eff(\X_i,R_i,R_i Y_i)\} + O_p(N^{1/2}h^m)\n\\
&=&N^{1/2} E[E\{\S^*\eff(\X_i,R_i,R_i Y_i)\mid \u_i\}] + O_p(N^{1/2}h^m)\n\\
&=&O_p(N^{1/2}h^m),\n
\ee
where the last equality uses Lemma
(1).

Consider $T_{332CB}$ as defined in  (\ref{ustat:t332cb}), applying the U-statistics property to it we get,
\be \label{ustat:t332cb2}
&&T_{332CB}\n\\
&=&N^{-3/2} \sumI  \sumJ \frac{\{1-r_i  \pi^{-1}(y_i,\u_i;\bb,g^*)\}C_i r_j K_{h} (\u_j-\u_i)  }{d^*(\x_i) D_i E(R_i\mid\u_i) f_{\U}(\u_i) } + O_p(N^{1/2}h^m)\n\\
&=&N^{-1/2} \sumI E\left[\frac{\{1-r_i  \pi^{-1}(y_i,\u_i;\bb,g^*)\}C_i R_j K_{h} (\U_j-\u_i)  }{d^*(\x_i) D_i E(R_i\mid\u_i) f_{\U}(\u_i) } \mid \x_i, r_i, r_i y_i\right] \n\\
&&+N^{-1/2} \sumJ E\left[\frac{\{1-R_i  \pi^{-1}(Y_i,\U_i;\bb,g^*)\}C_i r_j K_{h} (\u_j-\U_i)  }{d^*(\X_i) D_i E(R_i\mid\U_i) f_{\U}(\U_i) }  \mid \x_j, r_j, r_j y_j\right] \n\\
&&-N^{1/2} E\left[\frac{\{1-R_i  \pi^{-1}(Y_i,\U_i;\bb,g^*)\}C_i R_j K_{h} (\U_j-\U_i)  }{d^*(\X_i) D_i E(R_i\mid\U_i) f_{\U}(\U_i) } \right]+ O_p(N^{1/2}h^m)+O_p(N^{-1/2})\n\\
&=& T_{332CBA}+T_{332CBB}-T_{332CBC}+ O_p(N^{1/2}h^m)+O_p(N^{-1/2}).\n
\ee

Consider $T_{332CBA}$,
\be \label{ustat:t332cba}
&&T_{332CBA}\n\\
&=&N^{-1/2} \sumI E\left[\frac{\{1-r_i  \pi^{-1}(y_i,\u_i;\bb,g^*)\}C_i R_j K_{h} (\U_j-\u_i)  }{d^*(\x_i) D_i E(R_i\mid\u_i) f_{\U}(\u_i) } \mid \x_i, r_i, r_i y_i\right] \n\\
&=&N^{-1/2} \sumI \frac{\{1-r_i  \pi^{-1}(y_i,\u_i;\bb,g^*)\}C_i  }{d^*(\x_i) D_i}+ O_p(N^{1/2}h^m)\n\\
&=&N^{-1/2} \sumI \frac{A_i  }{d^*(\x_i)} + O_p(N^{1/2}h^m).\n
\ee

Consider $T_{332CBB}$,
\be \label{ustat:t332cbb}
&&T_{332CBB}\n\\
&=&N^{-1/2} \sumJ E\left[\frac{\{1-R_i  \pi^{-1}(Y_i,\U_i;\bb,g^*)\}C_i r_j K_{h} (\u_j-\U_i)  }{d^*(\X_i) D_i E(R_i\mid\U_i) f_{\U}(\U_i) }  \mid \x_j, r_j, r_j y_j\right] \n\\
&=&N^{-1/2} \sumJ r_j E\left[\frac{\{1-R_i  \pi^{-1}(Y_i,\U_i;\bb,g^*)\}C_i K_{h} (\u_j-\U_i)  }{d^*(\X_i) D_i E(R_i\mid\U_i) f_{\U}(\U_i) } \right] \n\\
&=&N^{-1/2} \sumJ r_j E\left[\frac{\{w^{-1}(\X_j)
-w^{*-1}(\X_j)\} C_j  }{d^*(\X_j) D_j  } \mid \u_j, 1\right] + O_p(N^{1/2}h^m)\n\\
&=&O_p(N^{1/2}h^m).\n
\ee

Consider $T_{332CBC}$,
\be \label{ustat:t332cbc}
&&T_{332CBC}\n\\
&=&N^{1/2} E\left[\frac{\{1-R_i  \pi^{-1}(Y_i,\U_i;\bb,g^*)\}C_i R_j K_{h} (\U_j-\U_i)  }{d^*(\X_i) D_i E(R_i\mid\U_i) f_{\U}(\U_i) } \right] \n\\
&=&N^{1/2} E\left[ \frac{\{1-R_i  \pi^{-1}(Y_i,\U_i;\bb,g^*)\}C_i  }{d^*(\X_i) D_i E(R_i\mid\U_i) f_{\U}(\U_i) } E\{R_j K_{h} (\U_j-\u_i) \}\right]  \n\\
&=&N^{1/2} E\left[\frac{\{1-R_i  \pi^{-1}(Y_i,\U_i;\bb,g^*)\}C_i  }{d^*(\X_i) D_i} \right]   + O_p(N^{1/2}h^m)\n\\
&=&N^{1/2} E\left\{\frac{A_i  }{d^*(\X_i)} \right\} + O_p(N^{1/2}h^m)\n\\
&=&O_p(N^{1/2}h^m),\n
\ee
where the last equality uses Lemma
(1).

Consider $T_{11}$ as defined in  (\ref{eq:t11}), applying the U-statistics property to it we get,
\be \label{ustat:t11}
&&T_{11}\n\\
&=& N^{-3/2} \sumI \sumJ \frac{A_i }{d^{*2}(\x_i)}  \frac{r_j
  K_{h}(\x_j-\x_i)}{E(R_i\mid\x_i)
  f_{\X}(\x_i)}\{e^{-h(y_j)}+e^{-g^*(\u_i)}e^{-2h(y_j)}\}\n\\
&=& N^{-3/2} \sumI \sumJ \frac{\{1-r_i  \pi^{-1}(y_i,\u_i;\bb,g^*)\}C_i r_j
  K_{h}(\x_j-\x_i)\{e^{-h(y_j)}+e^{-g^*(\u_i)}e^{-2h(y_j)}\} }{d^{*2}(\x_i) D_i E(R_i\mid\x_i) f_{\X}(\x_i)} \n\\
&=&N^{-1/2} \sumI E\left[  \frac{\{1-r_i  \pi^{-1}(y_i,\u_i;\bb,g^*)\}C_i R_j
  K_{h}(\X_j-\x_i)\{e^{-h(Y_j)}+e^{-g^*(\u_i)}e^{-2h(Y_j)}\} }{d^{*2}(\x_i) D_i E(R_i\mid\x_i) f_{\X}(\x_i)} \right]  \n\\
&&+N^{-1/2} \sumJ E\left[ \frac{\{1-R_i  \pi^{-1}(Y_i,\U_i;\bb,g^*)\}C_i r_j
  K_{h}(\x_j-\X_i)\{e^{-h(y_j)}+e^{-g^*(\U_i)}e^{-2h(y_j)}\} }{d^{*2}(\X_i) D_i E(R_i\mid\X_i) f_{\X}(\X_i)} \right] \n\\
&&-N^{1/2} E\left[\frac{\{1-R_i  \pi^{-1}(Y_i,\U_i;\bb,g^*)\}C_i R_j
  K_{h}(\X_j-\X_i)\{e^{-h(Y_j)}+e^{-g^*(\U_i)}e^{-2h(Y_j)}\} }{d^{*2}(\X_i) D_i E(R_i\mid\X_i) f_{\X}(\X_i)}\right] \n\\
&&+ O_p(N^{1/2}h^m)+O_p(N^{-1/2})\n\\
&=& T_{11A}+T_{11B}-T_{11C}+ O_p(N^{1/2}h^m)+O_p(N^{-1/2}).\n
\ee

Consider $T_{11A}$,
\be \label{ustat:t11a}
&&T_{11A}\n\\
&=&N^{-1/2} \sumI E\left[  \frac{\{1-r_i  \pi^{-1}(y_i,\u_i;\bb,g^*)\}C_i R_j
  K_{h}(\X_j-\x_i)\{e^{-h(Y_j)}+e^{-g^*(\u_i)}e^{-2h(Y_j)}\} }{d^{*2}(\x_i) D_i E(R_i\mid\x_i) f_{\X}(\x_i)} \right]  \n\\
&=&N^{-1/2} \sumI \frac{\{1-r_i  \pi^{-1}(y_i,\u_i;\bb,g^*)\}C_i  }{d^*(\x_i) D_i} + O_p(N^{1/2}h^m)\n\\
&=&N^{-1/2} \sumI \frac{A_i  }{d^*(\x_i)} + O_p(N^{1/2}h^m).\n
\ee

Consider $T_{11B}$,
\be \label{ustat:t11b}
&&T_{11B}\n\\
&=&N^{-1/2} \sumJ E\left[ \frac{\{1-R_i  \pi^{-1}(Y_i,\U_i;\bb,g^*)\}C_i r_j
  K_{h}(\x_j-\X_i)\{e^{-h(y_j)}+e^{-g^*(\U_i)}e^{-2h(y_j)}\} }{d^{*2}(\X_i) D_i E(R_i\mid\X_i) f_{\X}(\X_i)} \right] \n\\
&=&N^{-1/2} \sumJ r_j E\left[ \frac{\{1-R_i  \pi^{-1}(Y_i,\U_i;\bb,g^*)\}C_i
  K_{h}(\x_j-\X_i)\{e^{-h(y_j)}+e^{-g^*(\U_i)}e^{-2h(y_j)}\} }{d^{*2}(\X_i) D_i E(R_i\mid\X_i) f_{\X}(\X_i)} \right] \n\\
&=&N^{-1/2} \sumJ \frac{r_j  \{w^{-1}(\x_j)
-w^{*-1}(\x_j)\}C_j
  \{e^{-h(y_j)}+e^{-g^*(\u_j)}e^{-2h(y_j)}\} }{d^{*2}(\x_j) D_j }
+ O_p(N^{1/2}h^m)\n\\
&=&N^{-1/2} \sumI  \frac{ r_i  \{w^{-1}(\x_i)
-w^{*-1}(\x_i)\}C_i
  \{e^{-h(y_i)}+e^{-g^*(\u_i)}e^{-2h(y_i)}\} }{d^{*2}(\x_i) D_i }
+ O_p(N^{1/2}h^m).
\ee

Consider $T_{11C}$,
\be \label{ustat:t11c}
&&T_{11C}\n\\
&=&N^{1/2} E\left[\frac{\{1-R_i  \pi^{-1}(Y_i,\U_i;\bb,g^*)\}C_i R_j
  K_{h}(\X_j-\X_i)\{e^{-h(Y_j)}+e^{-g^*(\U_i)}e^{-2h(Y_j)}\} }{d^{*2}(\X_i) D_i E(R_i\mid\X_i) f_{\X}(\X_i)}\right] \n\\
&=&N^{1/2} E\left(\frac{\{1-R_i  \pi^{-1}(Y_i,\U_i;\bb,g^*)\}C_i E[R_j
  K_{h}(\X_j-\x_i)\{e^{-h(Y_j)}+e^{-g^*(\u_i)}e^{-2h(Y_j)}\}] }{d^{*2}(\X_i) D_i E(R_i\mid\X_i) f_{\X}(\X_i)}\right) \n\\
&=&N^{1/2} E\left[\frac{\{1-R_i  \pi^{-1}(Y_i,\U_i;\bb,g^*)\}C_i  }{d^*(\X_i) D_i}\right]+ O_p(N^{1/2}h^m)\n\\
&=&N^{1/2} E\left\{\frac{A_i}{d^*(\X_i)}\right\}+ O_p(N^{1/2}h^m)\n\\
&=&O_p(N^{1/2}h^m),\n
\ee
where the last equality uses Lemma
(1).

Consider $T_{12B}$ as defined in  (\ref{ustat:t12b}), applying the U-statistics property to it we get,
\be \label{ustat:t12b2}
&&T_{12B}\n\\
&=&N^{-3/2} \sumI \sumJ \frac{A_i r_j K_{h}
  (\x_j-\x_i) }{d^*(\x_i) E(R_i\mid\x_i) f_{\X}(\x_i)}
+ O_p(N^{1/2}h^m)\n\\
&=&N^{-3/2} \sumI \sumJ \frac{\{1-r_i  \pi^{-1}(y_i,\u_i;\bb,g^*)\}C_i r_j K_{h}
  (\x_j-\x_i) }{d^*(\x_i) D_i E(R_i\mid\x_i) f_{\X}(\x_i)}
+ O_p(N^{1/2}h^m)\n\\
&=&N^{-1/2} \sumI E\left[  \frac{\{1-r_i  \pi^{-1}(y_i,\u_i;\bb,g^*)\}C_i R_j K_{h}
  (\X_j-\x_i) }{d^*(\x_i) D_i E(R_i\mid\x_i) f_{\X}(\x_i)} \mid \x_i, r_i, r_i y_i\right]  \n\\
&&+N^{-1/2} \sumJ E\left[ \frac{\{1-R_i  \pi^{-1}(Y_i,\U_i;\bb,g^*)\}C_i r_j K_{h}
  (\x_j-\X_i) }{d^*(\X_i) D_i E(R_i\mid\X_i) f_{\X}(\X_i)} \mid \x_j, r_j, r_j y_j\right] \n\\
&&-N^{1/2} E\left[\frac{\{1-R_i  \pi^{-1}(Y_i,\U_i;\bb,g^*)\}C_i R_j K_{h}
  (\X_j-\X_i) }{d^*(\X_i) D_i E(R_i\mid\X_i) f_{\X}(\X_i)}\right]+ O_p(N^{1/2}h^m)+O_p(N^{-1/2})\n\\
&=& T_{12BA}+T_{12BB}-T_{12BC}+ O_p(N^{1/2}h^m)+O_p(N^{-1/2}).\n
\ee
Consider $T_{12BA}$,
\be \label{ustat:t12ba}
&&T_{12BA}\n\\
&=&N^{-1/2} \sumI E\left[  \frac{\{1-r_i  \pi^{-1}(y_i,\u_i;\bb,g^*)\}C_i R_j K_{h}
  (\X_j-\x_i) }{d^*(\x_i) D_i E(R_i\mid\x_i) f_{\X}(\x_i)} \mid \x_i, r_i, r_i y_i\right]  \n\\
&=&N^{-1/2} \sumI \frac{\{1-r_i  \pi^{-1}(y_i,\u_i;\bb,g^*)\}C_i  }{d^*(\x_i) D_i} + O_p(N^{1/2}h^m)\n\\
&=&N^{-1/2} \sumI \frac{A_i  }{d^*(\x_i) } + O_p(N^{1/2}h^m).\n
\ee
Consider $T_{12BB}$,
\be \label{ustat:t12bb}
&&T_{12BB}\n\\
&=&N^{-1/2} \sumJ E\left[ \frac{\{1-R_i  \pi^{-1}(Y_i,\U_i;\bb,g^*)\}C_i r_j K_{h}
  (\x_j-\X_i) }{d^*(\X_i) D_i E(R_i\mid\X_i) f_{\X}(\X_i)} \mid \x_j, r_j, r_j y_j\right] \n\\
&=&N^{-1/2} \sumJ r_j E\left[ \frac{\{1-R_i  \pi^{-1}(Y_i,\U_i;\bb,g^*)\}C_i  K_{h}
  (\x_j-\X_i) }{d^*(\X_i) D_i E(R_i\mid\X_i) f_{\X}(\X_i)} \right]\n\\
&=&N^{-1/2} \sumJ \frac{ r_j \{w^{-1}(\x_j)
-w^{*-1}(\x_j)\}C_j   }{d^*(\x_j) D_j} + O_p(N^{1/2}h^m)\n\\
&=&N^{-1/2} \sumI \frac{ r_i \{w^{-1}(\x_i)
-w^{*-1}(\x_i)\}C_i   }{d^*(\x_i) D_i} + O_p(N^{1/2}h^m).
\ee
Consider $T_{12BC}$,
\be \label{ustat:t12bc}
&&T_{12BC}\n\\
&=&N^{1/2} E\left[\frac{\{1-R_i  \pi^{-1}(Y_i,\U_i;\bb,g^*)\}C_i R_j K_{h}
  (\X_j-\X_i) }{d^*(\X_i) D_i E(R_i\mid\X_i) f_{\X}(\X_i)}\right] \n\\
&=&N^{1/2} E\left[\frac{\{1-R_i  \pi^{-1}(Y_i,\U_i;\bb,g^*)\}C_i E\{R_j K_{h}
  (\X_j-\x_i)\} }{d^*(\X_i) D_i E(R_i\mid\X_i) f_{\X}(\X_i)}\right] \n\\
&=&N^{1/2} E\left[\frac{\{1-R_i  \pi^{-1}(Y_i,\U_i;\bb,g^*)\}C_i }{d^*(\X_i) D_i}\right] + O_p(N^{1/2}h^m)\n\\
&=&N^{1/2} E\left\{\frac{A_i}{d^*(\X_i)}\right\}+ O_p(N^{1/2}h^m)\n\\
&=&O_p(N^{1/2}h^m),\n
\ee
where the last equality uses Lemma
(1).

Consider $T_{2211CC}$ as defined in  (\ref{ustat:t2211cc}) and $T_{2221C}$ as defined in  (\ref{ustat:t2221c}),
\be \label{ustat:t2211cc_t2221c}
&&T_{2211CC}-T_{2221C}\n\\
&=&N^{-1/2} \sumI \frac{\{1-r_i
  \pi^{-1}(y_i,\u_i;\bb,g^*)   \} \Delta_{1i}}{d^*(\x_i) D_i}   E\left[\{ \Delta_{1i} \Delta_{3i}/d^* (\X_i)\} \mid \u_i,1  \right]\n\\
&&-N^{-1/2} \sumI  \frac{\{1-r_i
  \pi^{-1}(Y_i,\u_i;\bb,g^*)\} \Delta_{3i} }{d^*(\x_i) }+ O_p(N^{1/2}h^m)\n\\
&=&N^{-1/2} \sumI \frac{\{1-r_i
  \pi^{-1}(y_i,\u_i;\bb,g^*)   \} }{d^*(\x_i)}  \left( \frac{E\left[\{ \Delta_{1i} \Delta_{3i}/d^* (\X_i)\} \mid \u_i,1  \right]\Delta_{1i}}{D_i}- \Delta_{3i} \right) + O_p(N^{1/2}h^m)\n\\
&=&N^{-1/2} \sumI \frac{\{1-r_i
  \pi^{-1}(y_i,\u_i;\bb,g^*)   \} }{d^*(\x_i)}  \left\{ \a^*(\u_i)\Delta_{1i}- \Delta_{3i} \right\} + O_p(N^{1/2}h^m)\n\\
&=&N^{-1/2} \sumI \frac{A_i   }{d^*(\x_i)} + O_p(N^{1/2}h^m).\n
\ee

$T_{2211CAC}$ is defined in  (\ref{ustat:t2211cac}) and $T_{2221AC}$ is defined in  (\ref{ustat:t2221ac}),
\be \label{ustat:t2211cac_t2221ac}
&&T_{2211CAC}-T_{2221AC}\n\\
&=&N^{1/2} E \left(\frac{\{1-R_i
  \pi^{-1}(Y_i,\U_i;\bb,g^*)\} \Delta_{1i}  }{d^*(\X_i) D_i }    E\left[\{ \Delta_{1i} \Delta_{3i}/d^* (\X_i)\} \mid \U_i,1  \right] \right)\n\\
&&-N^{1/2} E\left( \frac{\{1-R_i
  \pi^{-1}(Y_i,\U_i;\bb,g^*)\} \Delta_{3i} }{d^*(\X_i)
 }  \right)+ O_p(N^{1/2}h^m)\n\\
&=&N^{1/2} E \left\{\frac{\{1-R_i
  \pi^{-1}(Y_i,\U_i;\bb,g^*)\} \left\{ \a^*(\u_i)\Delta_{1i}- \Delta_{3i} \right\}}{d^*(\X_i) } \right\}+ O_p(N^{1/2}h^m)\n\\
&=&N^{1/2} E \left\{\frac{A_i }{d^*(\X_i) } \right\}+ O_p(N^{1/2}h^m).\n\\
&=&O_p(N^{1/2}h^m),\n
\ee
where the last equality uses Lemma
(1).

Putting all the terms together (cancelling and collecting similar terms) and rewriting (\ref{eq:final4}) we get the following,
\be \label{eq:final5}
&&\frac{1}{\sqrt{N}}\sumI\wh\S^*\eff(\x_i,r_i,r_i y_i,\bb)\n\\
&=& T_{2211ABB}+T_{2211BBB} +T_{2211CA}-T_{2211CC}+T_{2211AC}-2T_{2212ABBBB}
-T_{2221A}\n\\
&&+T_{2221C}-2T_{2222BB}-T_{2222C}+2T_{2223BB}
-T_{321}+T_{332CB}-T_{11}+T_{12B}\n\\
&&+O_p(N^{1/2}h^{2m}+ N^{-1/2}h^{-p}+N^{-1/2}+N^{1/2}h^m)\n\\
&=& \frac{1}{\sqrt{N}}\sumI \frac{A_i}{d^*(\x_i)}
-T_{2211ABC}+T_{2211ABBB}+T_{2211BBBB}+T_{2211CAB}-T_{2221AB}+T_{2222BC}-2T_{2222BBB}\n\\
&&-T_{11B}+T_{12BB}+O_p(N^{1/2}h^{2m}+ N^{-1/2}h^{-p}+N^{-1/2}+N^{1/2}h^m)
\ee
where
$T_{2211ABC}$ is defined in  (\ref{ustat:t2211abc}),
$T_{2211ABBB}$ is defined in  (\ref{ustat:t2211abbb}),
$T_{2211BBBB}$ is defined in  (\ref{ustat:t2211bbbb}),
$T_{2211CAB}$ is defined in  (\ref{ustat:t2211cab}),
$T_{2221AB}$ is defined in  (\ref{ustat:t2221ab}),
$T_{2222BC}$ is defined in  (\ref{ustat:t2222bc}),
$T_{2222BBB}$ is defined in  (\ref{ustat:t2222bbb}),
$T_{11B}$ is defined in  (\ref{ustat:t11b}), and
$T_{12BB}$ is defined in  (\ref{ustat:t12bb}).

Consider $T_{2211ABC}$ as defined in  (\ref{ustat:t2211abc}), and
$T_{2211ABBB}$ as defined in  (\ref{ustat:t2211abbb}),
\be \label{ustat:t2211abc_t2211abbb}
&&-T_{2211ABC}+T_{2211ABBB}\n\\
&=& -N^{-1/2} \sumI \frac{r_i  \Delta_{1i} \Delta_{3i}\{e^{-g(\u_i)}
-e^{-g^*(\u_i)} \} }{ d^*(\x_i)   } \n\\
&&+N^{-1/2} \sumI   \frac{ r_i   e^{-h(y_i)}\Delta_{3i} }{ d^*(\x_i) } \{e^{-g(\u_i)}
-e^{-g^*(\u_i)} \}
+ O_p(N^{1/2}h^m)\n\\
&=& N^{-1/2} \sumI \frac{r_i \Delta_{3i}\{ e^{-h(y_i)}-\Delta_{1i}\}  }{ d^*(\x_i)   }  \{e^{-g(\u_i)}
-e^{-g^*(\u_i)} \} + O_p(N^{1/2}h^m).
\ee

Combining (\ref{ustat:t2211abc_t2211abbb}) and (\ref{ustat:t2222bbb}) we get,
\be \label{ustat:t2211abc_t2211abbb_t2222bbb}
&&-T_{2211ABC}+T_{2211ABBB}-2T_{2222BBB}\n\\
&=& N^{-1/2} \sumI \frac{r_i \Delta_{3i}\{ e^{-h(y_i)}-\Delta_{1i}\}  }{ d^*(\x_i)   }  \{e^{-g(\u_i)}
-e^{-g^*(\u_i)} \} \n\\
&&-2N^{-1/2} \sumI \frac{r_i e^{-h(y_i)} \Delta_{1i} \a^*(\u_i)}{ d^*(\x_i) } \{e^{-g(\u_i)}-e^{-g^*(\u_i)}\}
+ O_p(N^{1/2}h^m)\n\\
&=&N^{-1/2} \sumI \frac{r_i \{e^{-g(\u_i)}-e^{-g^*(\u_i)}\} }{ d^*(\x_i)  }  \{\Delta_{3i}e^{-h(y_i)}-\Delta_{3i}\Delta_{1i}  -2 e^{-h(y_i)} \Delta_{1i}\a^*(\u_i)\}
+ O_p(N^{1/2}h^m)\n\\
&=&N^{-1/2} \sumI \frac{r_i \{e^{-g(\u_i)}-e^{-g^*(\u_i)}\} }{ d^*(\x_i)  }
\left\{-\Delta_{3i}e^{-h(y_i)}-\Delta_{3i}\Delta_{1i}   -
  2e^{-h(y_i)}\frac{C_i}{D_i} \right\}\n\\
&&+ O_p(N^{1/2}h^m).
\ee

Combining (\ref{ustat:t2211abc_t2211abbb_t2222bbb}), (\ref{ustat:t2211bbbb}) and
(\ref{ustat:t2222bc}), we get
\be \label{ustat:combine}
&&-T_{2211ABC}+T_{2211ABBB}+T_{2211BBBB}-2T_{2222BBB}+T_{2222BC}\n\\
&=&N^{-1/2} \sumI \frac{r_i \{e^{-g(\u_i)}-e^{-g^*(\u_i)}\} }{ d^*(\x_i)  }
\left\{-\Delta_{3i}e^{-h(y_i)}-\Delta_{3i}\Delta_{1i}   -
  2e^{-h(y_i)}\frac{C_i}{D_i} \right\} \n\\
&&+N^{-1/2} \sumI  \frac{ r_i  e^{-h(y_i)}\h'_\bb(y_i;\bb)  \Delta_{1i} }{ d^*(\x_i) } \{e^{-g(\u_i)}-e^{-g^*(\u_i)}\}\n\\
&&+N^{-1/2} \sumI \frac{  r_i \a^*(\u_i) \Delta_{1i}^2 \{e^{-g(\u_i)}-e^{-g^*(\u_i)}\} }{ d^*(\x_i)  } + O_p(N^{1/2}h^m)\n\\
&=&N^{-1/2} \sumI \frac{r_i\{e^{-g(\u_i)}-e^{-g^*(\u_i)}\} }{ d^*(\x_i)  }  \n\\
 && \times
\left\{\a^*(\u_i)   \Delta_{1i}^2 -\Delta_{3i}e^{-h(y_i)}-\Delta_{3i}\Delta_{1i}   -2e^{-h(y_i)}\frac{C_i}{D_i}+e^{-h(y_i)}\h'_\bb(y_i;\bb)  \Delta_{1i}  \right\} + O_p(N^{1/2}h^m)\n\\
&=&N^{-1/2} \sumI \frac{r_i \{e^{-g(\u_i)}-e^{-g^*(\u_i)}\} }{ d^*(\x_i)  }  \n\\
 &&\times \left[\Delta_{1i}\left\{\frac{C_i}{D_i}+ \Delta_{3i}\right\} -\Delta_{3i}e^{-h(y_i)}-\Delta_{3i}\Delta_{1i}   -2e^{-h(y_i)}\frac{C_i}{D_i}+e^{-h(y_i)}\h'_\bb(y_i;\bb)  \Delta_{1i}  \right] \n\\
&&+ O_p(N^{1/2}h^m)\n\\
&=&N^{-1/2} \sumI \frac{r_i\{e^{-g(\u_i)}-e^{-g^*(\u_i)}\}  }{ d^*(\x_i)  } \left[ \{\Delta_{1i}-2e^{-h(y_i)}\}\frac{C_i}{D_i} -\Delta_{3i}e^{-h(y_i)} +e^{-h(y_i)}\h'_\bb(y_i;\bb)  \Delta_{1i}  \right] \n\\
&&+ O_p(N^{1/2}h^m).\n
\ee

Consider $T_{2211CAB}$ as defined in  (\ref{ustat:t2211cab}) and $T_{2221AB}$ as defined in  (\ref{ustat:t2221ab}),
\be \label{ustat:t2211cab_t2221ab}
&&T_{2211CAB}-T_{2221AB}\n\\
&=&N^{-1/2} \sumI   \frac{\{w^{-1}(\x_i) -w^{*-1}(\x_i)\} r_i e^{-h(y_i)} }{d^*(\x_i) D_i }    E\left[\{ \Delta_{1i} \Delta_{3i}/d^* (\X_i)\} \mid \u_i,1  \right]\n\\
&&-N^{-1/2} \sumI \frac{\{w^{-1}(\x_i) -w^{*-1}(\x_i)\} r_i e^{-h(y_i)} \h'_\bb(y_i;\bb)}{d^*(\x_i)} + O_p(N^{1/2}h^m)\n\\
&=&N^{-1/2} \sumI   \frac{\{w^{-1}(\x_i) -w^{*-1}(\x_i)\} r_i e^{-h(y_i)} }{d^*(\x_i) D_i }  \left[E\left\{\left( \frac{\Delta_{1i} \Delta_{3i}}{d^*(\X_i)}\right)\mid \u_i,1  \right\}- \h'_\bb(y_i;\bb)D_i\right] \n\\
&&+ O_p(N^{1/2}h^m).
\ee

Consider $T_{12BB}$ as defined in  (\ref{ustat:t12bb}) and $T_{11B}$ as defined in  (\ref{ustat:t11b}),
\be \label{ustat:t12bb_t11b}
&&T_{12BB}-T_{11B}\n\\
&=&N^{-1/2} \sumI \frac{ r_i \{w^{-1}(\x_i)
-w^{*-1}(\x_i)\}C_i   }{d^*(\x_i) D_i}\n\\
&&-N^{-1/2} \sumI  \frac{ r_i  \{w^{-1}(\x_i)
-w^{*-1}(\x_i)\}C_i
  \{e^{-h(y_i)}+e^{-g^*(\u_i)}e^{-2h(y_i)}\} }{d^{*2}(\x_i) D_i }  +O_p(N^{1/2}h^m)\n\\
&=&N^{-1/2} \sumI  \frac{ r_i  \{w^{-1}(\x_i)
-w^{*-1}(\x_i)\}C_i
   }{d^{*2}(\x_i) D_i } \{d^*(\x_i)-e^{-h(y_i)}-e^{-g^*(\u_i)}e^{-2h(y_i)}\}\n\\
   &&+O_p(N^{1/2}h^m).
\ee
Combining (\ref{ustat:t2211cab_t2221ab}) and (\ref{ustat:t12bb_t11b}), we get
\be \label{ustat:t2211cab_t2221ab_t12bb_t11b}
&&T_{2211CAB}-T_{2221AB}+T_{12BB}-T_{11B}\n\\
&=&N^{-1/2} \sumI   \frac{\{w^{-1}(\x_i) -w^{*-1}(\x_i)\} r_i e^{-h(y_i)} }{d^*(\x_i) D_i }  \left[E\left\{\left(\frac{\Delta_{1i} \Delta_{3i}}{d^*(\X_i)}\right) \mid \u_i,1  \right\}- \h'_\bb(y_i;\bb)D_i\right] \n\\
&&+N^{-1/2} \sumI  \frac{ r_i  \{w^{-1}(\x_i)
-w^{*-1}(\x_i)\}C_i
   }{d^{*2}(\x_i) D_i } \{d^*(\x_i)-e^{-h(y_i)}-e^{-g^*(\u_i)}e^{-2h(y_i)}\}  \n\\
&&+O_p(N^{1/2}h^m)\n\\
&=&N^{-1/2} \sumI  \frac{ r_i  \{w^{-1}(\x_i)
-w^{*-1}(\x_i)\}}{d^{*2}(\x_i) D_i }\n\\
&&\times [C_i\{d^*(\x_i)-e^{-h(y_i)}-e^{-g^*(\u_i)}e^{-2h(y_i)}\}+d^*(\x_i) D_i e^{-h(y_i)}\{\a^*(\u_i)-\h'_\bb(y_i;\bb)\}]  \n\\
&&+O_p(N^{1/2}h^m)\n\\
&=&N^{-1/2} \sumI  \frac{ r_i  \{w^{-1}(\x_i)
-w^{*-1}(\x_i)\}}{d^{*2}(\x_i) D_i }\n\\
&&\times [C_i\{d^*(\x_i)-e^{-h(y_i)}\pi^{-1}(y_i,\u_i;\bb,g^*)\}+d^*(\x_i) D_i e^{-h(y_i)}\{\a^*(\u_i)-\h'_\bb(y_i;\bb)\}]  \n\\
&&+O_p(N^{1/2}h^m).\n
\ee

Putting all the terms together we can rewrite
(\ref{eq:final5}) as following,
\be \label{eq:final6}
&&\frac{1}{\sqrt{N}}\sumI\wh\S^*\eff(\x_i,r_i,r_i y_i,\bb)\n\\
&=& \frac{1}{\sqrt{N}}\sumI \frac{A_i}{d^*(\x_i)}
-T_{2211ABC}+T_{2211ABBB}+T_{2211BBBB}+T_{2211CAB}-T_{2221AB}+T_{2222BC}-2T_{2222BBB}\n\\
&&-T_{11B}+T_{12BB}+O_p(N^{1/2}h^{2m}+ N^{-1/2}h^{-p}+N^{-1/2}+N^{1/2}h^m)\n\\
&=& \frac{1}{\sqrt{N}}\sumI \frac{A_i}{d^*(\x_i)}\n\\
&&+N^{-1/2} \sumI \frac{r_i\{e^{-g(\u_i)}-e^{-g^*(\u_i)}\}  }{ d^*(\x_i)  } \left[ \{\Delta_{1i}-2e^{-h(y_i)}\}\frac{C_i}{D_i} -\Delta_{3i}e^{-h(y_i)} +e^{-h(y_i)}\h'_\bb(y_i;\bb)  \Delta_{1i}  \right] \n\\
&&+N^{-1/2} \sumI  \frac{ r_i  \{w^{-1}(\x_i)
-w^{*-1}(\x_i)\}}{d^{*2}(\x_i) D_i }\n\\
&&\times [C_i\{d^*(\x_i)-e^{-h(y_i)}\pi^{-1}(y_i,\u_i;\bb,g^*)\}+d^*(\x_i) D_i\{e^{-h(y_i)}\a^*(\u_i)- e^{-h(y_i)}\h'_\bb(y_i;\bb)\}]  \n\\
&&+O_p(N^{1/2}h^{2m}+ N^{-1/2}h^{-p}+N^{-1/2}+N^{1/2}h^m)\n\\
&=& \frac{1}{\sqrt{N}}\sumI \frac{A_i}{d^*(\x_i)}\n\\
&&+\frac{1}{\sqrt{N}} \sumI \frac{r_i\{e^{-g(\u_i)}-e^{-g^*(\u_i)}\}[\a^*(\u_i)
E\{e^{-h(Y_i)}\mid\x_i,1\}
-E\{e^{-h(Y_i)}\h'_\bb(Y_i;\bb) \mid\x_i,1\}] }{ d^*(\x_i)}\n\\
&&\times\left[2E\{e^{-h(Y_i)}\mid\x_i,1\}-e^{-h(y_i)}-\frac{E\{e^{-h(Y_i)}\mid\x_i,1\} \{e^{-h(y_i)}+e^{-g^*(\u_i)}e^{-2h(y_i)}\}}{d^{*}(\x_i)} \right] \n\\
&&+O_p(N^{1/2}h^{2m}+ N^{-1/2}h^{-p}+N^{-1/2}+N^{1/2}h^m)\n\\
&=& \frac{1}{\sqrt{N}}\sumI \{\S\eff^*(\x_i,r_i,r_iy_i,\bb)+ r_i \k(\x_i, y_i,\bb,g^*)\}\n\\
&&+O_p(N^{1/2}h^{2m}+N^{-1/2}h^{-p}+N^{-1/2}+N^{1/2}h^m),
\ee
where
\bse
\k(\x_i, y_i,\bb,g^*)&=& \frac{\{e^{-g(\u_i)}-e^{-g^*(\u_i)}\}[\a^*(\u_i)
E\{e^{-h(Y_i)}\mid\x_i,1\}
-E\{e^{-h(Y_i)}\h'_\bb(Y_i;\bb) \mid\x_i,1\}] }{ d^*(\x_i)}\n\\
&&\times\left[2E\{e^{-h(Y_i)}\mid\x_i,1\}-e^{-h(y_i)}-\frac{E\{e^{-h(Y_i)}\mid\x_i,1\} \{e^{-h(y_i)}+e^{-g^*(\u_i)}e^{-2h(y_i)}\}}{d^{*}(\x_i)} \right] \n.
\ese
\qed

\section{Proof of Lemma 2}
We write
\bse
&&N^{1/2}[\wh E\{\bz(\X,Y)\}-
E\{\bz(\X,Y)\}]\\
&=&N^{-1/2} \sumI
\left( r_i \bz(\x_i,y_i)+ (1-r_i)\frac{\wh
    E[\bz(\x_i,Y)\exp\{-h(Y,\wh\bb)\}\mid \x_i,1]}{\wh
    E[\exp\{-h(Y,\wh\bb)\}\mid \x_i,1]}
-
E\{\bz(\X,Y)\}\right)\\
&=&T_1+T_2+T_3,
\ese
where
\bse
T_1&=&N^{-1/2} \sumI
\left( r_i \bz(\x_i,y_i)+ (1-r_i)\frac{
     E[\bz(\x_i,Y)\exp\{-h(Y,\bb)\}\mid \x_i,1]}{
    E[\exp\{-h(Y,\bb)\}\mid \x_i,1]}
-
E\{\bz(\X,Y)\}\right),\\
T_2&=&N^{-1/2} \sumI
(1-r_i) \left( \frac{\wh
    E[\bz(\x_i,Y)\exp\{-h(Y,\bb)\}\mid \x_i,1]}{\wh
    E[\exp\{-h(Y,\bb)\}\mid \x_i,1]} -
\frac{
    E[\bz(\x_i,Y)\exp\{-h(Y,\bb)\}\mid \x_i,1]}{
    E[\exp\{-h(Y,\bb)\}\mid \x_i,1]}
\right),\\
T_3&=&N^{-1/2} \sumI
 (1-r_i) \left(\frac{
    \wh E[\bz(\x_i,Y)\exp\{-h(Y,\wh\bb)\}\mid \x_i,1]}{
    \wh E[\exp\{-h(Y,\wh\bb)\}\mid \x_i,1]}
-\frac{
    \wh E[\bz(\x_i,Y)\exp\{-h(Y,\bb)\}\mid \x_i,1]}{
    \wh E[\exp\{-h(Y,\bb)\}\mid \x_i,1]}
\right)\\
&=&E\left(
\frac{\partial}{\partial\bb\trans}\frac{
    E[\bz(\X,Y)\exp\{-h(Y,\bb)\}\mid \X,1]}{
    E[\exp\{-h(Y,\bb)\}\mid \X,1]} \right)
N^{1/2}(\wh\bb-\bb)+o_p(1),\\
&=&E\left(
\frac{\partial}{\partial\bb\trans}\frac{
    E[\bz(\X,Y)\exp\{-h(Y,\bb)\}\mid \X,1]}{
    E[\exp\{-h(Y,\bb)\}\mid \X,1]} \right)
N^{-1/2}\sumI\bphi_\bb(\x_i,r_i,r_iy_i,\bb,g^*)+o_p(1).
\ese
Here $\bphi_\bb(\x_i,r_i,r_iy_i,\bb,g^*)$ is defined in Theorems
1, 2 or 3,
corresponding to the estimator used as $\wh\bb$. To analyze $T_2$, we
distinguished the three cases in Theorems
1, 2 or 3
separately. When the true expectations are used, $T_2=\0$. When
parametric models involving $\ba$ are used,
\bse
T_2&=&E\left(\frac{\partial}{\partial\ba\trans}\frac{
    E[\bz(\X,Y)\exp\{-h(Y,\bb)\}\mid \X,1,\ba]}{
    E[\exp\{-h(Y,\bb)\}\mid \X,1,\ba]}\right)N^{1/2}(\wh\ba-\ba)+o_p(1)\\
&=&E\left(\frac{\partial}{\partial\ba\trans}\frac{
    E[\bz(\X,Y)\exp\{-h(Y,\bb)\}\mid \X,1,\ba]}{
    E[\exp\{-h(Y,\bb)\}\mid \X,1,\ba]}\right)N^{-1/2}\sumI\bphi_\ba(\x_i,r_i,r_iy_i)+o_p(1).
\ese
When nonparametric methods are used,
\bse
T_2&=&N^{-1/2} \sumI
(1-r_i) \left( \frac{\wh
    E[\bz(\x_i,Y)\exp\{-h(Y,\bb)\}\mid \x_i,1]
}{
    E[\exp\{-h(Y,\bb)\}\mid \x_i,1]} \right.\\
&&\left.-
\frac{
    E[\bz(\x_i,Y)\exp\{-h(Y,\bb)\}\mid \x_i,1]
 \wh E[\exp\{-h(Y,\bb)\}\mid \x_i,1]
}{
    (E[\exp\{-h(Y,\bb)\}\mid \x_i,1])^2}
\right)+o_p(1)\\
&=&N^{-1/2}\sumI
(1-r_i) \left( \frac{n^{-1}\sumJ r_jK_h(\x_j-\x_i)
    \bz(\x_i,y_j)\exp\{-h(y_j,\bb)\}
}{
    E[\exp\{-h(Y,\bb)\}\mid \x_i,1]n^{-1}\sumJ r_jK_h(\x_j-\x_i)
    } \right.\\
&&\left.-
\frac{
    E[\bz(\x_i,Y)\exp\{-h(Y,\bb)\}\mid \x_i,1]
n^{-1}\sumJ r_jK_h(\x_j-\x_i)
    \exp\{-h(y_j,\bb)\}
}{
    (E[\exp\{-h(Y,\bb)\}\mid \x_i,1])^2 n^{-1}\sumJ r_jK_h(\x_j-\x_i)}
\right)+o_p(1)\\
&=&N^{-1/2}\sumI
(1-r_i) \left( \frac{n^{-1}\sumJ r_jK_h(\x_j-\x_i)
    \bz(\x_i,y_j)\exp\{-h(y_j,\bb)\}
}{
    E[\exp\{-h(Y,\bb)\}\mid \x_i,1]f_{\X\mid R}(\x_i,1)
    } \right.\\
&&\left.-
\frac{
    E[\bz(\x_i,Y)\exp\{-h(Y,\bb)\}\mid \x_i,1]
n^{-1}\sumJ r_jK_h(\x_j-\x_i)
    \exp\{-h(y_j,\bb)\}
}{
    (E[\exp\{-h(Y,\bb)\}\mid \x_i,1])^2 f_{\X\mid R}(\x_i,1)  }
\right)+o_p(1)\\
&=&T_{21}+T_{22}-T_{23}+o_p(1).
\ese
Here
\bse
T_{21}&=&
N^{-1/2}\sumI
(1-r_i) \left( \frac{E[ RK_h(\X-\x_i)
    \bz(\x_i,Y)\exp\{-h(Y,\bb)\}]
}{
    E[\exp\{-h(Y,\bb)\}\mid \x_i,1]f_{\X,R}(\x_i,1)
    } \right.\\
&&\left.-
\frac{
    E[\bz(\x_i,Y)\exp\{-h(Y,\bb)\}\mid \x_i,1]
E[ RK_h(\X-\x_i)
    \exp\{-h(Y,\bb)\}]
}{
    (E[\exp\{-h(Y,\bb)\}\mid \x_i,1])^2 f_{\X, R}(\x_i,1)  }
\right)+o_p(1)\\
&=&N^{-1/2}\sumI
(1-r_i) \left( \frac{E[
    \bz(\x_i,Y)\exp\{-h(Y,\bb)\}\mid \x_i,1]
}{
    E[\exp\{-h(Y,\bb)\}\mid \x_i,1]
    } \right.\\
&&\left.-
\frac{
    E[\bz(\x_i,Y)\exp\{-h(Y,\bb)\}\mid \x_i,1]
E[ \exp\{-h(Y,\bb)\}\mid\x_i,1]
}{
    (E[\exp\{-h(Y,\bb)\}\mid \x_i,1])^2   }
\right)+o_p(1)\\
&=&o_p(1),\\
T_{22}&=&N^{-1/2}\sumJ
E \left\{(1-R) \left(\frac{r_jK_h(\x_j-\X)
    \bz(\X,y_j)\exp\{-h(y_j,\bb)\}
}{
    E[\exp\{-h(Y,\bb)\}\mid \X,1]f_{\X,R}(\X,1)
    } \right.\right.\\
&&\left.\left.-
\frac{
    E[\bz(\X,Y)\exp\{-h(Y,\bb)\}\mid \X,1]
r_jK_h(\x_j-\X)
    \exp\{-h(y_j,\bb)\}
}{
    (E[\exp\{-h(Y,\bb)\}\mid \X,1])^2 f_{\X, R}(\X,1)  }
\right)\right\}+o_p(1)\\
&=&N^{-1/2}\sumJ \frac{r_j\{1-E(R\mid\x_j)\} \exp\{-h(y_j,\bb)\}}{
E[\exp\{-h(Y,\bb)\}\mid \x_j,1]E(R\mid\x_j)}
\left(
    \bz(\x_j,y_j)
-
\frac{
    E[\bz(\X,Y)\exp\{-h(Y,\bb)\}\mid \x_j,1]
}{
    E[\exp\{-h(Y,\bb)\}\mid \x_j,1]  }
\right)\\
&&+o_p(1),
\ese
and
$T_{23}=E(T_{21})+o_p(1)=o_p(1)$.
Thus, summarizing the above results, we get the expansion
\bse
&&N^{1/2}[\wh E\{\bz(\X,Y)\}-
E\{\bz(\X,Y)\}]\\
&=&N^{-1/2} \sumI
\left\{ r_i \bz(\x_i,y_i)+ (1-r_i)\frac{
     E[\bz(\x_i,Y)\exp\{-h(Y,\bb)\}\mid \x_i,1]}{
    E[\exp\{-h(Y,\bb)\}\mid \x_i,1]}
-
E\{\bz(\X,Y)\}\right.\\
&&\left.+E\left(
\frac{\partial}{\partial\bb\trans}\frac{
    E[\bz(\X,Y)\exp\{-h(Y,\bb)\}\mid \X,1]}{
    E[\exp\{-h(Y,\bb)\}\mid \X,1]} \right)
\bphi_\bb(\x_i,r_i,r_iy_i,\bb,g^*)+\k_\ba(\x_i,r_i,r_iy_i,\bb)\right\}+o_p(1).
\ese
where
$\k_\ba(\x,r,ry,\bb)=\0$
 when the conditional expectations are known as in Theorem
1,
\bse
\k_\ba(\x,r,ry,\bb)=E\left(\frac{\partial}{\partial\ba\trans}\frac{
    E[\bz(\X,Y)\exp\{-h(Y,\bb)\}\mid \X,1,\ba]}{
    E[\exp\{-h(Y,\bb)\}\mid \X,1,\ba]}\right)N^{-1/2}\sumI\bphi_\ba(\x,r,ry)
\ese
 when the conditional expectations are parametrically estimated  as in Theorem
2, and
\bse
\k_\ba(\x,r,ry,\bb)=
\frac{r\{1-E(R\mid\x)\} \exp\{-h(y,\bb)\}}{
E[\exp\{-h(Y,\bb)\}\mid \x,1]E(R\mid\x)}
\left(
    \bz(\x,y)
-
\frac{
    E[\bz(\X,Y)\exp\{-h(Y,\bb)\}\mid \x,1]
}{
    E[\exp\{-h(Y,\bb)\}\mid \x,1]  }
\right)
\ese
when the conditional expectations are nonparametrically estimated  as in Theorem
3. These results directly lead to $N^{1/2}[\wh E\{\bz(\X,Y)\}-
E\{\bz(\X,Y)\}]\to N(\0,\V_\bz)$.\qed

\section{Proof of Theorem 4}

From $\sumI \wh E\{\bz(\X_i, Y_i,\wh\bt)\}=0$, we have
\bse
\0&=&N^{1/2} \wh E\{\bz(\X_i, Y_i,\wh\bt)\}\\
&=&N^{1/2}\wh E\{\bz(\X_i, Y_i,\bt)\}
+\frac{\partial E\{\bz(\X_i, Y_i,\bt)\}}{\partial\bt\trans} N^{1/2}(\wh\bt-\bt)+o_p(1).
\ese
Therefore,
\bse
N^{1/2}(\wh\bt-\bt)&=&
-\left[\frac{\partial E\{\bz(\X_i,
      Y_i,\bt)\}}{\partial\bt\trans}\right]^{-1}
N^{-1/2}\sumI \wh E\{\bz(\X_i, Y_i,\bt)\}+o_p(1).
\ese
Inserting the results from Lemma 2, we complete the
proof.
\qed